\documentclass[a4paper,12pt,times,numbered,print,index]{Classes/PhDThesisPSnPDF}

\usepackage{epigraph}
\usepackage{setspace}

\usepackage{hyperref}
\hypersetup{
   pdftitle={PhD Thesis Shailesh Kumar RSS2018001},
   colorlinks=true,
   citecolor=blue,
   linkcolor=blue,
   urlcolor=black
}

\input{Preamble/preamble}

\ifdefineAbstract
\fi

\ifdefineChapter
\fi

\begin{document}

\frontmatter


\begin{titlepage}	
	\centering 	
	{\LARGE \bfseries Gravitational Memory Effect for Near-Horizon} 
    \vspace{0.4\baselineskip}
    	
	{\LARGE \bfseries Asymptotic Symmetries}
	
	\vspace{1.5\baselineskip}
	
	\emph{\large A Thesis Submitted}
	\vspace{0.1\baselineskip}
	
	\emph{\large In partial fulfillment of the requirements for the degree of}

	\vspace{1.5\baselineskip}
	
	{\bf{\Large Doctor of Philosophy}
	
	\vspace{0.1\baselineskip}

	{\Large in}
	
	\vspace{0.1\baselineskip}
	
	{\Large Physics}}

	\vspace{0.8\baselineskip}
	
	\emph{\large by}
	
	\vspace{0.8\baselineskip}
	
	{\Large \bfseries Shailesh Kumar}

	\vspace{0.4\baselineskip}
	
	{\Large \bfseries (RSS2018001)}

	\vspace{1.0\baselineskip}
	
	\emph{\large  Under the supervision of }

	\vspace{0.5\baselineskip}
	
	{\Large \bfseries Dr. Srijit Bhattacharjee}

	\vspace{1.0\baselineskip}
	
	\includegraphics[scale=0.40]{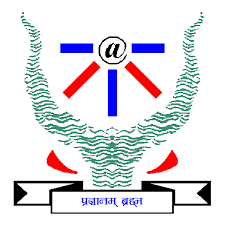}

	\vspace{1.0\baselineskip}
	
	\bf{{\large Department of Applied Sciences}

	
	{\large Indian Institute of Information Technology Allahabad}

	{\large	Prayagraj, Uttar Pradesh, India, 211015}

	{\large October, 2021}    }
	
\end{titlepage}		
\thispagestyle{empty}
\null\vfill
\noindent\copyright~2021 by Shailesh Kumar.\\
 All rights reserved.


\chapter*{\centering Acknowledgement}
\addcontentsline{toc}{chapter}{Acknowledgement}

\spacing{1.5}

It is with immense pleasure that I convey my gratitude to everyone who has assisted and supported me in my journey to IIIT-Allahabad to pursue my PhD. There are a number of people whom I would like to thank for their contribution to this thesis both directly and indirectly. I highlight a few names below; however, the list is so long that it is difficult for me to fit everyone in a single acknowledgement.  

First and foremost, I would like to express my heartfelt gratitude and profound appreciation to my supervisor, Dr. Srijit Bhattacharjee, whose expertise and invaluable ability kept the thesis work novel and level scientifically vivid. His insightful feedback, informative remarks and constructive criticism have pushed me to improve my skills and thoughts, as well as my research efforts that collectively brought up the research work to an advanced or higher level. I appreciate and respect his patience during the discussion of various research topics having relevance to the thesis. I am extremely grateful to him for constantly keeping an eye on the thesis-writing and suggestions. I am indebted for his relentless guidance and encouragement without which I would not have been able to finish
this dissertation.

I would also like to express my deep sense of thanks to Doctoral committee members Prof. Pavan Chakraborty and Prof. Anjan Ananda Sen for their valuable comments and suggestions during the colloquiums that led to addressing a question on observational signatures of asymptotic symmetries in the last chapter of the thesis.

I greatly appreciate the collaboration with Dr. Arpan Bhattacharyya in my projects, and sincerely thank him for stimulating ideas and insightful comments during various discussions. I would also like to thank my colleague and friend Subhodeep Sarkar for his contribution to the last chapter of the thesis and thoughtful discussions on many occasions and topics. Our friendly competition and joyful companionship reside beyond the horizon.   

I am grateful to our Director, Prof. P. Nagabhushan, for his continuous motivation and inspiration. Further, I humbly thank and acknowledge the technical and infrastructural support from IIIT-Allahabad. I thank the campus Library for providing the academic support and resources throughout the program. I would also like to thank the Ministry of Human Resource Development (MHRD), Government of India for providing the research fellowship during my PhD.

It has been a privilege to spend several years in the Department of Applied Sciences, and its members will always remain dear to me. I'm grateful to all the past Heads of our department, Prof. Pritish Varadwaj, Dr. Akhilesh Tiwari, Dr. Ashutosh Mishra, and to the current Head, Dr. Ratan K. Saha for their support and help. Special thanks to Mr. Sanjiv Kumar, who is no longer with us, for the assistance given by him.

I would further like to acknowledge the love and support given by my friends Sanjay Sharma, Subhodeep Sarkar and Ankur Gogoi, and seniors Ankit Singh, Rahul Maurya, Yogendra Singh, Dr. Faizan Ahmad and Dheeraj Chitara. The time spent with these people has taken my campus-journey to a different level and will not only be remembered by me but also remain as an eidetic memory. 

Last but not the least, I am eternally beholden to my family for their immutable and unconditional love, care and support. This thesis is dedicated to them. I would like to especially thank my parents, Uncle, Aunt and my brothers who have given me complete freedom, support and inspiration over the years. Last of all, I also thank everyone who has been a part of my journey at IIIT-Allahabad, Prayagraj.


\chapter*{\centering Abstract}
\addcontentsline{toc}{chapter}{Abstract}

\spacing{1.3}

Gravitational memory effect has emerged as a new window and opened up several intriguing avenues in the field of gravitational wave astronomy together with its inter-connection to asymptotic symmetries. The recent developments in this direction have drawn considerable attention from theoretical as well as observational perspectives. It has been shown that such effects can also be recovered near the horizon of black holes which might play a crucial role in understanding the information loss puzzle. With this motivation, we have studied the gravitational memory effect near the horizon of black holes analogous to the one obtained at asymptotic null infinity. 

We first study how the asymptotic symmetries emerge near the horizon of black hole spacetimes and how they can be detected via some ideal detectors as some permanent change of configurations (Memory) once gravitational wave passes through the setup. The emergence of asymptotic symmetries is studied from two different perspectives. First, how asymptotic symmetries emerge as soldering freedoms when two black holes are being glued along a common null surface. Secondly, in the form of asymptotic symmetries that preserve the asymptotic form of the near horizon metric of a black hole. We have established an analogous form of the displacement memory near the horizon of non-extreme and extreme black holes for both these perspectives. We also discuss the detection prospects of the supertranslation memory or hair. In this direction, other than the memory effect, we have explored the possibility of detecting a supertranslated black hole through standard tests of General Relativity. We propose a scenario in which such a configuration may be distinguished from ordinary black holes by examining the photon trajectories near the horizon. As we know that the gravitational memory has not been detected yet, in this direction, our study might serve as a useful contribution towards investigation of the signatures of asymptotic symmetries in the future.

\chapter*{\centering List of Publications}
\spacing{1.3}
\subsection*{Publications related to the thesis}
\begin{itemize}
\item[1.] S. Bhattacharjee, \textbf{S. Kumar} and A. Bhattacharyya, \textit{Memory effect and BMS-like symmetries for impulsive gravitational waves}, \href{https://journals.aps.org/prd/abstract/10.1103/PhysRevD.100.084010} {\color{blue}Phys. Rev. D 100, 084010 (2019)}.
\item[2.] S. Bhattacharjee, \textbf{S. Kumar}, \textit{Memory effect and BMS symmetries for extreme black holes}, \href{https://journals.aps.org/prd/abstract/10.1103/PhysRevD.102.044041}{\color{blue}Phys. Rev. D 102, 044041 (2020)}.
\item[3.] S. Bhattacharjee, \textbf{S. Kumar} and A. Bhattacharyya, \textit{Displacement memory effect near the horizon of black holes}, \href{https://link.springer.com/article/10.1007/JHEP03(2021)134}{\color{blue}J. High Energ. Phys. 2021, 134 (2021)}.
\item[4.] S. Sarkar, \textbf{S. Kumar}, S. Bhattachrajee, \textit{Can we detect a supertranslated black hole?}, \href{https://arxiv.org/abs/2110.03547}{\color{blue}arXiv:2110.03547 [gr-qc]}. [communicated]
\end{itemize}
\subsection*{Publication not related to the thesis}
\begin{itemize}
\item[1.] S. Bhattacharjee, \textbf{S. Kumar}, and S. Sarkar, \textit{Mass inflation and strong cosmic censorship in a nonextreme BTZ black hole}, \href{https://journals.aps.org/prd/abstract/10.1103/PhysRevD.102.044030}{\color{blue}Phys. Rev. D 102, 044030 (2020)}.\\
\end{itemize}

\section*{Conference Proceedings}
\begin{itemize}
\item[1.] \textbf{S. Kumar}, \textit{Displacement memory andn BMS symmetries}, Sixteenth Marcel Grossmann Meeting (MG16), \href{https://arxiv.org/abs/2109.13082}{\color{blue}arXiv: 2109.13082 [gr-qc]}. [Accepted].
\end{itemize}


\setcounter{tocdepth}{3}
\setcounter{secnumdepth}{3}
\tableofcontents

\listoffigures




\mainmatter

\chapter{Introduction}\label{CH1}  



\ifpdf
    \graphicspath{{Chapter1/Figs/Raster/}{Chapter1/Figs/PDF/}{Chapter1/Figs/}}
\else
    \graphicspath{{Chapter1/Figs/Vector/}{Chapter1/Figs/}}
\fi

\section{Gravitational Waves} 
\spacing{1.5}
The direct detection of gravitational waves (GW), which confirmed the fundamental predictions made by Einstein for the existence of gravitational waves almost a century ago, has set a compelling evidence and bolstered Einstein's general theory of relativity. It is the first direct observation of a binary black hole (BBH) system merging to form a single black hole. Since then researchers have been increasingly involved in exploring various aspects of compact binary systems as these have been very promising astrophysical candidates of gravitational waves. The compact binaries such as binary black hole mergers and binary neutron stars have been the most effective sources for generating gravitational waves.

The heart of gravitational wave generation lies in the theory of \textit{general relativity} (GR) \cite{Einstein, 1937FrInJ.223...43E, PhysRevLett.116.061102, PhysRevLett.125.101102}. The spacetime of general relativity is a generalization to the Minkowskian spacetime where we anticipate getting light cones arranged in irregular patterns. As an extension of \textit{special relativity} which deals with the relation between space and time for the particles or objects moving comparable to the speed of light ($c$) and does not include the effect of gravity, Einstein introduced a new understanding of the gravitation theory. After spending almost a decade on the more general picture, Einstein came up with the general theory of relativity which suggests that massive sources deform the spacetime geometry. General relativity has the ability to comprehend how gravity influences the fabric of spacetime. The theory is ruled by one of the most celebrated and highly non-linear equations, known as the \textit{Einstein field equation} originally established by Einstein. It is written as
\begin{align}
G_{\mu\nu} \equiv R_{\mu\nu} - \frac{1}{2}g_{\mu\nu}R = \frac{8\pi G}{c^{4}}T_{\mu\nu}
\end{align} 
where the Greek indices depict spacetime components of the given quantities and $G_{\mu\nu}$ is called \textit{Einstein tensor}. For four-dimensional spacetime coordinates, it can be written as $x^{\mu}= (x^{0}, x^{i})$ where $x^{0}$ denotes the temporal coordinate and $x^{i}$ denotes the spatial coordinate. $R_{\mu\nu}$ is called as \textit{Ricci tensor} and denotes the curvature of the spacetime geometry; it is written as contracted \textit{Riemann tensor} ($R^{\alpha}{}_{\beta\mu\nu}$), i.e., $R_{\mu\nu}=R^{\alpha}{}_{\mu\alpha\nu}$. The left-hand side of the field equation represents the spacetime geometry whereas the right-hand side describes the energy-momentum tensor. Thus the equation explains how spacetime geometry is coupled with matter or energy. In John Wheeler's terms \cite{1973grav.bookM}:

\textit{`` Matter tells spacetime how to curve, and curved spacetime tells matter how to move.''}

Einstein's general relativistic field equation, which has been validated numerous times, is now the most precise technique to predict gravitational events and interactions. As mentioned that the Einstein field equation is highly non-linear, it does not appear to be a simple differential equation with many known properties at all because of the geometric language and abbreviations utilized in expressing it. The best way to look at it is to apply weak field approximation. In this consideration, one can make expansion of the field equation for the given Minkowskian metric ($\eta_{\mu\nu}$) with perturbation ($h_{\mu\nu}$). We can keep the linear order term in $h_{\mu\nu}$ by avoiding the mathematical complications. This important formalism is known as the \textit{linearized theory of gravity}. We try to take the small linear perturbation ($h_{\mu\nu}$) to the Minkowskian metric ($\eta_{\mu\nu}$) and denote the resultant spacetime metric as $g_{\mu\nu}$. The metric components can be written as
\begin{align}
g_{\mu\nu} = \eta_{\mu\nu} + h_{\mu\nu} ; \hspace{3mm} \vert h_{\mu\nu} \vert << 1.
\end{align}
We know that when gravity is absent, spacetime is flat; therefore, a weak gravitational field is one in which spacetime is linearly or almost flat. The analysis produces the linearized Einstein field equation which can be written in compact form as follows
\begin{align}
\square \bar{h}_{\mu\nu} = -\frac{16\pi G}{c^{4}} T_{\mu\nu} \label{Gw}
\end{align}
where $\eta_{\mu\nu}$ is a simple diagonal metric ($ -1, +1, +1, +1 $). The $\bar{h}_{\mu\nu}$ is defined in terms of linear perturbation $h_{\mu\nu}$ and its trace $h$.
\begin{align}
\bar{h}_{\mu\nu} = h_{\mu\nu}-\frac{1}{2}\eta_{\mu\nu}h \hspace{3mm} ; \hspace{3mm} \square = -\frac{\partial^{2}}{\partial t^{2}}+\vec{\nabla}^{2}
\end{align}
We use the fact that as a result of gauge transformations generated by a vector, the theory remains invariant, and also we can always use the Lorentz gauge condition, $\partial^{\mu}\bar{h}_{\mu\nu} = 0 $. It is now obvious that the Eq.(\ref{Gw}) will express wave-like solutions with and without sources. If we consider the propagation of waves in empty space or vacuum, it can be regarded as the superposition of plane waves. The linearized field equation in vacuum case can be written as, $\square \bar{h}_{\mu\nu} = 0$. Physically this can be understood as a small perturbation with amplitude propagating as a wave at the speed of light ($c$) in the linearized theory of gravity. This is known as \textit{gravitational waves} in the literature. The plane wave solution for linearized wave equation in vacuum is written as, $\bar{h}_{\mu\nu}(x) = A_{\mu\nu}e^{ik_{\alpha}x^{\alpha}}$; where wave-vector $k^{\alpha}=(\omega,\vec{k})$ determines the direction of the wave-propagation together with its frequency, and $A_{\mu\nu}$ is the \textit{polarization tensor}. In general, there are ten independent components of $\bar{h}_{\mu\nu}$. By checking the orthogonal relation, i.e. $A_{\mu\nu}k^{\mu}=0$, and  \textit{transverse-traceless} gauge (TT gauge: $\partial_{\mu}\bar{h}^{\mu\nu}=0$, $\bar{h}^{\mu}_{\mu}=0$), one is finally left with two polarizations.  These are two distinct polarization states or degree of freedom plus ($h_{+}$) and cross ($h_{\times}$). 

\begin{figure}[h!]\centering
    \includegraphics[scale=0.57]{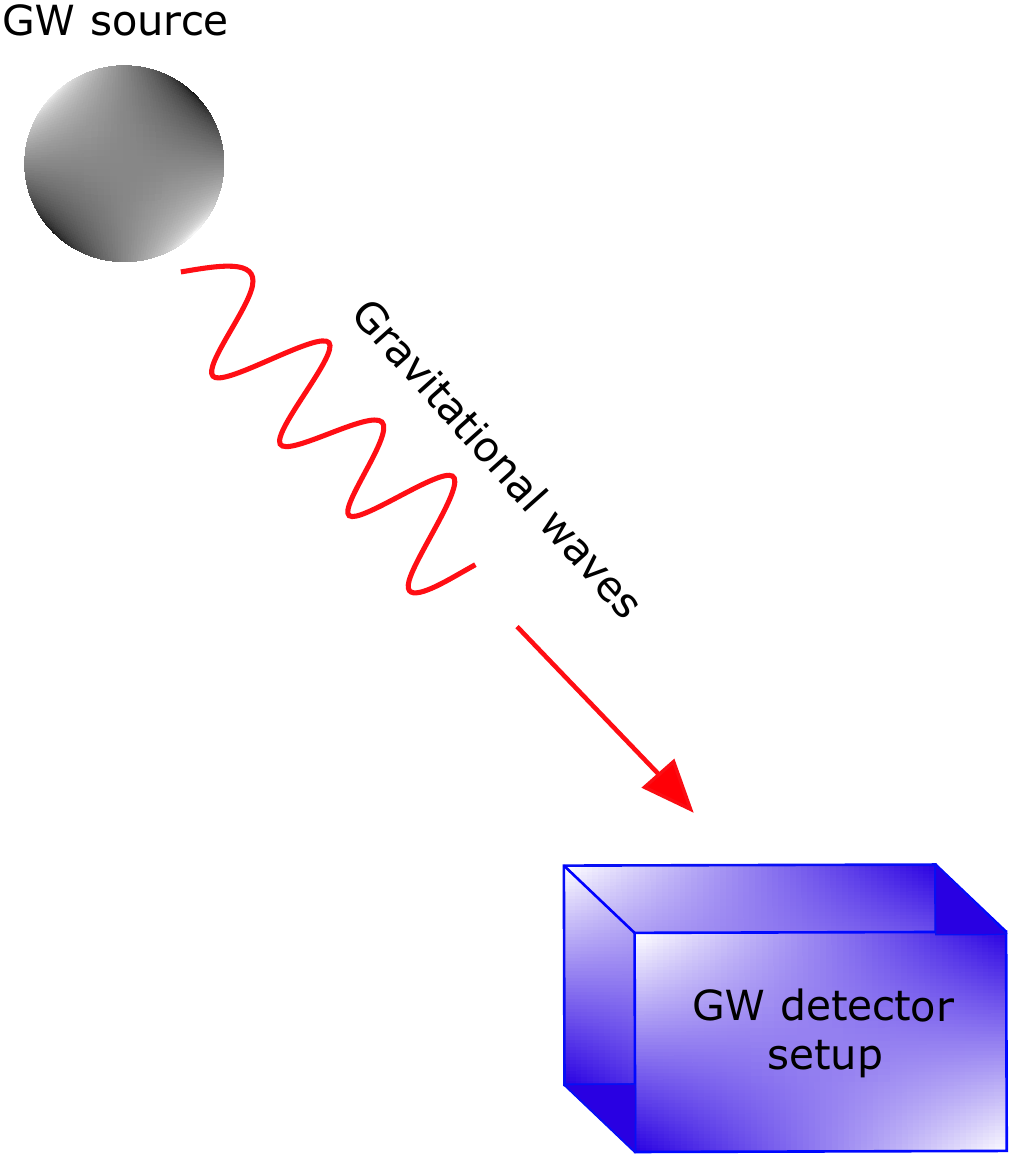} 
    \caption{Generation of  GW from a source and its interaction with earth based detectors.}
\end{figure}\label{pic1}

Gravitational waves are ripples or distortions in the fabric of spacetime that propagate at the speed of light. Gravitational waves carry the information and properties of the source. The moving compact astrophysical sources can generate gravitational waves as sketched in Fig(\ref{pic1}). To show typical \textit{chirp} waveforms with two polarizations for non-spinning compact binary system up to 2 post-Newtonian (2PN) order\footnote{The explicit details of the calculations are contained in \textit{Gravitational waveforms from inspiralling compact binaries to second-post-Newtonian order}, Luc Blanchet et al. 1996 Class. Quantum Grav. 13 575.} have been plotted in Fig.(\ref{fig:irrfdad}) \cite{Blanchet_1996}.   
\begin{figure}[!htb]\centering
\includegraphics[width=7.8cm, height=6.3cm]{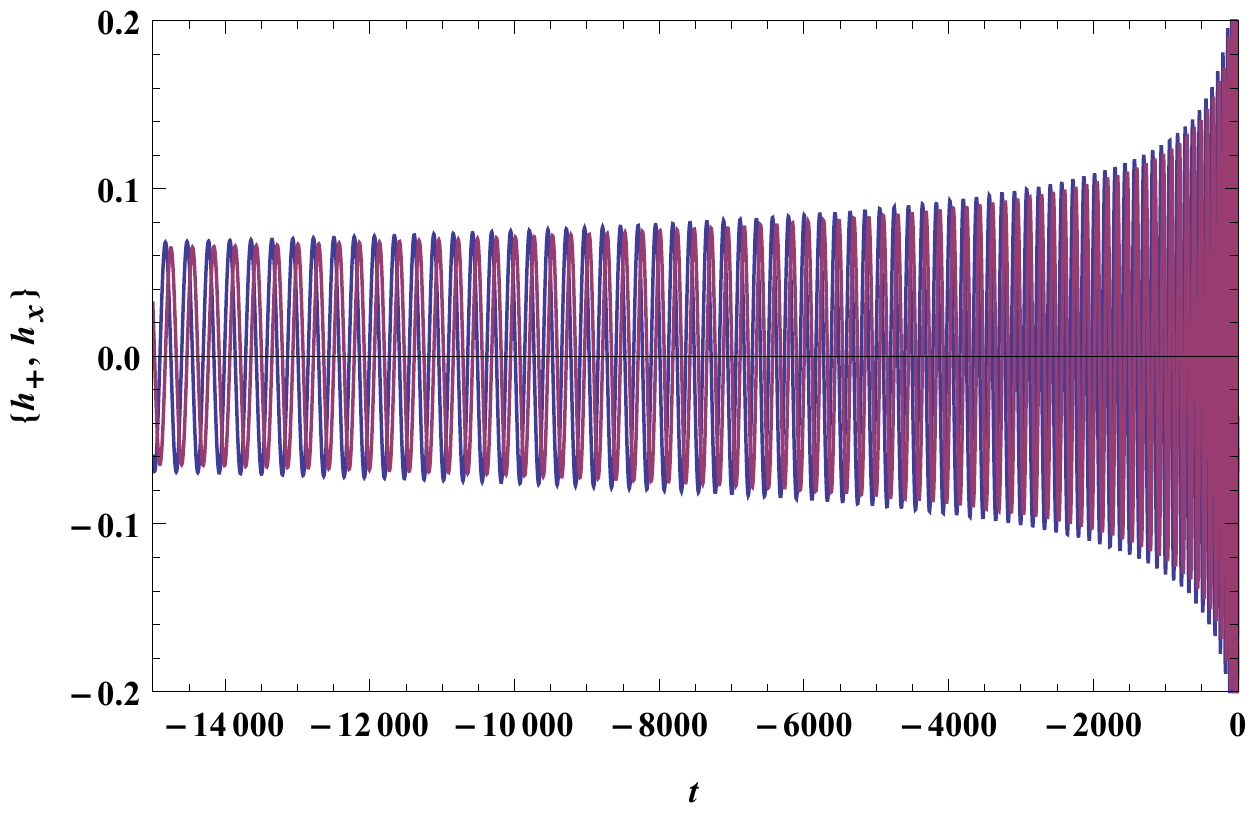}\includegraphics[width=7.65cm, height=6.3cm]{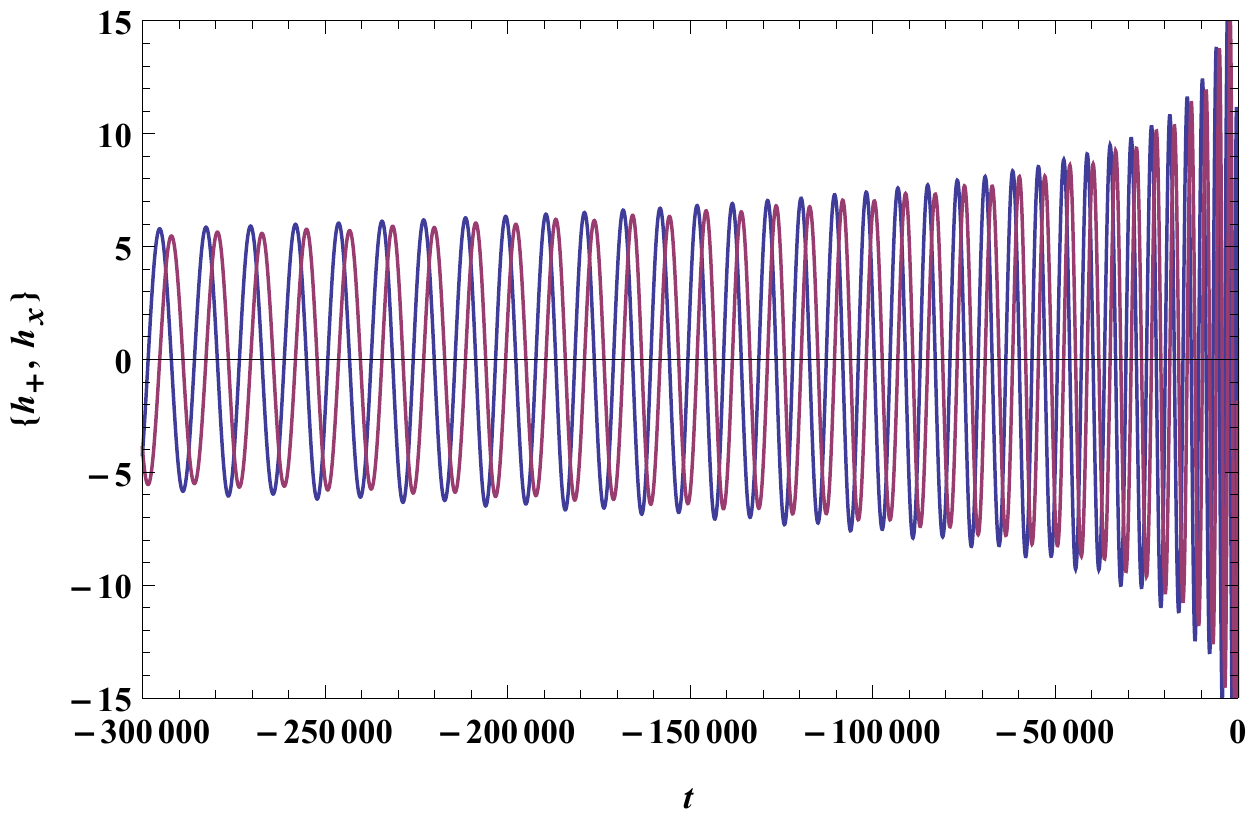} 
\caption{Waveforms for a binary inspiral: It is a typical chirp waveform for inspiralling, non-spinning, compact binaries. In each graph, blue is for \textit{plus}-polarization ($h_{+}$) and maroon is for \textit{cross}-polarization ($h_{\times}$). The waveforms have been plotted up to 2 post-Newtonian (2PN) order. The left graph is for half-half solar mass binaries whereas the right graph is for 29 and 36 solar masses.}
\end{figure}\label{fig:irrfdad}
There are several stages associated with the motion of massive binaries. As compact binary objects are one of the most effective gravitational wave sources, there are three phases involved in the method of transforming a binary system into a single system, known as \textit{Inspiral, Merger} and \textit{Ringdown}. If we consider a compact binary as a two black hole system orbiting each other, one can estimate the rate at which general relativity computes a binary will lose energy by measuring the quadrupole moment for the motion of the system. In this mechanism, the size of the orbit gets shrunk and the orbital period is also reduced significantly. This process is termed the \textit{inspiral phase}. In the next stage, the two orbiting black holes will eventually merge into one larger source, called the \textit{merger phase}. Immediately after the highly unstable second stage, the final phase \textit{ringdown} continues to generate gravitational waves until the source achieves the final stable stage. The perturbative techniques are being used for examining the inspiral and ringdown phases whereas numerical relativity is required for investigating and analysing the merger stage. However, as mentioned earlier, the formulation of quadrupole moment suggests an indirect way of detecting gravitational waves by measuring the change in the motion of a compact binary system due to the loss of energy. The size of the orbit shrinks as the system loses energy, and the period of the orbit also decreases. The change of time rate is then estimated using the quadrupole formula for energy loss. This is the technique that Hulse and Taylor adopted for a system of a binary pulsar and measured the rate change of period and showed the prediction of general relativity to be accurate, in 1975 \cite{1975ApJ195L51H}. This discovery opened up new avenues for research in the special and general relativistic physics for compact binaries. In the last 40 years, gravitational wave theory has made significant advances in experimental and theoretical fields, the most well-known of which is the direct detection of gravitational waves generated from binary black holes merger \cite{PhysRevLett.116.061102}. 

The experimental theme of GW-detection starts in the 1960s with Weber's resonant mass detector setup which was followed by an international network of cryogenic resonant detectors.  Over the time, experiments were improved and interferometers were introduced and a large number of observatories were established to improve the noise and performance of gravitational wave signal \cite{PhysRevD.82.022003, drever1983interferometric, Gertsenshtein:1962kfm, Moss:71, PhysRev.117.306, Weiss}. A number of detector setups in Japan, Germany, USA and Italy were ready by the beginning of the twenty first century. These observatories made simultaneous observations from 2002 to 2011. The detectors were further improved, as a result, LIGO with modified techniques in 2015 became the first sensitive network of specialized detectors to start making observations \cite{2015, Acernese_2014, Affeldt_2014,PhysRevD.88.043007}. The GW150914 was the first direct observation by LIGO in Sept 2015 and was announced in Feb 2016. LIGO, a Laser Interferometer Gravitational-Wave Observatory, is a large-scale experimental setup to study gravitational waves in order to test Einstein's general theory of relativity and various properties of GW sources. LIGO is consist of two interferometers, each with two 4 km (2.5 miles) long arms arranged in an \textit{L} shape. The modified interferometer setup is called advanced LIGO or aLIGO which uses the length difference between its orthogonal arms to calculate gravitational wave strain. Each arm contains two mirrors that serve as test masses and are separated by $L=L_{x}=L_{y}=4 km$. The length between the mirrors in both arms contracts and expands during the passage of GW through the detector. As a result, it changes the lengths of the arms by $\delta L_{x}$ and $\delta L_{y}$ also written as $\delta L(t) = \delta L_{x}-\delta L_{y} = h(t) L $; where $h(t)$ is the GW strain amplitude projected onto the detector setup. In other words, the strain generated as a result of the mirrors' motion can be described as a linear combination of two polarizations, i.e., $h(t)=f_{+}h_{+}(t)+f_{-}h_{-}(t)$; where $f_{+}$ and $f_{-}$ are coefficients of two degrees of freedom which store the information of detectors' orientation with respect to the GW source. The phase difference between the two light rays returning to the beam splitter produces an interference pattern and provides us a way to measure the amplitude $h(t)$ of the pattern. As the GW signal is very small ($\sim 10^{-19}m.$ amplitude), it is extremely important to have simultaneous observations of the signal. In this respect, LIGO together with its sister observatories have played a crucial and significant role in gravitational wave astronomy. These observatories are situated at United States \cite{LIGO}, Virgo in Italy \cite{Virgo}, GEO 600 in Germany \cite{GEO} and KAGRA in Japan \cite{Kagra}. Making the observatory a larger global network, India has also planned a joint India-US detector, named as LIGO-India or IndIGO. 

The various detection techniques are being employed in search of weak gravitational wave signals by minimizing noise transients \cite{Blackburn_2008, Abbott_2009, Aasi:2012wd, Cornish_2015}. The amount of complication of noise and signal models are not defined in advance; rather estimated by the \textit{Bayesian model}. The detection techniques employed in search of GW signals are usually motivated by analytical approaches assuming Gaussian noise, then the impacts of non-Gaussianity are taken into consideration. In \cite{Cornish_2015}, Cornish et al. discuss the identification of weak GW amplitude signals in the presence of non-stationary and non-Gaussian noise. It is a central challenge for unmodeled transient signals to separate the GW signal from noise because it has been noticed that the data from initial LIGO-Virgo searches for GW have exhibited the non-Gaussian aspects which affect the ability to implicate weak signals. Authors in \cite{Cornish_2015} have adopted the Bayesian technique which explicitly explains the wider range of non-stationarity and non-Gaussianity in the LIGO data. After the direct detection of gravitational waves, more recently, other computational techniques such as machine learning and deep learning have emerged as powerful sources in astrophysics especially in search of GW signals. In \cite{Cuoco_2020}, a detailed review of machine learning applications to analyze the LIGO data and techniques for improving the aLIGO-Virgo sensitivity has been provided. In addition to this, researchers have also started implementing deep learning concepts with convolutional neural networks to identify and characterise the GW signals efficiently \cite{2018PhLB77864G, morales2020deep}. The next generation gravitational wave detectors are aimed at observing compact binaries with a high signal to noise ratio across the universe \cite{Evans:2016mbw}. The challenges in analyzing the GW data efficiently and reducing the noise significantly have opened the GW astronomy era to a wider community and welcome researchers from various backgrounds from theoretical and experimental domains.

The direct detection has enabled the scientific community involved in GW astronomy to investigate numerous aspects of gravitation theory; recently, researchers pointed out a major physical effect called as \textit{gravitational memory effect} which has not been detected yet experimentally. This might help us in understanding the minute and hidden properties of black hole spacetimes and gravitational waves. It has also a connection with asymptotic symmetries and soft theorem. In the next section, we provide a disjoint study of gravitational memory and asymptotic symmetries, and subsequently, we shall also discuss the triangular relation in memory effect, asymptotic symmetries and soft theorem.  
\section{Gravitational Memory and BMS Symmetries} 

\subsection{Gravitational Memory}

A feature of gravitational wave signal that can implicate a permanent displacement in the LIGO test masses is yet to be observed. The direct detection of gravitational waves has enabled researchers to look for various aspects of black hole spacetimes; \textit{gravitational memory} (GW memory) is one of such physical effects which could uncover many new and intriguing findings of the compact massive sources like population of binary black hole mergers, super-massive black hole mergers etc. Early on, gravitational wave signals have a small amplitude, rise to a maximum value, and then fall to zero. Technically, the GW memory is defined as a gravitational wave signal that does not always decay to zero but eventually approaches a nonzero finite value. The gravitational wave passes through the detector setup and distorts the spacetime geometry. This induces a relative change in the position of the detectors imparting a memory to the configuration after the passes of gravitational waves. The term \textit{memory} implies the information or properties of spacetimes from where it is being generated and carried by gravitational waves.
\begin{figure}[h!]\centering
    \includegraphics[scale=0.65]{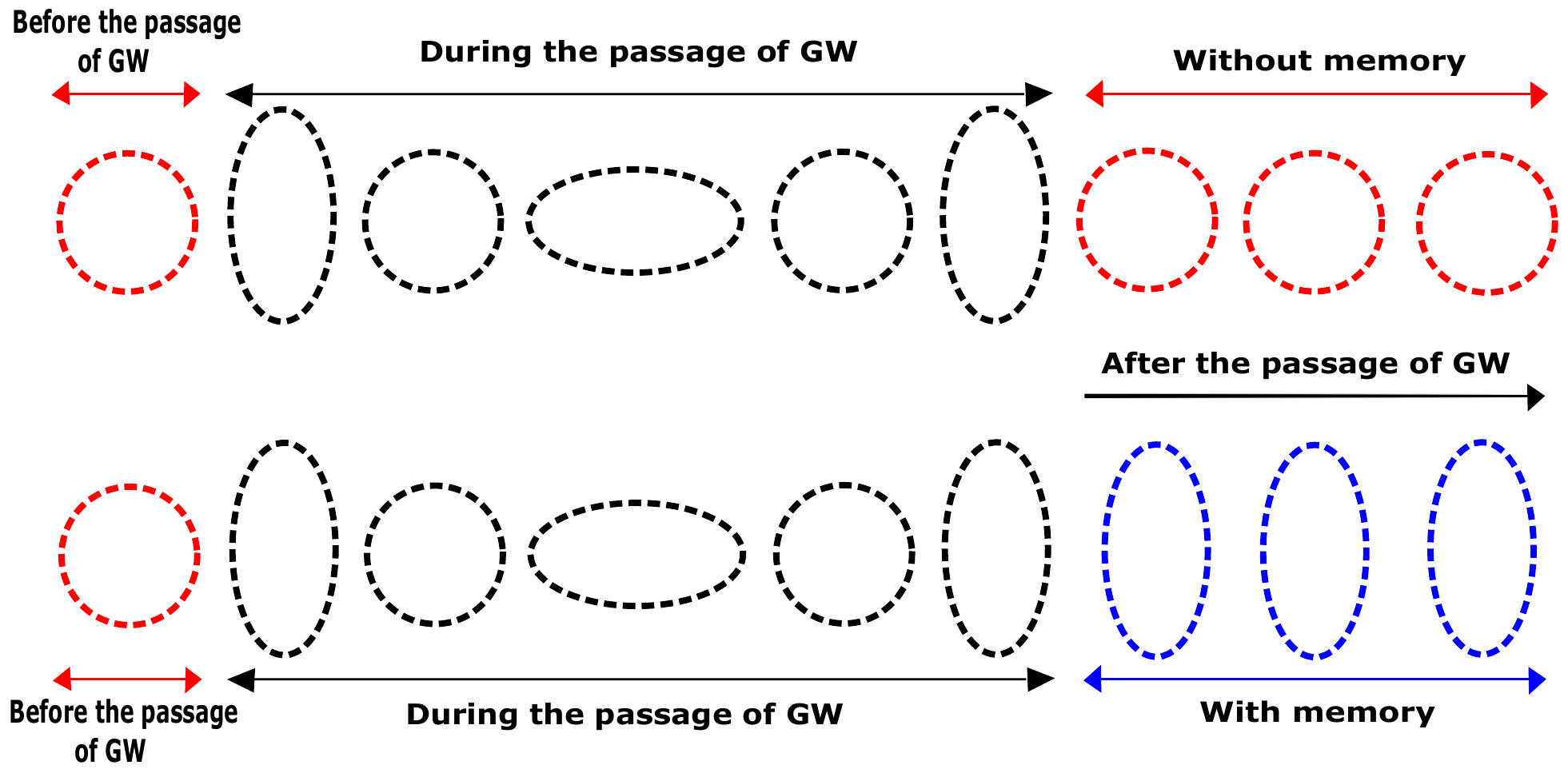} 
    \caption{It shows the effect of GW which propagate along the z-axis through a ring of test masses. It is describing the resulting effect on the ring of test masses, with and without memory, before and after
the passage of GW.}\label{pic2}
\end{figure}

Let us consider our setup to be a ring of test masses and let the gravitational waves generated from some violent astrophysical phenomena pass through the configuration as depicted in Fig(\ref{pic2}). As a result, the interaction between setup and gravitational waves changes the orientation of the ring of test masses during the passage of gravitational waves due to distortion generated in the fabric of spacetime. There are two possibilities associated in the process of interaction between detector setup and gravitational waves, i.e., when gravitational wave is \textit{not carrying memory} and when gravitational wave is \textit{carrying memory}. In the first case, when gravitational wave is not carrying memory, we can see in Fig(\ref{pic2}) that there is no relative change in the orientation of ring of test masses after the passage of gravitational waves. The setup contracts and expands during the passage of gravitational waves and attains its initial orientation after the passage of gravitational waves. Whereas in the second case when gravitational wave is carrying memory, we find that there is a relative change in the orientation of ring of test masses after the passage of gravitational waves (in blue colour) and the memory is imparted to the setup. This change in orientation of the ring of test masses is a permanent change and can be understood as gravitational memory effect. 

There are two kinds of memory known in the literature: \textit{linear memory} \cite{1974SvA1817Z, Braginsky:1986ia, Braginsky1987} and \textit{nonlinear memory} \cite{Christodoulou:1991cr, PhysRevD.46.4304}. The change in the second time derivative of the quadrupole moment of the source gives rise the linear memory. Hyperbolic orbits, two-body scattering, and explosions are well-known candidates of unbounded systems that produce linear memory. The story begins with Y. B. Zeldovich and A. G. Polnarev who explored the prediction for pulses of gravitational waves, and how to improve the instrument's sensitivity in order to have a fair chance of recording the radiation from these remarkable processes \cite{1974SvA1817Z}. They have provided a rough estimate to have the enhanced sensitivity in the instrumental design and showed the analysis for achieving the \textit{linear memory} for a pair of free bodies by computing the change in the linearly perturbed metric (h) before and after the interaction with gravitational radiation. This idea of Zel’dovich and Polnarev was further elaborated by Braginsky and Grishchuk \cite{Braginsky:1986ia}, who coined the term \textit{memory effect}. They came up with some possible applications from experimental perspectives for the detection of gravitational radiation from binary stars with the enhancement in the sensitivity of detectors. The abstract of Braginsky et al. \cite{Braginsky:1986ia} states that 

\textit{``In the memory effect, the distance between a pair of bodies is different from the initial distance in the presence of a gravitational radiation pulse.''} 

Further, Braginsky and Thorne, in a study on a class of gravitational waves which experimentalists expect to detect, reported that the transverse-traceless component of the linearly perturbed metric $h^{TT}_{ij}$ rises from zero and oscillates for a few cycles before settling down to a nonzero final value of $\delta h^{TT}_{ij}$ \cite{Braginsky1987}. This class of gravitational wave is known as \textit{burst with memory} (BWM). BWM could be one of the earliest types of gravitational waves to be detected, albeit it is very unlikely. However, when devising search techniques and doing data analysis, investigators should also consider them. 

As a remarkable progress in 1991, Christodoulou disproved the assumption that gravitational waves produced by astrophysical sources, and reaching the Earth's vicinity would necessitate the use of a linearized gravity theory \cite{Christodoulou:1991cr}. When studying the generation of gravitational waves from strong sources, it is widely acknowledged that the non-linearity of Einstein's field equation must be fully considered. In this respect, Christodoulou showed an additional memory due to the energy radiated in gravitational radiation and proved that a permanent displacement of test masses after the passage of gravitational wave is a signature of the nonlinear memory effect. In other words, the Christodoulou memory is a nonlinear type of memory that emerges from the contribution of the gravitational wave stress–energy tensor which acts as a source for additional GWs. It is generated by gravitational waves that are released as a result of the waves that have already been emitted. Christodoulou and Blanchet et al. independently formulated the nonlinear memory effect. A follow up on Christodoulou's work was given by Thorne who provided a more physical interpretation with experimental strategies and detector sensitivities \cite{Thorne:1992sdb}. Whereas Blanchet et al. took a different approach and termed the nonlinear memory as the \textit{Hereditary effect}. They could compute the gravitational waveform explicitly including hereditary contributions up to quadratically nonlinear order $\mathcal{O}(\frac{v}{c})^{4}$ using an extension of existing multipole-moment wave generation formalism \cite{PhysRevD.46.4304}. The potential candidates for nonlinear memory are bounded binary black hole mergers, a population of binary black hole mergers and binary neutron stars etc.

As GW memory might contain minute information of black hole spacetimes, the field has made a significant progress recently. In the context of post-Newtonian (PN) approximation, the Christodoulou memory is derived from 2.5 post-Newtonian order multipole interactions and has an effect on the waveform at leading order \cite{PhysRevD.80.024002}. Earlier, the memory was estimated as a burst (unmodeled) during the phase-inspiral \cite{Thorne:1992sdb, PhysRevD.50.3587}. The first practical computation of memory's evolution was done by Marc Favata where he accounted for all phases of binary black hole coalescence that is inspiral, merger, ringdown \cite{PhysRevD.80.024002, Favata:2008ti, Favata:2009ii, Favata:2010zu}. He showed that the peak sensitivity of LIGO and LISA detectors might capture the effect of reciprocal of the memory's rise time during the merger and ringdown phases of a BBH. At least one of the polarizations for late and early time values will deviate from zero. For $h_{+}$ and $h_{\times}$ polarizations, the memory in general for an observer at time $t$ can be written as \cite{Favata:2010zu}
\begin{align}
\bigtriangleup h_{+,\times} = \lim_{t\rightarrow +\infty}h_{+,\times}(t) - \lim_{t\rightarrow -\infty}h_{+,\times}(t)
\end{align} 
where $\bigtriangleup$ depicts the change in the late and early time values of GW polarizations. The analysis is based on the expansion of GW polarizations in spin-weighted spherical harmonics basis and radiative mass, current multipole moments also expanded in the basis of spherical harmonics. In order to relate this with family of source multipole moments, one can use the PN formalism \cite{Blanchet2006} which gives a non-linear and iterative approach written in the form of integrals over the stress pseudo-tensor and source's gravitational field. 

On the other hand, Bieri and her collaborators have provided an extensive study on the nonlinear memory effect \cite{2013arXiv13083100B, 2013CQGra30s5009B, 2014PhRvD90d4060T, 2015arXiv150505213B, 2016PhRvD94f4040B, 2017APSAPRC14001B, 2017CQGra34u5002B, 2018PhRvD98l4038B, 2021PhRvD103b4043B}. In \cite{2013arXiv13083100B}, Bieri et al. showed the nonlinear memory for neutrino radiation characterized by a source like supernova explosion or binary neutron star merger. Further in \cite{2013CQGra30s5009B}, the authors again showed a gravitational wave memory analogue in the electromagnetic realm. They examined the change or shift to an electromagnetic radiation detector after the wave has passed rather than finding distortion in the detector. Her group also provided the study of memory for the sources in expanding cosmology with a simplified version for de Sitter cosmological spacetime. A finding of memory effect which is similar to the one obtained for asymptotically flat spacetimes in $\Lambda$CDM cosmology where the universe is highly inhomogeneous has been considered in \cite{2016PhRvD94f4040B, 2017CQGra34u5002B}. Recently, Bieri has shown a growing magnetic memory for asymptotically flat spacetimes which is contributed by the magnetic part of the curvature. This study motivates us to explore the various properties of gravitational wave sources and the possible detectability of dark matter via gravitational waves etc \cite{2021PhRvD103b4043B}. 

Memory effect has made much greater progress from an observational standpoint through advanced detectors like LIGO or LISA. Researchers are focusing on the measurements on ensemble of BBH mergers per year through advanced detectors and trying to explore general relativistic effects. GW memory will be detectable with dozens of such occurrences using a global network of ground-based GW detectors. In this direction, recently there have been several implications of detecting GW memory in GW150914 through LIGO detector \cite{PhysRevLett.117.061102, PhysRevD.101.083026}. As PD Lasky et al. proposed to search for GW memory associated with the population of single BBH mergers; they showed that 100 BBH merger events like GW150914 are required to have the strong observation of memory effect by avoiding the degeneracies of some extrinsic parameters, excluding mass and spin of the black hole, in response of the detector setup to the gravitational waves. They adopted the technique for detecting the memory which relies on the coherent summation of an ensemble of sub-threshold signals because incoherent summation might cancel the memory. Though one can still measure the sum but signal will grow much slower. We know that the amplitude of the memory signal and the sign are both modulated by polarization. Thus for individual detection, it is important to have the knowledge of memory's sign. The mode of the oscillatory signal $h_{lm} = h_{22}$ , which is typically employed in parameter estimation, cannot be used to infer the sign of the memory because $l=2$ and $m=2$ does not break the degeneracy for certain extrinsic parameters \cite{PhysRevLett.116.241102, Aasi_2014}. Therefore, for parameter estimation of BBH mergers, Lasky et al. emphasized the need of including higher order gravitational wave modes. They took $l=2, 3$ and $\vert m \vert > 0$ to show degeneracy breaking parameter $\Delta h_{lm}\neq 0$, which helps in determining the sign of memory. As a result, they estimated that aLIGO could detect memory with a signal-to-noise ratio of $3 (5)$ using $35 (90)$ events with masses and distances close to GW150914 \cite{PhysRevLett.117.061102}.

A detailed analysis of the work \cite{PhysRevLett.117.061102} has also been carried out recently in \cite{PhysRevD.101.083026} where authors estimate the amount of time taken by aLIGO and Virgo detectors to detect the memory signal in the population of BBH mergers. Their study showed that the signal-to-noise ratio for the GW memory will be near the threshold ~3 after spending a five year observational period in the population of BBH mergers provided that the detectors are operating at their design sensitivities. A study in \cite{PhysRevD.101.023011} develops a Bayesian framework for detecting GW memory through advanced detectors. Their analysis for ten BBH mergers in LIGO/Virgo’s first transient gravitational wave catalog does not show any evidence of memory signal for the detectors providing consistency with the expectations. Further commented that a population of $\mathcal{O}(2000)$ BBH mergers observations can provide us a strong proof of GW memory.    

There have been several studies on gravitational memory effect for plane gravitational waves \cite{Penrose:1972xrn, Hogan2003, Hogan2013, 1998gr.qc9054S, Luk:2012hi, PhysRevD.101.064022} also important from asymptotic symmetry point of view \cite{PhysRevD.96.064013, ZHANG2017743, Zhang_2018, PhysRevD.99.024031}. P.-M Zhang et al. have provided a study on memory for plane gravitational waves and showed how this enables us to have the exact description of soft gravitons which are important ingredients for information paradox. The possible detectability of soft gravitons can be made by observing the motion of test detectors or freely falling particles \cite{PhysRevD.96.064013, ZHANG2017743, Zhang_2018}. Moreover, long back Zel'dovich and Polnarev \cite{1974SvA1817Z} had shown that the detectors will possess a very negligible relative velocity before and after the passage of gravitational waves. This differs from Bondi and Pirani's later findings \cite{1988Natur.332..212B, 1989RSPSA.421..395B}. However, the study of Zhang et al.  on \textit{velocity memory effect} supports Bondi and Pirani's analysis and took the side of Braginsky, Grishchuk, Thorne, and Polnarev \cite{Braginsky:1986ia,  Braginsky1987, Grishchuk:1989qa, lasenby2017black}. As a conclusion, their results showed the velocity memory effect as the nonvanishing asymptotic relative velocity. Another study \cite{PhysRevD.101.064022} has shown the computation of kinematic variables like expansion, shear and rotation, and provided the study on memory effect for the exact, vacuum plane parallel gravitational waves. Plane gravitational waves in the context of stitching of two spacetimes have also been the interest of memory effect \cite{PhysRevD.99.024031}. Memory in the context of stitching of two black hole spacetimes has also been one of the goals of this thesis; we have separately contributed two chapters for this focusing on the memory for extreme and nonextreme black holes and its relation with asymptotic symmetries \cite{PhysRevD.100.084010, Bhattacharjee2021}. 

As we now know that memory effect implies the permanent change or shift in the orientation of ring of test masses or more precisely change in the relative displacement of LIGO test masses which are nothing but geodesics, the direct intuition suggests us to implement the idea of geodesic deviation equation (GDE). This computes the deviation vector between two timelike geodesics or LIGO test masses which directly restores the memory. 
\begin{figure}[h!]\centering
    \includegraphics[scale=0.4]{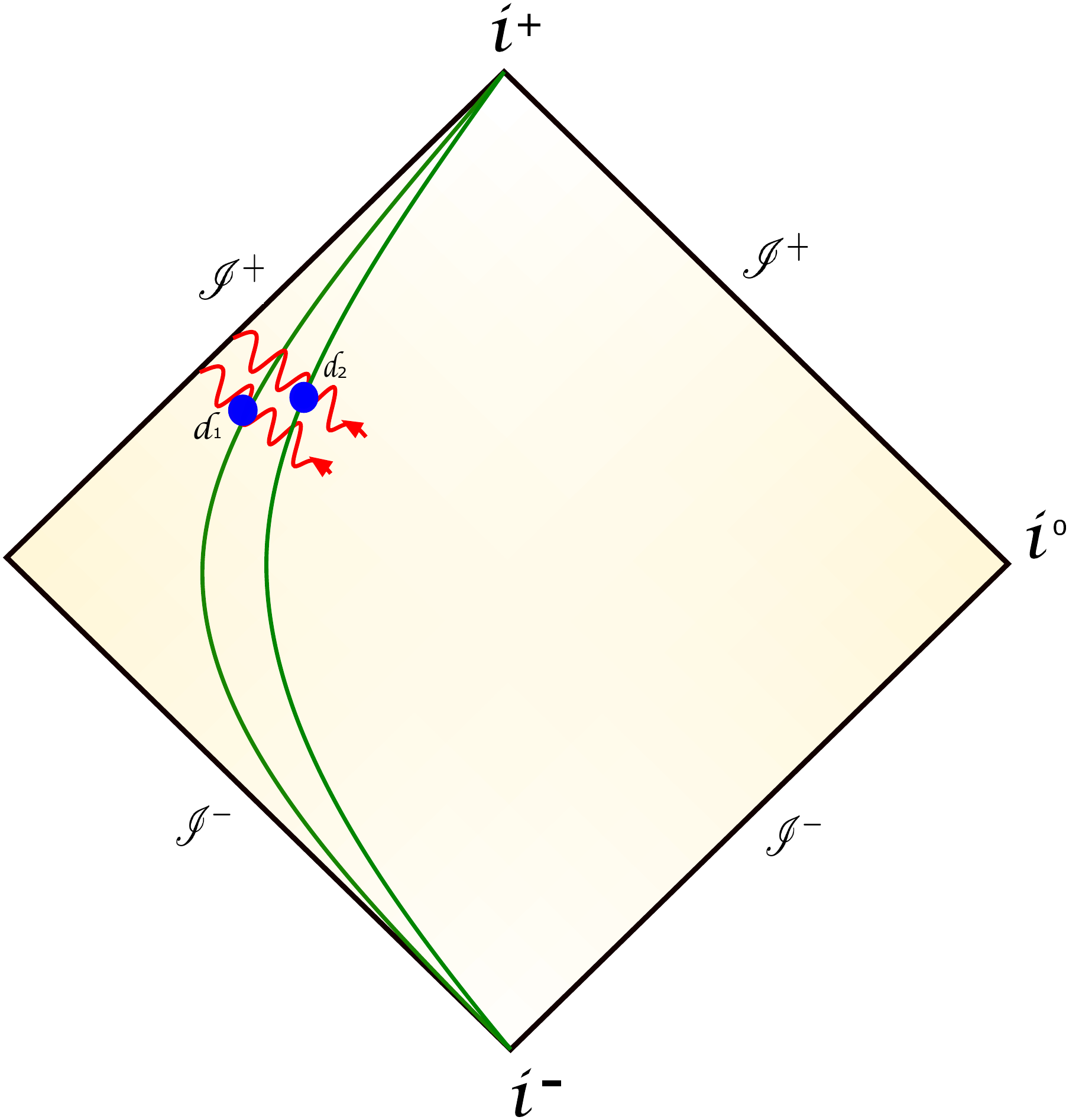} 
    \caption{An schematic diagram showing two detectors $d_{1}$ and $d_{2}$ stationed at future null infinity will capture the memory effect due to the interaction between detector setup and gravitational waves generated at late time $\mathcal{I}^{-}$.}\label{pic3} 
\end{figure}
Briefly, if we have two nearby timelike geodesics or inertial detectors described by the tangent vector $T^{\mu}$ together with a deviation vector $s^{\mu}$, and let us position them at the future null infinity ($\mathcal{I}^{+}$). The evolution of the deviation vector before and after the interaction with gravitational waves will be captured in the GDE:
\begin{align}
\frac{D^{2}s^{\mu}}{d\tau^{2}} = -R^{\mu}{}_{\delta\sigma\lambda}T^{\delta}T^{\lambda}s^{\sigma} \label{gde}
\end{align}
The solution of the GDE will give us a permanent change in the deviation vector $s^{\mu}$ which implicates the achievement of displacement memory effect. This is also known as conventional memory effect which is written to the linear order of $\frac{1}{r}$, where $r$ denotes a radial coordinate. Recently, it has been shown, the change in the deviation vector can be realized by the asymptotic symmetry (supertranslation) which gives rise the same shift between the detectors \cite{strominger2018lectures, Strominger2016}. The detailed analysis and its relation with asymptotic symmetries together with soft theorem have been carried out in \cite{strominger2018lectures, Strominger2016}.


\subsection{Asymptotic symmetries: The Bondi-Metzner-Sachs group}  
\label{section1.3}
 Symmetry considerations have aided many physical systems including spacetime geometry. If we do not take gravity into the picture, i.e. for flat spacetimes, we can define Poincar\'e or inhomogeneous Lorentz group which leaves the Minkowskian metric invariant under some symmetry transformations. Those symmetry transformations are nothing but Lorentz transformations and translations. Thus the isometry group of Minkowskian spacetime is the \textit{Poincar\'e group}. This itself motivates us to look for a generalized concept of isometry which can be implemented for curved spacetimes. An isometry is a diffeomorphism that leaves the metric invariant. Let us now put an extremely small source of gravity into the picture, under this consideration, one is confronted with an unusual predicament. First of all, in a general curved spacetime, one does not recover many symmetries. If we are in an asymptotically flat (AF) spacetime, then one may expect that the symmetry group at the infinities would resemble that of a Minkowski spacetime. However, in practice the group becomes much rich. This was analysed a long ago by Bondi, van der Burg, Matzner and Sachs \cite{Bondi, Sachs:1962zza}. The  inhomogeneous (Poincare) transformations somehow get disappeared. It is not trivial and also difficult to accept the disappearance of inhomogeneous group in the presence of dynamically small gravitational fields because its survival implies the conservation laws. As an other possibility, one also needs to think about if Lorentz transformations are \textit{approximate symmetry} transformations. With this, it is sensible to consider the concept of approximate symmetry in general relativity by introducing \textit{asymptotic symmetry} in an asymptotically flat spacetime. One can again ask what we actually mean by asymptotically flat spacetime, and what are the appropriate boundary conditions which are being preserved by unknown coordinate transformations. The answer to this question has been provided by Arnowitt et al. and Bondi in \cite{PhysRev.122.997, Bondi:1962px}. Both approaches in terms of basic assumptions are quite similar. The initial approach to the problem is to construct the most appropriate boundary conditions and then search for the asymptotic symmetries. As a first attempt Bondi, van der Burg, Metzner and Sachs expected to recover Lorentz transformations as asymptotic symmetry transformations. But surprisingly, they found additional symmetry transformations differing from Lorentz transformations. These transformations in particular are \textit{asymptotic symmetry transformations} \cite{Bondi, Sachs:1962zza, Penrose:1962ij, Newman:1966ub}. The asymptotic symmetries (AS), symmetries that preserve the asymptotic form of the metric of a spacetime asymptotically, are also called \textit{BMS-transformations}. These can be divided into two parts: \textit{Supertranslation} and \textit{Superrotation}, collectively they are called as \textit{Supertransformations}. These are diffeomorphisms of spacetime that preserve the asymptotic behaviour of the metric. Therefore, it is asymptotically flat spacetime, curved spacetimes whose geometry becomes indistinguishable from flat spacetimes at asymptotic infinities, which forms a more general isometry group, i.e. \textit{asymptotic symmetry group} (ASG) also called as \textit{BMS group}. It is an infinite-dimensional group, and it turns out that the Poincar\'e group is the subgroup of this \textit{BMS group}.

Now if we take Minkowskian spacetime, the Poincar\'e group can be recovered as an ASG with the conserved mass or energy; and the group consists of spacetime translations together with Lorentz transformations whereas for asymptotically flat spacetimes, the ASG is nothing but BMS group: semi-direct product of spacetime translations and Lorentz group.
\begin{align*}
Poincare = Lorentz \ltimes Translations \hspace{0.3cm} ; \hspace{0.3cm}
BMS = Lorentz \ltimes Supertranslations
\end{align*} 
where $`\ltimes'$ is the symbol of semi-direct product which means that elements of the BMS group is made up of Lorentz transformations and supertranslations in pairs.

The BMS group has an infinite-dimensional abelian subgroup composed of supertranslations \cite{Bondi,Sachs:1962zza, Newman:1966ub, Newman:1962cia, PhysRevD.95.044002}. These are angle-dependent translations. An extended BMS symmetry superrotations have recently been recovered at null infinities of asymptotically flat spacetimes \cite{Barnich_2007, PhysRevLett.105.111103, Barnich:2011ct}. It is a local conformal transformation of the spatial slice of the metric at null infinities \cite{PhysRevLett.105.111103, Barnich:2011ct}. These are transformations or diffeomorphism acting on the asymptotic null infinities. The asymptotic symmetries have also been recovered near the spatial infinity \cite{Henneaux2018, 2020n}. Recently, in the context of black hole horizon, one could recover the near-horizon BMS symmetries by examining the asymptotic symmetries preserving the near-horizon structure of black holes \cite{PhysRevLett.116.091101, Donnay2016}. In the context of soldering of two spacetimes, say black holes, one can also recover supertranslation and superrotations near the horizon of black holes \cite{Blau:2016juv, Blau2016, PhysRevD.98.104009}. More discussion on BMS symmetries, and their connection with GW memory and soft theorem is provided in Sec.(\ref{s4}).


\subsection{The Triangular Relation}\label{s4}
\label{section1.4}

\begin{figure}[h!]\centering
    \includegraphics[scale=0.55]{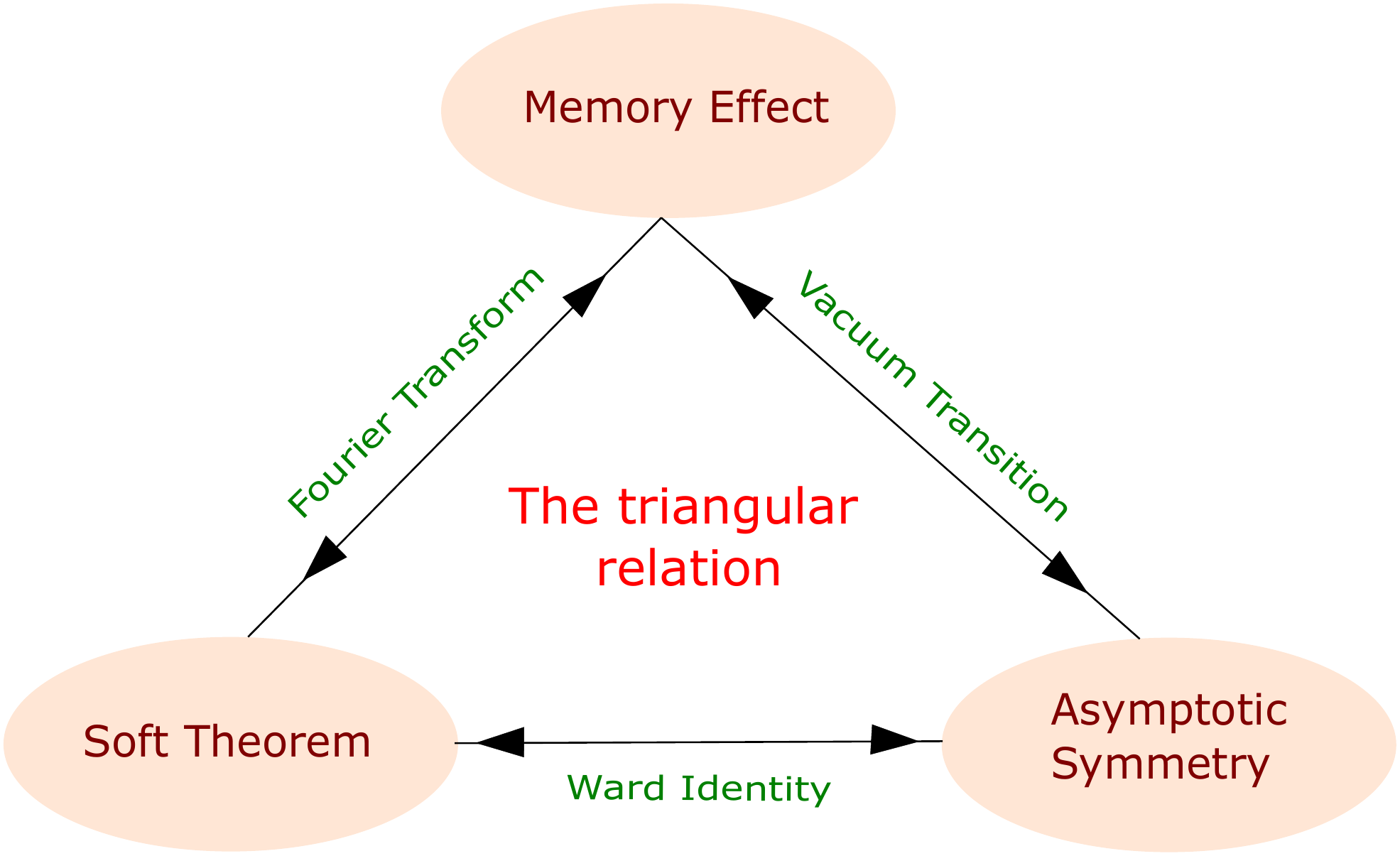} 
    \caption{The infrared triangle: representing the triangular inter-connection between memory effect, asymptotic symmetries and soft theorem.}\label{memm2}
\end{figure}

More recently, it has been established that the infrared structure of gravity and asymptotic symmetries are closely linked. This has brought researchers to look for various intriguing properties for black hole spacetimes. The infrared triangle or the triad contains very informative elements of the topics which are inter-connected with each other. Each of the topics of the triangular relation has been explored intensively in an independent way for decades. The recent developments in this direction establish a connection among these ingredients of the triad. As it can be seen in Fig.(\ref{memm2}) that these three elements are nothing but- \textit{soft theorem}, \textit{asymptotic symmetry} and \textit{GW memory effect}. 

It is known that these three topics seemingly unrelated are essentially one and the same, albeit approached from very different angles and stated in quite distinct notations. This connection was first noticed by Strominger and his collaborators while working on gravitation theory \cite{Strominger2016, Strominger2014}. Let us start corner by corner and see how they are related with each other. The story begins from the \textit{first corner}: Soft theorems which were discovered with Bloch and Nordsieck's work in quantum electrodynamics in 1937, and were extensively refined by Low and others \cite{PhysRev.52.54, PhysRev.96.1428, PhysRev.96.1433, PhysRev.110.974, Kazes1959, YENNIE1961379}. Later, Weinberg expanded this to gravity in 1965 \cite{PhysRev.140.B516}.

In general relativity, the local energy defined by the Bondi mass aspect ($m_{B}$) at each angle on the conformal sphere remains conserved. This local energy is clearly conserved by a single free massless particle in Minkowski space that enters at past null infinity ($\mathcal{I}^{-}$) and exits at future null infinity ($\mathcal{I}^{+}$)  without changing propagation angle. A pair of interacting massless particles incoming from $\mathcal{I}^{-}$, on the other hand, can scatter and travel at right angles to $\mathcal{I}^{+}$. As a result, the particle's contributions to local energy cannot be conserved at all angles in this scenario. \textit{Soft gravitons}, which have zero total energy, are created as a result of this process. However, these can now provide positive or negative localized contributions to the energy. But these are dispersed in such a way that energy conservation is ensured from every viewpoint. When a massless external particle becomes soft, soft theorems describe universal features of Feynman diagrams and scattering amplitudes. More precisely, matrix elements of conservation laws in quantum gravity provide an infinite number of accurate relationships between scattering amplitudes. Weinberg, using Feynman diagrammatics, had already found these relationships \cite{PhysRev.140.B516} in 1965 which is known as the soft graviton theorem. The statement can be reversed also: from the soft graviton theorem, one can infer both the infinity of conservation rules and the supertranslation symmetry of gravitational scattering. This has also set a connection between BMS symmetry and the soft graviton theorem. This exact equivalence has opened up new insights on infrared structure of gravitational theories.
\begin{figure}[h!]\centering
    \includegraphics[scale=0.65]{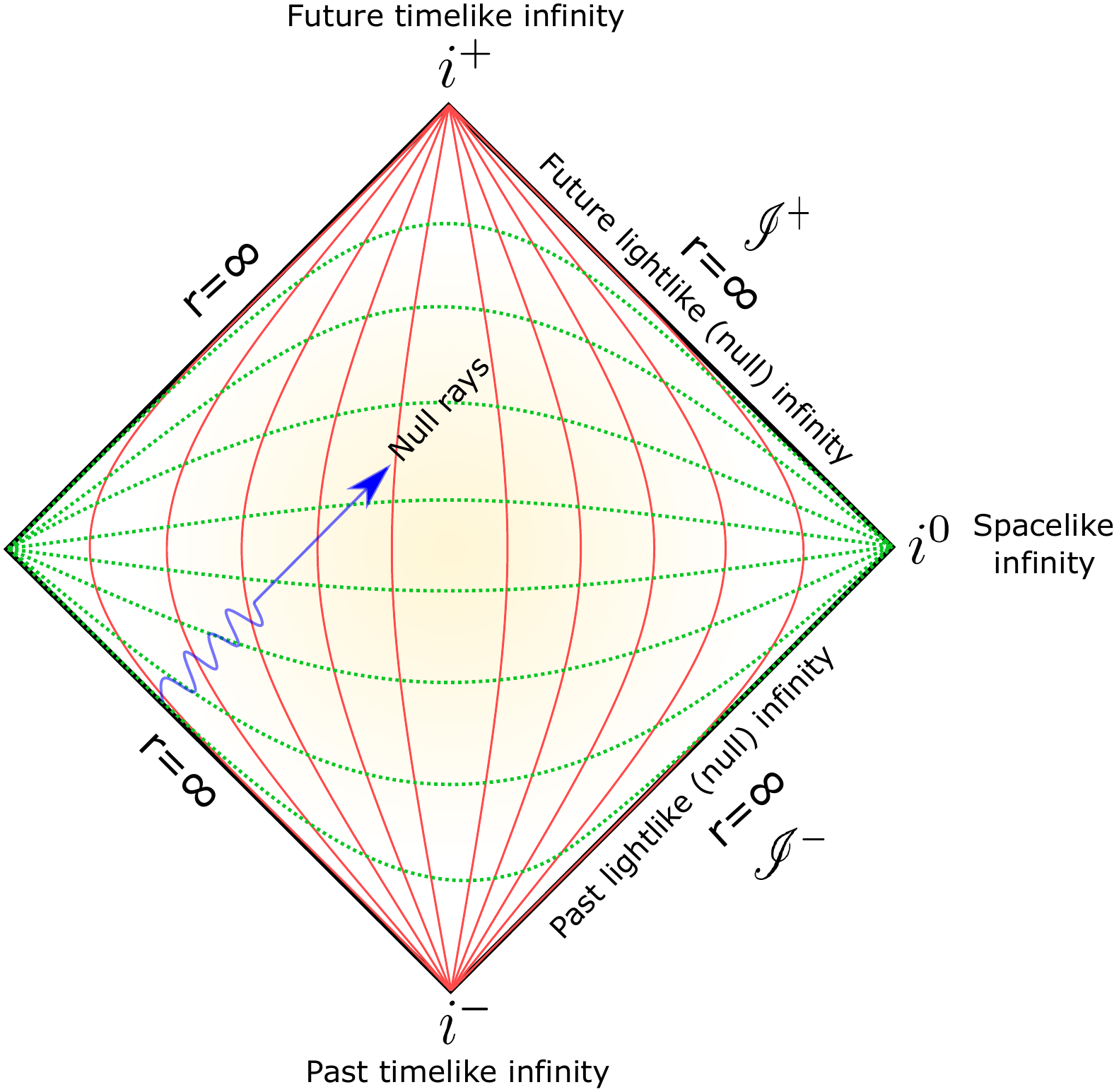} 
    \caption{Minkowski space-Penrose diagram: green curves are constant $t$ surfaces, and red curves are constant $r$ surfaces. Blue line depicts the trajectory of null rays.}
\end{figure}\label{mem1}

The \textit{second corner} of the triangle is \textit{asymptotic symmetry}. As mentioned earlier that this concept is related to the formation of the BMS group (semi-direct product of Lorentz group and supertranslations); Bondi-van der Burg-Metzner-Sachs (BMS) showed new symmetries for asymptotically flat spacetimes: Supertranslations and more recently the concept of extended BMS symmetry Superrotations. It is of course different from Poincare symmetries. Recently in the context of black hole information paradox, the asymptotic symmetries and soft theorem have played a crucial role in order to have a better understanding of gravity at quantum level. A black hole can contain three measurable parameters: mass, charge, and angular momentum. One can describe a black hole if these three parameters are known. The \textit{No-hair theorem} argues that if an item collapses into a black hole, its material attributes become unmeasurable. Materialistic qualities are referred as \textit{hair} in this context. The term ``A black hole has no hair'' was given by John Wheeler. At first glance, it appears that if a black hole is characterised just by $M$, $Q$, and $J$, it will have zero temperature and give us zero entropy, violating thermodynamic laws. When the quantum field theory is applied to black hole descriptions, we discover that black holes have temperatures that are related to their mass. As a result, Hawking's black hole can be defined by its mass, rotation, and charge. It simply states that the no-hair theory holds true in quantum black holes as well. The question arises when an object crosses the event horizon boundary; what happens to the object or the information that enters into the black hole. The physical information that enters a black hole is never lost from quantum theory perspectives whereas it is lost forever according to GR and the No-hair theorem. A black hole without information will violate the law that preserve the information. This contradiction shows one of the most celebrated issues of black hole physics and called as the \textit{Hawking information paradox}.

The pioneering study was initiated by Hawking \cite{hawking2015information} where he claimed, ``How does the information of the quantum state of infalling particles re-emerge in the outgoing radiation?'' For the answer to this question, he proposed that the information is kept on the event horizon, not in the black hole's interior. He turned down his old study on ``Breakdown of Predictability in Gravitational Collapse'' which says that the final quantum state of the completely evaporated black hole will be lost \cite{PhysRevD.14.2460}. Almost after forty years, the initial conclusion of information being lost has become largely seen as improbable. However, during this period, no universal defect in the original assertion was provided which could question the underlying assumptions. On the other hand, string theory has aided in the understanding of the problem of black hole information puzzle. From string theory point of view, a holographic plate at the horizon stores all of the information about a black hole's quantum state. In addition, the storage capacity was discovered to be exactly what the Hawking-Bekenstein area entropy law anticipated \cite{STROMINGER199699}. Recently, new findings on the infrared structure of quantum gravity in asymptotically flat spacetimes have provided such an a priori ground for information loss puzzle. It is directly related  to BMS symmetries originally discovered in early sixties of the twentieth century. In asymptotically flat spacetimes, supertranslation symmetries entail an infinite number of conservation laws for all gravitational theories. Strominger and Hawking's latest discovery uses the asymptotic symmetries of BMS group to prove that information is not lost rather stored in something called as a soft particle. One can construct conserved currents using these soft particles. They have no or zero total energy and may have negative local energy, hence the name \textit{soft hairs} which refers to the hairs in the no-hair theorem. Therefore, soft hair or low-energy quantum excitations may be carried by a black hole and leak information when it evaporates. These principles necessitate a substantial amount of soft supertranslation hair being carried by black holes. 

The classical vacuum in general relativity, according to BMS, is substantially degenerate, and these different vacua are related by supertranslation. For flat spacetime, the vacuum is transformed into a physically inequivalent zero-energy vacuum through supertranslations. Hence, the supertranslation symmetry is spontaneously disrupted or broken because the vacuum is not invariant. Strominger and Zhiboedov provided an explicit form for the supertranslation relating initial and final vacua \cite{Strominger2016}. Furthermore, Hawking, Perry and Strominger (HPS) have shown an explicit treatment of soft hair in terms of soft gravitons or photons. They examined that at the future edge of horizon, the complete information about their quantum state is preserved on a holographic plate \cite{PhysRevLett.116.231301}. These are the recent attempts to tackle the information loss puzzle.

Further, the triad shows a connection between soft theorem and asymptotic symmetries via Ward identity. It is a quantity that connects scattering amplitudes of symmetry-related states and turns out to be nothing but the soft theorems \cite{strominger2018lectures, Strominger2014, He2015, He2014}. It has received considerable attention recently after the derivation of BMS-charges \cite{Barnich2011} and the Ward identities corresponding to BMS charge conservation were proven to be equivalent to Weinberg’s soft graviton theorem and subleading soft graviton theorem \cite{He2015, cachazo2014evidence, WHITE2014216}. Strominger conjectured that the quantum gravity S-matrix has an exact symmetry given by a certain infinite-dimensional diagonal subgroup of the supertranslation symmetries. He provided a follow up work on this with his collaborators and showed that the universal soft graviton theorem of \cite{PhysRev.140.B516, Weinberg:1995mt} is simply the Ward identity following from the diagonal BMS supertranslation symmetry of \cite{Strominger2014}. In \cite{He2015}, authors derive a Ward identity by the S-matrix elements between states with incoming and outgoing particles at in and out points on the conformal sphere at null infinity. S-matrix elements with and without soft graviton current insertions are related by the supertranslation Ward identity and the Weinberg’s soft graviton theorem is the Ward identity following from diagonal supertranslation invariance \cite{He2015}. This implies a connection between soft theorem and BMS symmetries via Ward identity.

\begin{figure}[h!]\centering
    \includegraphics[scale=0.6]{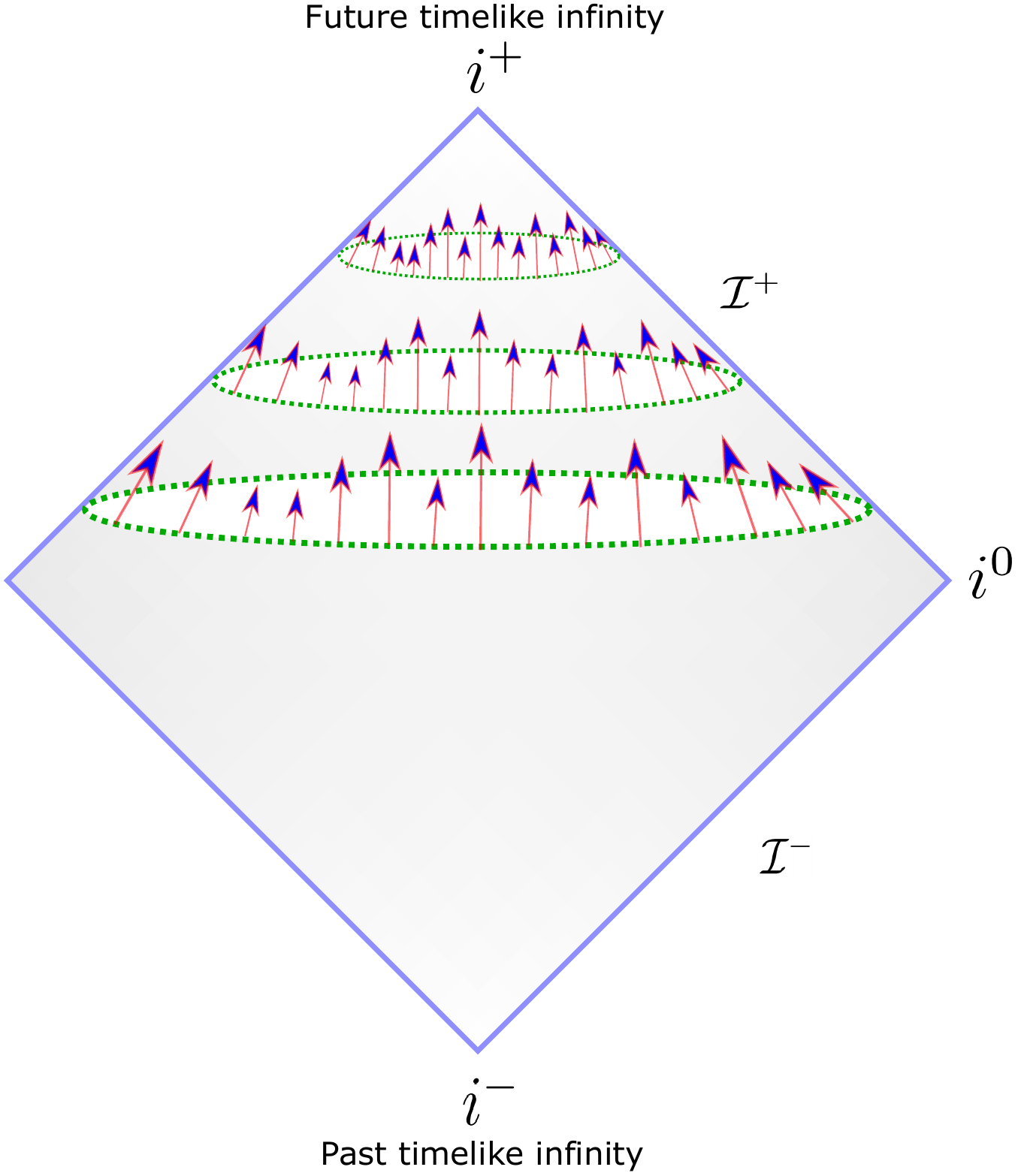} 
    \caption{A supertranslation shifts retarded time $u$ individually at each angle on $\mathcal{I}$.}\label{memmn}
\end{figure}

Now we come to the \textit{third} and last component of the triad: \textit{memory effect}. The precise ingredients soft theorem, asymptotic symmetries and memory effect of the triad are inter-connected with each other. In classical picture, for a given spacetime geometry, BMS transformations produce an infinite class of space-time metrics that are physically unique or distinct. Let us assume that the BMS transformations act on a given metric $g_{\mu\nu}(x^{\mu})$ with $x^{\mu} = (x^{0}, x^{i})$, i.e., one time and three spatial coordinates. The action of the BMS transformations on the metric can result in a permanent relative change in the metric.
\begin{align}
g_{\mu\nu}(x^{\mu})  \xrightarrow{\text{BMS transformation}} \tilde{g}_{\mu\nu}(x^{\mu}) 
\end{align}     
The metric $g_{\mu\nu}(x^{\mu})$ and $\tilde{g}_{\mu\nu}(x^{\mu})$ are distinct and this relative change implies the generation of \textit{GW memory}, and also motivates us to seek for a connection between memory and BMS symmetries.  This change can be understood in the following way- GWs generated from a black hole spacetime and carrying information or properties in terms of BMS parameters would interact with the detector setup placed at the asymptotic null infinity, this would induce a permanent relative change in the initial configuration of the setup.  This is what we are referring as GW memory. Similar set up can also be considered at a place near to the horizon of black holes. A persistent effect similar to that of null-infinity may again be observed. In this thesis this is examined. It provides a physical meaning to the inter-connection between memory and asymptotic symmetries emerging near the horizon of black holes. Technically, $g_{\mu\nu}(x^{\mu})$ can be thought of as the metric of a given asymptotically flat spacetime and $\tilde{g}_{\mu\nu}(x^{\mu})$ is the resultant metric appears as a consequence of the interaction between GWs and detectors which implies a net relative change in the configuration and gives a definition to the memory. The Fig.(\ref{memmn}) also explains the transformation of a geometry into a new one by the action of supertranslation.  

While describing the quantum treatment, we consider the metric ($g_{\mu\nu}$) description as a particular state ($\vert g_{\mu\nu}>$) in the Hilbert space. The transformation now is elevated to an operator with the corresponding charges denoted as $\hat{Q}_{BMS}$ \cite{lust2018black}.
\begin{align}
\hat{Q}_{BMS}\vert g_{\mu\nu}> = \vert\tilde{g}_{\mu\nu}>
\end{align}	  
It describes a degenerate set of vacua, and vacua are distinguished by a spacetime metric. It explains a degenerate collection of vacua each of which is defined by a metric. The entropy of a particular spacetime geometry is therefore connected to the degeneracy of the vacuum that appears from the BMS symmetries. The BMS transformation turns one metric into another, resulting in the spontaneous breakage of the BMS symmetry group. This spontaneous breaking of the BMS symmetry produces the \textit{Goldstone} modes which are known as \textit{soft gravitons} \cite{lust2018black}. This appears in any gravitational scattering amplitude theory. The BMS Goldstone modes implicate the generation of soft gravitons by the action of $\hat{Q}_{BMS}$.

This is related to soft theorem. If we consider four-point graviton amplitude with momenta $\{p_{i}=p_{1},p_{2},p_{3},p_{4}\}$. These amplitudes tend to IR diverging as $\{p_{i}\}\rightarrow 0$. Such divergences can be avoided by adding the infinite number of gravitons, and an infinite cloud of soft particles appear in the final state. This is also a description of \textit{soft theorem}. It is related by the BMS Ward identity and which was first recognised by Strominger. 

Let us briefly sum up the connection for soft theorem, BMS symmetries and memory effect. Soft theorems, Ward identities linked with BMS symmetries, and the memory effect in gravity and gauge theories have recently been discovered and found mathematically to be equivalent. The connection among the three ingredients of the triad can be understood in the following way: it has been shown that a Fourier transform of the soft graviton factor yields a formula for the gravitational memory. It implies that the classical outcome is a special instance or limiting case of the quantum outcome. The Ward identity of the BMS supertranslation, which is claimed to be a spontaneously broken symmetry of the quantum gravity S-matrix, is exactly equivalent to the soft graviton theorem. The permanent change in the metric components as an effect of gravitational radiation is the Fourier transform of the Ward identity which is satisfied by the $S$-matrix of the infrared sector of quantum gravity corresponding to the supertranslations. For the GW memory effect, the initial and final vacua of the gravitational field are associated by the specific BMS supertranslation, and GW memory can be seen as a transition between two inequivalent vacua of the gravitational field. This is how these three fields are inter-linked and no longer different from each other. This triad relation recently has gained considerable attention and has been extended to electromagnetism \cite{He2014, Pasterski2017}, Yang-Mills theory \cite{He2016} and other theories \cite{Campiglia2015, PhysRevD.97.046002}.  A good review of the asymptotic symmetries and displacement memory can be found in \cite{compere2019advanced}. In this thesis, we investigate the possible connection between the corners- asymptotic symmetries and the memory effect of the infrared triangle (\ref{memm2}) near the horizon of black holes \cite{strominger2018lectures, Strominger2016}. 

\section{Near-Horizon Structure}

From theoretical perspectives, the far region analysis has been extensively carried out by the research community. Recently, the near-horizon structure of black hole spacetime is becoming one of the most appealing area of research \cite{PhysRevD.101.124010, doi:10.1142/S0218271820430063, PhysRevLett.124.041601, Adami2020, Adami2021, adami2021null}. This is opening up the window for three nodes of the IR triangle, i.e., soft theorem, BMS symmetries and GW memory not only from theoretical perspectives but also from experimental point of view. The study of recovering BMS symmetries for asymptotically flat spacetimes has recently got a considerable attention for near the horizon of black holes \cite{ PhysRevLett.116.091101, Donnay2016, PhysRevD.95.104053, PhysRevD.98.124016}. These studies  have found the diffeomorphisms that preserve the near-horizon asymptotic structure of black hole spacetimes. The existence of symmetry implies the presence of a conserved quantity. In the context of asymptotic symmetries, the existence of soft hair on black holes is necessary for charge conservation of supertranslation and superrotation \cite{strominger2018lectures, PhysRevLett.116.231301}. As the conservation principles are derived from the long-distance behaviour of fields close to spatial infinity, the presence of black holes should have no effect on them. We know that the conserved charges can be expressed as bulk integrals over any cauchy surface. A contribution from the future event horizon should be taken into account for conserved charges as $\mathcal{I^{+}}$ is no longer a Cauchy surface in the presence of a classical black hole. This brings a direct motivation for the emergence of asymptotic symmetries near the horizon of black holes from conservation perspectives.

There are two methods to recover BMS symmetries near the horizon of a black hole spacetime. The study of near-horizon BMS symmetries had started in the new millennium  \cite{PhysRevD.64.124012, Hotta_2001}, and as a recent progress,  Donnay et al. showed one way of obtaining BMS symmetries near the horizon of a stationary black hole \cite{PhysRevLett.116.091101}. Defining suitable choice of boundary conditions in the near-horizon region, the asymptotic Killing vectors are computed which preserve such boundary conditions. In three and four dimensions, they showed that for the appropriate choice of boundary conditions, the near-horizon region of a stationary black hole induces supertranslations including semi-direct sum with extended BMS symmetry superrotations which is being represented by Virasoro algebra. Both supertranslations and superrotations appear near the horizon in this way. The extended analysis of this study was further provided by the same authors where they considered time-dependent boundary conditions; as a result of asymptotic isometries, supertranslation was again recovered for near-horizon region of non-extremal black holes. Here, authors ensured that these time-dependent boundary conditions are preserved by asymptotic form of the Killing vectors. Furthermore, they have looked at the scenario of stationary black holes and Rindler horizons and showed that the black hole entropy is gathered by the zero modes of the charges. The detailed study of asymptotic symmetry group and charge algebra can be seen in \cite{Donnay2016}. These infinite-dimensional symmetries near black hole horizon have been studied in various contexts, one is the membrane paradigm \cite{PhysRevD.64.124012, Hotta_2001, Carlip_1999, PhysRevD.66.124021}. Therefore, this indicates that one can recover BMS symmetries by preserving the near-horizon asymptotic structure of black holes which would mimic the ones obtained at asymptotic null infinity \cite{PhysRevLett.116.091101, Donnay2016}. 

Recently, in the context of gluing of two spacetimes, the memory effect and its connection with BMS symmetries have been given a considerable attention. This is our second approach which deals with the soldering of two spacetimes across a common null hypersurface \cite{Blau2016, Blau:2016juv, PhysRevD.98.104009}. Let us consider two spacetimes, say black holes, are being patched along a null hypersurface and one of the null hypersurfaces happens to be the horizon of the black holes, we obtain \textit{horizon shell}. To ensure that the analysis is consistent with the theory of general relativity, we need to satisfy junction conditions. It has been shown that we can stitch them in infinite ways by demanding that the induced metric remains invariant under the translations generated by the null generators of the shell \cite{Blau2016, PhysRevD.98.104009}. The freedom for the choice of the intrinsic coordinates on null hypersurface in the null-direction is termed as \textit{soldering freedom}. The sandwiched spacetime generates a certain kind of GW, called as \textit{impulsive gravitational wave} (IGW). 
\begin{figure}[h!]\centering
    \includegraphics[scale=1.0]{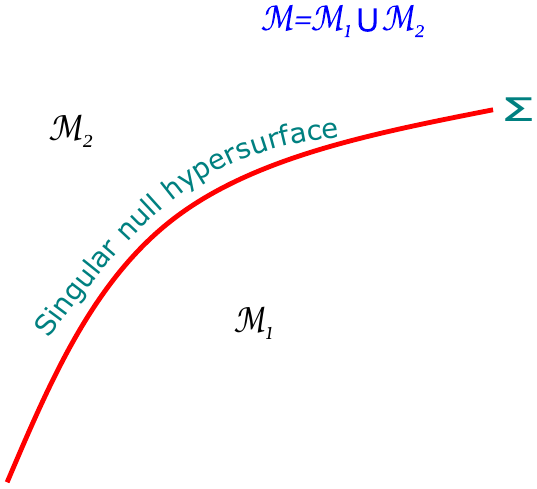} 
    \caption{Null hypersurface $\Sigma$ separating manifolds $\mathcal{M}_{1}$ and $\mathcal{M}_{2}$ each with different metric. The whole spacetime will be the union on both manifolds, $\mathcal{M}=\mathcal{M}_{1} \cup \mathcal{M}_{2}$.} \label{memff}
\end{figure}
The null hypersurface, representing the history of impulsive lightlike signals, separates spacetime manifold into two parts ($\mathcal{M}_{1}, \mathcal{M}_{2}$) as depicted in the Fig.(\ref{memff}) each with a different metric. There are two methods known to describe this scenario \cite{PhysRevD.43.1129}. One way is to introduce a common coordinate system which covers  both sides of the singular null hypersurface . In other words, this approach is based on distributional algorithm where the spacetime metric in a common coordinate system is written as a union of two metrics. The other method of IGW generation is based on the Penrose's cut and paste formalism \cite{Penrose:1972xrn, Aliev_2001, Nutku} which uses the freedom of performing the general coordinate transformation to the one side of the metric and re-attach with the other side of the spacetime along a common null hypersurface. Both these approaches have their own importance. However, both of them can be used alternatively. 



The soldering analyses for two spacetimes generate a singular term in the Einstein field equation supported on the null hypersurface. This singular term is proportional to delta distribution function, when the distributional algorithm is employed. The study of this singular term induces IGWs, which is constituted by the transverse-traceless part of the induced metric on the shell. In fact this singular term modifies the expression of energy-momentum tensor, and also carries the delta function which would depict the thin layer or shell. Hence, a singular null hypersurface in general would carry thin shell or thin distribution of lightlike matter (e.g. neutrino fluid) together with IGWs. It can also contain only lightlike matter or pure IGWs \cite{Hogan2003, PhysRevD.43.1129, MSM_1927__25__1_0, Israel1966, Poisson2002ARO}. But its history will remain the null hypersurface. Such impulsive signals are usually produced during violent astrophysical phenomena like supernova explosion or a coalescence of black holes. One can also study the intrinsic properties, such as surface current, surface pressure and surface energy density, of the null hypersurface situated at the horizon by analysing the energy-momentum tensor. 

Interestingly, a class of soldering symmetries mimic the BMS symmetries. This implies, there are infinite number of ways to solder two manifolds across a null shell as long as the null generators of the shell keep the induced metric invariant. Naturally, the appearance of BMS like symmetries at the horizon (null shell) of black holes would provide a very good platform to see the effect of such symmetries on test particles. In this way, this paves the window of studying the memory effect for IGW on detectors stationed near the horizon. 
The supertranslation-like symmetry is recovered as a soldering freedom, which acts as a shift of the null coordinate on the shell (installing a null analogue of Gaussian normal coordinates across the shell)
\begin{align*}
Supertranslation: \hspace{2mm} V \rightarrow V + T(x^{A}).
\end{align*}  
 $T(x^{A})$ is nothing but supertranslation parameter depending on the spatial or angular coordinates. Further, one recovers superrotation as a soldering freedom near the horizon mimicking the one obtained at null infinity, if one allows local conformal transformations to the spatial slice of the metric, $z\rightarrow f(z)$ $\&$ $\bar{z}\rightarrow f(\bar{z})$, and compensate that by a corresponding shift in null coordinate. This induces extended BMS symmetry known as superrotation near the black hole horizon \cite{PhysRevD.98.104009}. This will discussed in the Sec.(\ref{BMSsymmsuptr}).

A detailed study of horizon shell and soldering freedom for stitching of two black hole spacetimes has been considered in \cite{Blau2016, PhysRevD.98.104009}. Blau et al. \cite{Blau2016} showed supertranslation as a soldering freedom. Authors provided their analysis for Schwarzschild black hole by explicitly obtaining the intrinsic properties of the horizon shell. They obtained the shell intrinsic properties by employing an off-shell extension of the soldering transformations. The extension of the Blau's work was carried out by Bhattacharjee et al. \cite{PhysRevD.98.104009} where authors provided the results for intrinsic formulation of horizon shell in terms of BMS parameters for rotating black holes and reinterpreted the results. The authors also provided a detailed study on the superrotation-like soldering symmetries and showed that one can not recover superrotations at the black hole event horizon instead one can recover them for the null shell placed just outside the event horizon. The authors further showed that the soldering freedom can also be recovered for pure de Sitter black holes and Rindler horizons. These studies certainly show a rich structure of black hole horizon. 

Now, a pertinent  question arises, how to establish the connection between memory effect and BMS-like symmetries near the horizon of black holes? The study of memory effect and its connection with near-horizon BMS symmetries has been carried out in \cite{PhysRevD.100.084010, PhysRevD.102.044041}. It is also the major goal of this thesis. We now know that the BMS symmetries can be recovered near the BH horizon as well. In this respect, our study touch upon the emergence of asymptotic symmetries in the context of soldering of two spacetimes, and show some measurable effects together with the displacement in the deviation vector in terms of BMS parameters for the test detectors placed near the horizon of black holes. So our natural task would be to study some measurable effects of IGWs carrying BMS like parameters on test particles or test detectors or test geodesics, timelike and null. The interaction between horizon shell (the null shell placed at the horizon of a black hole) and two nearby null geodesics would amount us to locally estimate some measurable effects like expansion and shear on and near the horizon of the black hole. These physical parameters (optical tensors) suffer a permanent jump before and after the passage of the IGWs (with and without matter). This depicts a particular form of the memory known as \textit{B-memory}. Using the optical parameters one can further compute the $B-$tensor which describes the failure of two nearby null geodesics being remain parallel. Similarly, one can also study the interaction between timelike geodesics and null shell placed at the horizon. This requires to solve the geodesic deviation equation and find the change in the components of the deviation vectors before and after the passage of the IGWs. This also depicts memory effect induced by IGWs on timelike geodesics. 

Leaving the method of spacetime-soldering behind, which involves studying the memory on the horizon itself, a slightly different picture would be to estimate memory effect near the horizon of black holes using asymptotic analysis. In the far region case, some gravitational waves approaching the past null infinity of an asymptotically flat spacetime would interact with the test detector placed near the future null infinity. This interaction will generate a permanent relative displacement or deviation in the test detector which signifies a net non-vanishing or non-zero change in the deviation vector before and after the passage of gravitational waves. This quantifies the memory effect at asymptotic null infinity for asymptotically flat spacetimes. In order to relate it with BMS symmetries, one can obtain a supertranslation which would give rise the same change in the deviation vector ensuring the memory effect is obtained in terms of supertranslation parameter. The challenge is to establish a similar study on memory effect and its relation with BMS-like symmetries near the horizon of black holes. As mentioned earlier that the BMS symmetries and its extended version \cite{PhysRevLett.105.111103, Barnich:2011ct} can be recovered at the horizon of black holes by studying the asymptotic symmetries preserving the near-horizon structure of black holes \cite{PhysRevLett.116.091101, Donnay2016, PhysRevD.98.124016}, we provide an explicit analysis for the \textit{displacement memory effect for near-horizon BMS symmetries} \cite{Bhattacharjee2021}, i.e., we compute the displacement memory effect near the horizon of black holes and its possible connection with supertranslation and superrotation-like symmetries. It is understandable that the excitations in the vicinity of the black hole spacetimes would affect the initial separation vector for the detector setup placed near the horizon and would leave an imprint on the test detector. This depicts memory effect near the horizon of black holes. However, the asymptotic fall offs or boundary conditions are preserved by the asymptotic Killing vectors. To establish the connection with BMS symmetries, we again look for the transformation(s) of metric parameter(s) contributing in the memory effect along the Killing direction. This establishes the connection between displacement memory and BMS symmetries for near the horizon of black holes. The explicit computations have been carried out for extreme and non-extreme black holes (fixed temperature configurations) in \cite{Bhattacharjee2021} and will be discussed in the separate chapter on displacement memory. This study certainly provides a good model to study GW memory in terms of supertranslation and superrotation-like symmetries. It is not only  promising from theoretical perspectives but also from experimental point of views as it might help to relate the soft hairs that may give a direction in resolving information puzzle as conjectured by Hawking, Perry and Strominger.

So far we have considered only theoretical models to investigate the signatures of the asymptotic symmetries near the black holes. A detailed review of these studies can be found in \cite{kumar2021displacement}. These considerations have to be improved to say something concretely on the possibility of detection of such signatures. Therefore, it should be important to examine the possibility of finding the signatures of asymptotic symmetries in real astrophysical observations, e.g. \cite{PhysRevD.103.024031}. For this, we have studied the observational signatures of supertranslation through the behaviour of photon sphere for a dynamically evolving supertranslated black hole. This study provides a useful detection device in examining the role of asymptotic symmetries for future observations. This analysis shows first and direct motivation in the direction of detecting the supertranslation hair through standard tests of general relativity. The detailed analyses can be found in the final chapter of the thesis.

\section{Thesis in a nutshell}
\label{section1.4}
As the gravitational memory signal has not been observed yet, it is one of the fascinating fields to investigate the theoretical connections of asymptotic symmetries in gravitational wave data astronomy. It will certainly help us to probe the near horizon structure of black holes in order to examine various aspects of gravitation theory not only from classical perspectives but also from quantum regime. This sets a simultaneous motivation for theoreticians and GW data analysts to study different signals and their connections with asymptotic symmetries.

The goal of the thesis is to study GW memory effect near the horizon of black holes and its possible connection with near-horizon BMS symmetries. In chapter (\ref{CH2}), we provide the necessary ingredients to develop the stage for computing the BMS memory in the context of soldering of two spacetimes. We provide the review for formation of horizon shell, intrinsic properties of the null shell and its connection with BMS-like symmetries. Next, chapter (\ref{CH3}) deals with BMS memory effect for non-extreme black holes where we study the interaction of IGWs with timelike and null geodesics. We show that the null geodesic congruence will exhibit a finite and non-zero change in the measurable quantities like shear and expansion between two nearby null geodesics. Further, the interaction between IGWs and timelike geodesics leads us to solve the geodesic deviation equation, and show that the deviation vector between two nearby timelike geodesics suffer a finite jump before and after the passage of gravitational waves. Both these studies on null and timelike geodesics will carry the BMS parameters, i.e., supertranslations and superrotations. The next chapter (\ref{CH4}) is based on the similar approach but deals with extreme black holes. We again provide the study of interaction between timelike and null geodesics before and after the passage of the gravitational waves together with the measurable effects in terms of BMS parameters. Thus this would conclude the BMS memory effect near the horizon of non-extreme and extreme black holes in the context of stitching of two black hole spacetimes. These studies might serve as a model to help from the observational perspectives in the detection of BMS-like symmetries. 

The chapter (\ref{CH5}) is an analogous study of the conventional memory effect which was obtained at asymptotic null infinity for asymptotically flat spacetimes. Here, we provide a similar study for near-horizon asymptotic form of a metric. We compute the displacement memory effect near the horizon of black holes in three and four dimensions for extreme and non-extreme black holes. We also highlight the major differences by providing the comparison with the far region analysis. Further, we also provide an explicit connection with near-horizon BMS symmetries. As a result, we show that the amount of change in the deviation vector can be given by the supertranslation or superrotation parameters.    

In chapter (\ref{CH6}), we highlight the memory tensor together with a new gravitational memory, spin memory, near the horizon of black holes analogous to the one obtained at null infinity. We address the difficulties involved in the detection of asymptotic symmetries. As an alternative approach, we study the behaviour of photon sphere carrying supertranslation field for a dynamically evolving supertranslated black hole. This chapter sets a strong motivation for determining the black hole shadows and inspires to have the observational signatures of such symmetries through standard tests of general relativity. Finally, in chapter (\ref{CH7}), we discuss and summarize our major findings of the studies together with its future prospects which might play a crucial role in understanding the GW memory from observational standpoints. The calculation details can be seen in the appendices.

\nomenclature[z-DEM]{DEM}{Discrete Element Method}
\nomenclature[z-FEM]{FEM}{Finite Element Method}
\nomenclature[z-PFEM]{PFEM}{Particle Finite Element Method}
\nomenclature[z-FVM]{FVM}{Finite Volume Method}
\nomenclature[z-BEM]{BEM}{Boundary Element Method}
\nomenclature[z-MPM]{MPM}{Material Point Method}
\nomenclature[z-LBM]{LBM}{Lattice Boltzmann Method}
\nomenclature[z-MRT]{MRT}{Multi-Relaxation 
Time}
\nomenclature[z-RVE]{RVE}{Representative Elemental Volume}
\nomenclature[z-GPU]{GPU}{Graphics Processing Unit}
\nomenclature[z-SH]{SH}{Savage Hutter}
\nomenclature[z-CFD]{CFD}{Computational Fluid Dynamics}
\nomenclature[z-LES]{LES}{Large Eddy Simulation}
\nomenclature[z-FLOP]{FLOP}{Floating Point Operations}
\nomenclature[z-ALU]{ALU}{Arithmetic Logic Unit}
\nomenclature[z-FPU]{FPU}{Floating Point Unit}
\nomenclature[z-SM]{SM}{Streaming Multiprocessors}
\nomenclature[z-PCI]{PCI}{Peripheral Component Interconnect}
\nomenclature[z-CK]{CK}{Carman - Kozeny}
\nomenclature[z-CD]{CD}{Contact Dynamics}
\nomenclature[z-DNS]{DNS}{Direct Numerical Simulation}
\nomenclature[z-EFG]{EFG}{Element-Free Galerkin}
\nomenclature[z-PIC]{PIC}{Particle-in-cell}
\nomenclature[z-USF]{USF}{Update Stress First}
\nomenclature[z-USL]{USL}{Update Stress Last}
\nomenclature[s-crit]{crit}{Critical state}
\nomenclature[z-DKT]{DKT}{Draft Kiss Tumble}
\nomenclature[z-PPC]{PPC}{Particles per cell}

\chapter{Thin-shell formalism and near-horizon asymptotic symmetries}\label{CH2}



\ifpdf
    \graphicspath{{Chapter2/Figs/Raster/}{Chapter2/Figs/PDF/}{Chapter2/Figs/}}
\else
    \graphicspath{{Chapter2/Figs/Vector/}{Chapter2/Figs/}}
\fi

In this chapter, the mathematical ingredients needed for the analyses of the near horizon asymptotic symmetries and their detection are discussed. In general relativity, a shell is a geometric configuration that can be used to investigate the propagation of thin distribution of null matter (e.g. neutrino) or impulsive gravitational waves (IGWs). The thin surface layer of null matter together with impulsive waves is precisely referred as \textit{thin-shell} or \textit{thin null shell}. The basic construction of this formalism, relying on the junction conditions, was first provided in pioneering works by C. Lanczos and N. Sen \cite{lanczos1922bemerkungen, lanczos1924flat, sen1924grenzbedingungen}. The formalism 
was historically developed by Darmois and Israel in the literature \cite{MSM19272510, Israel:1966rt}, and is being widely utilized in general relativity \& cosmology \cite{PhysRevD.43.1129, Israel:1966rt, doi:10.1063/1.531740}. Darmois initially formulated the geometric criteria for the thin-layer to be treated as a boundary of two separate manifolds joined together at the hypersurface \cite{MSM19272510}. In this direction, the seminal finding of Israel \cite{Israel:1966rt} has been one of the core inputs for the further investigation of a variety of topics such as gravitational collapse, domain wall configuration, bubble dynamics, interior structure of black holes and wormholes \cite{doi:10.1103/PhysRev.56.455, PhysRevD.35.1747, VISSER1989203, PhysRevD.53.3215}. These are some recent applications of thin-shell formalism in general relativity. The major motivation for various applications of the formalism comes from a famous study of 1939 on Oppenheimer-Snyder collapse where authors showed the initial insights about the nature of gravitational collapse to a black hole \cite{doi:10.1103/PhysRev.56.455}.

\begin{figure}[h!]\centering
    \includegraphics[height=9.0cm, width=9.0cm]{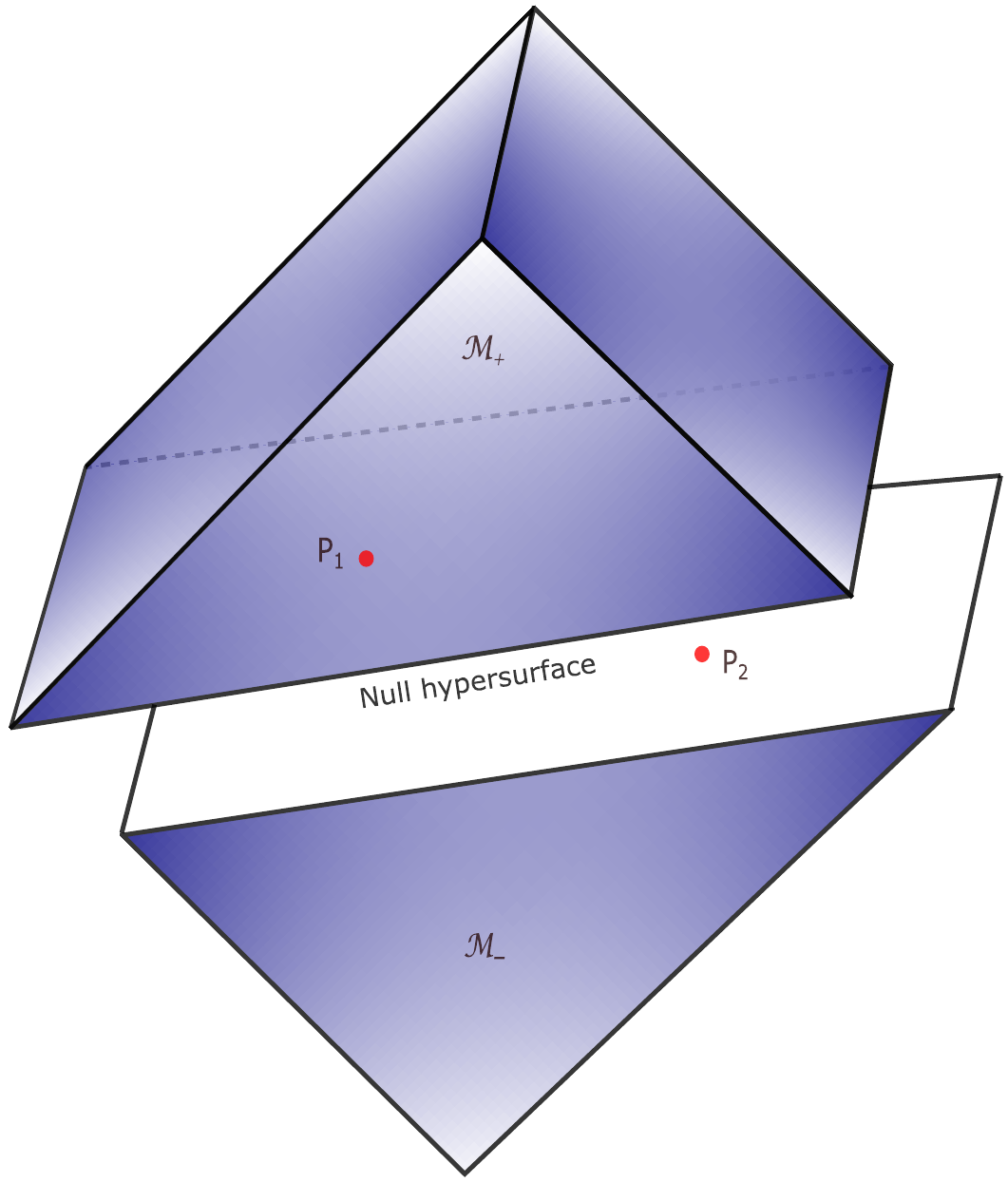} 
    \caption{A schematic diagram of Minkowski spacetime which is being cut into two parts $\mathcal{M}_{-}$ and $\mathcal{M}_{+}$ along a plane null hypersurface ($\Sigma$). $\mathcal{M}_{\pm}$ are re-attached along the common null hypersurface, and points are relocated along the null-generators of the cut and found (denoted as $P_{1}$ and $P_{2}$). The whole spacetime will be the union of both manifolds, $\mathcal{M}=\mathcal{M}_{+} \cup \mathcal{M}_{-}$. The methodology generates impulsive gravitational waves in flat spacetime. This process can be generalized to black holes and generation of IGWs can be understood in the similar way.}\label{solderingn}
\end{figure}

Technically, the Darmois-Israel junction requirements specify the proper boundary conditions for a singular hypersurface supported by a localized energy-momentum source with a Dirac delta distribution. In addition to the junction conditions, one also needs to make sure the metric continuity across the singular hypersurface. The formalism expresses the hypersurface properties in terms of jump in extrinsic curvature as a function of intrinsic coordinates.  The concept relies on the construction of spacetime geometries using the \textit{cut-and-paste} approach across the singular null hypersurface. Penrose demonstrated a geometrical cut-and-paste methodology for generating impulsive waves in a flat spacetime background \cite{Penrose:1972xrn}; where an impulsive gravitational wave intuitively can be thought of as a suitable class of sandwich waves whose profiles approach the Dirac delta distribution. The cut-and-paste, also known as \textit{scissors-and-paste}, technique works by slicing spacetime along a null hypersurface and then re-attaching the two halves along that common null hypersurface \cite{Penrose:1972xrn, PhysRevD.100.024040, PhysRevD.96.064043} as it can be seen in the Fig(\ref{solderingn}). 
There is an alternative way of representing such spacetimes by using continuous or distributional metrics \cite{Hogan2003, doi:10.1142/9789812776938_0007}. The null shells are important and simultaneously difficult to investigate because normal vector becomes the null surface generator. A null-shell placed at the horizon is termed as a \textit{horizon shell} \cite{Blau2016}. As the near horizon structure might contain very minute and key information of black hole spacetimes, it is crucial to dig into the local properties of horizon shells. To avoid the difficulty of normal vector to the null hypersurface becoming tangent, we need an auxiliary normal which can help in capturing the intrinsic properties of the shell. However, because of the difficulties involved in the procedure, the study of lightlike or null shells has been neglected and remains little understood. The overcoming contribution was made by Barrab$\grave{e}$s and Israel in his pioneering work \cite{PhysRevD.43.1129} where authors introduced a \textit{transversal} or \textit{cross vector} on the null hypersurface $\Sigma$. 
This helped in resolving the ambiguities involved in analyzing the null shells.  

On the other hand, as a recent development, there has been considerable attention from asymptotic symmetry perspective. It has been shown that there can be a substantial freedom in \textit{soldering} two spacetime geometries along a given null hypersurface. However, one has to maintain the Israel junction condition that the induced metric is continuous across the null surface. For an isolated example, there exists a study on \textit{Dray `t Hooft shell} focusing on separating or joining two equal mass Schwarzschild black holes along their horizon. Despite the fact that there have been existing studies on this subject, there appears to have been no comprehensive follow-up investigations. In this respect, a systematic take on was performed by Blau and O'Loughlin in 2016 \cite{Blau2016} where authors demonstrated the circumstances in which non-trivial soldering transformations can occur. They showed that the resulting soldering group is infinite dimensional which one can achieve by preserving the induced null hypersurface metric along the killing direction, i.e., the translations along the null generators of the null hypersurface. Later this analysis was extended for rotating black holes in \cite{PhysRevD.98.104009}. These analyses imply the existence of asymptotic symmetries at the black hole horizon. We shall look at the methodologies in greater details in subsequent sections of this chapter.

In this chapter, we would mainly focus on the basic understanding required for soldering of two spacetimes (say black holes) which results the generation of IGWs and thin surface layer or thin distribution of null matter (e.g. neutrino fluid). This chapter provides a basic building block for the emergence of asymptotic symmetries at the horizon of black holes when two spacetimes are stitched together, and intrinsic properties of null hypersurface of the formulation. Further, in subsequent chapters, we shall analyze the interaction between timelike or null test particles and horizon shells, and try to investigate some measurable effects on the test particles or geodesics in terms of asymptotic symmetries (BMS parameters). This attempt would establish a connection between measurable quantities and asymptotic symmetries near the horizon of black holes. 

\section{Thin-shell formalism and impulsive gravitational waves}\label{isrl}

In general relativity, the fundamental development for the formation of shock waves, hypersurfaces and thin shells is devised in which the intrinsic properties of a hypersurface are mathematically specified by the extrinsic curvature of the embedding spacetime. This necessitates a central problem to establish some suitable junction conditions at the surfaces of discontinuity. Because the extrinsic curvature is constricted as a carrier of transverse geometrical information, the standard prescription fails for lightlike surfaces. The underlying reason for this is that the normal to the null hypersurface becomes the tangent, hence, can not have a suitable description of extrinsic curvature, and stops us to examine the surface properties as we move along the null hypersurface. The introduction of transversal extrinsic curvature in terms of an auxiliary normal vector addresses the resolution to this issue and helps us to analyze the intrinsic properties of a null surface. The formalism includes a detailed description of the Einstein field equation when two spacetimes are soldered with some appropriate junction conditions. 

\subsection{Junction condition, IGW and intrinsic properties of null shells}\label{sisrl}

We look at a case where two spacetimes, for our consideration- black holes, are being stitched along a common null hypersurface. Carrying out the computation for Einstein field equation, one obtains a Ricci tensor with a term proportional to a Dirac delta distribution function sustaining on the null hypersurface when soldering the two spacetime metrics. This singularity is linked to a thin distribution of null matter or impulsive gravitational waves (IGWs) or a combination of both supported on a null hypersurface. This is what we call a \textit{thin-shell}, or \textit{thin-null shell}. A thin shell is termed as a \textit{horizon shell} when a null hypersurface across which two spacetimes are stitched becomes the event horizon of a black hole. We aim to determine its stress-energy tensor in terms of the geometric attributes of the null hypersurface, as provided by Barrab$\grave{e}$s-Israel and others \cite{PhysRevD.43.1129, Israel1966, Poisson2002ARO}. This framework provides a basic identification of stress-energy tensor in terms of surface energy density, surface current and surface pressure. This also enables us to look for the possibilities of having freedom to solder two black hole spacetimes with the implication of asymptotic symmetries in this context, and how one can obtain some measurable effects on a family of freely-moving observers crossing the null surface. We shall now try to show and understand the basic construction of the formalism. 

As a convention, the Greek letters will denote the spacetime components of a given quantity, and the lower Latin cases will denote the spatial components. In other words, we shall frequently use $\mu, \nu, \alpha, \beta$ etc in order to denote spacetime components, and $a, b, c, i, j, k, l$ for denoting the spatial components. In general for four dimensional case, one can write a quantity $x^{\mu} = (x^{0}, x^{i})$; where the first component $x^{0}$ represents the temporal part and $x^{i}$ implies the spatial part. Here, Greek index runs as $\mu = 0, 1, 2, 3$ depicting spacetime components whereas lower Latin index as, $i = 1, 2, 3$ depicting spatial components. The capital Latin indices will represent last two component of the spatial part denoted as $A, B$ or $C = 2, 3$. 

\begin{figure}[h!]\centering
    \includegraphics[height=5.5cm, width=10.0cm]{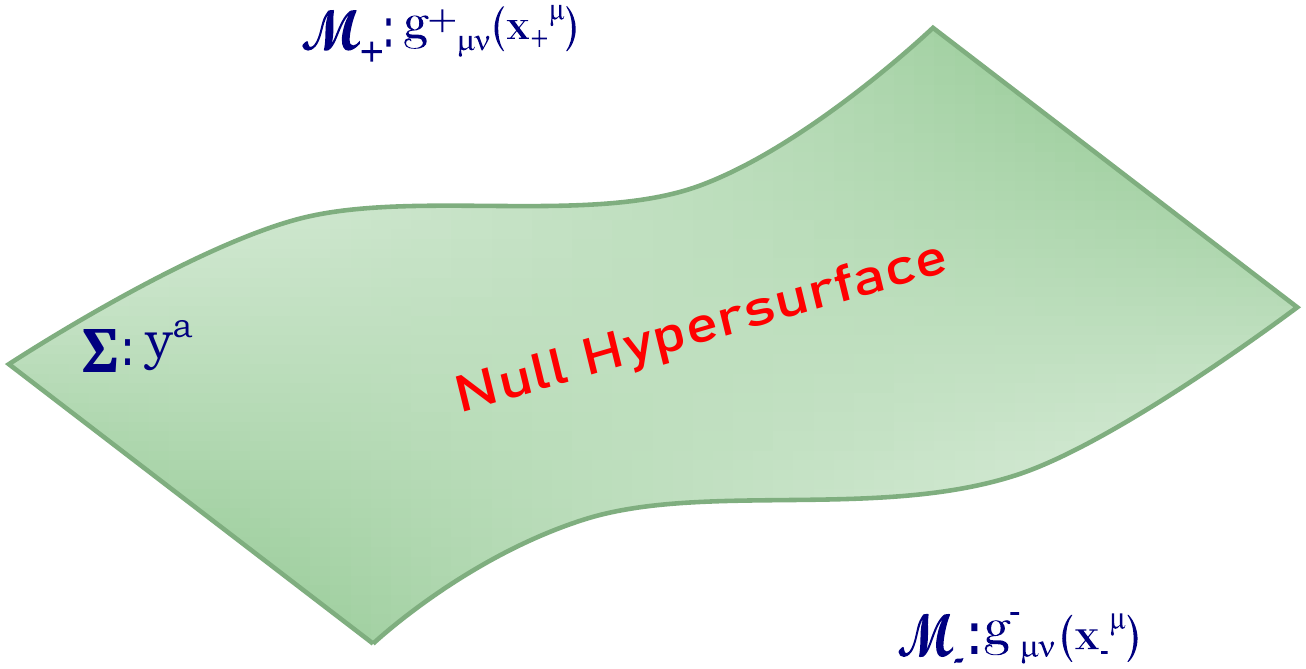} 
    \caption{A schematic diagram representing soldering of two spacetimes $\mathcal{M}_{-}$ and $\mathcal{M}_{+}$ along a common null hypersurface ($\Sigma$) defined with their respective coordinates.}\label{stching}
\end{figure}

Now if we consider a null hypersurface $\Sigma$, the intrinsic coordinates to the hypersurface can be written as $y^{a}$. In Kruskal coordinates, it can be written as $y^{a}={(V, y^{A})}$. The two sides of the hypersurface is depicted in the Fig(\ref{stching}) as $\mathcal{M}_{\pm}$ defined with their respective coordinates $x^{\mu}_{\pm}$. Let us consider an arbitrary parameter $\lambda$ on the null generators of the hypersurface. Thus we can write intrinsic coordinates to the $\Sigma$ as $y^{a}=(\lambda, \Xi^{A})$ with $A = (2, 3)$. We also assume that the hypersurface coordinates $y^{a}$ are same on both sides of the $\Sigma$. $\Xi^{A}$ is used to label the generator of the $\Sigma$ and each generator is being assigned by these two coordinates $\Xi^{A}$. Further a third coordinate $\lambda$ is assigned to each event on a specified generator. When $\lambda$ is changed while $\Xi^{A}$ remains constant, a displacement is produced along a single generator whereas a displacement across generators is produced by varying $\Xi^{A}$ while keeping $\lambda$ constant. As it can be seen from manifolds $\mathcal{M}_{\pm}$, the hypersurface is described by a set of parametric relations: $x_{\pm}^{\mu}(y^{a})$. One can write tangent vectors on the hypersurface $\Sigma$ as, $e^{\mu}_{a}=\frac{\partial x^{\mu}}{\partial y^{a}}$. Since we have the displacement along the null generator $dx^{\mu}=n^{\mu}d\lambda$, hence the tangent to the null generators can be given by $n^{\mu}=\frac{dx^{\mu}}{d\lambda}$. 
Thus we can separately write down the null vector and two spacelike vectors, satisfying the conditions $n_{\mu}e^{\mu}_{A} = 0 = n^{\mu}k_{\mu}$, in the following way,
\begin{align}
e^{\mu}_{1}\equiv n^{\mu} = \Big(\frac{\partial x^{\mu}}{\partial \lambda}\Big)\Big\vert_{\Xi^{A}} \hspace{2mm} ; \hspace{2mm} e^{\mu}_{A} = \Big(\frac{\partial x^{\mu}}{\partial \Xi^{A}}\Big)\Big\vert_{\lambda}.
\end{align}  \label{e1}
The hypersurface metric can be written as
\begin{align}\label{e2}
ds_{\Sigma}^{2} = \sigma_{AB}d\Xi^{A}d\Xi^{B} \hspace{2mm} ; \hspace{2mm} \sigma_{AB}=g_{\mu\nu}e^{\mu}_{A}e^{\nu}_{B}.
\end{align}
\begin{figure}[h!]\centering
    \includegraphics[height=2.5cm, width=10.0cm]{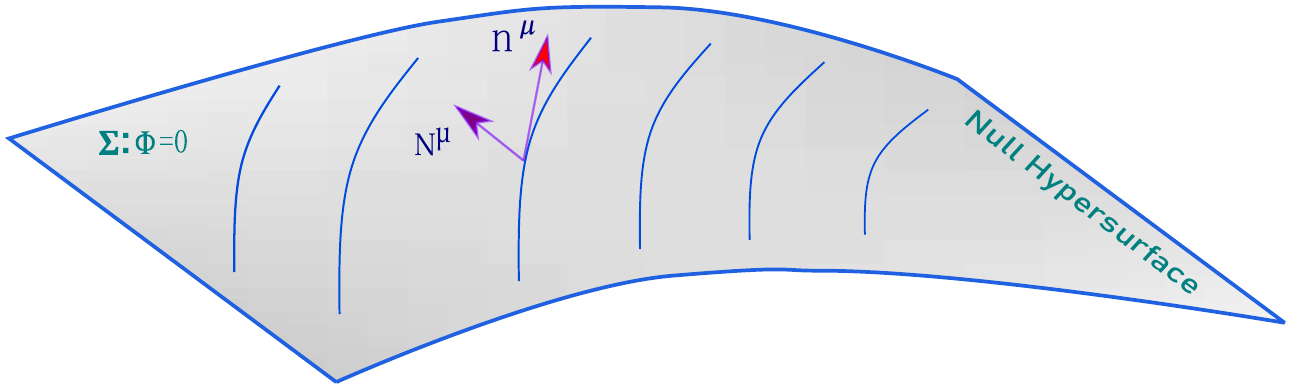} 
    \caption{A null hypersurface ($\Sigma$) defined by an equation $\Phi =0 $. $n^{\mu}$ is normal to the hypersurface which ultimately turns out to be a tangent to the $\Sigma$ itself. Hence $n^{\mu}$ is a generator of the null hypersurface. $N^{\mu}$ is a transverse or auxiliary normal which is not tangent to the null surface and satisfies $n\cdot N=-1$.}\label{stching2}
\end{figure}
As mentioned earlier for a null surface that the normal vector to the $\Sigma$ is also a tangent vector depicted in the Fig.(\ref{stching2}) so it does not deliver information about the null surface. In this respect, we must define an auxiliary normal vector $N^{\mu}$ distinct from the usual normal of the hypersurface $n^{\mu}$, and it is not tangent to $\Sigma$. This should satisfy certain normalization conditions given as, $n^{\mu}N_{\mu}=-1$, $n^{\mu}n_{\mu}=0$ and $N_{\mu}e^{\mu}_{A}=0$. Therefore, one can finally write down the inverse metric or the completeness relation for the metric in the following way, and can also be cross-verified by computing all possible inner products among $N^{\mu}$, $e^{\mu}_{A}$ and $n^{\mu}$,
\begin{align}
g^{\mu\nu} = -N^{\mu}n^{\nu} - N^{\nu}n^{\mu} + \sigma^{AB}e^{\mu}_{A}e^{\nu}_{B}.
\end{align}\label{e3}
This further requires two conditions to satisfy
\begin{align}
\delta^{a}_{b}-n^{a}N_{b} = g_{*}^{ac}g_{cb} \hspace{2mm} ; \hspace{2mm} n^{a}(N\cdot N)+g_{*}^{ab}N_{b} = 0 , 
\end{align}
where $g_{*}^{ab}$ is a pseudo inverse of the induced degenerate metric $g_{ab}$. One of the normalization conditions in general can be written as $N\cdot N=p$ with $p$ equals 0 and -1 for null and timelike cases respectively. The transverse vector on the hypersurface is written as: $N\cdot e_{(a)}\vert_{\Sigma}\equiv N_{a}$. 
We also consider that the induced metric is same on both sides of the hypersurface $\Sigma$, i.e., $[g_{\mu\nu}]e^{\mu}_{a}e^{\nu}_{b}=0$
\begin{align}\label{jncs}
[ g_{ab} ] = g_{ab}^{+} - g_{ab}^{-} = 0.
\end{align}  
Here the notation `[ ]' depicts a change or jump in a given quantity. For a quantity $Q$, $[Q] = Q(\mathcal{M}_{+})\vert_{\Sigma}-Q(\mathcal{M}_{-})\vert_{\Sigma}$. We also make sure that $[e^{\mu}_{a}]=0$ and transverse vector is same on both sides of the hypersurface, i.e., $[N_{\mu}]=0$; it ensures that the vector remains continuous across the surface. This is coordinate invariant condition which allows us to have a well-behaved geometry, and known as the \textit{first junction condition}.

Further, we establish a continuous coordinate system locally in the neighborhood of the $\Sigma$ which is distinct from the respective manifold's coordinates $x^{\mu}_{\pm}$ and cover both spacetimes $\mathcal{M}_{\pm}$. The benefit of this particular coordinate system is that it overlaps with both the manifolds and also simplifies our computation; it can be used to describe both sides of the $\Sigma$. This overlap $\mathcal{M}_{+}\cap \mathcal{M}_{-}$ should contain the hypersurface $\Sigma$. In this common coordinate system, we have
\begin{align}
[N^{\mu}] = [n^{\mu}] = [e^{\mu}_{a}] = 0. 
\end{align}
In addition to this, the Eq.(\ref{e2}) should also follow the continuity across $\Sigma$, i.e., $[\sigma_{AB}]=0$. 

Let us now assume that the hypersurface is given by $\Phi(x^{\mu}) = 0$ where $\Phi$ is a smooth function written in common coordinates $x^{\mu}$. Given a hypersurface, one can define a normal vector to be $n^{\mu}=g^{\mu\nu}\partial_{\nu}\Phi(x^{\mu})$. The glued metric along the $\Sigma$ is written as
\begin{align}\label{spc}
g_{\mu\nu}(x^{\mu}) = g^{+}_{\mu\nu}(x^{\mu}) \mathcal{H}(\Phi) + g^{-}_{\mu\nu}(x^{\mu}) \mathcal{H}(-\Phi),
\end{align}
where $\mathcal{H}(\Phi)$ is a Heaviside step function. This particular Eq.(\ref{spc}) is implying the phrase \textit{soldering} or \textit{stitching} or \textit{gluing of two spacetimes} $\mathcal{M}_{+}$ and $\mathcal{M}_{-}$ across a common null hypersurface $\Sigma$. We consider that the metric in this common coordinate system is continuous across both sides of the hypersurface $\Sigma$, i.e., $[g_{\mu\nu}] = 0$. With this form of the metric, we try to investigate the behavior of Einstein field equation. We start from the Christoffel symbol which gets decomposed in two separate connection coefficients for the respective manifolds. The glued Christoffel symbol takes the following form
\begin{align}\label{chrst}
\Gamma^{\mu}_{\rho\nu} = \frac{1}{2}g^{\mu\sigma}(g_{\rho\sigma,\nu}+g_{\nu\sigma,\rho}-g_{\rho\nu,\sigma})
					   = \Gamma^{+\rho}_{\rho\nu} \mathcal{H}(\Phi)+\Gamma^{-\rho}_{\rho\nu}\mathcal{H}(-\Phi).
\end{align}
Where the derivative of the joint metric can be written as
\begin{align}\label{dm}
\partial_{\rho}g_{\mu\nu} = \partial_{\rho}g^{+}_{\mu\nu}\mathcal{H}(\Phi)+\partial_{\rho}g^{-}_{\mu\nu}\mathcal{H}(-\Phi)+[g_{\mu\nu}](\partial_{\rho}\Phi)\delta(\Phi).
\end{align}
Given the derivative of the metric, we can write down the Christoffel symbols. $\Gamma^{\pm\rho}_{\mu\nu}$ denotes the ordinary connections for $\mathcal{M}_{+}$ and $\mathcal{M}_{-}$ respectively. This will be carried by Riemann tensor expression which contains a singular term. The final expression of the same is given by
\begin{align}\label{rmn}
R^{\mu}_{\nu\lambda\rho} = R^{+\mu}{}_{\nu\lambda\rho}+R^{-\mu}{}_{\nu\lambda\rho}+Q^{\mu}{}_{\nu\lambda\rho}\delta(\Phi),
\end{align}
where the non-vanishing last term is, $Q^{\mu}{}_{\nu\lambda\rho}=-\Big([\Gamma^{\mu}{}_{\nu\rho}]n_{\lambda}-[\Gamma^{\mu}{}_{\nu\lambda}]n_{\rho}\Big)$; it is nothing but the Riemann tensor associated with the hypersurface. It is proportional to Dirac delta distribution function whose analysis plays a crucial role in understanding the null shell and implicates the generation of impulsive gravitational waves. Further computation leads us to have a singular term in Einstein field equation. In general, we can write
\begin{align}\label{soldeinst}
G_{\mu\nu} = G^{+}_{\mu\nu}\mathcal{H}(\Phi)+G^{-}_{\mu\nu}\mathcal{H}(-\Phi)+G_{\mu\nu}\delta(\Phi).
\end{align}
In order to satisfy the Einstein field equation, for non-zero stress-energy tensor, the stress tensor will take the following form,
\begin{align}\label{sett}
T_{\mu\nu} = T^{+}_{\mu\nu}\mathcal{H}(\Phi)+T^{-}_{\mu\nu}\mathcal{H}(\Phi)+S_{\mu\nu}\delta(\Phi),
\end{align}
where the last of the Eq.(\ref{sett}) is written as: $S_{\mu\nu}=\frac{1}{8\pi}\Big(Q_{\mu\nu}-\frac{1}{2}Qg_{\mu\nu}\Big)$. This particular expression of $S_{\mu\nu}$ contribute from the singular part of the Einstein field equation which is proportional to $\delta(\Phi)$. The last term of the Eq.(\ref{sett}) can be thought of as a thin distribution of stress-energy. In general, this can carry impulsive gravitational waves or null matter or mixture of both. In the next subsection, we shall look at the decomposition of IGWs and matter part which can further be projected on the null surface, given by $\Sigma$ 
\begin{align}
S_{ab} = S_{\mu\nu}e^{\mu}_{a}e^{\nu}_{b}.
\end{align}
It is to note here that the condition $[g_{\mu\nu}]=0$ ensures that the tangential derivative of the metric is also continuous, i.e., $n^{\alpha}[\partial_{\alpha}g_{\mu\nu}]=0$. The possible discontinuity will be contained in the transverse derivative of the metric ($N^{\alpha}\partial_{\alpha}g_{\mu\nu}$). Hence, we consider a tensor field $\gamma_{\mu\nu}$ such that 
\begin{align}\label{gkappa}
[\partial_{\alpha}g_{\mu\nu}] = -\gamma_{\mu\nu}n_{\alpha}
\end{align}
One can write the singular term of the Eq.(\ref{rmn}) in terms of $\gamma_{\mu\nu}$ tensor field:
\begin{align}
Q^{\mu}{}_{\nu\lambda\rho} = \frac{1}{2}\Big(\lambda^{\mu}_{\rho}n_{\nu}n_{\lambda}+\gamma_{\nu\lambda}n^{\mu}n_{\rho}-\gamma_{\nu\rho}n^{\mu}n_{\lambda}-\gamma^{\mu}_{\lambda}n_{\nu}n_{\rho}\Big)
\end{align}
The stress-energy tensor becomes
\begin{align}\label{sttn}
S^{\mu\nu} = \frac{1}{16\pi}\Big(n^{\mu}\gamma^{\nu}_{\rho}n^{\rho}+n^{\nu}\gamma^{\mu}_{\rho}n^{\rho}-\gamma^{\rho}_{\rho}n^{\mu}n^{\nu}-\gamma_{\rho\lambda}n^{\rho}n^{\lambda}g^{\mu\nu}\Big)
\end{align}
The expression (\ref{sttn}) gets more simplified when we write it in the basis $(n^{\mu}, e^{\mu}_{A}, N^{\mu})$.  Therefore, we define following projections
\begin{align}
\gamma_{A}\equiv \gamma_{\mu\nu}e^{\mu}_{A}n^{\nu} \hspace{2mm} ; \hspace{2mm} \gamma_{AB} \equiv \gamma_{\mu\nu} e^{\mu}_{A}e^{\nu}_{B}.
\end{align} 
Under these projections, the stress tensor takes the following form
\begin{align}\label{smunu}
S^{\mu\nu} = \mu n^{\mu}n^{\nu}+j^{A}(n^{\mu}e^{\nu}_{A}+e^{\mu}_{A}n^{\nu})+p\sigma^{AB}e^{\mu}_{A}e^{\nu}_{B}.
\end{align}
where $\mu$, $j^{A}$ and $p$ are the intrinsic quantities of the shell termed as surface energy density, surface current and isotropic surface pressure. 
\begin{align}
\mu \equiv -\frac{1}{16\pi} \sigma^{AB}\gamma_{AB} \hspace{2mm} ; \hspace{2mm} j^{A}\equiv \frac{1}{16\pi} \sigma^{AB}\gamma_{B} 
\hspace{2mm} ; \hspace{2mm} p \equiv -\frac{1}{16\pi}\gamma_{\mu\nu}n^{\mu}n^{\nu}.
\end{align}
We may adopt a slightly different notation in order to express the Eq.(\ref{smunu}) as a quantity to the hypersurface. We consider following definitions,
\begin{align}
\gamma^{\dagger} = \gamma_{ab}n^{a}n^{b} = \gamma_{a}n^{a} \hspace{2mm} ; \hspace{2mm} \gamma_{a} = \gamma_{ab}n^{b} \hspace{2mm} ; \hspace{2mm} \gamma^{a} = g_{*}^{ab}\gamma_{b} 
\hspace{2mm} ; \hspace{2mm} \gamma^{*} = g_{*}^{ab}\gamma_{ab} = \sigma^{AB}\gamma_{AB}.
\end{align}
where $g_{*}^{ab}=\delta^{a}_{A}\delta^{b}_{B}g^{AB}$; it is inverse of the spatial part of the hypersurface metric. The intrinsic stress tensor of the shell is given by
\begin{align}\label{inst}
S^{ab} = \frac{1}{16\pi}\Big(\mu n^{a}n^{b}+j^{a}n^{b}+j^{b}n^{a}+pg_{*}^{ab}\Big).
\end{align}
with
\begin{align}\label{stn1}
\mu = -\frac{1}{16\pi}\gamma_{ab}g_{*}^{ab} \hspace{2mm} ; \hspace{2mm} j^{a} = \frac{1}{16\pi}g_{*}^{ac}\gamma_{c} \hspace{2mm} ; \hspace{2mm} p = -\frac{1}{16\pi}\gamma^{\dagger}.
\end{align}
This is an alternative representation of the intrinsic quantities to the null surface. We can further simplify the expression (\ref{smunu}) and (\ref{inst}) and can make the equation independent of coordinates $x^{\mu}$. Therefore, we can remove the need to have the dependence of common coordinate system in Eq.(\ref{smunu}) and Eq.(\ref{inst}). As a result, this can be expressed in any coordinates.

Since our discussion is purely based on null hypersurfaces and null shells so we nowhere mentioned any details regarding timelike and spacelike shells. However, the analyses for timelike and spacelike shells has explicitly been carried out in the literature \cite{PhysRevD.43.1129, Poisson2002ARO, Poisson:2009pwt}. In that case the important ingredients were the continuity of induced metric $g_{ab}$ and discontinuity in extrinsic curvature $[\mathcal{K}_{ab}]$. Our goal is to provide a similar study for null shells by removing continuous coordinates for computing intrinsic properties. Our expectation is that the intrinsic quantities would contain the extrinsic curvature similar to timelike and spacelike shells. We are also aware of the major issue involved in expressing extrinsic curvature $\mathcal{K}_{ab}$ that the normal vector $n^{\mu}$ becomes the tangent to the null hypersurface $\Sigma$. So the definition $\mathcal{K}_{ab}=\frac{1}{2}(\mathcal{L}_{n}g_{\mu\nu})e^{\mu}_{a}e^{\nu}_{b}$ no longer holds for null shells. This object has no transverse properties; however, it is valid for timelike and spacelike hypersurfaces because the normal $n^{\mu}$ has correct meaning of transverse which is pointing away from the surface. The ultimate solution to this problem is to introduce a transverse curvature which carries the transverse derivative of the metric. Hence, the modified definition takes the following form
\begin{align}
\mathcal{K}_{ab} = \frac{1}{2}\mathcal{L}_{N}g_{\mu\nu}e^{\mu}_{a}e^{\nu}_{b} = -N_{\mu}e^{\mu}_{a;\nu}e^{\nu}_{b}.
\end{align}
This equation assumes that the $N_{\mu}e^{\mu}_{a}$ is a constant and also uses one identity $e^{\nu}_{b}\nabla_{\nu}e^{\mu}_{a}=e^{\nu}_{a}\nabla_{\nu}e^{\mu}_{b}$. This property ensures that the tensor $\mathcal{K}_{ab}$ is a symmetric tensor. The jump in the extrinsic curvature can be written as \begin{align}\label{extcur}
[\mathcal{K}_{ab}] = [N_{\mu;\nu}]e^{\mu}_{a}e^{\nu}_{b} = \frac{1}{2}\gamma_{\mu\nu}e^{\mu}_{a}e^{\nu}_{b}.
\end{align}
Using Eq.(\ref{gkappa}), one can further write
\begin{align}\label{gmaab}
\gamma_{ab} = N^{\mu}[\partial_{\mu}g_{ab}] = 2[\mathcal{K}_{ab}].
\end{align}
Therefore, the intrinsic quantities can be written in the following form,
\begin{align}\label{intrinsic}
\mu = -\frac{1}{8\pi}\sigma^{AB}[\mathcal{K}_{AB}] \hspace{2mm} ; \hspace{2mm} j^{A} = \frac{1}{8\pi}\sigma^{AB}[\mathcal{K}_{\lambda B}] \hspace{2mm} ; \hspace{2mm}
p = -\frac{1}{8\pi}[\mathcal{K}_{\lambda\lambda}].
\end{align}
This is known as the \textit{transverse curvatures algorithm} for determining the intrinsic properties of a given null shell. The approach reduces the algebra quite significantly. The expressions of the null shell quantities depend on the induced metric $\sigma^{AB}$ and jump in the extrinsic curvature. Here, we have taken the hypersurface coordinates $(\lambda, x^{A})$. This completes our understanding of null shell formalism in terms of estimating intrinsic properties. One can further show the conservation properties of a null shell. However, we are not going into that detail. The explicit computation for the same can be referred from the study of Barrab$\grave{e}$s and Israel \citep{PhysRevD.43.1129}. In an abbreviation, the conservation equation depends on the arithmetic mean of connection coefficients and extrinsic curvatures. The relevant expression under the absence of matter source is given by
\begin{align}
N^{a}(\partial_{b}+\tilde{\Gamma}_{b})S^{ab}-S^{ab}\tilde{\mathcal{K}}_{ab} = 0,
\end{align}
where 
\begin{align}
\tilde{\Gamma}_{b} = \frac{1}{2}(\Gamma^{+}_{b}+\Gamma^{-}_{b}) \hspace{2mm} ; \hspace{2mm} \tilde{\mathcal{K}}_{ab} = \frac{1}{2}(\mathcal{K}^{+}_{ab}+\mathcal{K}^{-}_{ab}).
\end{align}
$\tilde{\Gamma}_{b}$ and $\tilde{\mathcal{K}}_{ab}$ indicate the arithmetic mean of $\Gamma^{\pm\alpha}{}_{\alpha b}$ and $\mathcal{K}^{\pm}_{ab}$ respectively.
Here $\Gamma^{\pm}_{b}$ is nothing but $\Gamma^{\pm\alpha}{}_{\alpha b}$. So far we have analyzed the stress-tensor of a null shell and interpreted the intrinsic quantities of the shell. Next, we shall discuss how we can study the gravitational wave part of the impulsive lightlike signal. 

An alternative way to compute the intrinsic quantities is to consider the off-shell extension of the soldering transformations which would give rise the same result as appears from the transverse curvatures algorithm. It is a transformation of soldering coordinates off the horizon shell locally in the vicinity of the horizon. The intrinsic quantities will be estimated using off-shell extended metric via Eq.(\ref{stn1}). The details of these approaches can be found in \cite{Hogan2003, Blau2016, PhysRevD.98.104009, PhysRevD.43.1129}. We have used both the techniques for the analysis which can be seen in the Chapter (3) and Chapter (4).

\subsection{Decomposition of null matter and IGW}

The singular component of the stress-energy tensor was formerly interpreted as a thin distribution of null matter, e.g. neutrino fluid, or impulsive gravitational wave or a combination of both. In this subsection, we provide a brief study on how we decompose matter part and impulsive gravitational waves. In order to examine gravitational wave degree of freedom, one needs to analyze Weyl tensor components of the spacetime $\mathcal{M}=\mathcal{M}_{+}\cup \mathcal{M}_{-}$. As we already have a singular term in the Riemann tensor, hence the Weyl tensor also exhibits a term proportional to delta distribution function. 
\begin{align}
C_{\kappa\lambda\mu\nu} = \tilde{C}_{\kappa\lambda\mu\nu}+\hat{C}_{\kappa\lambda\mu\nu}\delta(\Phi),
\end{align}
where the tilde part of the Weyl tensor is sum of the individual Weyl tensors of both the manifolds while the singular term is given by

\begin{align}
\hat{C}_{\kappa\lambda\mu\nu} = 2n_{[\mu}\gamma_{\nu][\kappa}n_{\lambda]}-8\pi\lbrace S_{\kappa[\mu}g_{\nu]\lambda}-S_{\lambda[\mu}g_{\nu]\kappa}\rbrace+\frac{16\pi}{3}S^{\beta}_{\beta}g_{\kappa[\mu}g_{\nu]\lambda}.
\end{align}
The analysis for intrinsic expression of stress tensor ensures that there is a part of $\gamma_{ab}$ which does not contribute to stress-energy tensor. We denote this non-contributing part as $\hat{\gamma}_{ab}$. In general, one can write down $\gamma_{ab}$ containing both matter and gravitational wave parts.
\begin{align}
\gamma_{ab} = \hat{\gamma}_{ab}+\bar{\gamma}_{ab}
\end{align} 
where $\hat{\gamma}_{ab}$ is the gravitational wave degree of freedom and $\bar{\gamma}_{ab}$ is the null matter part.

For any timelike or spacelike shells, there are six independent components of energy-momentum tensor whereas for null shell we have four independent components of energy-momentum tensor. However, $\gamma_{ab}$ has six independent components. As a result, one can characterize the transverse-traceless components of $\gamma_{ab}$ denoted as $\hat{\gamma}_{ab}$:
\begin{align}\label{TT}
\hat{\gamma}_{ab}n^{b} = 0 \hspace{2mm} ; \hspace{2mm} g_{*}^{ab}\hat{\gamma}_{ab} = 0.
\end{align} 
$\hat{\gamma}_{ab}$ does not have any impact on the shell's matter content. This reduces to have only two independent components of $\hat{\gamma}_{ab}$. In particular, the gravitational wave degree of freedom with extracted $\gamma_{ab}$ can be written as
\begin{align}\label{gwdf}
\hat{\gamma}_{ab} = \gamma_{ab}-\frac{1}{2}\gamma^{*}g_{ab}+2\gamma_{(a}N_{b)}+\Big(N_{a}N_{b}-\frac{1}{2}N\cdot Ng_{ab}\Big)\gamma^{\dagger}.
\end{align} 
Using this, we can easily check conditions (\ref{TT}) are satisfied. This in general represents two degrees of freedom of polarization present in an impulsive gravitational wave signal. The matter part of $\gamma_{ab}$ is given by
\begin{align}
\bar{\gamma}_{ab} = 16\pi\Big(g_{ac}S^{cd}N_{d}N_{b}+g_{bc}S^{cd}N_{d}N_{a}-\frac{1}{2}g_{cd}S^{cd}N_{a}N_{b}-\frac{1}{2}g_{ab}S^{cd}N_{c}N_{d}\Big).
\end{align}
Therefore, whenever $\bar{\gamma}_{ab}$ vanishes, we obtain pure impulsive gravitational wave $\hat{\gamma}_{ab}$. If it is non-vanishing, it would contain null matter part together with IGWs. The further details related to $\hat{\gamma}_{ab}$, $\bar{\gamma}_{ab}$ and $S^{ab}$ will be discussed in the next two chapter. This sets our understanding on how one can decompose impulsive lighlike signals into gravitational wave degree of freedom and null matter part. Let us now focus on how asymptotic symmetries arise in the null shell formalism.

\section{Soldering freedom \& BMS transformations}\label{solrtrnssuptr}

We shall turn our discussion on emergence of asymptotic symmetries of spacetimes for null shells when two spacetimes are being glued along a common null hyerpsurface. It is to note here that the analysis for flat spacetimes has already been carried out in \cite{PhysRevD.99.024031}. Hence, our discussion will be oriented towards the asymptotic symmetries on black hole horizon shells. It is already known and also discussed in Chapter(I) that the asymptotic symmetries were originally discovered by Bondi-van der Burg-Metzner-Sachs (BMS) in early sixties and recently it has gained considerable attention due to its possible connection with gravitational memory and soft theorem. 
Our goal in this section is to provide a study on possible soldering freedoms in patching of two spacetimes along a common null surface. This is nothing but a freedom of choosing intrinsic coordinates on the hypersurface. This is usually termed as \textit{soldering freedom}. These soldering freedoms turn out to be the BMS-like transformations comprising \textit{supertranslation} and \textit{superrotation}, collectively called as \textit{BMS-transformations} \cite{Blau2016, PhysRevD.98.104009}.

We start with a thorough examination of the circumstances that allow for non-trivial soldering transformations. As a result of this study, the soldering group that results is infinite-dimensional. 
We can construct an infinity number of physically unique or distinct shells on this hypersurface; we shall refer to them all together as \textit{horizon shells}. Each of the shells will be parameterized by an arbitrary function on the horizon which is nothing but turns out to be a BMS-like transformation. Every shell in general carries stress-energy tensor comprising surface energy density, current and pressure together with impulsive gravitational waves residing on the horizon shell. Let us discuss the emergence of BMS-like symmetries in this context.

\subsection{Soldering transformation analysis}\label{BMSsymmsuptr}

Here, we shall be providing the detailed analysis for obtaining the freedom in the choice of intrinsic coordinates along the null direction. This soldering freedom of stitching the two spacetimes along a common null surface provides us BMS-like transformations on the horizon shell. It emerges as a coordinate transformation which preserves the induced metric on the hypersurface. However, one needs to keep in mind that the analysis is consistent with the junction condition (\ref{jncs}). Therefore, the first question is to ask about the permissible coordinate transformations on both sides of the shell while maintaining the junction condition. This implies us to figuring out the Killing vectors of the hypersurface metric in an appropriate coordinate system. Therefore, our computation is to deal with the Lie derivative of the induced metric along the Killing direction (say $Z^{a}\partial_{a}$ with components $Z^{a}$) or Killing equation.
\begin{figure}[h!]\centering
    \includegraphics[height=4.2cm, width=10.5cm]{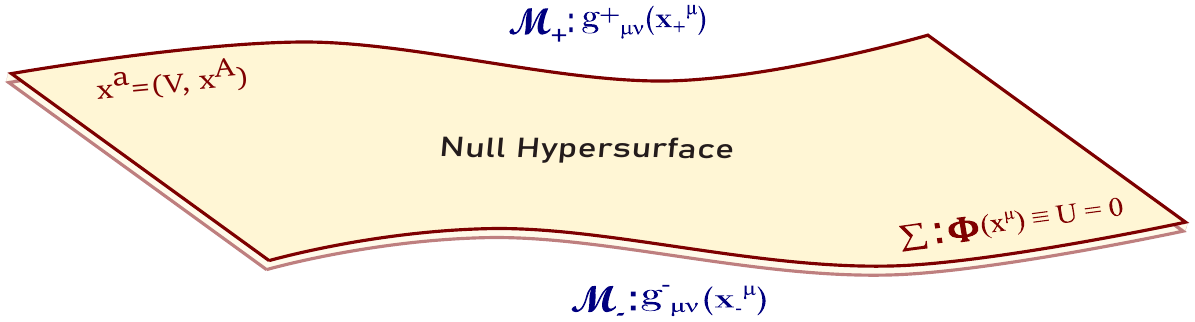} 
    \caption{A horizon shell with a thin surface layer of null matter and impulsive light-like signal which appear as a result of gluing two spacetimes. The geometry is described in Kruskal coordinate system with a hypersurface defined as $\Phi$.}\label{IGW}
\end{figure}
We consider Kruskal coordinates $(U, V, x^{A})$ with coordinates $(V, x^{A})$ on the null hypersurface $\Sigma$ as can be seen in the Fig.(\ref{IGW}). $V$ is the responsible parameter along the hypersurface for generating null congruence. One can also use Eddington-Finkelstein (EF) coordinates to carry out the analysis. We do not consider EF coordinates particularly here because Kruskal coordinates are suitable choice for performing memory computation as the horizon is positioned at $U=0$. However, in some cases for the computation of intrinsic properties, we shall also be looking at EF coordinates. We shall of course mention about this wherever it is required.

Since the analysis is based on finding the Killing vectors as a metric preserving coordinate transformations, let us start with the Killing equation,
\begin{align}\label{killing}
\mathcal{L}_{Z}g_{ab} = 0,
\end{align}
where $\mathcal{L}_{Z}$ denotes the Lie derivative along the Killing direction $Z$. The vector field on the null hypersurface $\Sigma$ is given by
\begin{align}
Z = Z^{a}\partial_{a} = Z^{V}\partial_{V} + Z^{A}\partial_{A}.
\end{align}
One simply use the standard formula of the Lie derivative of $2$-rank covariant tensor in order to expand it.
\begin{align}\label{kexp}
\mathcal{L}_{Z}g_{ab} = 0 \hspace{0.3cm} \Longrightarrow \hspace{0.3cm} Z^{c}\partial_{c}g_{ab} + (\partial_{a}Z^{c})g_{c b}+(\partial_{b}Z^{c})g_{ac}=0.
\end{align}
This consideration will also hold for timelike or spacelike shells because it is just a Killing equation. The ambiguity gets involved for a null shell with its degenerate metric with metric components, $g_{aV}=0$, and non-degenerate $g_{AB}$. Apparently, the scenario has the potential to get more interesting. The Eq.(\ref{kexp}) takes the following form
\begin{align}Z^{c}\partial_{c}g_{ab}+(\partial_{a}Z^{A})g_{Ab}+(\partial_{b}Z^{A})g_{aA} = 0.
\end{align}
Since we are taking metric with $g_{aV}=0$ and performing Lie derivative on $g_{aV}$, as a result we obtain $V$-independent transformations of $Z^{A}$. 
\begin{align}\label{kilzc}
\mathcal{L}_{Z}g_{aV} = 0 \hspace{0.3cm} \Longrightarrow \hspace{0.3cm} \partial_{V}Z^{A} = 0.
\end{align}
Further if we consider the Lie derivative of spatial coordinates, it implies
\begin{align}\label{kilspa}
\mathcal{L}_{Z}g_{AB} = 0 \hspace{0.3cm} \Longrightarrow \hspace{0.3cm} Z^{V}\partial_{V}g_{AB}+Z^{C}\partial_{C}g_{AB}+(\partial_{A}Z^{C})g_{CB}+(\partial_{B}Z^{C})g_{AC} = 0.
\end{align}
Now we shall try to examine the special cases when the metric $g_{AB}$ has $V$ dependence and and independence. We expect to get some non-trivial soldering transformations \cite{Blau2016, Nutku}, known as BMS-like transformations. Let us look at the details section-wise below.

\subsubsection{Emergence of Supertranslation}\label{sup}

The first special case is when metric does not depend on $V$ parameter. This induces a new type of translation which has angle dependent notion, termed as \textit{Supertranslation}. It is similar to the one obtained at asymptotic null infinity for asymptotically flat spacetimes. To see this, let us start inspecting the Eq.(\ref{kilspa}). The first term of the equation generates a coordinate freedom along $Z$-direction when the metric $g_{AB}$ is independent of $V$ parameter. Hence we are free to choose transformation along $V$ as an arbitrary function of $F(V, x^{A})$.  
\begin{align}\label{sup1n}
\partial_{V}g_{AB} = 0 \hspace{0.3cm} \Longrightarrow \hspace{0.3cm} Z^{V} = F(V, x^{A})
\end{align}
The effect of this isometry transformation on the normal vector can be investigated for the Killing horizon. The direction of the normal to the shell is preserved through isometry transformation which implies that Lie transportation of normal should be proportional to itself. We would resume the investigation for normal in a moment. However, Eq.(\ref{sup1n}) shows that the $Z^{V}$ remains unconstrained and provides us a freedom to apply arbitrary coordinate transformation on $V$. The allowed coordinate transformation as a soldering freedom $V \rightarrow F(V, x^{A})$ can be written as
\begin{align}
(V, x^{A}) \hspace{0.3cm} \longrightarrow \hspace{0.3cm} \Big((F(V, x^{A}), x^{A}\Big)
\end{align}
Because the coordinate transformation includes an arbitrary function $F(V, x^{A})$, it indicates the formation of infinite dimensional soldering (\textit{SOL}) or isometry group $SOL(\Sigma)$ which is a group of coordinate transformations.
\begin{align}
SOL (\Sigma, g_{AB}) = \lbrace V \longrightarrow F(V, x^{A})\rbrace.
\end{align}
Physically it means that one can solder two black hole spacetimes in an infinite number of ways along a common null hypersurface. The only required condition is to maintain the junction condition through out the analysis. 

Let us examine the impact of the $Z$-generated transformations on the null normal $n^{a}$ of $\Sigma$ for the Killing horizon. We know that for $\partial_{V}g_{ab}$ the metric is translation invariant along the null generator $n^{a}$ of $\Sigma$. This is equivalent to say: $n^{c}\partial_{c}g_{ab} \hspace{0.15cm} \Longleftrightarrow \hspace{0.15cm} \partial_{V}g_{ab} = 0$. Thus, we can write
\begin{align}
g_{ab}n^{b} = 0 \hspace{0.2cm} \xrightarrow{\text{Operating Lie derivative}} \hspace{0.2cm} \mathcal{L}_{Z}n^{a}\sim n^{a}
\end{align}
Alternatively, this is clear from a given normal $n=\partial_{V}$ and using Eq.(\ref{kilzc}). 
\begin{align}\label{lzn1}
\mathcal{L}_{Z}n^{a} =& -\partial_{V}Z^{a} = -(\partial_{V}Z^{a})\delta^{a}_{V} \hspace{0.3cm} \Longrightarrow \hspace{0.3cm} \mathcal{L}_{Z}n^{a} = -(\partial_{V}Z^{V})n^{A},
\end{align}
where $\delta^{a}_{V}$ denotes component of null normal, i.e., $n^{a}=\delta^{a}_{V}$. Now we would impose a further restriction on soldering transformations which preserve the null normal. For horizon shells, we have $\mathcal{L}_{Z}n^{a}=0$. This implies
\begin{align}\label{lzn2}
\partial_{V}Z^{V} = 0 \hspace{0.3cm} \Longrightarrow \hspace{0.3cm} V \hspace{0.2cm} \longrightarrow \hspace{0.2cm} V+F(x^{A}) \hspace{0.2cm} : \hspace{0.2cm} BMS \hspace{0.15cm} Supertranslation
\end{align} 
This is instantly indentified as a supertranslation in the literature. For $x^{A}$ with $x^{A}=(\theta, \phi)$, it is interpreted as an angle dependent translation, hence named \textit{supertranslations}-like transformation or supertranslation-like symmetries. The soldering group that keeps this structure preserved is still infinite dimensional. We shall use these metric preserving transformations while implementing the intrinsic formulation of null shell where we have a seed metric and perform soldering transformation on other side of the null hypersurface $\Sigma$, and finally stitch them along a common null hypersurface $\Sigma$. The intrinsic quantities will carry the supertranslation parameter. So this is all about the emergence of supertranslation-like symmetries as a soldering freedom which would help us in examining intrinsic properties of null shells. Next, we shall discuss about the appearance of extended asymptotic symmetry- superrotation which arises as a result of $\partial_{V}g_{AB}\neq 0$. 

\subsubsection{Emergence of Superrotation}\label{suprr}

A new type of symmetry labeled as \textit{superrotation} has just been discovered in an extended form of BMS symmetries near the horizon of black holes which mimics the one obtained at asymptotic null infinity \cite{PhysRevLett.105.111103, PhysRevLett.116.091101, Blau2016, PhysRevD.98.104009}. It is a local conformal transformation of the spatial slice of the metric, or local conformal transformation of celestial sphere at null infinity \cite{strominger2018lectures, Barnich:2011ct}. Let us now turn our discussion towards the another possibility where metric depends on the $V$ parameter, i.e., $\partial_{V}g_{AB}\neq 0$. As a result, one can obtain extended BMS symmetries known as \textit{superrotation}-like symmetries. The analysis begins with the Eq.(\ref{kilspa}) in search of possible non-trivial soldering freedoms. Allowing the conformal transformations to the spatial sector of the metric such that
\begin{align}
Z^{C}\partial_{C}g_{AB}+(\partial_{A}Z^{C})g_{CB}+(\partial_{B}Z^{C})g_{CA} = \Omega(x^{A})g_{AB},
\end{align} 
where $\Omega(x^{A})$ is the conformal factor and it is independent of $V$. It can generally be a function of spatial coordinates $x^{A}$ and will be constrained by above equation. If one performs the conformal transformation in spatial coordinates represented in complex coordinates via $z \longrightarrow f(z)$ and $\bar{z} \longrightarrow f(\bar{z})$ such that the Eq.(\ref{kilspa}) can be written in the following way, 
\begin{align}\label{suprreq}
Z^{V}\partial_{V}g_{AB} + \Omega(x^{A})g_{AB} = 0,
\end{align}
where $x^{A}=(z, \bar{z})$. We have the equation that we may seek for a feasible solution which will appear as a soldering freedom. Since the first term of the Eq.(\ref{suprreq}) involves single derivative of the metric $g_{AB}$ with respect to $V$ coordinate and our assumption is that the spatial slice depends on $V$ parameter, hence, we shall consider a solution which is at least quadratic in $V$. This will serve our purpose of examining extended BMS symmetries. One may obviously think that this particular solution is non-generic, and it is also true in principle. As a initial guess to the Eq.(\ref{suprreq}), we investigate our analysis with a solution having $V^{2}$ factor with the spatial slice in order to ensure the existence of the first term. One can also take any regular function of $V$ instead of $V^{2}$ which would be the most general case discussed in \cite{PhysRevD.98.104009}. However, keeping in mind the conformal transformations to the spatial slice of the metric, we also perform a compensating transformation along $V$ direction so that the analysis is consistent with the junction condition and also satisfies Eq.(\ref{kilspa}). Let us consider the following solution of the Eq.(\ref{suprreq}),
\begin{align}\label{solkill}
g_{AB} = V^{2}\tilde{g}_{AB},
\end{align}  
where $\tilde{g}_{AB}$ exhibits the dependence of spatial coordinates $x^{A}$. Put the solution into Eq.(\ref{suprreq}),
\begin{align}
2VZ^{V}\tilde{g}_{AB}+\Omega(x^{A})V^{2}\tilde{g}_{AB} = 0
\end{align}
This provides us a freedom along $V$ direction as a compensation of conformal transformation of spatial slice of the metric. The $Z^{V}$ component is given by
\begin{align}
Z^{V} = -\frac{V \Omega(x^{A})}{2}
\end{align} 
Therefore, our consideration $\partial_{V}g_{AB}\neq 0 \Longrightarrow r^{2}(U,V)\tilde{g}_{AB}(x^{A})$ gives a feasible solution to $Z^{V}$  which compensates the conformal transformation. This analysis gives rise metric preserving extended BMS transformations known as \textit{superrotation}-like symmetries. This completes our discussion on emergence of superrotation-like transformations near the horizon of black holes. Hence, as a result, we recover supertranslation and superrotation like symmetries mimicking the ones originally obtained at null infinity. This ends our investigation on the emergence of asymptotic symmetries appearing as a soldering freedom in the context of stitching of black hole spacetimes.

This concludes our theoretical framework of horizon shell formalism, and how supertranslation and superrotation-like near-horizon symmetries can emerge as a soldering freedom in the context of stitching of two black hole spacetimes. In the next two chapters, we shall implement the intrinsic formulation of horizon shell to black hole spacetimes- Schwarzschild and extreme Reissner Nordstr$\ddot{o}$m (ERN) black holes. We shall also try to understand the interaction of timelike and null geodesics with a horizon shell and what all measurable quantities can be extracted. 
We would examine these analyses for both the considerations, i.e., non-extreme (Schawarzschild) and extreme RN black holes, and try to address the difficulties involved in the process.


\chapter{Memory effect and BMS symmetries for non-extreme black holes}\label{CH3}

\ifpdf
    \graphicspath{{Chapter3/Figs/Raster/}{Chapter3/Figs/PDF/}{Chapter3/Figs/}}
\else
    \graphicspath{{Chapter3/Figs/Vector/}{Chapter3/Figs/}}
\fi

As we know that the study of null shells was initiated after the mid of the twentieth century \cite{Penrose:1972xrn, PhysRevD.43.1129, Israel1966, Synge:1957zz}, recently, the field has gained considerable interest in the context of gravitational memory effect \cite{PhysRevD.100.084010, Blau2016, PhysRevD.98.104009}. Let us briefly recap the idea of null shell. Null shells appear as impulsive lightlike signals and represent the mixture of impulsive gravitational waves (IGWs) and lightlike matter. Impulsive gravity waves are generated during violent astrophysical phenomena such as supernovae explosion, merger of heavy black holes etc. Therefore, it would be interesting to study the measurable effects of such signals on test detectors. There have been several studies of IGWs on test geodesics in different considerations \cite{1998gr.qc9054S, PhysRevD.100.024040, PhysRevD.96.064043, doi:10.1142/9789812776938_0007, Podolsk__1999, PODOLSKY19991, Ortaggio_2002, PhysRevD.67.064013} where authors take a distributional metric and investigate the effect of impulsive signals on test particles. Here, we adopt a local coordinate system in which the metric is continuous; however, its first derivative is discontinuous across the null hypersurface. As a result, this singular term or discontinuity appears in the stress-energy tensor and gives rise to the notion of null shells. When the null shell is being placed at the horizon of a black hole, it is termed as horizon shell. The study of memory effect for null shells has been carried out extensively for plane gravitational waves \cite{PhysRevD.96.064013, ZHANG2017743, Zhang_2018,  PhysRevD.99.024031}. Further, the study of plane gravitational waves and its relation with GW memory and soft theorem has been a well-studied area \cite{PhysRevD.96.064013}. Such analyses have been extensively examined near null infinity. Recently, there has been growing interest near the horizon of black holes as well since such effects and connection with BMS symmetries can also be recovered near the horizon of black holes which exactly mimics the one obtained at asymptotic null infinity \cite{Blau2016, PhysRevD.98.104009}. In this chapter, we focus on the emergence of asymptotic symmetries when two Schwarzschild black holes are soldered along a common null hypersurface. Our analyses extends the impact of plane waves on test geodesics to spherical waves, i.e., we consider spherically symmetric static black holes in order to investigate the imprints of asymptotic symmetries on test geodesics \cite{PhysRevD.100.084010}.

The basic construction of horizon shell formalism and ingredients to determine the appearance of asymptotic symmetries has already been discussed in chapter (\ref{CH2}). We adopt the basic formulation and perform the study on Schwarzschild black hole which is nothing but a non-extreme black hole as its surface gravity is non-vanishing. It can be considered as finite temperature configuration. Since, we already have the basic set up, let us start with the emergence of asymptotic symmetries in the context of gluing formalism of two Schwarzschild black holes along a common null surface.

\section{Emergence of BMS symmetries}
\subsection{Supertranslation-like freedom on Killing horizon:}

We take Kruskal form ($U, V, x^{A}$) of the Schwarzschild metric with the coordinates for the  horizon  $\Sigma$, $\zeta^{a}=\{V,x^{A}\},$ where $V$ is the parameter along the hypersurface generating null congruences.  In this coordinate, the normal to the surface becomes, 
$ \label{non1} n^{\alpha}=(\partial_{V})^{\alpha}$. The metric is written as, 
\be \label{smet}
ds^2= -2 G(r) dU dV +r^2(U,V)(d\theta^2+\sin^2(\theta) d\phi^2),
\ee
where 
\be \label{gshw}
G(r)=\frac{16\, m^3}{r} e^{-r/2m} \hspace{3mm} ; \hspace{3mm} U V= -\Big(\frac{r}{2m}-1\Big) e^{r/2m}.
\ee
The Kruskal coordinates ($U, V$) are related to Eddington-Finkelstein and Tortoise coordinates in the following way,
\begin{align}
    U = -e^{-u/4m} \hspace{2mm} ; \hspace{2mm} V = e^{v/4m} \hspace{5mm} and \hspace{5mm} u = t-r* \hspace{2mm} ; \hspace{2mm} v = t+r*,
\end{align}
with Tortoise coordinate, $r* = r+2m\hspace{1mm}ln\Big\vert \frac{r}{2m}-1\Big\vert$; where $u$ and $v$ are the retarded and advanced Eddington-Finkelstein coordinates respectively. The null shell is located at Killing horizon $U=0.$  It follows from the Eq.(\ref{kilspa}) that $Z^V=F(V,\theta,\phi)$. Further, from Eq.(\ref{lzn1}) and Eq.(\ref{lzn2}),
\be \label{killing3}
\partial_{V} Z^{V}=0.
\ee
This will further restrict the form of $Z^V$ as
$
Z^{V}= T(\theta,\phi).
$
Clearly, $Z$ generates the following symmetry, \be \label{trans}
V\rightarrow  V+ T (\theta,\phi).
\ee
This is strikingly similar to the supertranslation found at null infinity of asymptotically flat spacetimes \cite{Bondi,Sachs:1962zza}. This construction has been extended for rotating black holes in \cite{PhysRevD.98.104009}.

\subsection{Superrotation-like soldering freedom}

Superrotation-like transformations can also be recovered  when a null surface situated just a little away from the horizon of a black hole \cite{PhysRevD.98.104009}. Using complex or stereographic coordinates the 2-dimensional spatial slice of Schwarzschild black hole becomes $r^2 \frac{d\zeta d\bar \zeta}{(1+\zeta\,\bar \zeta)^2}$, with 
\begin{align}
r^2(U, V)= 4m^2-\frac{8 m^2}{e}U\, V+ \mathcal{O} (U\, V)^2,
\end{align}
where $e$ appears from the exponential form of the Eq.(\ref{gshw}), known as Euler's number. A null shell is located at $U=\epsilon$ and we notice that $\partial_{V}g_{AB}\neq 0$. We obtain soldering transformations,
\begin{align}
\begin{split} \label{trans2}
V\rightarrow V(1-\tilde \Omega(\zeta,\bar \zeta) )-\frac{a\,\tilde \Omega(\zeta,\bar \zeta) }{b \,\epsilon} +\mathcal{O}(\epsilon) \hspace{2mm} ; \hspace{2mm}
\zeta\rightarrow \zeta+ h(\zeta) \hspace{2mm} ; \hspace{2mm} \bar \zeta \rightarrow \bar \zeta+ \bar h (\zeta).
\end{split}
\end{align}

Hence, when we glue a supertranslated or superrotated metric with a seed metric, these BMS-like symmetries emerge in the shell's intrinsic quantities. This will be discussed in detail in the next section. 

\section{Off-shell extension of soldering freedom and Intrinsic properties of horizon shell}
Here, we provide necessary expressions useful for computation of stress tensor (\ref{inst}) on null shell. 
The Scwarzschild Killing horizon case has been presented in \cite{Blau2016}, we briefly review that here.  Let us consider the case of Schwarzschild black hole mentioned in (\ref{smet}). The null shell is located at $U=0.$ On the $+$ side, we can perform the following transformations,
\be
V_{+}= F(V, \theta,\phi) \hspace{2mm} ; \hspace{2mm} \theta_{+}=\theta \hspace{2mm} ; \hspace{2mm} \phi= \phi_{+}.
\label{on-shell}\ee
We choose the intrinsic coordinates $x^{\mu}$ to be coordinates on the $-$ side. Introducing the following ansatz \cite{Blau2016},
\begin{align}
\begin{split} \label{kerr4}
V_{+}&=F(V,\theta,  \phi)+U A(V,\theta,\phi) \hspace{2mm} ; \hspace{2mm} U_{+}=U C(V,\theta, \phi),\\& \theta_{+}=\theta+U B^{\theta}(V,\theta,\phi) \hspace{2mm} ; \hspace{2mm} \phi_{+}= \phi+U B^{ \phi} (V,\theta, \phi),
\end{split}
\end{align}
and demanding the continuity of spacetime metric across the junction at leading order in $U$, we get,
 \begin{align}
\begin{split} \label{kerr5}
C=&\frac{1}{\partial_V F} \\ A=&\frac{e}{4}\partial_{V} F \Big((\frac{2}{e} \frac{\partial_{\theta}F}{\partial_V F})^2+\sin(\theta)^2(\frac{2}{e}\frac{1}{\sin(\theta)^2}\frac{\partial_{\tilde \phi}F}{\partial_{V} F})^2\Big),\\
B^{\theta}=&\frac{2}{e} \frac{\partial_{\theta}F}{\partial_V F} \hspace{2mm} ; \hspace{2mm}
B^{\phi}= \frac{2}{e}\frac{1}{\sin(\theta)^2}\frac{\partial_{ \phi}F}{\partial_{V} F}.
\end{split}
\end{align}
We then expand the tangential components of the metrics of both sides $\mathcal{M}_{+}$ and $\mathcal{M}_{-}$ to the linear order in $U$ and read off the components $\gamma_{ab}$.
For $\mathcal{M}_{-}$  we have,
\begin{align}
\begin{split}
g^{-}_{ab}dx^{a} dx^{b}=g^{0}_{AB} dx^{A} dx^{B}- U V\frac{8 m^2}{e} \Big( d\theta^2+ \sin(\theta)^2 d  \phi^2\Big).
\end{split}
\end{align}
For $\mathcal{M}_{+}$ we have, 
\begin{align}
\begin{split}
g^{+}_{ab}dx^{a} dx^{b} =& g^{0}_{AB} dx^{A} dx^{B}+8 m^2 U \Big(-\frac{2}{e}dA dF+\sigma_{AB} dx^A(dB^B-(F/eF_V) dx^{B})+ \\
& \sin(\theta)\cos(\theta)\,B^{\theta}\,\, d\phi^2 \Big).
\end{split}
\end{align}
using Eq.(\ref{gmaab}),
$
\gamma_{ab}=N^{\alpha}[\partial_{\alpha} g_{ab}]= N^{U}[\partial_{U} g_{ab}]
$
with,
$
N^{U}=\frac{e}{8 m^2}.$
Using (\ref{kerr5}), $\mathcal{M}_{+}$, we get 
\begin{align}
\begin{split}
g^{+}_{ab}dx^{a} dx^{b}=& g^{0}_{AB} dx^{A} dx^{B}+8 m^2 U \Big(\frac{2}{e}\Big( \frac{\partial_{V}\partial_{a} F}{\partial_{V} F }dV dx^a\Big)+\frac{2}{e}\frac{\partial_{A}\partial_{B} F}{\partial_{V} F}dx^{A} dx^{B}-\frac{4}{e}\cot(\theta)\frac{\partial_{\phi} F}{\partial_{V} F} d\theta d\phi \\
& -\sigma_{AB} dx^A(F/eF_V) dx^{B}+ \frac{2}{e}\sin(\theta)\cos(\theta)\frac{\partial_{\theta} F}{\partial_{V} F}  d\phi^2\Big).
\end{split}
\end{align}
We write down different components of $\gamma_{ab}$ tensor,
\begin{align}
\begin{split} \label{kerr6}
\gamma_{V a }=& 2\frac{\partial_{V}\partial_{a} F}{\partial_{V} F } \\ \gamma_{\theta \phi }=& 2\Big(\frac{\nabla_{\theta}^{(2)} \partial_{\phi} F}{\partial_{V} F  }\Big) \\
 \gamma_{\theta \theta }=& 2\Big(\frac{\nabla_{\theta}^{(2)} \partial_{\theta} F}{\partial_{V} F }-\frac{1}{2}\Big(\frac{F}{\partial_{V} F}-V\Big)\Big), \\
 \gamma_{\phi \phi }=& 2\Big(\frac{\nabla_{\phi}^{(2)} \partial_{\phi} F}{\partial_{V} F }-\frac{1}{2}\sin(\theta)^2\Big(\frac{F}{\partial_{V} F}-V\Big)\Big).
\end{split}
\end{align}
Using (\ref{kerr6}), we can identify shell-intrinsic properties as follows.\begin{align}
\begin{split} \label{kerr7}
\mu =&-\frac{1}{32 m^2 \pi \partial_{V} F}\Big(\nabla^{(2)} F-F +V\partial_{V}F\Big)\\
j^{A}=&\frac{1}{32 m^2 \pi}\sigma^{AB}\frac{\partial_{B}\partial_{V}F}{\partial_{V} F}\\
p=&-\frac{1}{16 \pi} \gamma_{VV}=-\frac{1}{8\pi}\frac{\partial_{V}^2 F}{\partial_{V} F}.
\end{split}
\end{align}
For supertranslation we need to replace $F(V,\theta,\phi)$ by $V+ T(\theta,\phi).$ Then only $\mu$ turns out to be nonzero,
\be
\mu =-\frac{1}{32 m^2 \pi}\Big(\nabla^{(2)} T(\theta,\phi)-T(\theta,\phi) \Big).
\label{mu}\ee
This shows that the non-vanishing surface energy density $\mu$ carries the supertranslation parameter $T(\theta,\phi)$. Next, we investigate the effects of horizon shell on timelike congruence.

\section{Memory Effect: Timelike congruence} \label{sec3}

In this section, we study interaction between IGW and test particles following timelike geodesics.  We must find some observable effects on the relative motion of neighbouring test particles. To see this, we consider two test particles whose worldlines are passing through a null hypersurface supporting IGW. The effect of IGW on the separation vector of these nearby geodesics can be captured by the use of geodesic deviation equation (GDE) and junction conditions. It should be noted, the treatment presented here is based on the fact that a consistent $C^0$ matching of the metrics can be done maintaining $C^1$ regularity of the geodesics across the null shells \cite{2014}. Here, we closely follow the construction given in \cite{Hogan2003, doi:10.1142/S0218271801001098} and review the necessary parts.\par
Let us consider a congruence having  $T^{\mu}$ to be the tangent vector, where $T^{\mu}$ is a unit time-like vector field.
\begin{equation}
g_{\mu\nu}T^{\mu}T^{\nu} = -1
\end{equation} 
The integral curves of $T^{\mu}$ happens to be time-like geodesics. Therefore,
\begin{equation}
\dot{T}^{\mu} = T^{\nu} \nabla_{\nu}T^{\mu}= 0.
\end{equation}
Dot denotes the covariant derivative of a tensor field along $T^{\mu}$ direction. This congruence is passing through the null shell located at $\Sigma.$ Now we introduce a separation vector $X^{\mu}$ satisfying $g_{\mu\nu}T^{\mu}X^{\nu}=0$ such that,
\begin{equation}
\dot{X}^{\mu} =X^{\nu}\nabla_{\nu} T^{\mu} .
\end{equation} 
This vector $X^{\mu}$ satisfies the geodesic deviation equation,
\begin{equation}
\ddot{X}^{\mu} = -R^{\mu}_{\lambda\sigma\rho}T^{\lambda}X^{\sigma}T^{\rho}.
\end{equation}
$R^{\mu}_{\lambda\sigma\rho}$ is Riemann tensor of full space-time $\mathcal{M}$. To be consistent with the jump in transverse derivative of the quantity (\ref{gmaab}), which produces a Delta function term in Riemann tensor, we can assume the jumps in partial derivatives of $T^{\mu}$ and $X^{\mu}$ across $\Sigma$, given by,
\begin{equation} \label{t1}
[T^{\mu}_{,\lambda}] \footnote{`,' means derivative.}= P^{\mu}n_{\lambda};\,\,\, [X^{\mu}_{,\lambda}] = W^{\mu}n_{\lambda},
\end{equation}
for some vectors $P$ and $W$ defined on $\Sigma.$ The vectors $P, \,W$ are not necessarily tangential to $\Sigma.$
Let $\lbrace E_{a}\rbrace$ be a triad of vector fields defined along the timelike geodesics tangent to $T^{\mu}$ by parallel transporting $\lbrace e_{a}\rbrace$ along these geodesics. Therefore,
\begin{equation}
\dot{E}^{\mu}_{a} = 0,
\end{equation}
where, $E_{a} \big|_{\Sigma}=e_{a}$. Consequently the jump in the partial derivatives of $E^{\mu}_{a}$ should take the following form for some $F^{\mu}_{a}$ defined on $\Sigma$,
\begin{equation}
[E^{\mu}_{a,\lambda}] = F^{\mu}_{a}n_{\lambda}.
\end{equation}
Let $X^{\mu}_{(0)}$ is the vector $X^{\mu}$ evaluated on $\Sigma$, i.e., the value of separation vector $X^{\mu}$ at the intersection point of the null hypersurface with the timelike congruence. Similarly, let $T^{\mu}_{(0)}$ denote the tangent vector $T^{\mu}$ evaluated on $\Sigma$. We can write $X^{\mu}_{(0)}$ in terms of components along the orthogonal and the tangential vectors to the hypersurface in the following way, \begin{equation} \label{t2}
X^{\mu}_{(0)} = X_{(0)}T^{\mu}_{(0)}+X^{a}_{(0)}e^{\mu}_{a},
\end{equation}
for some function $X_{(0)}$, and $X^{a}_{(0)}$ is the $X^{a}$ evaluated at $\Sigma$. Now if $\Sigma$ is located at $U=0$ (this happens typically in Kruskal-like coordinates) , then using (\ref{t1}), and expanding around $U=0$ (keeping only the linear term)  we get, 
\begin{equation} \label{t3}
X^{\mu} = X^{-\mu} +U\, {\cal{H}} (U)W^{\mu},
\end{equation}
with $X^{-\mu}$ in general having dependence on $U$ (in the $-$ side) such that when $U=0$, $X^{-\mu}=X^{\mu}_{0}$. It is convenient to calculate $X^{\mu}$ in the basis $\lbrace E^{\mu}_{a},T^{\mu}\rbrace$. Its non-vanishing components are
\begin{equation}
X_{a} = g_{\mu\nu}X^{\mu}E^{\nu}_{a}.
\end{equation} 
Using (\ref{t2}) and (\ref{t3}), we calculate the above expression for small $U>0$ (in $+$ side),
\begin{equation} \label{t6}
X_{a} = \Big(\tilde{g}_{ab}+\frac{1}{2}\,U\, \gamma_{ab}\Big)X^{b}_{(0)}+U\,V_{(0)a}^{-}
\end{equation}
where $V_{(0)a}^{-}=\frac{dX_{a}^{-}}{du} \big |_{U=0}$ and $\tilde{g}_{ab}$  given by (using the definition of $\gamma_{ab}$ in Eq.(\ref{gmaab})),
\begin{equation} \label{t7}
\tilde{g}_{ab} = g_{ab} + (T_{(0)\mu}e^{\mu}_{a})(T_{(0)\nu}e^{\nu}_{b}).
\end{equation}
The last term in Eq. (\ref{t6}) denotes the relative displacement of test particles for small $U$ when no signal were present. 
As the $\tilde{g}_{ab}$ is non-degenerate we can invert it,
\begin{equation}\label{tt6}
X_{a} = \tilde{g}_{ab}X^{b},\qquad X^a=\tilde{g}^{ab}X_b.
\end{equation}
To see the effect of wave and shell part of the signal separately, we decompose $\gamma_{ab}$ into a transverse-traceless part.
\begin{equation}
\gamma_{ab} = \hat{\gamma}_{ab} + \bar{\gamma}_{ab}
\end{equation}
with,
\begin{equation} \label{barmet}
\bar{\gamma}_{ab} = 16\pi\, \Big(g_{ac}S^{cd}N_{d}N_{b}+g_{bc}S^{cd}N_{d}N_{a}-\frac{1}{2}g_{cd}S^{cd}N_{a}N_{b}-\frac{1}{2}g_{ab}S^{cd}N_{c}N_{d}\Big),
\end{equation}
where $S^{ab}$ is defined in (\ref{inst}), and we have also used the definitions (\ref{intrinsic}). 
Here
\begin{equation}\label{hgamma}
\hat{\gamma}_{ab} = \gamma_{ab}-\frac{1}{2}g_{*}^{cd}\gamma_{cd}g_{ab}-2 n^{d}\gamma_{d(a}N_{b)}+\gamma^{\dagger}N_{a}N_{b},
\end{equation}
and $g_{*}^{ab}$ is the pseudo-inverse of $g_{ab}$ defined as \cite{Blau2016},
\begin{equation}
g_{*}^{ab}g_{bc} = \delta^{a}_{c}-n^{a}N_{c},\,\, g_{*}^{cd}\gamma_{cd}= g^{AB}\gamma_{AB}.	
\end{equation}
Also, $\gamma^{\dagger}=\gamma_{ab}n^{a}n^{b}$ and $\gamma_{ab}$ is defined via Eq.(\ref{gmaab}).
$\hat{\gamma}_{ab}$ encodes the pure gravitational wave degrees of freedom as $\hat{\gamma}_{ab}n^b=0=g_*^{ab}\hat{\gamma}_{ab}$. On the other hand, $\bar \gamma_{ab}$ encodes the degrees of freedom corresponding to the null matter of the shell. If $\bar \gamma_{ab}$ vanishes then we will have a pure impulsive gravitational wave. Otherwise, the IGW will be accompanied by some null-matter. \par
Next, from (\ref{barmet}) and using (\ref{intrinsic}) for Kruskal type coordinates, we get
\begin{align} \label{t5}
\bar \gamma_{VB}=\bar \gamma_{BV}= 16\pi g_{B C}S^{VC} \hspace{2mm} ; \hspace{2mm} \bar{\gamma}_{AB} =& -8\pi\,S^{VV}g_{AB}.
\end{align}
Using (\ref{tt6}), (\ref{t7}), (\ref{barmet}) and (\ref{hgamma})  in (\ref{t6}) we get,
\begin{align}
\begin{split}
&X_{V} = X_{(0) V}+\frac{1}{2}\, U\, \bar{\gamma}_{VB}X^{B}_{(0)}+U\, V_{(0) V}^{-}, \\&
 X_{A} = g_{AB}X^{B}_{(0)}+\frac{1}{2}\, U\, \gamma_{AB}X^{B}_{(0)}+U\, V_{(0)A}^{-}
\end{split}
\end{align}
We assume before the arrival of signal the test particles reside at the spacelike 2-dimensional surface. Therefore, we must have: $X_{(0) V}= V_{(0)V}^{-}=0$, leading to (using Eq.(\ref{t5})),
\begin{align}
\begin{split}
&X_{V} = \frac{1}{2}\, U\, \bar{\gamma}_{VB}X^{B}_{(0)}= 8\pi \, U\, g_{B C}S^{VC} X^{B}_{(0)},\\&
 X_{A} = (1-4\pi\,U\,S^{VV})(g_{AB}+\frac{U}{2}\, \hat \gamma_{AB}\, ) X^{B}_{(0)}+U\, V_{(0)A}^{-}.
\label{dev}\end{split}
\end{align}
Note that, if the surface current $S^{V C}\neq0$, then $X_{V}\neq0,$ which indicates that test particles will be displaced off the 2-dimensional surface. Since $\hat{\gamma}_{ab}$ does not affect $X^V$, this component of the deviation vector can only have non-vanishing change if the IGW becomes the history of a null shell containing current. Note that, for plane fronted wave we can have shell without any matter, in which this part will be zero \cite{Hogan2003}. However, for spherical-shells (such as horizon-shell in Schwarzschild spacetime), we can't have a pure gravitational wave without matter \cite{Blau2016, PhysRevD.98.104009}. For shells having  $S^{VC}=0$, we only have the spatial parts of the deviation vector non-vanishing,
\begin{align}
\begin{split}
X_{A} =  (1-4\pi\,U\,S^{VV})(g_{AB}+\frac{U}{2}\, \hat \gamma_{AB}\, ) X^{B}_{(0)}.
\label{dev1}\end{split}
\end{align}
The terms involving $\hat{\gamma}_{AB}$ in  (\ref{dev1}) describe the usual distortion effect of the wave part of the signal on the test particles in the signal front. Whereas there is an overall diminution factor in the first bracket of the above expression due to presence of the null shell. We now examine the expression of deviation vector components for different cases.
 
\subsection{ Memory effect and supertranslations at the black hole horizon}
\begin{figure}[h!]\centering
    \includegraphics[height=4.5cm, width=12.2cm]{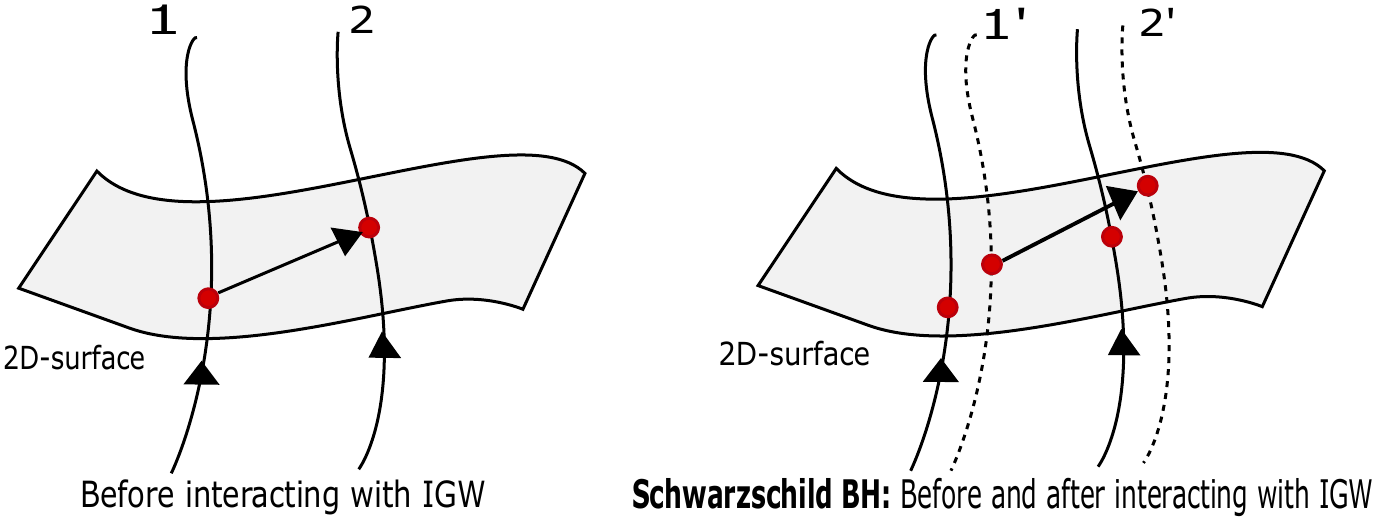} 
    \caption{Timelike geodesics 1 \& 2 get displaced upon interacting with IGW, depicted as 1’ \& 2’ with a new relative displacement vector residing within the 2-dimensional surface.}\label{IGWtimelike}
\end{figure}

We first consider horizon shell of Schwarzschild metric (\ref{smet}). For this case, following \cite{Blau2016} we have, 
 \begin{align}
 \begin{split}
\hat{\gamma}_{\theta\phi} = 2\frac{\nabla_{\theta}^{(2)}\partial_{\phi} F}{F_V} \hspace{2mm} ; \hspace{2mm}
\hat{\gamma}_{\theta\theta} = \frac{1}{2}\left(\gamma_{\theta\theta}-\frac{1}{\sin^2\theta}\gamma_{\phi\phi}\right)
\end{split}
\end{align}
If we focus on the case  (\ref{trans}) for which we get supertranslation from the  horizon soldering freedom, we get a shell having zero pressure and zero surface current, i.e. $S^{VA}=0=p$ \cite{Blau2016}.
Then we get,
\be
\hat{\gamma}_{\theta\phi}=2\nabla_{\theta}^{(2)}\partial_{\phi}T (\theta,\phi) \hspace{2mm} ; \hspace{2mm}
 \hat{\gamma}_{\theta\theta}=2\left(\nabla_{\theta}^{(2)}\partial_{\theta}T (\theta,\phi)-\frac{1}{\sin^2\theta}\nabla_{\phi}^{(2)}\partial_{\phi}T(\theta,\phi)\right).
 \ee
Also the energy density becomes,
 \be
S^{VV}=-\frac{1}{32 m^2\,\pi}(\nabla^2 T(\theta,\phi)-T(\theta,\phi)).
 \ee
Since the surface current is zero, there is no contribution of $X_{V}$ component of the deviation vector. This shell allows impulsive gravitational waves along with some matter and the neighbouring test particles will encode this into the $X^A$ components of the deviation vector (\ref{dev}),
\begin{align}
\begin{split}
X_{\theta}=& \Big(1+\frac{U}{8 m^2}\, (\nabla^2 T(\theta,\phi)-T(\theta,\phi))\Big)\Big[\Big(4 m^2+U\,(\nabla_{\theta}^{(2)}\partial_{\theta}T (\theta,\phi)- \\
& \frac{1}{\sin^2\theta}\nabla_{\phi}^{(2)}\partial_{\phi}T(\theta,\phi))\Big)X^{\theta}_{(0)}+U\, \nabla_{\theta}^{(2)}\partial_{\phi}T (\theta,\phi) X^{\phi}_{(0)}
\Big]
\end{split}
\end{align}
Similarly the $X^{\phi}$ component can also be computed. Thus, test particles will exhibit a relative change in the deviation vector residing within the 2-dimensional surface. It can be seen in the Fig.(\ref{IGWtimelike}). 

Clearly, there is a distortion in the relative position of the particles after encountering the impulsive gravitational wave and the distortions are determined by the supertranslation parameter $T(\theta,\phi)$. The displacement is confined to 2-surface and for physical matter having a positive energy density,  there will be a diminishing effect on the test particles.  Hence, this is a reminiscence of BMS memory effect at future null infinity. We may integrate this expression with respect to the parameter of the geodesics to get a shift in the spatial direction and obtain the displacement memory effect. For a slowly rotating spacetime, we can get similar kind of memory effect following \cite{PhysRevD.98.104009}. Also, the memory effect and soldering freedoms for constant curvature spacetimes have also been carried out in \cite{PhysRevD.100.084010}.

\section{Memory Effect on Null geodesics}\label{nullgeo3}
Here, we consider a null congruence crossing a null hypersurface $\Sigma$ orthogonally, and study the change in the optical parameters like shear, expansion etc. We compute these changes for a congruence defined by the tangent vector $\mathcal{N}$ as it crosses the IGW. To do so, we first take a null congruence $\mathcal{N}_+$  to the $`+'$ side of the null shell and evaluate the components of it in a common coordinate system installed across the shell. Further, the components of the same vector are evaluated to the $`-'$ side using the matching (soldering) conditions. In this procedure, the geodesic congruence suffers a jump and so as the expansion and shear corresponding to the congruence \cite{PhysRevD.99.024031}. The setup is shown in the Fig.~(\ref{IGW}). 
\begin{figure}[h!]\centering
    \includegraphics[height=4.5cm, width=6.0cm]{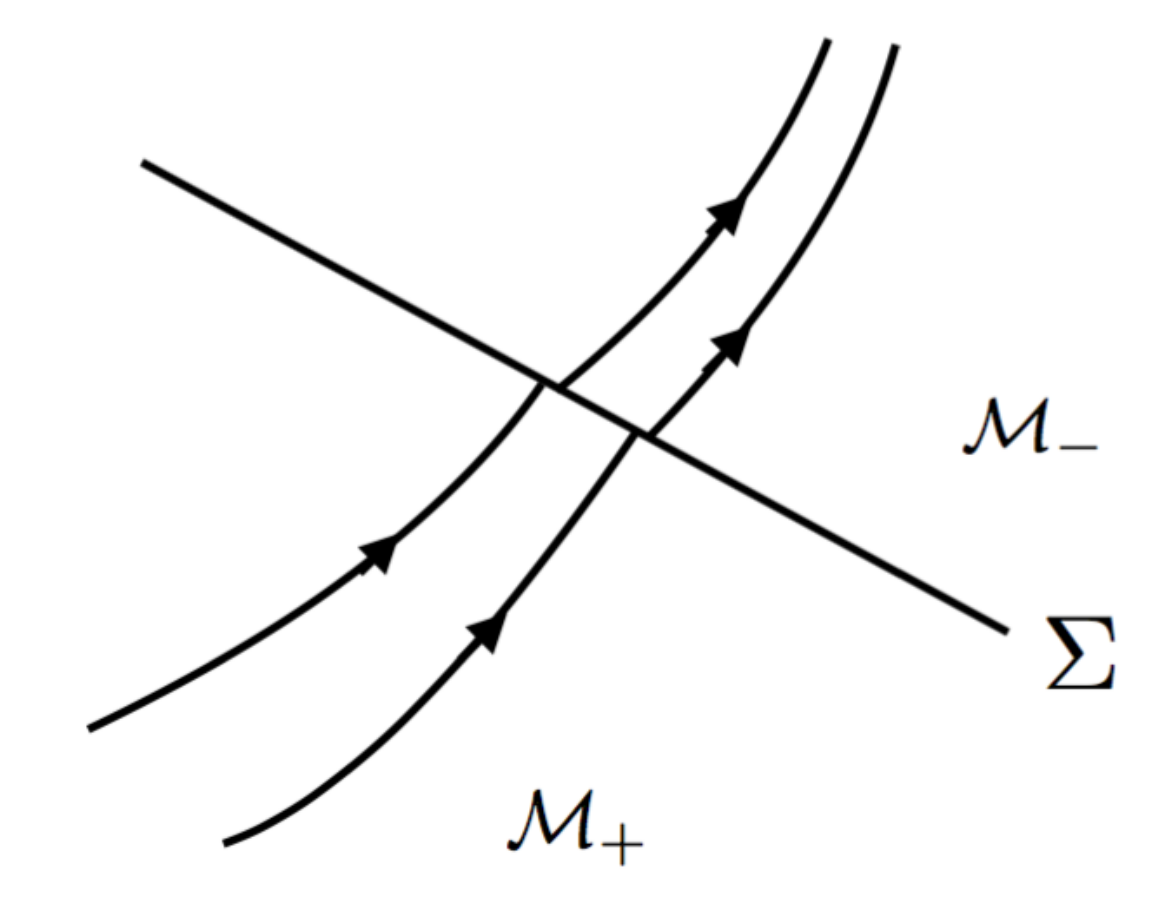} 
    \caption{Null geodesics crossing a null hypersurface $\Sigma$ from $\mathcal{M}_{+}$ to $\mathcal{M}_{-}$. Geodesics experience a jump upon crossing the hypersurface $\Sigma$ represented by soldering transformations.}\label{IGW}
\end{figure}

Let us consider null congruence with tangent vector $\mathcal{N}$ crossing the null surface $\Sigma$ orthogonally with normalization conditions: $n\cdot \mathcal{N}=-1, n\cdot n=0, \mathcal{N}\cdot e=0$. $\mathcal{N}$ can be thought of as the tangent vector of some hypersurface orthogonal (locally with the null analogue of Gaussian normal coordinates) to null geodesics across $\Sigma$. We denote $\mathcal{N}_-$ to be the tangent to the congruence to the $-$ side of $\Sigma$. Mathematically, the effect of the congruence $\mathcal{N}_+$ crossing the null shell is obtained by applying a coordinate transformation between the coordinates $x^{\mu}_+$ to a coordinate $x^{\mu}$ installed locally across the shell \cite{PhysRevD.99.024031}\footnote{One could have done the transformation in the $'-'$ side equivalently.}. These relations between coordinates carry the soldering parameters producing memory effect. 


Let $\mathcal{N}_0$ denotes the vector $\mathcal{N}_+$ transformed in the common coordinate system and evaluated at the hypersurface $\Sigma$. We take coordinates $x^{\mu}_-$ coinciding with $x^{\mu}$.  For expressing $\mathcal{N}_0$ in a common coordinate system, one needs the following transformation relation,
\begin{align}
\mathcal{N}_{0}^{\alpha}(x) = \Big(\frac{\partial x_{+}^{\beta}}{\partial x^{\alpha}}\Big)^{-1}\mathcal{N}_{+}^{\beta}\Big\vert_{\Sigma}.
\label{coordtt}
\end{align}
The inverse Jacobian matrix $\Big(\frac{\partial x_{+}^{\beta}}{\partial x^{\alpha}}\Big)^{-1}$ is to be evaluated using the (``on-shell") soldering transformations. We call soldering freedom confined on the null hypersurface to be ``on-shell" see (\ref{on-shell}). The ``off-shell'' version can be obtained from the ``on-shell'' version by extending the coordinate transformations off-the surface in the transverse direction, see (\ref{kerr4}). The vector in the $'-'$ side is obtained by considering components of $\mathcal{N}_0$ as initial values. Next, we compute the ``failure'' tensor $B$ for the vector $\mathcal{N}_0$ by taking the covariant derivative of it and projecting it on the hypersurface. This tensor encodes the amount of failure for the congruence to remain parallel to each other. 
\begin{align}
\begin{split}
B_{AB} = e^{\alpha}_{A}e^{\beta}_{B}\nabla_{\beta}\mathcal{N}_{0\alpha}. 
\end{split}
\end{align}
Kinematical decomposition of this tensor leads to two measurable quantities (assuming zero twist for hypersurface orthogonal congruence):
\begin{align}
\textrm{Expansion}:\,\, \Theta=\gamma^{AB}B_{AB} \hspace{3mm} ; \hspace{3mm}
\textrm{Shear}:\,\, \Sigma_{AB}=B_{AB}-\frac{\Theta }{2}\gamma_{AB},
\end{align}
where $\gamma_{AB}$ is the induced metric on the null shell. Then memory of the IGW or shell is captured  through the jump of these quantities across $\Sigma$. Note that even if we use ``on-shell'' version of the soldering transformations, jump in these quantities will be captured for a geodesic crossing the shell originating from $+$ side of the shell. 
\begin{align}
\begin{split} \label{jump}
&[\Theta]=\Theta  \big |_{\Sigma_+}-\Theta \big |_{\Sigma_-},\\&
[\Sigma_{AB}]=\Sigma_{AB} \big |_{\Sigma_+}-\Sigma_{AB}\big |_{\Sigma_-}.
\end{split}\end{align}
Next, we let the test geodesics to travel off-the shell, and we get additional contributions in jumps of the expansion and shear. This is achieved by transforming $B$ tensor further to the `-' side of the shell via the following relation that is obvious from hypersurface orthogonal nature of the congruence:
\be\label{ofsxts}
x^{\mu}_-\equiv x^{\mu}=x^{\mu}_0 + U \mathcal{N}_0^{\mu}(x^{\mu})+\mathcal{O}(U^2).
\ee
 Note that all the additional contributions in expansion and shear are proportional to $U$, the parameter of off-shell extension, and would coincide with the jumps when we restrict the transformations on the shell, i.e. for $U=0$. After determining these off shell transformations we can evaluate the failure tensor. For determining $B$ tensor off the shell, we first compute $B_{AB} = e^{\alpha}_{A}e^{\beta}_{B}\nabla_{\beta}\mathcal{N}_{0\alpha}$ and then pull it back to a point infinitesimally away from the shell. 
 \begin{align}\label{b-tensor}
\tilde{B}_{AB}(x^{\mu}) = \frac{\partial x_0^{M}}{\partial x^{A}}\frac{\partial x^{N}_0}{\partial x^{B}}B_{MN}(x^{\mu}_0).\end{align} 
In the far region of the shell, $B$ tensor is evaluated using $\mathcal{N}_+=\lambda \partial_U$, and clearly it will be different from the expression we get from the above equation leading to the jumps in its different components. Next, we show these jumps for shells in Schwarzschild spacetime.

\subsection{ Memory for supertranslation-like transformations} \label{result1}
We consider the case of the Schwarzschild black hole in Kruskal coordinate (\ref{smet}). For plane fronted waves, a similar kind of memory effect has been reported in \cite{PhysRevD.99.024031}. Here we extend that for spherical waves. The details of off-shell transformations are given in appendix (\ref{appen1}). We first compute the failure tensor off-the shell using the inverse Jacobian of transformations given in  
(\ref{Jacobian-1}),
\be
 \mathcal{N}_{0}=\frac{e}{8 m^2} \Big(\partial_{U}+ \frac{1}{4F_{V}^2m^{2}e}\Big(F_{\theta}^{2}+\frac{F_{\phi}^{2}}{\sin^{2}\theta}\Big)\partial_{V}-\frac{F_{\theta}}{2 F_{V}m^{2}e }\partial_{\theta}-\frac{F_{\phi}}{2 F_{V} m^{2}e\sin^{2}\theta}\partial_{\phi}\Big)\Big\vert_{\Sigma}. 
\ee

With this and using (\ref{Jacobian4}) one can evaluate $\tilde{B}_{AB}$ off-the shell:
\begin{align}
\tilde{B}_{AB} =  \frac{1}{det(J)^{2}}\left(
\begin{array}{cc}
 \Big(1+\frac{2U}{e}T_{\phi\phi}\Big)^{2}+\frac{4U^{2}}{e^{2}}T_{\theta\phi}^{2} & -\frac{2U}{e}T_{\theta\phi}\Big(2+\frac{2U}{e}T_{\phi\phi}+\frac{2U}{e}T_{\theta\theta}\Big) \\
 -\frac{2U}{e}T_{\theta\phi}\Big(2+\frac{2U}{e}T_{\phi\phi}+\frac{2U}{e}T_{\theta\theta}\Big) &  \Big(1+\frac{2U}{e}T_{\theta\theta}\Big)^{2}+\frac{4U^{2}}{e^{2}}T_{\theta\phi}^{2}
\end{array}
\right) B_{AB},
\end{align}

The failure tensor $B_{AB}$ for the congruence with $\mathcal{N}_+=\lambda \partial_U$\footnote{ $\lambda$ is arbitrary and only equal to $\frac{e}{8 m^2}$ on the shell} yields,
\begin{align}
B_{CD} =& e_{C}^{\alpha}e_{D}^{\beta}\bigtriangledown_{\beta}\mathcal{N}_{\alpha} = - \Gamma^{V}_{DC}\mathcal{N}_{V}.
\end{align}
 We find the $B$-tensor before and after the null congruence crosses the shell to be different which means there is a jump. Clearly for $U=0$, the off-shell failure tensor $\tilde{B}$ reduces to $B$ evaluated at the shell $\Sigma$. Now focusing particularly on supertranslation, and using (\ref{trans}) and (\ref{Jacobian-2}) one gets the following jumps across the shell: 
\begin{align}
 \begin{split}
&[\Theta ]=\frac{1}{(4 m^2)^2}\Big(T_{\theta\theta}+\csc(\theta)^2 T_{\phi\phi}+T_{\th}\, \cot(\theta)\Big),\\
&[\Sigma_{\theta\theta}]=\frac{1}{8 m^2}\Big(T_{\theta \theta }-T_{\phi \phi} \csc ^2(\theta )-T_{\th} \cot (\theta )\Big),\\&
[\Sigma_{\phi\phi}]=\frac{1}{ 8m^2}\Big(T_{\phi \phi }-T_{\theta \theta } \sin ^2(\theta )+\frac{T_{\th} \sin (2\theta )}{2}\Big),\\&
[\Sigma_{ \theta \phi}]=\frac{1}{(2m)^2}\Big(T_{\theta\phi}- T_{\phi} \cot (\theta )\Big).
 \end{split}
 \end{align}
 It is apparent from the above expressions that the jumps in different components of $B$ are related to supertranslation parameter $T(\theta,\phi)$, and we get a memory corresponding to this. We could have got jumps using the off-shell tensor $\tilde{B}$, in that case the jumps will differ from this by the terms proportional to $U$. Such memories associated with null geodesics are termed as $B$-memory. 




\subsection{Memory for null shell near a black hole horizon }
 In this case, the null surface is situated at $U=\epsilon$ for Schwarzschild black hole. Here, we use transformation relations mentioned in (\ref{trans2}). The inverse Jacobian and other details are displayed in appendix (\ref{appen1}). Using (\ref{Jb3}) the tangent vector of past congruence can be written as,
\begin{align}
 \mathcal{N}_{0} =\frac{e-2 V \epsilon }{8 m^2} \Big(\partial_{U}-B\partial_{\zeta}-\bar{B}\partial_{\bar{\zeta}}\Big)+\mathcal{O}(h^2,\bar h^2, \epsilon\, h, \epsilon\, \bar h)\Big\vert_{\Sigma}.
\end{align}
 
Performing similar analysis as done in section (\ref{result1})  we get,
\begin{align}
[\Theta]=& \frac{4 e^{2-m^2} \left((1+\zeta \bar \zeta) \left(h'(\zeta)+\bar h'(\bar \zeta)\right)-2 \bar \zeta h(\zeta)-2 \zeta \bar h (\bar \zeta)\right)}{1+\zeta \bar\zeta}+\mathcal{O}(h^2,\bar h^2, \epsilon\, h, \epsilon\, \bar h),\\
[\Sigma_{\zeta\zeta}] =&  -\frac{e}{4\,(1+\zeta\bar{\zeta})^{2}}\partial_{\zeta}\bar{B}+\mathcal{O}(h^2,\bar h^2, \epsilon\, h, \epsilon\, \bar h)\\
[\Sigma_{\bar{\zeta}\bar{\zeta}}] =& -\frac{e}{4\,(1+\zeta\bar{\zeta})^{2}}\partial_{\bar\zeta}B+\mathcal{O}(h^2,\bar h^2, \epsilon\, h, \epsilon\, \bar h), \\
[\Sigma_{\zeta\bar{\zeta}}] =& \mathcal{O}(h^2,\bar h^2, \epsilon\, h, \epsilon\, \bar h)\hspace{2mm} ; \hspace{2mm}
[\Sigma_{\bar{\zeta}\zeta}] = \mathcal{O}(h^2,\bar h^2, \epsilon\, h, \epsilon\, \bar h).
\end{align}

$B$ and $\bar B$ are defined in (\ref{new4}). This shows that the $B$-memory effect can be obtained for superrotation-like symmetries as well.

\section{ Discussions}
In violent astrophysical phenomena, gravitational wave pulses with considerable strength are generated.  We have modelled such waves as null shells which represent history of such impulsive gravitational waves. The motivation of this work is to find BMS memory effect near the vicinity of a black hole horizon. For this, we have reviewed the theory of null-shells endowed with BMS-like symmetries.  It is well known that these shells, in general, carry non-zero stress tensor along with IGW. We have studied the interaction of such IGW and stress tensor components on test particles. Let us summarize the findings of Schwarzschild consideration.


First, we have studied the horizon shell of Schwarzschild black hole carrying IGW and evaluated its effect on the deviation vector connecting two nearby timelike geodesics crossing the shell. In this case, where no surface current is present, the effect of IGW on test particles, initially at rest at the 2-dimensional spatial surface, is to displace them relative to each other within the surface. This displacement is characterized by a supertranslation parameter indicating the supertranslation memory effect. This work is an extension of earlier works on similar effects for planar gravitational waves \cite{1998gr.qc9054S, PhysRevD.100.024040, PhysRevD.96.064043, doi:10.1142/9789812776938_0007, Podolsk__1999, PODOLSKY19991, Ortaggio_2002, PhysRevD.67.064013}. We have also considered the effect of superrotations on the deviation vector again on timelike congruence. In this case, due to the presence of non-zero surface current, the test particles initially situated at rest on the 2-dimensional spatial surface, will get displaced off the surface. A similar effect is expected to be exhibited by a null shell situated near to the event horizon of a black hole. This superrotation-like memory effect near the horizon has not been considered earlier. The analyses can be referred from \citep{PhysRevD.100.084010}. Furthermore, we turned our attention on null geodesics. We have shown the expansion and shear parameters of a null geodesic congruence encounter jumps characterized by supertranslation parameter upon crossing a null shell situated at the Schwarzschild horizon. Finally, we have shown that a null shell placed just outside of a black hole horizon exhibits superrotation-like memory effect on the observers in a similar fashion. An analogous approach can be adopted for non-extreme Reissner-Nordstr$\ddot{o}$m black hole which qualitatively will not differ much; however, the results will be a bit lengthier. It will not provide any significant differences from the study of Schwarzschild black hole.

The memory effect near the horizon of a black hole may offer many minute details of the symmetries appearing at the horizon region. We hope the more advanced versions of present day detectors might capture the effects considered here. Relation between BMS-like memories and the Penrose limit for exact impulsive wave spacetimes can offer interesting findings \cite{Shore2018}. As there exists some modified version of the BMS algebra on a null-surface situated at a finite location of a manifold \cite{Chandrasekaran2018}, it will be interesting to investigate the role of those symmetries on the gravitational memory effect. The role of BMS-like symmetries in the study of quantum fluctuations of these spherical IGW background could be a fascinating area of research \cite{Hortacsu:1990ku, Hortacsu:1998sw}. It will also be interesting to see the implications of our results in the context of holography (AdS/CFT correspondence). Finally, understanding the connection between the quantum vacuum structure of gravity and these memories will be an active field of study.  



\chapter{Memory effect and BMS symmetries for extreme black holes}\label{CH4}

\ifpdf
    \graphicspath{{Chapter3/Figs/Raster/}{Chapter3/Figs/PDF/}{Chapter3/Figs/}}
\else
    \graphicspath{{Chapter3/Figs/Vector/}{Chapter3/Figs/}}
\fi


Extreme black holes are important to the general relativity, and String theory community, primarily because of its relevance in the calculation of black hole entropy \cite{STROMINGER199699}. Extreme black holes allow conformal symmetry when we zoom in to its near-horizon geometry, and this played a crucial role not only in the study of quantum black holes but also in finding new kinds of hairs in extreme black holes \cite{PhysRevLett.121.131102}. As it has been studied that almost 70\% of astrophysical black holes are near extremal and many supermassive black holes are also near extremal \cite{2014n, 2005, 2006, 2006n, 2011n}, more attention needs to be paid in exploring the properties of these black holes. Furthermore, Strominger and Vafa derived the Bekenstein-Hawking area-entropy relation for a extreme black hole \cite{STROMINGER199699}. So extreme black holes are important from experimental as well as theoretical perspectives.

The basic framework of this chapter is similar to the last chapter. In particular, here we deal with the emergence of asymptotic symmetries in the context of stitching of two extreme black holes along a common null surface, and its effects on test geodesics. The results turn out to be quite interesting and oppose the Schwarzschild consideration discussed in chapter (\ref{CH3}). Let us start with the shell-intrinsic properties and emergence of asymptotic symmetries for extreme black holes.

\section{Shell-intrinsic properties and BMS symmetries}

In this section, we consider the extreme Reissner Nordstr$\ddot{o}$m (ERN) metric and examine the supertranslation-like symmetries together with BMS intrinsic quantities. We follow the recipe indicated in \cite{PhysRevD.100.084010, PhysRevD.98.104009}. We have a seed ERN metric which is identified as manifold $\mathcal{M}_{-}$ and consider a metric for $\mathcal{M}_{+}$ manifold on which we do a supertranslation type coordinate transformation. The ERN metric in Eddington-Finkelstein (EF) coordinates for $\mathcal{M}_{-}$ manifold is given by
\begin{equation}
ds^{2} = -\Big(1-\frac{M}{r}\Big)^{2}dv^{2}+2dv dr+r^{2}(d\theta^{2}+\\sin^{2}\theta)d\phi^{2} \label{RN},
\end{equation}
with $v=t + r_*=t+\int \frac{dr}{(1-M/r)^2}$. It is evident from the metric that the 2-sphere metric will again give rise to supertranslation-like transformations. The supertranslation can be written similar to Schwarzschild case, i.e., $v\longrightarrow v+T(\theta,\phi)$.

Further, we perform this supertranslation type transformation on $v_+$ coordinate and keep other coordinates unaltered.
\begin{align}
v_{+}=v+T(\theta,\phi) \hspace*{3mm} ; \hspace*{3mm} r_{+} = r \hspace*{3mm} ; \hspace*{3mm} \theta_{+}=\theta \hspace*{3mm} ; \hspace*{3mm} \phi_{+}=\phi. \label{supertranslation}
\end{align}

Under these transformations, the metric takes the following form
\begin{align}
ds_{+}^{2} =& -\Big(1-\frac{M}{r}\Big)^{2}(dv+T_{\theta}(\theta ,\phi)d\theta + T_{\phi}(\theta ,\phi)d\phi)^{2} \nonumber \\
& + 2(dv+T_{\theta}(\theta ,\phi)d\theta + T_{\phi}(\theta ,\phi)d\phi)dr+r^{2}d\Omega_{2}^{2},
\end{align} 
where the horizon is situated at $r=M$ and $T_{\theta}(\theta,\phi)=\partial_{\theta}T(\theta.\phi)$, $T_{\phi}(\theta,\phi)=\partial_{\phi}T(\theta.\phi)$. We choose the transversal (or the auxiliary null normal) vector $N_{\mu}=(-1, 0, 0, 0)$. We use Eq.(\ref{extcur}) to compute the extrinsic curvature $\mathcal{K}_{\theta\theta}$, $\mathcal{K}_{\phi\phi}$ and $\mathcal{K}_{\theta\phi}$ on the horizon. Finally using Eq.(\ref{gmaab}), we get expressions for $\gamma_{ab}$,
\begin{itemize}
\item[(i)] $\gamma_{\theta\theta} = 2\nabla_{\theta}T_{\theta}(\theta,\phi) $
\item[(ii)] $\gamma_{\phi\phi} = 2\nabla_{\phi}T_{\phi}(\theta,\phi)$
\item[(iii)] $\gamma_{\theta\phi} = \gamma_{\phi\theta} = 2\nabla_{\theta}T_{\phi}(\theta,\phi).$
\end{itemize}
Collectively, one can write $\gamma_{AB}=2\nabla_{A}T_{B}$. $\gamma_{AB}$ is symmetric in $A, B$. Notation `$\nabla$' denotes the covariant derivative of $T(x^{A})$ with respect to the unit 2-sphere metric.  Now using Eq.(\ref{gwdf}), one can also compute transverse traceless part of $\gamma_{ab}$
\begin{itemize}
\item[(i)] $\hat{\gamma}_{\theta\theta}  = \nabla_{\theta}T_{\theta}(\theta,\phi)-\frac{1}{\sin^{2}\theta}\nabla_{\phi}T_{\phi}(\theta,\phi) $
\item[(ii)] $\hat{\gamma}_{\phi\phi}  = \nabla_{\phi}T_{\phi}(\theta,\phi)-\sin^{2}\theta \nabla_{\theta}T_{\theta}(\theta,\phi)$
\item[(iii)] $\hat{\gamma}_{\theta\phi} = \hat{\gamma}_{\phi\theta} = 2\nabla_{\theta}T_{\phi}(\theta,\phi). $
\end{itemize}

With the help of extrinsic curvature expressions, we directly compute the surface energy density of the shell as,
\begin{align}
\mu = -\frac{1}{8M^{2}\pi}\bigtriangleup^{(2)}T(\theta,\phi).
\label{edc}\end{align}
The surface current $J^{A}$ and pressure of the shell is given by,
\begin{align}
J^{A} = 0 \hspace*{4mm} ; \hspace*{4mm} p = 0.
\end{align}
Interestingly, the current turns out to be zero in EF coordinates unlike the Schwarzchild or non-extreme case \cite{Blau:2016juv, Blau2016}. It is easy to see from Eq.(\ref{edc}),  the energy density is conserved along the null direction of the shell i.e. $\partial_v \mu=0$. However, there is no non-zero charge corresponding to this supertranslation-like BMS translation, as it vanishes when evaluated on the spherical surface.

Now let us try to see if we can get a shell without matter supporting only gravitational waves. For this, one needs to see if there is any regular solution of the equation obtained by setting $\mu$ equal to zero. 
\begin{align}
\bigtriangleup^{(2)}T(\theta,\phi) = 0.
\end{align} 
This is a Laplace's equation on a sphere. We know this Laplacian has spherical harmonics $Y_l^m(\theta, \phi)$ as eigenfunctions with $-l(l+1)$ as eigenvalues. However, here we have only $l=0$ as a feasible solution which corresponds to a constant only. Therefore, the allowed shift in the $v$ direction is of the form-
\be v\to v+ c,\ee
where $c$ is a constant. Direct substitution of this into the components of $\hat{\gamma}_{ab}$ yields zero. So there can not be a shell supporting pure gravitational waves. A similar situation was also obtained for Schwarzschild case \cite{Blau:2016juv, Blau2016}. Next, we would like to see if conformal symmetries can be recovered as a soldering freedom in ERN spacetime. The basic technique is to glue two metrics across a null surface after performing a conformal transformation to the spatial part of the metric and a subsequent shift in the null direction on one side before attaching the other side \cite{Penrose:1972xrn, 2017}. In the following subsection, we wish to find conformal symmetries in ERN horizon shell from a similar kind of construction.
\subsection{Superrotation near the horizon of  ERN spacetime}\label{supr1}
As outlined in (\ref{suprreq}), superrotation-like soldering freedom can be obtained if one finds a metric that contains a $V$ dependent spatial part ($g_{AB}$) at some $U$ \cite{PhysRevD.98.104009}. Let us now examine this for a horizon shell in ERN spacetime. We introduce a null coordinate $U$ with the help of retarded null coordinate $u=t - r_*$, and write (\ref{RN}) as \cite{Lucietti2013}
\begin{equation}\label{ern-dn}
ds^2=-\frac {2f(r)}{f(M-U)}dUdv + r^2(U,v)d\Omega_2^2,
\end{equation}
where, 
\begin{equation}
f(r)=\left( 1-\frac{M}{r}\right)^2;\,\,r_*=r-M +2M\Big(ln\Big\vert\frac{r}{M}-1\Big\vert-\frac{M}{2(r-M)}\Big);\,u=-2r_*(M-U),
\end{equation}
and
\begin{equation}
2dU=\frac{du}{f(M-U)}.
\end{equation}
Using $r_*=(v-u)/2$ for small $U$ one finds $r(U,v)$ as
\begin{equation}\label{r-eqn}
r(U,v)=M-U+\frac{v}{2M^2}U^2 +\mathcal{O}(U^3)
\end{equation}

The metric (\ref{ern-dn}) is analytic at the event horizon where $U=0$. Now let us consider a null surface just outside the horizon.  For small $U\neq0$ the spatial section of the metric should now become a function of $v$. Therefore
as discussed in (\ref{suprr}), we can recover superrotation-like symmetries on the shell situated close to horizon. To understand this clearly, we reparametrize the metric of unit sphere in terms of complex coordinates $z,\bar{z}$ and write
$d\Omega_2^2=\frac{2dzd\bar{z}}{(1+z\bar{z})^2}$. Now we perform the following transformations in the (say) $+$ side of the shell placed at $U=\epsilon$ (with $\epsilon$ small) 
\begin{equation}
z\to z+ f(z); \bar{z}\to\bar{z}+ \bar{f}(\bar{z});v\to v(1-\tilde{\Omega}(z,\bar{z}) ) -\frac{M^2}{\epsilon^2}\tilde{\Omega}(z,\bar{z}) )+\frac{2M^2}{\epsilon}\tilde{\Omega}(z,\bar{z}) )+\mathcal{O}(\epsilon),
\end{equation}
where $f(z)$ and $\bar{f}(\bar{z})$ are holomorphic and anti-holomorphic functions. These transformations will induce an infinitesimal conformal transformations on the unit two-sphere satisfying the equation (\ref{suprreq}). The conformal factor $\tilde{\Omega}(z,\bar{z})$ is expressed in terms of $z$, $\bar{z}$, $f(z)$ and $\bar{f}(\bar{z})$. As long as $\epsilon\neq0$, the transformation is valid and if one sets $\epsilon=0$ then one must also set $\tilde{\Omega}(z,\bar{z})$ equal to zero indicating at the horizon these transformations do not exist. Therefore, the shell placed near the horizon of ERN black hole gives rise to superrotation-like soldering transformations \cite{PhysRevD.98.104009}. The intrinsic properties of this shell can be obtained by similar manner as described earlier. 


\section{Memory effect: Extreme RN black hole}\label{sec4}
In this section, we shall study the memory effect for ERN black hole on timelike geodesics. To study memory effect, one needs to calculate geometric quantities and their derivatives at the horizon. We also need to extend components of geodesic tangents and  deviation vectors in a near vicinity of the horizon shell orthogonally. A natural choice for this should be a Kruskal-like coordinate system that is regular at the event horizon. In \cite{CARTER1966423}, a maximal analytic extension of ERN black hole was constructed by overlapping two sets of double null coordinates. The construction is useful for examining global features of the spacetime but due to presence of trigonometric functions, it provides less analytical control at the event horizon (for example, the metric is not $C^1$ at the event horizon). In Kruskal type coordinates, one uses an exponential mapping from usual advanced or retarded null coordinates which provides a better analytical control. However, for extreme case since the horizon is degenerate (surface gravity $\kappa=0$) the Kruskal coordinates (eg. $u=-{1\over \kappa}\ln({-U}))$ are constant for any value of advanced or retarded coordinates. This can be remedied using a slightly modified version of Kruskal coordinates as presented in \cite{PhysRevD.62.024005}. Therefore, we shall adopt a Kruskal extension that unambiguously places the shell at $U=0$, and also better suited for memory effect analysis. 
 
Recall $u=t-r_*;\,v=t+r_*$ and the tortoise coordinate is given by
\begin{align}\label{pole}
r_{*} = r+2M\Big(ln(r-M)-\frac{M}{2}\frac{1}{(r-M)}\Big)+constant \,.
\end{align}
Although $r_*$ can be made continuous at the event horizon ($r=M$)  by taking the $Q^2\to M^2$ limit from the $r_*$ of non-extreme case (here $Q$ is the charge parameter of RN black hole) \cite{PhysRevD.62.024005}, but the Kruskal transformations become divergent exactly at $r=M$ as they are related by
\[u=-{1\over \kappa}\ln({-U}); \,v={1\over \kappa}\ln({V}),\]
and $\kappa=0$ at the horizon. Therefore the following transformations for the metric (\ref{RN}) are to be made \cite{PhysRevD.62.024005}
\begin{align}
u = -\psi(-U) \hspace{5mm} ; \hspace{5mm} v = \psi(V),
\end{align}
where we consider $\psi(V)$ to be of the form\footnote{In general for any $\xi$, $\psi(\xi)=4M\Big(ln \xi-\frac{M}{2\xi}\Big)$.}
\begin{align}
\psi(V)= 4M\Big(ln V-\frac{M}{2V}\Big).
\end{align}
Near the horizon
\begin{align}
r_{*}\sim\frac{1}{2}\psi(r-M).
\end{align}
Under this assumption, metric is given by,
\begin{align}
ds^{2} = -\frac{(r-M)^{2}}{r^{2}}\psi(-U)^{'}\psi(V)^{'} dU dV +r^{2}d\Omega_{2}^{2},
\end{align}
where prime denotes the derivative of function $\psi(-U)$ with respect to $U$ and derivative of function $\psi(V)$ with respect to $V$. The transformations are not well defined if the metric is degenerate on the horizon. However, we construct the asymptotic form of the metric as one can have, in the asymptotic limit, $t\sim r_{*}$, and $u\sim -2r_{*} \sim -\psi(r-M)$. Therefore, the inverse transformation is
\begin{align}
U = -\psi^{-1}(-u)\sim -\psi^{-}(\psi(r-M)) = -(r-M),
\end{align}
together with
\begin{align}
\psi(-U)^{'} \sim \frac{4M}{r-M}+\frac{2M^{2}}{(r-M)^{2}}.
\end{align}
$\psi(V)^{'}$ is regular as it is finite and non-zero everywhere. Thus we have a set of Kruskal coordinates that are well defined on the horizon. The metric is written as \cite{PhysRevD.62.024005}
\begin{align}
ds^{2} = -\frac{2M^{2}}{r^{2}}\psi(V)^{'} dU dV+r^{2}(U)d\Omega_{2}^{2}.\label{lib}
\end{align}
We would consider this metric and study the off-shell extension of the transformations and memory effect. The construction follows the one considered in \cite{PhysRevD.100.084010}. We must mention here, for an extreme-Kerr metric, similar kind of  $(U,V)$ coordinates can be obtained as the structure of $r_*$ is almost identical to (\ref{pole}). However, for obtaining shell's intrinsic properties and to study memory effect, one needs to find the extension of soldering freedom off the shell. For a full 4-dimensional rotating metric, this is not an easy task. 
\subsection{Off-Horizon Shell Extension of Soldering Transformations}
To determine the stress-tensor supported on the horizon shell, we extend the soldering transformations off the horizon shell by making an expansion to linear order in $U$  to one side of the shell. We shall call this as ``off-shell'' extension of soldering transformations. Of course, the soldering procedure should be consistent with the junction conditions so that the metric remains continuous across the horizon shell. We also find the generators for off-shell soldering transformations and determine the exact off-shell coordinate transformations to the linear order of $U$. More detailed description can be found in chapter (\ref{CH3}) and \cite{Blau2016}.

We extend the soldering transformations off the horizon shell in $\mathcal{M}_{+}$ side for extreme RN black hole. As $U=-(r-M)$, and $r=M$ is the horizon, $U=0$ depicts the horizon shell. The off-shell soldering transformations to the linear order in $U$ are given by
\begin{align}\label{ern-offc}
U_{+} = U C(V,x^{A}) \hspace*{4mm} ; \hspace*{4mm} V_{+} = F(V,x^{A})+UA(V,x^{A}) \hspace*{4mm} ; \hspace*{4mm} x_{+}^{A}=x^{A}+UB^{A}(V,x^{A}),
\end{align}
where $x^{A}$ denote $(\theta,\phi)$. We first need to determine the functions $A(V,x^{A})$ $C(V,x^{A})$ and $B(V,x^{A})$. For this, we take the transformed metric and compare $g_{U\alpha}$ components with the non-transformed metric. We determine
\begin{align}
C = \frac{\partial_{V}\psi (V)}{\partial_{V}\psi (F)} \hspace{2mm} ; \hspace{2mm}
A = \frac{M^{2}}{2}\frac{F_{V}}{\partial_{V}\psi(V)}\sigma_{AB} B^{A} B^{B} \hspace{2mm} ; \hspace{2mm}
B^{A} = \partial_{V}\psi(V)\frac{1}{M^{2}F_{V}}\sigma^{AB} F_{B},
\end{align}
where $\sigma_{AB}$ is unit 2-sphere metric. Now, we specialize our calculations for BMS case i.e. for $V\rightarrow V+T(\theta,\phi)$. We can explicitly write the metric components linear in $U$. 
\begin{align}
Ug_{ab}^{(1)+}dx^{a}dx^{b} =& 2M^{2}U\Big[\Big(-\frac{\psi(V)^{''}}{M^{2}}+\frac{\psi(V)^{'}}{M^{2}\psi(T)^{'}}\psi(T)^{''} \Big)dV^{2} \nonumber
 \\
 & +\frac{\psi(V)^{'}}{M^{2}}\Big(\frac{\partial_{A}\psi(T)^{'}}{\psi(T)^{'}}+T_{A}\frac{\psi(T)^{''}}{\psi(T)^{'}} \Big)dVdx^{A} \nonumber \\ 
& +\frac{\psi(V)^{'}}{M^{2}}\Big(T_{AB}+T_{B}\frac{\partial_{A}\psi(T)^{'}}{\psi(T)^{'}}-\frac{M\sigma_{AB}}{\psi(T)^{'}}\Big)dx^{A}dx^{B} \nonumber\\
& -\frac{2\psi(V)^{'}}{M^{2}}T_{\phi}\cot\theta d\theta d\phi+B^{\theta}\sin\theta\cos\theta d\phi^{2}\Big],
\end{align}
where $\psi(T)^{'}$ denotes the derivative of function $\psi(T)$ with respect to $T$ and $\psi(V)^{'}$ is derivative of function $\psi(V)$ with respect to $V$. For $\mathcal{M}_{-}$ manifold, the metric is
\begin{align}
Ug_{ab}^{(1)-}dx^{a}dx^{b} = -2MU\sigma_{AB}dx^{A}dx^{B}
\end{align}
 Recall the normalization conditions $n^{\mu}n_{\mu}=0$ and $n^{\mu}N_{\mu}=-1$.  The auxiliary normal is chosen to be $N^{\mu}=(1,0,0,0)$.
Now we can extract all components of $\gamma_{ab}$ using Eq.(\ref{gmaab}). The  components of $\gamma_{ab}$ are given by
\begin{align}
\gamma_{VV} =& 2\Big(-\psi(V)^{''}+\frac{\psi(V)^{'}}{\psi(T)^{'}}\psi(T)^{''}\Big) \\
\gamma_{VA} =& 2\frac{\psi(V)^{'}\psi(T)^{''}}{\psi(T)^{'}}T_{A}  \\
\gamma_{\theta\theta} =& 2\psi(V)^{'}\Big(T_{\theta\theta}+\frac{T_{\theta}^{2}\psi(T)^{''}}{\psi(T)^{'}}-\frac{M}{\psi(T)^{'}}+\frac{M}{\psi(V)^{'}}\Big) \label{gtt}\\
\gamma_{\phi\phi} =& 2\psi(V)^{'}\Big(T_{\phi\phi}+T_{\phi}^{2}\frac{\psi(T)^{''}}{\psi(T)^{'}}-M\frac{\sin^{2}\theta}{\psi(T)^{'}}+T_{\theta}\sin\theta\cos\theta+M\frac{\sin^{2}\theta}{\psi(V){'}}\Big) \label{gpp}\\
\gamma_{\theta\phi} =& \gamma_{\phi\theta} =  2\psi(V)^{'}\Big(\frac{T_{\theta}T_{\phi}}{\psi(T)^{'}}\psi(T)^{''}+T_{\theta\phi}-T_{\phi}\cot\theta\Big)	
\label{gm2}\end{align}
Next, we  directly compute the transverse traceless components $\hat{\gamma}_{ab}$
\begin{align}
\hat{\gamma}_{\theta\phi} =& \hat{\gamma}_{\phi\theta} = \gamma_{\theta\phi} \label{gamma1} \\
\hat{\gamma}_{\theta\theta} =& \frac{1}{2}\Big(\gamma_{\theta\theta}-\frac{\gamma_{\phi\phi}}{\sin^{2}\theta}\Big) \label{gamma2} \\
\hat{\gamma}_{\phi\phi} =& \frac{1}{2}\Big(\gamma_{\phi\phi}-\gamma_{\theta\theta}\sin^{2}\theta\Big). \label{gamma3}
\end{align}\label{hat}
Other components can also be calculated in the same way. We can estimate surface energy density in the following manner
\begin{align}
\mu =& -\frac{1}{16\pi M^{2}}\left(\gamma_{\theta\theta}+\frac{1}{\sin^{2}\theta}\gamma_{\phi\phi}\right).
\end{align}
Therefore
\begin{align}
\mu = -\frac{\psi(V)^{'}}{8\pi M^{2}}\Big(\bigtriangleup^{(2)}T(\theta,\phi)-2M\Big(\frac{1}{\psi(T)^{'}}-\frac{1}{\psi(V)^{'}}\Big)+\Big(T_{\theta}^{2}+\frac{T_{\phi}^{2}}{\sin^{2}\theta}\Big)\frac{\psi(T)^{''}}{\psi(T)^{'}}\Big).
\end{align}
The surface current and surface pressure have the following forms
\begin{align}
    J^{A} =& \frac{1}{8\pi M^{2}}\sigma^{AB} \Big(T_{B}\frac{\psi(T)^{''}}{\psi(T)^{'}}\Big) \label{J}\\
    p =& -\frac{1}{8\pi} \frac{1}{(\psi(V)^{'})^{2}} \Big(-\psi(V)^{''}+\frac{\psi(V)^{'}}{\psi(T)^{'}}\psi(T)^{''}\Big),
\end{align}
where $\sigma^{AB}$ is the inverse of the unit 2-sphere metric. Here, unlike the EF shell, we get non-vanishing current and pressure. 
\subsection{Memory Effect for Timelike Geodesics}\label{mtimelike}
In this section, we shall discuss the interaction between null shells supporting IGW and timelike geodesics. We consider two timelike geodesics crossing the ERN horizon shell supporting IGW. We determine the change in the deviation vector between two nearby geodesics passing the horizon shell. We work in a local coordinate system in which the metric is continuous but its first derivative is discontinuous across the null surface. The standard formulation of the mechanism has already be discussed in section (\ref{sec3}). Following the similar analyses, the deviation vectors upon interacting with impulsive lightlike signals turn out be
\begin{align}\label{X_A}
X_{V} =& \frac{1}{2}U\bar{\gamma}_{VB} X^{B}_{(0)} = 8\pi U g_{BC}S^{VC}X^{B}_{(0)} \\ 
X_{A} =& X_{A(0)}+\frac{U}{2}\gamma_{AB}X^{B}_{(0)}+UV^{-}_{(0)A}.
\label{X_B}\end{align}
The term involving $\gamma_{AB}$ represents the distortion effect of the wave on the test particle.  Note that, if the surface current is non-zero i.e. $S^{VC}\neq 0$,  then $X_{V}\neq 0$. This means the test particle will no longer reside on 2-dimensional surface. It gets displaced off the surface. We also observe that in the case under study, we have a non-zero current or non-zero $S^{VC}$. Hence, the effect of passage of impulsive wave is to deflect the particles off the 2-d surface. The spatial components can also be obtained from (\ref{X_B}). The expression for the $X_{\theta}$ component of the deviation vector
\begin{align}\label{dS}
X_{\theta} = X_{\theta(0)}+\frac{U}{2}\Big(\gamma_{\theta\theta}X^{\theta}_{(0)}+\gamma_{\theta\phi}X^{\phi}_{(0)}\Big) + UV^{-}_{(0)\theta}.
\end{align}
The $X_{\phi}$ component can also be recovered in the same way. One can now replace the requisite components of $\hat{\gamma}_{AB}$ from (\ref{gamma1}-\ref{gamma3}) and $\bar{\gamma}_{AB}$  from (\ref{t5}) into (\ref{dS}) to explicitly see the effect of IGW and matter part on the deviation vector.

Suppose the surface current is zero, i.e. $J^A=0$, then (\ref{X_A}) implies $X_V=0.$ Replacing the second expression of (\ref{t5}) into  (\ref{X_B}) yields
\begin{align}\label{dwave}
    X_{A} =(1-4\pi U S^{VV})(\delta_{AB}+\frac{1}{2}U\hat{\gamma}_{AB})X^B_{(0)} .
\end{align}
\begin{figure}[h!]\centering
    \includegraphics[height=4.5cm, width=12.2cm]{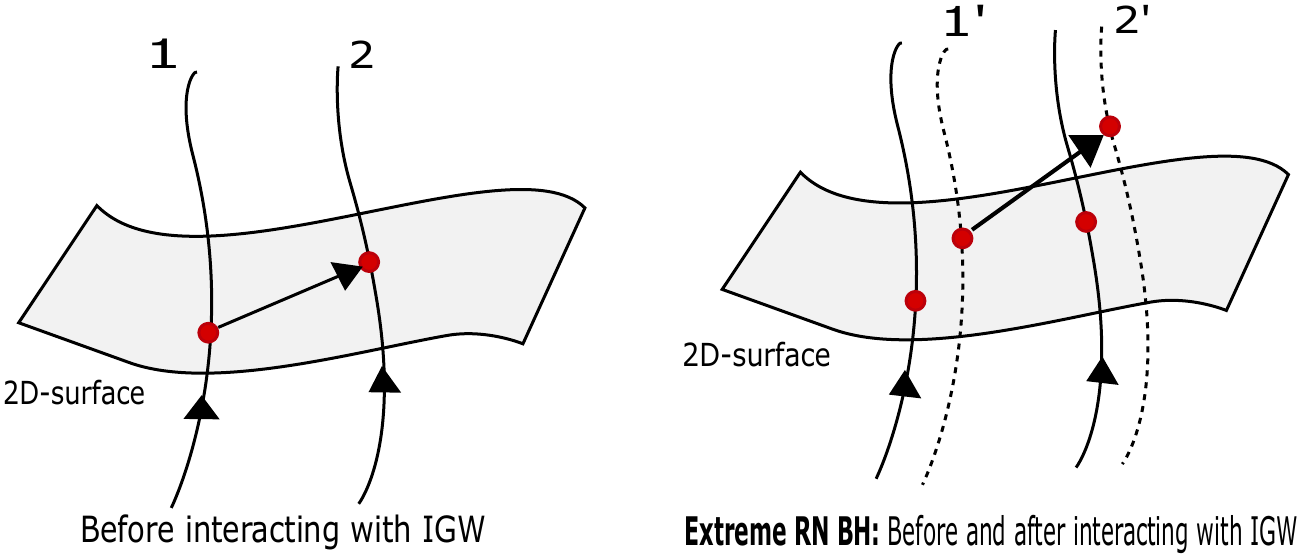} 
    \caption{Timelike geodesics 1 \& 2 get displaced upon interacting with IGW, depicted as 1’ \& 2’ with a new relative displacement vector lying within the 2-dimensional surface.}\label{IGWtimelikeern}
\end{figure}
From (\ref{J}), it is apparent that the anisotropic-stress or surface current of the shell becomes zero for a constant $T$, or constant shifts of $V$. In this case, from (\ref{gtt}), (\ref{gpp}), (\ref{gamma2}), and (\ref{gamma3}) it can be seen that the pure gravitational wave components are identically zero.  So for a shell without surface current we don't have any effect of gravity wave on the test particles as opposed to the case studied for a Schwarzschild black hole in \cite{PhysRevD.100.084010}.  Therefore it is impossible to have a relative displacement of the test particles confined purely in the 2-d plane for IGW supported at an ERN horizon shell. The particles will always be displaced from their initial plane as clearly depicted in the Fig(\ref{IGWtimelikeern}). Clearly, the deviations between two timelike geodesics are determined in terms of supertranslation parameter $T(\theta,\phi)$ contained in $\gamma_{AB}$ or the stress-tensor intrinsic to the shell.  We can integrate the expressions of the  components of deviation vectors and obtain the shift with respect to the parameter of the geodesics. This gives us ``displacement memory effect''. For physical shells having $S^{VV}>0$, we can also see there is a diminishing effect for the transverse components of the deviation vectors carrying the BMS-like memories. 

\section{B-tensor and Null geodesics}\label{mNull}
In this section, we shall study the effect of IGW on null geodesics passing orthogonally through the horizon shell. We would estimate the jump in optical parameters such as shear, expansion. Throughout this study, we  assume a continuous coordinate system is installed covering both sides of the null hypersurface or horizon, and the components of the geodesic tangent vector crossing the horizon are continuous in this coordinate system.  Let us consider a null congruence whose tangent vector is denoted by $\mathcal{N}$ and it is orthogonally crossing the hypersurface $\Sigma$ supporting IGW \footnote{The chosen congruence obeys hypersurface orthogonality \cite{PhysRevD.99.024031}. }. The normalization condition with the normal to the horizon will be as usual $n\cdot \mathcal{N}=-1,n\cdot n=0$ and $\mathcal{N}\cdot e=0$. The effect of null geodesic crossing the surface containing IGW from (say) ``-'' side to the ``+'' side is to apply a coordinate transformation from past coordinates to a continuous coordinate system $(x^{\mu})$ at $\Sigma$. This coordinate transformation then serves as the initial condition to develop it to the future or ``+'' side of the shell. This construction is similar as described in (\ref{sisrl}), only for assigning the past and future of the shell unambiguously, the off-the shell coordinate transformations are done in the ``-'' side. The major object of interest in our study is the failure tensor or the $B$-tensor with respect to the vector $\mathcal{N}_0$ projected onto the hypersurface $\Sigma$. The detailed theoretical construction has been discussed in section (\ref{nullgeo3}). Let us proceed directly with measuring the effect on null geodesics.



\subsection{B-memory Effect for Null Geodesics}
We consider extreme RN black hole in Kruskal coordinates given by (\ref{lib}). The tangent vector in past side to the null congruence is $\mathcal{N}_{-} = \lambda \partial_{U}$. Components of congruence $\mathcal{N}_{0}$ on hypersurface in continuous coordinates are obtained using (\ref{ofsxts}) and  calculating the inverse Jacobian of coordinate transformation  (see Appendix \ref{apnB})
\begin{align}\label{N_0}
\mathcal{N}_{0}^{\alpha} = \Big(\frac{\partial x^{\beta}_-}{\partial x^{\alpha}}\Big)^{-1} \mathcal{N}_{-}^{\beta}\Big\vert_{\Sigma}.
\end{align}
We first find the tangent vector at the hypersurface
\begin{align}
\mathcal{N}_{0} = \lambda\Big(\frac{F_{V} \psi (F)'}{\psi (V)'}\partial_{U}+\frac{\psi (F)'}{2 M^2 F_{V}}  \left(F_{\theta}^2+ \frac{F_{\phi}^2}{\sin^{2}\theta}\right) \partial_{V}-\frac{F_{\theta} \psi (F)'}{M^2}\partial_{\theta}-\frac{F_{\phi} \psi (F)'}{M^2 \sin^{2}\theta}\partial_{\phi}\Big)\Big\vert_{\Sigma}.
\end{align}
From $U$-component of $\mathcal{N}_{0}$, we determine: \hspace{1mm}$\lambda = \frac{\psi(V)^{'}}{\psi(F)^{'}F_{V}}$. Using Eq.(\ref{ofsxts}) and $\mathcal{N}_{0}$ expression, we get
\begin{align}
B_{AB} = 2\frac{\psi(V)^{'}}{F_{V}^{2}}F_{A}F_{BV}-\frac{\psi(V)^{'}}{F_{V}^{3}}F_{A}F_{B}F_{VV}-\frac{\psi(V)^{'}}{F_{V}}F_{AB}+\frac{F_{A}F_{B}}{F_{V}^{2}}\psi(V)^{''}-\Gamma^{\delta}_{AB}\mathcal{N}_{0\delta}.
\end{align}
We will consider now the expressions of $\Theta$ and $\Sigma_{AB}$ evaluated with respect to $\mathcal{N}_0$ and $\mathcal{N}_-$ to find the jumps in these optical tensors. If we specialize our case for BMS supertranslation, the non-vanishing change in expansion and shear on the shell (at $U=0$) are
\begin{align}
[\Theta] =& \frac{1}{M^{2}} \Big(-\psi(V)^{'} \bigtriangleup^{(2)}T(\theta,\phi) +\psi(V)^{''}\sigma^{AB}T_{A}T_{B}\Big)\\
[\Sigma_{\theta\theta}] =& \frac{\psi(V)^{'}}{2}\Big(-T_{\theta\theta}+\frac{T_{\phi\phi}}{\sin^{2}\theta}+T_{\theta}\cot\theta\Big)+\frac{\psi(V)^{''}}{2}\Big(T_{\theta}^{2}-\frac{T_{\phi}^{2}}{\sin^{2}\theta}\Big) \\ 
[\Sigma_{\phi\phi}] =& \frac{\psi(V)^{'}}{2}\Big(-T_{\phi\phi}+T_{\theta\theta}\sin^{2}\theta-\frac{T_{\theta}\sin 2\theta}{2}\Big)+\frac{\psi(V)^{''}}{2}\Big(T_{\phi}^{2}-T_{\theta}^{2}\sin^{2}\theta\Big) \\
[\Sigma_{\theta\phi}] =& -\psi(V)^{'}T_{\theta\phi}+T_{\theta}T_{\phi}\psi(V)''+\psi(V)^{'}T_{\phi}\cot\theta
\end{align}
where $\sigma^{AB}$ is inverse of unit 2-sphere metric. Here, we have non-vanishing jumps for expansion and shear comprising BMS parameter $T(\theta,\phi)$ and its derivatives. The jump in the expansion and shear are  determined by the shell stress-energy tensor and a combination of IGW and stress-tensor respectively. Here, the term \emph{B}-memory is used to recognise the fact that the interaction of the shell with test detectors give rise changes in optical tensors, and these changes are expressed via BMS soldering parameters. This is similar to the BMS memory effect that one obtains in the far region of asymptotically flat spacetimes.

\subsection{B-memory Effect in Off-Shell Extension of Null Congruence}
Now we compute the off-shell extended $B$-tensor for the soldering transformation of supertranslation type. To find the Jacobians of the transformation used in (\ref{b-tensor}), we write \cite{PhysRevD.99.024031, PhysRevD.100.084010}
\begin{align*}
P^M_A=\frac{\partial x_0^M}{\partial x^A}=(W^{-1})^M_A,\end{align*}
where $W^M_A=(\delta^{M}_{A}-U\frac{\partial \mathcal{N}^{M}}{\partial x^{A}})$, and compute the inverse of the $W$ matrix as
\begin{align}
 \frac{1}{det(W)}\left(
\begin{array}{cc}
 1+\frac{U\psi(V)^{'}}{M^{2}}\frac{T_{\phi\phi}}{\sin^{2}\theta} & -\frac{U\psi(V)^{'}}{M^{2}}T_{\theta\phi} \\
 -\frac{U\psi(V)^{'}}{M^{2}}\Big(\frac{T_{\theta\phi}}{\sin^{2}\theta}-\frac{2\cos\theta T_{\phi}}{\sin^{3}\theta}\Big) &  1+\frac{U\psi(V)^{'}}{M^{2}}T_{\theta\theta}
\end{array}
\right).
\label{Jacobian4}
\end{align}
where 
\begin{align*}
det(W) =\Big(1+\frac{U\psi(V)^{'}}{M^{2}}\frac{T_{\phi\phi}}{\sin^{2}\theta}\Big)\Big(1+\frac{U\psi(V)^{'}}{M^{2}}T_{\theta\theta}\Big) -\frac{U^{2}\psi(V)^{'2}}{M^{4}}T_{\theta\phi}\Big(\frac{T_{\theta\phi}}{\sin^{2}\theta}-\frac{2\cos\theta T_{\phi}}{\sin^{3}\theta}\Big).
\end{align*}
Here we also used the fact, $\mathcal{N}(x^{\mu})=\mathcal{N}_0(x_0^{a}(x^{\mu}))$\footnote{This follows from the hypersurface orthogonality of the congruence.}. The full expression of (\ref{b-tensor}) is quite huge to be written here as it will be multiplication of two $W^{-1}$ matrices. However, we  display the shortest component of the $B$-tensor up to linear order in $U$ here.
\begin{align}
B_{\theta\theta} =& (\psi ''(V) T_{\theta}^{2}-\psi '(V) T_{\theta\theta}-M)+\frac{2U}{M^{2}}\Big( -\cot (\theta ) \psi '(V)^2 T_{\phi} T_{\theta\phi}-\psi '(V) \psi ''(V) T_{\phi} T_{\theta} T_{\theta\phi}+ \nonumber \\
& M \psi '(V) T_{\theta\theta}+\psi '(V)^2 T_{\theta\phi}^2+\psi '(V)^2 T_{\theta\theta}^2-\psi '(V) \psi ''(V) T_{\theta\theta} T_{\theta}^2 \Big)\,+ {\cal{O}}(U^2).
\label{B-tt}\end{align}
If one sets $U=0$, the off-shell $B$-tensor reduces to on-shell $B$-tensor which we have already computed in the previous subsection. it is clear from (\ref{B-tt}) that the off-shell $B$-tensor also suffers jump ($U$ dependent) across the null shell and the jump is parametrized by BMS parameters. Thus, B-memory effect is again visible in the off-shell extension of the soldering transformation. An alternative approach to see the change in $B$-tensor for the null geodesics crossing the shell could be to consider the Lie derivative of $B$-tensor with respect to the vector field $\mathcal{N}_{0}$. This would also give rise a $B$-memory effect for null geodesics.   

\section{ Discussion}
The motivation of this work is to find the memory effect of IGW supported at a horizon shell of extreme black holes for timelike and null geodesics (or test detectors) crossing the null shell. Although the memory corresponding to BMS type symmetries discussed here is quite distinct from the memory being studied in the far region or asymptotic infinities of black holes, but this study may serve as a model for determining the effect of impulsive gravity waves (together with some thin layer of null matter like neutrino fluid), generated during violent astrophysical phenomena, on test particles. The appearance of BMS-like symmetries at the horizon or at any null surface situated at a finite distance of a spacetime, provides an intriguing possibility to investigate direct evidence of those symmetries. Our attempt here is to provide a theoretical model that can capture such symmetries. It would be interesting to see how our considerations can be related to recent studies of BMS symmetries on null hypersurfaces and local horizons \cite{Chandrasekaran2018, Ashtekar2018}. 

Although the mathematical frameworks used in this note are already applied to study BMS-memory effect for non-extreme black holes in \cite{PhysRevD.100.084010}, but several new features have also been obtained in this study for the extreme black holes. The nature of horizon-shells containing BMS memories for ERN black hole has non-vanishing surface current as opposed to the case of non-extreme (Schwarzschild) black holes \cite{PhysRevD.100.084010}. In fact one can not have a physical shell carrying BMS-like charges with a vanishing surface current for ERN case. This feature changes the way test (timelike) particles get deflected from their initial position after passage of IGW. There are also other novel features of this study which we summarize below in more detail.

We have shown how BMS supertranslation-like symmetry arises as soldering freedom for a horizon shell in ERN black hole. We started with estimating the intrinsic properties of the null shell for ERN black hole. We also study the BMS type soldering freedom for extreme BTZ case. The detailed study for BTZ black hole is shown in Appendix (\ref{apnA}).  We observe, for both the cases, there is no possibility of a shell where IGW  and the matter supported on the shell get decoupled. Therefore these horizon shells do not support pure IGW without matter. This is same as the case of non-extreme black holes. We also discussed how conformal symmetries may arise as soldering freedom when we place a null shell infinitesimally close to the event horizon of an ERN black hole.  We have related this with the Penrose's cut-paste construction. For an ERN black hole, introducing a double null coordinate system, we have demonstrated the appearance of superrotation-like symmetry near (but not on the) horizon. 

Next, we performed the off-shell extension of soldering transformations for ERN black holes. We used a Kruskal-like double null metric that is regular at the horizon and obtained the off-shell extension to the linear order of U. We then computed shell-intrinsic properties. The Kruskal shell for ERN has non-vanishing pressure and surface current. This is in contrast to the Kruskal shell in Schwarzschild black hole where both of these  intrinsic quantities vanish. Thereafter, we obtained the components of deviation vectors in terms of BMS-like parameters that are present in the shell's stress-energy tensor. We show that the test particles initially at rest get displaced from their initial plane after they interact with the shell supporting IGW and null mater. This corresponds to the memory effect for timelike geodesics crossing the null shell. We have  also  shown there can not be a deflection that will {\emph keep} the test particles on the initial 2-d surface (codimension 1 surface  of the shell) and only induce a relative displacement between them, as was seen for a non-extreme black hole \cite{PhysRevD.100.084010}. Further, we computed the memory effect for null geodesics passing orthogonally through the horizon shell placed in ERN spacetime. We observed, there is a non-vanishing change in optical tensors (the expansion and shear) which again shows a type of memory effect (or $B$-memory) for null geodesics crossing orthogonally to the null hypersurface. We also get a finite jump in the expansion and shear, for the congruence at points infinitesimally away from the shell. Also due to non-existence of pure IGW, the jumps in shear can't be attributed to the pure gravitational degree of freedom encoded in IGW as was found for flat space \cite{PhysRevD.99.024031}.

The constructions depicted here in principle can be generalized for Kerr and extreme Kerr spacetimes also. A double null or Kruskal type coordinate system, to study the memory effect, can be obtained very much similar manner as described in (\ref{sec4}), but due to the absence of spherical symmetry, the analytic calculation becomes much more challenging. We would like to consider the rotating metrics in $ 4$-dimensions in future. 

It is known that any spacetime metric can be reduced to a plane wave metric around a point of a null geodesic. This was shown by Penrose and the resulted metric is known as Penrose limit of a spacetime \cite{RevModPhys.37.215}. As Penrose limit produces plane wave spacetimes (pp-wave, plane symmetric, homogeneous waves etc.), it is comparatively much easier to study geodesic deviation vectors and optical tensors in those backgrounds.  These studies can provide some useful theoretical setups that may prove useful in the future detection schemes of memory effect.  Using Penrose limit, the conventional non covariant memory effect for many impulsive gravitational wave and shock wave metrics can be studied \cite{Shore2018}. The near horizon limit of ERN black hole has $AdS_2 \times S^2$ geometry. In static coordinates it reads \cite{Lucietti2013} 
\begin{align}
    ds^{2} \simeq -\lambda^{2} dt^{2}+\frac{r_{0}^{2}}{\lambda^{2}}d\lambda^{2}+r_{0}^{2}d\Omega^{2}, \label{penrose limit}
\end{align}
where $\lambda = \frac{r-r_{0}}{r_{0}} << 1$ and $r_{0}=M$ is the horizon of the black hole. Penrose limit of this metric produces a symmetric plane wave spacetime. It will be interesting to study the memory effect in such symmetric plane wave spacetimes. Another useful extension of this work is to study different flux-balance laws as depicted in \cite{Compre2020}. Studying quantum effects in such IGW spacetimes generated in extreme black holes could be another interesting study where semiclassical features of near horizon symmetries may show up \cite{1996}.
\chapter{Displacement memory effect for near-horizon BMS symmetries}\label{CH5}

\ifpdf
    \graphicspath{{Chapter3/Figs/Raster/}{Chapter3/Figs/PDF/}{Chapter3/Figs/}}
\else
    \graphicspath{{Chapter3/Figs/Vector/}{Chapter3/Figs/}}
\fi

Motivated by the proposal of Hawking, Perry and Strominger \cite{PhysRevLett.116.231301}, we provide a detailed analysis of computing the displacement memory effect near the horizon of black holes and find its connection with the near-horizon BMS symmetries. It mimics the conventional memory effect obtained at asymptotic null infinity. Our particular interest is to see the effect of near-horizon BMS symmetries \cite{PhysRevD.98.124016} on the test detectors placed near the horizon by computing their permanent shift upon passage of gravitational waves. To compute the displacement memory effect, one starts with the geodesic deviation equation (GDE) which measures the deviation or displacement between two nearby timelike geodesics. We focus on estimating the memory effect for the test detectors stationed near the black hole horizon in three and four dimensions and its possible connection with BMS symmetries. 

In order to understand the detection framework, consider two nearby timelike geodesics or test masses of GW detectors positioned near the horizon of a black hole spacetime with a tangent vector $T^{\mu}$ and deviation or separation vector $S^{\mu}$. The deviation vector $S^{\mu}$ between adjacent geodesics evolves according to GDE, given by
\begin{align}
    \frac{D^{2}}{d\tau^{2}}S^{\mu} = -R^{\mu}{}_{\alpha\beta\gamma}T^{\alpha}T^{\gamma}S^{\beta},\label{deviation}
\end{align}
where $\tau$ is the proper time, $S^{\mu}$ denotes the spatial separation between two nearby timelike geodesics, and $R^{\mu}{}_{\alpha\beta\gamma}$ is the Riemann tensor. Just to recap, we adopt Greek letters for spacetime index ($\mu = 0, 1, 2, 3$), Latin lower case letters for hypersurface index ($a = 1, 2, 3$) and Latin upper case letters for the spatial coordinates on the hypersurface, i.e., coordinates for the co-dimension one surface ($A = 2, 3$). The GDE, in a particular coordinate system, is given by \cite{2015arXiv150505213B}
\begin{align}\label{BGDE}
    \Ddot{S}^{i} = -R_{titj} S^{j}. 
\end{align}
In weak field slow motion approximation, the Riemann tensor depends on the time derivatives of the Quadrupole moment tensor ($Q_{ij}$), $R_{titj} = -\frac{1}{r}\mathcal{P}[\ddddot{Q}_{ij}]$.
Where $\mathcal{P}$ function gives ‘projected orthogonally to the radial direction and trace-free' part of the tensor on which it acts. The interaction between the detectors and GW should induce a permanent change in the spatial separation of the setup. This can be seen by integrating the Eq. (\ref{BGDE}) twice.
\begin{align}\label{BiSep}
    \Delta S^{i} = \frac{1}{r}S^{j}\mathcal{P}[\Delta\ddot{Q}_{ij}], 
\end{align}
where `$\Delta$' denotes the difference between the separation vector before and after the passage of GW. This spatial separation between the detectors, to the linear order of $\frac{1}{r}$, depicts the conventional memory effect. It should be noted here that the Eq. (\ref{BiSep}) is only valid for the linear approximation to general relativity and for slowly moving sources. Here, we focus on the near vicinity of the horizon, and should not consider the weak field slow-motion approximation. Nonetheless, we can still obtain an analogous effect given by a change in the metric parameter in the right-hand side of Eq. (\ref{BiSep}) in a suitable coordinate system. Near the horizon, the jump in the separation is governed by the change in the metric parameters instead of Quadrupole moment tensor,
\[\Delta S^i\sim \rho \Delta B_{ij}S^j,\] where $B$ denotes a function of the coordinates carried by the metric, and $\rho$ denotes the radial coordinate. Further, we show that the change in the separation of deviation vector can be realized by a combination of supertranslation and superrotation revealing the BMS memory effect near the horizon of black hole spacetimes. Let us first start with the brief review of the conventional displacement memory obtained at null infinity.
\section{Memory effect at null infinity in flat spacetimes} \label{2D}
\begin{figure}[h!]\centering
    \includegraphics[height=7.5cm, width=7.2cm]{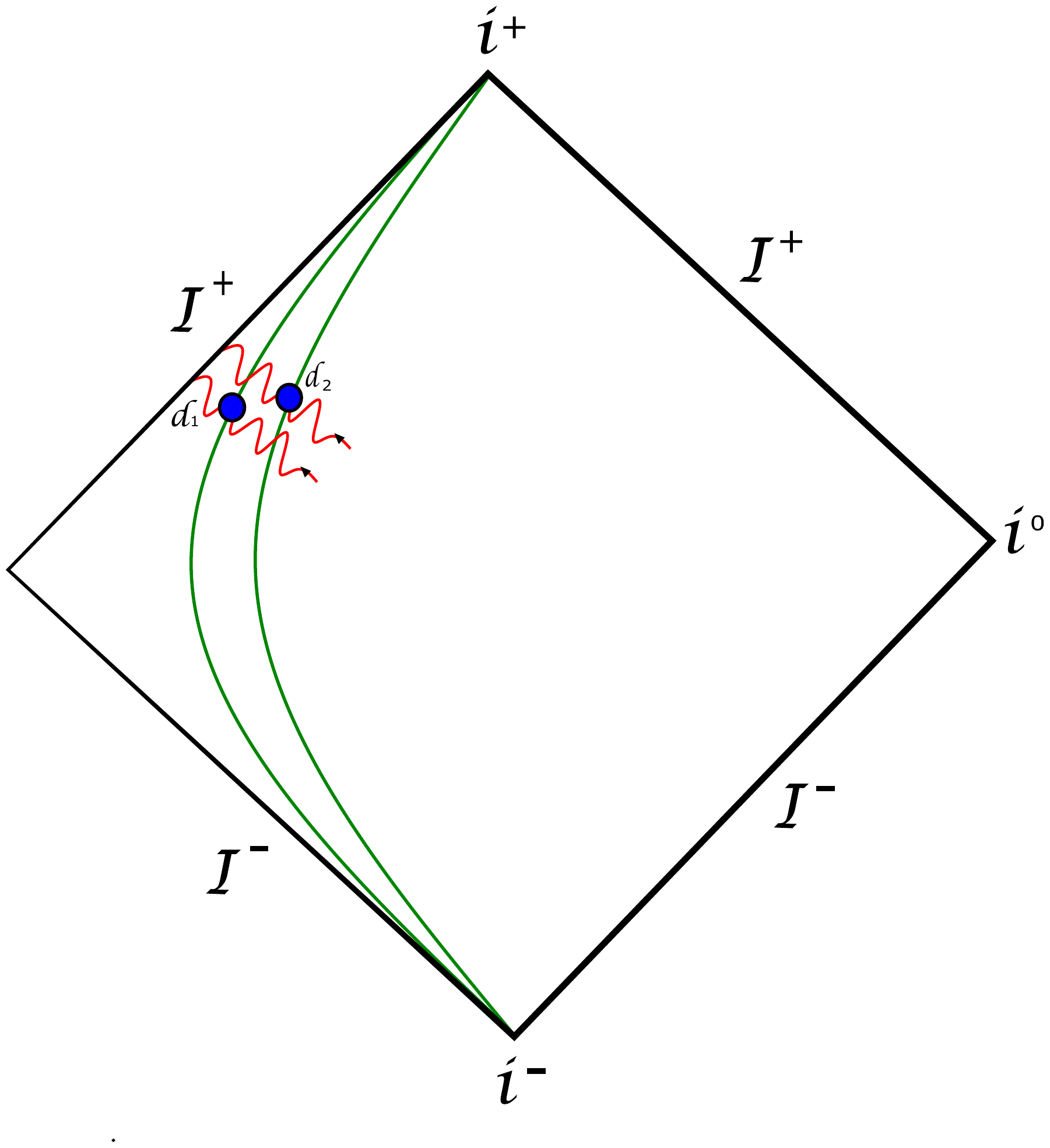} 
    \caption{A schematic diagram of the displacement memory effect at null infinity $I^{+}$. The pulses of gravitational radiation pass the detectors $d_{1}$ and $d_{2}$, and generate a permanent change in the relative separation.}\label{ds1}
\end{figure}

The gravitational memory effect for the asymptotically flat spacetimes near future null infinity has been a well-studied area. The inertial detectors positioned near the future null infinity get permanently displaced after the passage of GW as depicted in the Fig(\ref{ds1}). This is regarded as the conventional gravitational memory effect \cite{1974SvA1817Z, Braginsky:1986ia, Braginsky1987, Christodoulou:1991cr, 2014PhRvD90d4060T, strominger2018lectures, Strominger2014, 2017, PhysRevD.89.064008}. It has also been established that there is a direct relation between this displacement memory and the BMS symmetries \cite{strominger2018lectures, Strominger2014, 2017}. 
Recently, it has been explicitly shown in \cite{2017} how the displacement memory effect at the far region is related to a supertranslation. To see this, consider the general form of the asymptotically flat metric
\begin{align}
ds^{2} =& -du^{2}-2du dr+2r^{2}\gamma_{z\bar{z}}dz d\bar{z}+\frac{2m_{B}}{r}du^{2}+D^{z}C_{zz}du dz+ D^{\bar{z}}C_{\bar{z}\bar{z}}du d\bar{z}+ \nonumber \\ 
& rC_{zz}dz^{2}+rC_{\bar{z}\bar{z}}d\bar{z}^{2}+\frac{1}{r}\Big(\frac{4}{3}(N_{z}+u\partial_{z}m_{B})-\frac{1}{4}\partial_{z}(C_{zz}C^{zz})\Big) du dz+c.c.+..., \label{asymfd}
\end{align}
where $m_{B}$, $N_{z}$ in general depend on ($u,z,\bar{z}$); and they are also called as \textit{Bondi mass} and \textit{angular momentum aspect} respectively. Whereas $C_{zz}(u,z,\bar{z})$ describes GWs. It is, in fact, the free data available near asymptotic null infinity. It is related to the \textit{Bondi news tensor} ($N_{zz}$), written as $N_{zz}=\partial_{u}C_{zz}$. We can see that the first three terms in metric (\ref{asymfd}) represent the flat Minkowski spacetime and other terms represent leading order correction to the metric. ``...'' tells about the subleading terms at large $r$. For large $r$ fall-offs, the metric (\ref{asymfd}) components are
\begin{align}
g_{uu} =& -1+\mathcal{O}(\frac{1}{r}) \hspace{3mm} ; \hspace{3mm} g_{ur} = -1+\mathcal{O}(\frac{1}{r}) \hspace{3mm} ; \hspace{3mm} g_{uz} = \mathcal{O}(1) \nonumber \\
g_{zz} =& \mathcal{O}(r) \hspace{3mm} ; \hspace{3mm} g_{z\bar{z}} = r^{2}\gamma_{z\bar{z}}+\mathcal{O}(r) \hspace{3mm} ; \hspace{3mm} g_{rr} = g_{rz} = 0 \label{asymfd1}
\end{align} 
The geodesic deviation equation for the metric (\ref{asymfd}) takes the following form
\begin{align}
r^{2}\gamma_{z\bar{z}}\partial^{2}_{u}S^{\bar{z}} = -R_{uzuz}S^{z}, \label{asymr}
\end{align}
where $R_{uzuz} = -\frac{r}{2}\partial^{2}_{u}C_{zz}$. Therefore, the change in the displacement is given by
\begin{align}
\Delta S^{\bar{z}} = \frac{\gamma^{z\bar{z}}}{2r}\Delta C_{zz}S^{z}. \label{asymdev}
\end{align} 
From the $uu$-component of the Einstein field equation (EFE), and for a stress tensor with shockwave profile of the form
\begin{align}
T_{uu}(u,z,\bar{z})=\mu\delta(u-u_{rad})\frac{\delta^{2}(z-z_{rad})}{\gamma_{z\bar{z}}}, \label{str4dasmp}    
\end{align}
the leading order change in the $C_{zz}$ is given by
\begin{align}
\Delta C_{zz}(z,\bar{z}) =& C_{zz}(z,\bar{z})\vert_{u=u_{f}}-C_{zz}(z,\bar{z})\vert_{u=u_{0}} \nonumber \\
\Delta C_{zz}(z,\bar{z}) =& 2\mu D^{2}_{z}G(z,\bar{z};z_{rad},\bar{z}_{rad})-\frac{\mu}{2\pi}\int d^{2}z'\gamma_{z'\bar{z}'}D^{2}_{z}G(z,\bar{z};z',\bar{z}'),
 \label{asydel4d}
\end{align}
where $G(z,\bar{z};z_{rad},\bar{z}_{rad})$ is the Green's function, and can be found in \cite{strominger2018lectures}. The above expression contributes to the deviation equation to obtain the memory effect at the far region. 

Further, to relate the memory effect with BMS symmetry, consider the supertranslation of type $u\rightarrow u+f(z,\bar{z})$. One can find a $f(z,\bar{z})$ which would give rise the same change in $C_{zz}$ of (\ref{asydel4d}). For this, taking Lie derivative of $C_{zz}$ along the supertranslation parameter $f$ we get,
\begin{align}
\mathcal{L}_{f}C_{zz} = f N_{zz}-2D^{2}_{z}f. \label{lie4d1}
\end{align}
Eq. (\ref{lie4d1}) is directly related to change in the $C_{zz}$. If before and after the passage of GW the $N_{zz}$ is zero, then one obtains
\begin{align}
\Delta C_{zz} = -\mathcal{L}_{f}C_{zz} = 2D^{2}_{z}f. \label{lie4d2} 
\end{align}
Now one can choose
\begin{align}
f(z,\bar{z}) = \mu G(z,\bar{z};z_{rad},\bar{z}_{rad})-\frac{\mu}{4\pi}\int d^{2}z'\gamma_{z'\bar{z}'}D^{2}_{z}G(z,\bar{z};z',\bar{z}').
\end{align}
This choice of $f$ produces the same change in the $\Delta C_{zz}$ as appears in (\ref{asydel4d}). This establishes the connection between memory effect and BMS symmetries near the future null infinity for asymptotic flat spacetimes. 

Based on the analyses of conventional displacement memory, we expect to estimate a similar effect near the horizon of black holes and its explicit connection with near-horizon asymptotic symmetries. On the similar note, in the next and subsequent sections of the chapter, we provide necessary ingredients to develop the basic framework and estimate the BMS displacement memory effect near the horizon of non-extreme and extreme black holes in three and four dimensions. Let us begin with the near-horizon asymptotic form of the three dimensional black hole.
\section{Near-Horizon Memory for Three-Dimensional Black Holes}\label{3D}
\begin{figure}[h!]\centering
    \includegraphics[height=7.5cm, width=7.2cm]{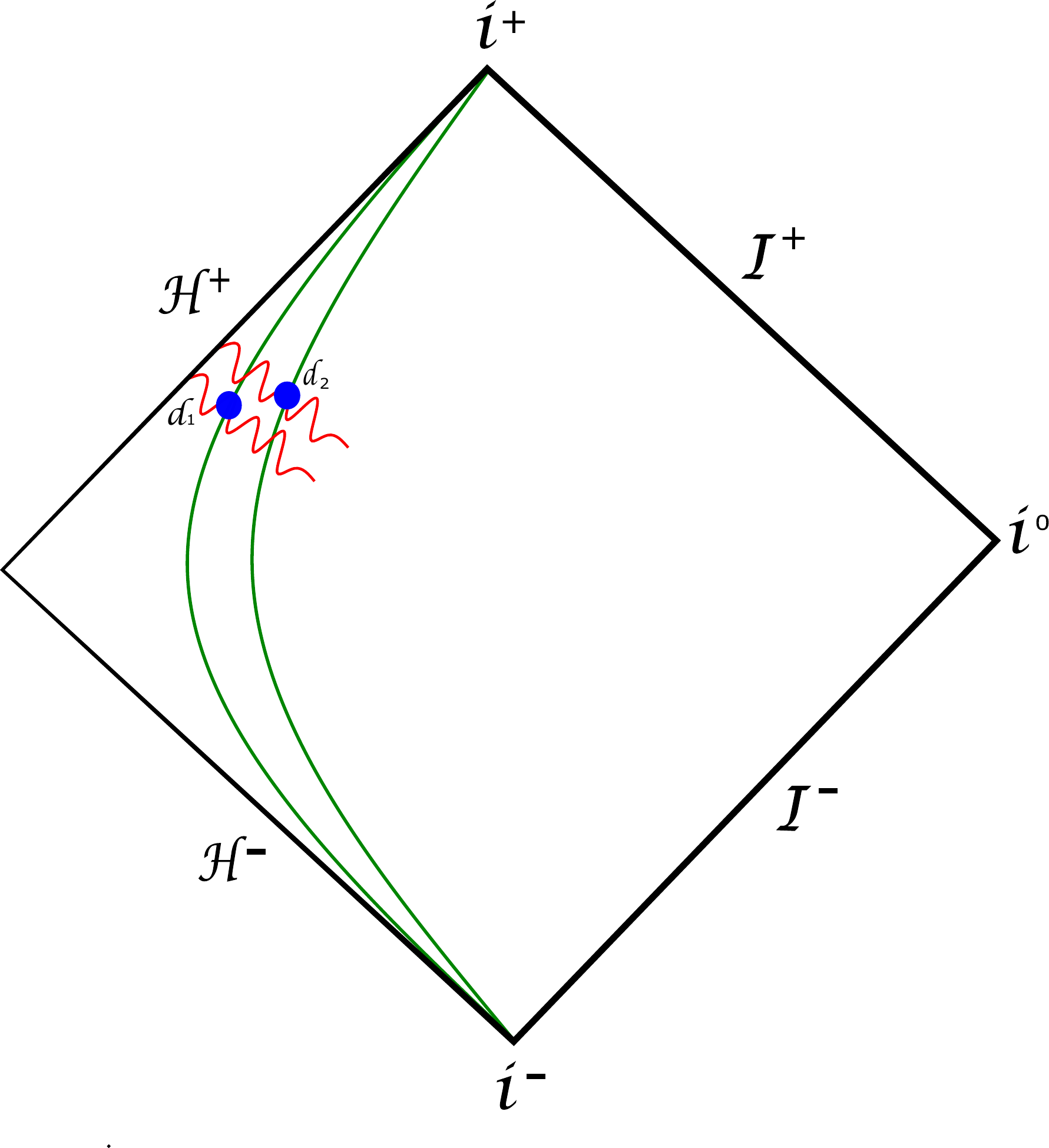} 
    \caption{A schematic diagram of the displacement memory effect near the horizon $\mathcal{H}^{+}$ of a black hole. The interaction between GWs and detector setup $d_{1}$ and $d_{2}$ generate a permanent change in the relative separation.}\label{dsern}
\end{figure}

Three-dimensional gravity theories sometimes give us a good hint of what should be expected in the four-dimensional scenario. In many cases, analysing three-dimensional models becomes easy. Therefore, to start with, we analyse the three-dimensional near-horizon metric for anti-de Sitter (AdS) spacetime, and compute the displacement memory effect. We are interested in measuring the change in the relative displacement or deviation vector between two nearby timelike geodesics or test detectors in the vicinity of black hole horizon, induced via the interaction with GWs generated from the black hole spacetime. The configuration can be understood from Fig.(\ref{dsern}). 

Before investigating the memory effect for a given near-horizon structure of a three-dimensional metric, let us provide a brief description for the origin of such a metric. The basic construction of studying a spacetime metric close to the horizon has started very recently \cite{Krishnan_2012, doi:10.1142/9789814425704_0004, PhysRevD.75.084019}. The term ``near-horizon’’ itself suggests that one needs to develop the mathematical computations by stepping off of the horizon and study the spacetime near the horizon.  Such a metric can be written in terms of horizon-based, ingoing Gaussian null coordinates. Further the metric can also be expanded in radial parameter in the form of a series. The whole idea of obtaining a near-horizon form of a spacetime metric has been elaborated very nicely in \cite{PhysRevD.87.024008}. For the permissible geometry of a horizon, the author provides a set of constraint equations and evolution equations which describe the changes for moving away from the horizon. 
After a study on geometry of deformations and given a horizon together with its foliation, one can construct the coordinates with affine parameter $\rho$ along the geodesics as the coordinates off the horizon and Lie-dragging the on-horizon coordinates to the level surface of $\rho$. The metric components can be calculated perturbatively by analysing the deformations generated by $\frac{\partial}{\partial\rho}$ vector field. As a result, we obtain the metric expanded in a series form where the authors have calculated the expressions of the series up to second order \cite{PhysRevD.87.024008}.

A simplified description and form of a near-horizon metric has been provided in \cite{PhysRevLett.116.091101,Donnay2016,PhysRevD.98.124016}. Let us consider the near-horizon metric for a three-dimensional black hole in Gaussian null coordinates.
\begin{align}
ds^{2}=\xi dv^{2}+2kdvd\rho +2h dvd\phi +R^{2}d\phi^{2}, \label{1}
\end{align}
where $v$ is the temporal coordinate, $\rho \geq 0$ represents the radial distance to the horizon and $\phi$ is the angular coordinate of period $2\pi$. Functions $\xi$, $k$, $h$, and $R$ are expected to comply with the following fall-off conditions near the horizon, which is at $\rho = 0$:
\begin{align}
\xi =& -2\kappa\rho +\mathcal{O}(\rho^{2}) \hspace{3mm} ; \hspace{8mm} k = 1+\mathcal{O}(\rho^{2}) \nonumber\\
h =& \theta (\phi)\rho + \mathcal{O}(\rho^{2}) \hspace{5mm} ; \hspace{8mm} R^{2} = \gamma(\phi)^{2}+\lambda(v,\phi)\rho + \mathcal{O}(\rho^{2}) \label{2}
\end{align}
where $\kappa(v,\phi)$, $\theta(\phi)$, $\gamma(\phi)$ and $\lambda(v,\phi)$ are arbitrary functions of the coordinates. 
The metric components $g_{\rho\phi}$ and $g_{\rho\rho}$ decay rapidly as $\mathcal{O}(\rho^{2})$. 

The asymptotic boundary conditions are being preserved by asymptotic Killing vectors, given by \cite{Donnay2016}
\begin{align}
Z^{v} =& f(v,\phi) \nonumber\\
Z^{\rho} =& -\partial_{v}f \rho + \partial_{\phi}f \frac{\theta}{2\gamma^{2}}\rho^{2}+\mathcal{O}(\rho^{3}) \label{killing3D} \\
Z^{\phi} =& Y(\phi)-\partial_{\phi}f\frac{\rho}{\gamma^{2}}+\partial_{\phi}f\frac{\lambda}{2\gamma^{4}}\rho^{2}+\mathcal{O}(\rho^{3}), \nonumber 
\end{align}
where $f(v,\phi)$ and $Y(\phi)$ are arbitrary functions, and prime denotes the derivative with respect to $\phi$. In addition to this, $Z^{\rho}$ may contain an $\mathcal{O}(1)$ term $\tilde{z}(v,\phi)$. However, we set such terms to be zero provided $Z=0, \partial_{v}Y=0$.  Under such transformations, the arbitrary functions transform along the Killing direction as follows
\begin{align}
\delta_{Z}\kappa =& Y\partial_{\phi}\kappa + \partial_{v}(\kappa f)+\partial^{2}_{v}f \hspace{2mm} ; \hspace{2mm}
\delta_{Z}\gamma =\partial_{\phi}(Y\gamma)+f\partial_{v}\gamma \nonumber\\
\delta_{Z}\theta =& \partial_{\phi}(Y\theta)+f\partial_{v}\theta-2\kappa\partial_{\phi}f-2\partial_{v}\partial_{\phi}f+2\partial_{\phi}f\frac{\partial_{v}\gamma}{\gamma} \label{trans3Dn1}\\
\delta_{Z}\lambda =& Y\partial_{\phi}\lambda +2\lambda\partial_{\phi}Y+2\theta\partial_{\phi}f-2\partial^{2}_{\phi}f+2\partial_{\phi}f\frac{\partial_{\phi}\gamma}{\gamma}+f\partial_{v}\lambda -\lambda\partial_{v}f . \nonumber 
\end{align}
Introducing the modified Lie bracket
\begin{align}
[Z_{1},Z_{2}] = \mathcal{L}_{Z_{1}}Z_{2}-\delta_{Z_{1}}Z_{z}+\delta_{Z_{2}}Z_{1}, \label{mbrcs3D}
\end{align}
One gets
\begin{align}
[Z(T_{1},X_{1},Y_{1}),Z(T_{2},X_{2},Y_{2})] = Z(T_{12},X_{12},Y_{12}), \label{alg3D}
\end{align}
where
\begin{align}
T_{12} =& Y_{1}\partial_{\phi}T_{2}-Y_{2}\partial_{\phi}T_{1} \nonumber \\  
X_{12} =& Y_{1}\partial_{\phi}X_{2}-Y_{2}\partial_{\phi}X_{1}-\kappa (T_{1}X_{2}-T_{2}X_{1})   \\
Y_{12} =& Y_{1}\partial_{\phi}Y_{2}-Y_{2}\partial_{\phi}Y_{1}. \nonumber
\end{align}
Here we considered fixed temperature configuration, i.e., fixed $\kappa$ for which 
\begin{align}
    f(v,\phi) = T(\phi)+e^{-\kappa v} X(\phi), \label{fxdT3D}
\end{align}
where $T(\phi)$ and $X(\phi)$ are two sets of supertranslation generators \cite{PhysRevLett.116.091101, Donnay2016}. 

One can further establish the algebra between generators $T_{n}$, $X_{n}$ and $Y_{n}$ by defining Fourier modes, $T_{n}=Z(e^{in\phi},0,0)$, $X_{n}=Z(0,e^{in\phi},0)$ and $Y_{n}=Z(0,0,e^{in\phi})$, given by
\begin{align}
    i[Y_{m},Y_{n}] =& (m-n)Y_{m+n} \hspace{3mm} ; \hspace{3mm}
    i[Y_{m},T_{n}] = -nT_{m+n} \nonumber \\
    i[Y_{m},X_{n}] =& -nX_{m+n} \hspace{9mm} ; \hspace{3mm} i[T_{m},X_{n}] = -\kappa X_{m+n}  \label{alg3D1}
\end{align}
The generator $T_{n}$ appears as a supertranslation from the symmetry of type $v\rightarrow v+T(\phi)$ and $Y_{n}$ causes superrotations. The conserved charges corresponding to the symmetries have also been found in \cite{PhysRevLett.116.091101,Donnay2016,PhysRevD.98.124016} and form a representation of (\ref{alg3D1}). The zero modes of the generators produce the Bekenstein-Hawking entropy of stationary black holes \cite{PhysRevLett.116.091101,Donnay2016}. 
In the extremal scenario for which $\kappa=0$, the charge associated with the zero mode of $X$ is the one that gives the entropy \cite{Donnay2016}.

\subsection{Geodesic Deviation Equation and Memory Effect}
Now, let us find the geodesic deviation equation for the metric (\ref{1}). As mentioned earlier that in general $\kappa$ depends on ($v,\phi$), the deviation equation (\ref{deviation}) takes the following form\footnote{Here, we have considered terms leading order in $\rho$ for the Riemann tensor.}
\begin{align}
\gamma^{2}(\partial_{v}^{2}S^{\phi}-\kappa(v,\phi)\partial_{v}S^{\phi}) = -R_{\phi v\phi v} S^{\phi},\label{4}
\end{align}
where the Riemann tensor component is given by
\begin{align}
R_{\phi v\phi v} =& \rho\Big(\frac{1}{2}\kappa \theta (\phi)^{2}+\kappa^{2}\lambda+\frac{\kappa\theta(\phi)\gamma(\phi)^{'}}{\gamma(\phi)}-\kappa\theta(\phi)^{'}-\frac{\gamma(\phi)^{'}}{\gamma(\phi)}\partial_{\phi}\kappa+\partial_{\phi}^{2}\kappa-\\ 
& \frac{1}{2}\lambda\partial_{v}\kappa+\frac{1}{2}\kappa\partial_{v}\lambda-\frac{1}{2}\partial^{2}_{v}\lambda\Big)+\mathcal{O}(\rho^{2}).\label{55}
\end{align}
prime denotes the derivative of the function with respect to $\phi$ coordinate, and $\lambda\equiv \lambda(v,\phi)$, $\kappa\equiv \kappa(v,\phi)$. However, as our motivation is to find the memory near the horizon of black holes, we consider only fixed temperature configurations i.e. cases where $\kappa$ is constant. We also set $\partial_v\gamma=0$. Under these assumptions, the Riemann tensor component becomes
\begin{align}
R_{\phi v\phi v} =& \rho\Big(\frac{1}{2}\kappa \theta (\phi)^{2}+\kappa^{2}\lambda+\frac{\kappa\theta(\phi)\gamma(\phi)^{'}}{\gamma(\phi)}-\kappa\theta(\phi)^{'}+\frac{1}{2}\kappa\partial_{v}\lambda-\frac{1}{2}\partial^{2}_{v}\lambda\Big)+\mathcal{O}(\rho^{2}),\label{5}
\end{align}
Here, only $\lambda$ depends on $(v,\phi)$ and $\kappa$ is fixed as mentioned above. where $\kappa = \frac{1}{l^{2}r_{+}}(r_{+}^{2}-r_{-}^{2})$; $r_{+}$ and $r_{-}$ denoting the outer and inner horizons respectively. $l$ is the length scale of AdS spacetime. 

\subsubsection{Extremal Case}
Above equations (Eq. \ref{4}, \ref{5}) considerably simplify in the case when $\kappa=0$.  In this case,  the $g_{vv}$ component of the asymptotic metric becomes, $g_{vv} = L(v,\phi)\rho^{2} + \mathcal{O}(\rho^{3})$ \cite{PhysRevD.98.124016}, and therefore does not contribute to the leading order. Now, for this extremal form of the metric, the geodesic deviation equation (\ref{deviation}) takes the following form
\begin{align}
\gamma(\phi)^{2}\partial^{2}_{v}S^{\phi}_{E} = \frac{\rho}{2}\partial^{2}_{v}\lambda S^{\phi}_{E},
\end{align}
where $S^{\phi}_{E}$ depicts the $\phi$ component of the deviation vector for extremal case. 

Now we can measure the change in the deviation vector which can be written as
\begin{align}
\Delta S^{\phi}_{E} =\frac{\rho}{2\gamma(\phi)^{2}}(\Delta \lambda) S^{\phi}_{E} + \mathcal{O}(\rho^{2}).\label{gd1}
\end{align}
Here we observe that our goal is to find the change in $\lambda(v,\phi)$ in order to estimate the jump in the deviation vector. For this, we solve the $vv$-component of the Einstein field equation, given by
\begin{align}
-\frac{\rho}{2\gamma(\phi)^{2}}\partial^{2}_{v}\lambda = 8\pi T^{M}_{vv}. \label{f1}
\end{align}
Further, defining $T_{vv}=\lim_{\rho \to 0}8\pi\frac{T_{vv}^{M}}{\rho}$ and considering a shock wave type profile for the stress tensor as 
\begin{align}
    T_{vv}(v,\phi) = s \delta(v-v_{0})g(\phi) ,\label{tvv}
\end{align} where $s$ is just a constant, the solution of the field equation becomes
\begin{align}
\lambda(v,\phi) = c_{1}(\phi)+\Big(c_{2}(\phi) -2s\gamma(\phi)^{2}g(\phi)\Big)v. \label{ful1}
\end{align}
To compute the change in $\lambda(v,\phi)$, i.e., $\Delta\lambda(\phi)$ for some initial and final $v$, we set the boundary conditions as following
\begin{align}
\lambda(v=v_i,\phi) = \lambda_{0}(\phi) \hspace{5mm} ; \hspace{5mm} \lambda(v=v_f,\phi) = \lambda_{f}(\phi).
\end{align}
Hence, the full solution can be written as
\begin{align}
\lambda(v,\phi) =C(\phi)+ F(\phi)v, \label{ld3d}
\end{align}
for some function $F(\phi)$ that contains the metric parameters only depending on $\phi$. Therefore, the change in $\lambda(v,\phi)$ can be written as
\begin{align}
\Delta\lambda = \lambda(v=v_{f},\phi)-\lambda(v=v_{i},\phi) = F(\phi)\Delta v, \label{lambda1}
\end{align}
where $\Delta v = v_{f}-v_{i}$. We plug this expression of $\Delta\lambda$ into (\ref{gd1}). This gives us the change in the  deviation vector $S^{\phi}_{E}$.
\begin{align}
\Delta S^{\phi}_{E} = \frac{\rho}{2\gamma(\phi)}F(\phi)\Delta v S^{\phi}_{E}+\mathcal{O}(\rho^{2}). \label{extreme3D}
\end{align}
This provides the shift in the deviation vector between the test geodesics near the horizon for three-dimensional extreme black holes. This is the leading order change in the deviation vector which depends on metric parameters. Here, the derivative of $\lambda$ behaves as a Bondi News tensor that appears in the far region case. The memory (\ref{extreme3D}) mimics the one originally obtained at asymptotic null infinity (\ref{asymdev}). The analysis implies a permanent change in the setup induced by the interaction with GWs and regarded as the displacement memory effect. Fig.(\ref{ds1}) provides an illustration of the situation we are examining. 



\subsubsection{Relation with BMS symmetry} \label{bms3d}

To see the connection between displacement memory effect and BMS symmetry, we consider the change in the parameter $\lambda$. For extreme case, the $v$ component of the Killing vector, generating two sets of supertranslations $T(\phi)$ and $X(\phi)$  becomes 
\begin{equation}
    f(v,\phi)=T(\phi)+v X(\phi).
\end{equation}
In general the $\lambda(v,\phi)$ can be written as $\lambda(v,\phi)=C(\phi)+v F(\phi)$, as shown in (\ref{ld3d}). If  $\Delta v$ is the difference in advanced-time before and after the passage  of GWs, then
\begin{align}
\Delta \lambda= F(\phi)\Delta v . \label{dlmbda1}
\end{align}
This is an equivalent way of writing (\ref{lambda1}). Now, we consider the change in $\lambda(v,\phi)$ induced by asymptotic Killing vector $Z$ \cite{PhysRevD.98.124016},
\begin{align}
{\cal L}_{Z}\lambda =2 \lambda \partial_{\phi}Y+Y\partial_{\phi} \lambda + 2\theta \partial_{\phi} f-2 \partial_{\phi}^2 f +2\partial_{\phi}f\frac{\partial_{\phi}\gamma}{\gamma(\phi)}+f\partial_{v}\lambda -\lambda \partial_v f . \label{sup1}
\end{align}
Our goal is to produce Eq. (\ref{lambda1}) like change via ${\cal L}_{Z}\lambda$. Here, to make the variation of $\lambda(v,\phi)$ along the Killing direction $Z$ to be independent of $v$, we set coefficient of $v$ as zero. This will make sure the whole right hand side of the Eq. (\ref{sup1}) matches with the displacement in $\lambda$ derived in (\ref{lambda1}). We obtain the following condition
\begin{align}
    2F\partial_{\phi}Y+Y\partial_{\phi}F+2\theta\partial_{\phi} X-2\partial^{2}_{\phi}X+2\partial_{\phi}X\frac{\partial_{\phi}\gamma (\phi)}{\gamma(\phi)} = 0, \label{3d-cndn}
\end{align}
where $F$, $X$, $Y$ and $\theta$ are functions of $\phi$. We observe the condition of $\Delta \lambda(v,\phi)$ to be $v$ independent involves both $Y(\phi)$ and $X(\phi)$. The remaining terms in Eq. (\ref{sup1}) are functions of $\phi$ only and can be arranged in such a way that they become equal to $\Delta v F(\phi)$. To see the feasibility of such BMS transformation, we may consider only superrotation to be acting on the metric parameters, in that case Eq. (\ref{3d-cndn}) offers a solution $Y(\phi)=F^{-1/2}.$ On the other hand in the absence of superrotations the above equation provides a solution for $X(\phi)\propto \int d\phi e^{-\int A(\phi) d\phi}.$ Where $A(\phi)$ is a function of metric parameters $\theta$, $\gamma$ etc. In general one may also get a solution when both supertranslation and superrotation are present by choosing an appropriate $g(\phi)$ sitting in the expression of energy momentum tensor (\ref{tvv}). All these results confirm that the memory can be induced by BMS parameters superrotation $Y(\phi)$ and supertranslation $X(\phi)$.

\subsection{Non-Extremal Case}\label{nextml1}
For nonzero but constant $\kappa$ configuration, we again need to find the change in $\lambda(v,\phi)$. This can  obtained by solving the $vv$-component of the Einstein's field equation to the leading order in $\rho$, 
\begin{align}
    -\frac{\rho}{2l^{2}\gamma^{3}}\Big(-4\kappa\gamma^{3}+2l^{2}\kappa\theta\gamma^{'}+l^{2}\gamma(\kappa\theta^{2}+2\kappa^{2}\lambda-2\kappa\theta^{'}+3\kappa\partial_{v}\lambda+\partial^{2}_{v}\lambda)\Big) = 8\pi T^{M}_{vv}. \label{6}
\end{align}
Here, $\lambda\equiv \lambda(v,\phi)$. We shall again define $T_{vv}=\lim_{\rho \to 0}8\pi\frac{T^{M}_{vv}}{\rho}$ and consider the terms other than $\lambda(v,\phi)$ as a function  $\kappa\, \mathcal{F(\phi)}$, defined by,
\begin{align}\mathcal{F} (\phi)=\rho\Big(\frac{2}{l^{2}}-\frac{1}{\gamma(\phi)^{2}}(\frac{\theta(\phi)\gamma(\phi)^{'}}{\gamma(\phi)^{2}}+\frac{\theta(\phi)^{2}}{2}-\theta(\phi)^{'}) \Big). \end{align}
Since we are interested to solve the field equation (\ref{6}) with respect to $v$ coordinate, $\phi$-dependent functions would be unaffected while solving the equation. Now Eq. (\ref{6}) becomes
\begin{align}
-\frac{1}{2\gamma(\phi)^{2}}\frac{\partial^{2}}{\partial v^{2}}\lambda(v,\phi)-\frac{3\kappa}{2\gamma(\phi)^{2}}\frac{\partial}{\partial v}\lambda(v,\phi)-\frac{\kappa^{2}}{\gamma(\phi)^{2}}\lambda(v,\phi)+\kappa\,\mathcal{F}(\phi)- T_{vv}(v,\phi) = 0. \label{7}
\end{align}
Further, we consider a shock wave type stress tensor, i.e., $T_{vv}(v,\phi) = s \delta(v-v_{0})g(\phi)$. Where $s$ is a constant or can also be a function of $\phi$.  The situation still describes a fixed temperature but not a stationary black hole configuration. 

For non-stationary background, let us first try to solve (\ref{4}) together with (\ref{5}) perturbatively in $\kappa$ to gain insight into the problem. Upto leading order in $\kappa$, we get from (\ref{4}) and (\ref{5}),
\begin{align}
\gamma^{2}(\partial_{v}^{2}S^{\phi}-\kappa\partial_{v}S^{\phi}) =- \rho\Big(\frac{1}{2}\kappa \theta (\phi)^{2}+\frac{\kappa\theta(\phi)\gamma(\phi)^{'}}{\gamma(\phi)}-\kappa\theta(\phi)^{'}+\frac{1}{2}\kappa\partial_{v}\lambda-\frac{1}{2}\partial^{2}_{v}\lambda\Big) S^{\phi}+\mathcal{O}(\kappa^2).
\end{align}
Integrating this twice, we get
\begin{align}
\Delta S^{\phi}(1-\kappa\,\Delta v)=\frac{\rho}{\gamma^2}\Big(\frac{1}{2}\,\Delta\lambda-\frac{1}{2}\,\kappa\,\Delta v\, \Delta\lambda-\kappa (\Delta v)^2 \Theta(\phi)\Big)S^{\phi}+\mathcal{O}(\kappa^2),
\end{align}\footnote{$\Delta v=v_f-v_i$ as defined earlier.}
where
\begin{align}
    \Theta(\phi)=\frac{\theta(\phi)^2}{2}+\frac{\theta(\phi)\gamma(\phi)^{'}}{\gamma(\phi)}-\theta(\phi)^{'}.
\end{align}
Finally, we get
\begin{align}
    \Delta S^{\phi}=\frac{\rho}{\gamma^2}\Big(\frac{1}{2}\,\Delta\lambda-\kappa (\Delta v)^2 \Theta(\phi)\Big)S^{\phi}+\mathcal{O}(\kappa^2),\label{7b}
\end{align}
Now from (\ref{7}) up to $\mathcal{O}(\kappa)$, integrating twice, we obtain,
\begin{align}
\Delta\lambda\simeq 2\gamma^2\Big[-s\,\Delta v\, g(\phi)+\kappa\,(\Delta v)^2\, (\mathcal{F}+3\,s\,g(\phi) )\Big]. \label{7a}
\end{align}
Inserting Eq. (\ref{7a}) in (\ref{7b}) describes an expression of a shift in the deviation vector due to passage of a gravitational wave pulse near the asymptotic horizon of a non-extreme black hole. This implies the equivalent expression of memory as obtained for extreme consideration (\ref{extreme3D}) and mimics the one originally obtained at asymptotic null infinity (\ref{asymdev}).

Now let us look at the connection with BMS symmetries for non-extreme case. We shall adopt the similar approach as the one which has already been shown for extreme case. The general solution for $\lambda(v,\phi)$ can be written as
\begin{align}
    \lambda(v,\phi)=c(\phi)+\tilde{g}(\phi)v+\kappa\mathcal{\tilde{F}}(\phi)v\Delta v,
\end{align}
The $v$ component of the Killing vector for the fixed temperature configuration is given by Eq.(\ref{fxdT3D}). With these considerations, the variation of $\lambda(v,\phi)$ along Killing direction gives a condition generated from the $v$ coefficient. This makes sure that the ${\cal L}_{Z}\lambda$ remains independent of $v$. The condition is given by
\begin{align}
    (2\tilde{g}+\kappa\mathcal{\tilde{F}}\Delta v)\partial_{\phi}Y+(Y\partial_{\phi}\tilde{g}+\kappa\Delta v \partial_{\phi}\mathcal{\tilde{F}})-2\kappa\theta\partial_{\phi}X+2\kappa\partial^{2}_{\phi}X-2\kappa\frac{\partial_{\phi}\gamma(\phi)}{\gamma(\phi)}\partial_{\phi}X = 0 , \label{3dnxt2}
\end{align}
where $\tilde{g}$, $\mathcal{\tilde{F}}$, $Y$, $\theta$ and $X$ are functions of $\phi$. This condition also involves the superrotation $Y$ and supertranslation $X$. Further, since $X$ does not appear in the charge of BMS symmetries \cite{Donnay2016}, then it can be considered to be zero. This simplifies the Eq. (\ref{3dnxt2}), and reduces to
\begin{align}
    (2\tilde{g}+\kappa\mathcal{\tilde{F}}\Delta v)\partial_{\phi}Y+(Y\partial_{\phi}\tilde{g}+\kappa\Delta v \partial_{\phi}\mathcal{\tilde{F}}) = 0 , \label{3dnxtn2}
\end{align}
However, except the $v$ coefficient, the rest of the terms in Eq. (\ref{sup1}) would contain only functions of $\phi$. In this way a connection between BMS parameters and displacement memory can be established.

\section{Near-Horizon Memory for Four-Dimensional Black Holes}\label{4D}
\begin{figure}[h!]\centering
    \includegraphics[height=7.5cm, width=7.2cm]{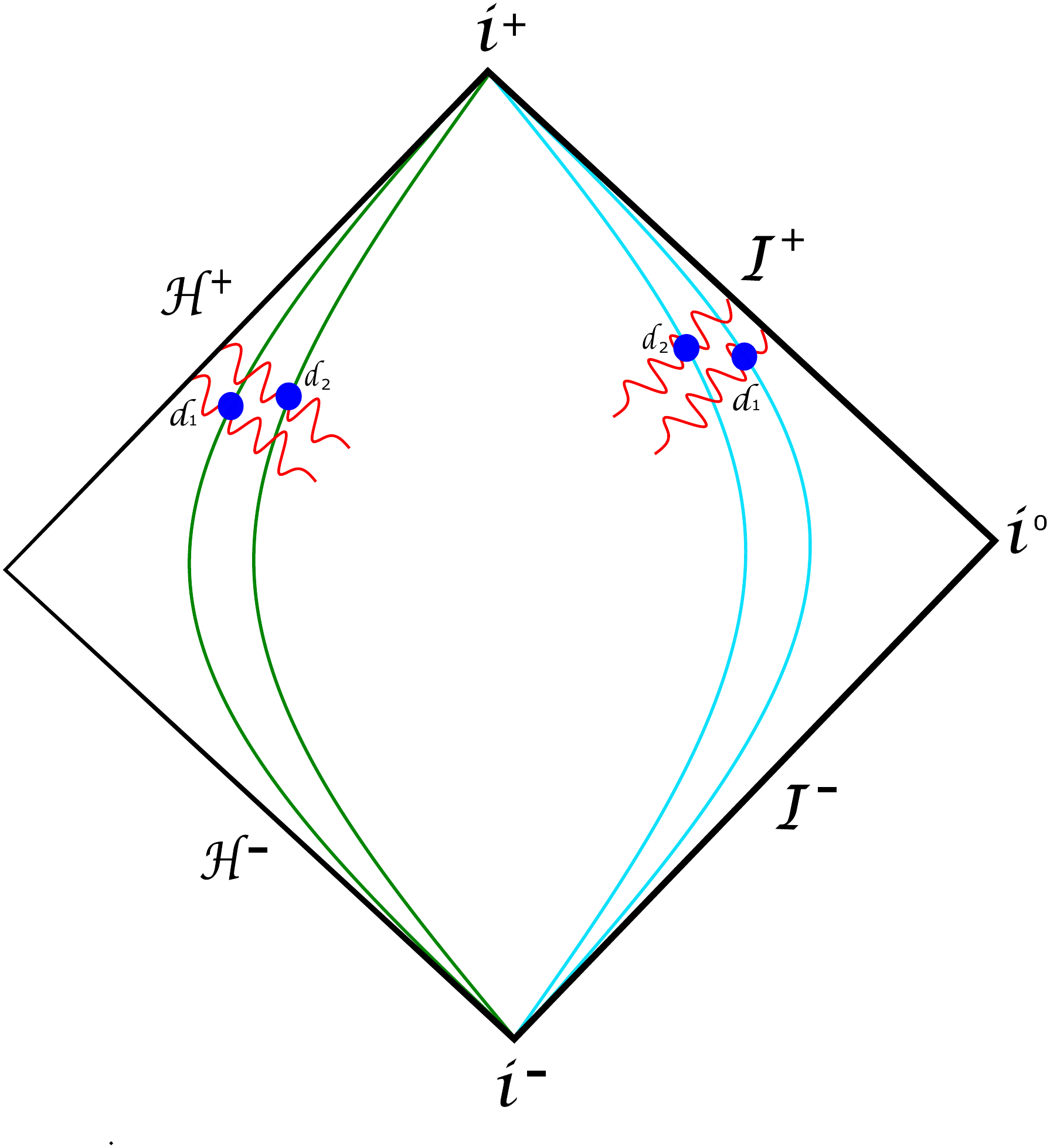} 
    \caption{A schematic diagram representing the detectors setup near the horizon $\mathcal{H}^{+}$ of a black hole and null Infinity $I^{+}$. The interaction between GWs and detectors near the horizon of a black hole generates a permanent relative change in the displacement vector which analogous to the one obtained at null infinity $I^{+}$.}\label{dmfa}
\end{figure}

In this section, we consider a realistic scenario by extending our analysis to four-dimensional black holes. We shall again adopt a similar strategy as implemented in section (\ref{3D}). We consider the same configuration, i.e., two nearby timelike geodesics or test detectors are being positioned near the black hole horizon. The schematic diagram (\ref{dmfa}) shows a relative setups at $I^{+}$ and $\mathcal{H}^{+}$. 
The basic construction of a near-horizon four-dimensional black hole metric carries the similar concepts as discussed in section (\ref{3D}). Let us start by writing the  general form of the near-horizon four-dimensional metric \cite{PhysRevLett.116.091101,Donnay2016,PhysRevD.98.124016}
\begin{align}
ds^{2} =& g_{vv} dv^{2}+2k dv d\rho +2g_{vA} dv dx^{A}+g_{AB}dx^{A}dx^{B}, \label{m}
\end{align}
with following fall-off conditions for the horizon $\rho = 0 $:
\begin{align}
    g_{vv} =& -2\kappa\rho + \mathcal{O}(\rho^{2}) \hspace{3mm} ; \hspace{5mm} k = 1+\mathcal{O}(\rho^{2}) \nonumber\\
    g_{vA} =& \rho \theta_{A}+\mathcal{O}(\rho^{2}) \hspace{8.3mm} ; \hspace{5mm} g_{AB} = \Omega \gamma_{AB}+\rho \lambda_{AB}+\mathcal{O}(\rho^{2}) \nonumber
\end{align}
where $\theta_{A}$, $\Omega$ and $\lambda^{AB}$ are functions of ($v,x^{A}$). We shall consider $\Omega$ to be independent of $v$. $\gamma_{AB}$ represents the 2-sphere metric. In stereographic coordinates, $x^{A}=(\zeta, \bar{\zeta})$, the spherical part of the metric is $\gamma_{AB} dx^{A} dx^{B} = \frac{4}{(1+\zeta\bar{\zeta})^{2}} d\zeta d\bar{\zeta}$.
While metric components $g_{\rho A}$ and $g_{\rho\rho}$ fall as $\mathcal{O}(\rho^{2})$.  
The asymptotic Killing vectors preserving fall-off boundary conditions are given by 
\begin{align}
\chi^{v} =& f(v,x^{A}) \nonumber \\
\chi^{\rho} =& -\partial_{v}f \rho + \frac{1}{2}g^{AB}\theta_{A}\partial_{B}f\rho^{2}+\mathcal{O}(\rho^{3}) \label{killing4D}\\
\chi^{A} =& Y^{A}(x^{A})+g^{AC}\partial_{C}f \rho+\frac{1}{2}g^{AD}g^{CB}\lambda_{DB}\partial_{C}f \rho^{2}+\mathcal{O}(\rho^{3}), \nonumber 
\end{align}
where $f(v,x^{A})$ and $Y^A(x^{A})$ are arbitrary functions, and $g^{AB}$ is inverse of $g_{AB}$. $Y^{A}$ is a function of $x^{A}$ only, i.e. $Y^{\zeta} = Y^{\zeta}(\zeta)$ and $Y^{\bar{\zeta}} = Y^{\bar{\zeta}}(\bar{\zeta})$. One can find the transformation of the functions $\theta_{A}, \lambda_{AB}, g_{AB}$ and $\kappa$, and can establish the algebra between supertranslation and superrotation as an  extension of three-dimensional analysis. Further, for $\kappa$ constant together with $\partial_{v}\Omega =0$, the conserved charge at the horizon turns out to be
\begin{align}
    \mathcal{Q}(T,Y^{A}) = \frac{1}{16\pi G}\int d\zeta d\bar{\zeta}\sqrt{\gamma}\Omega  \Big(2\kappa T-Y^{A}\theta_{A}\Big). \label{chrg1d4}
\end{align}
They close under the Poisson bracket given by
\begin{align}
    \{\mathcal{Q}(T_{1},Y^{A}_{1}),\mathcal{Q}(T_{2},Y^{A}_{2})\} = \mathcal{Q}(T_{12},Y^{A}_{12}). \label{4dxtnn}
\end{align}
Similarly, one also obtains for $\kappa = 0$
\begin{align}
    \mathcal{Q}(X,Y^{A}) = \frac{1}{16\pi G}\int d\zeta d\bar{\zeta}\sqrt{\gamma}\Big(2X\Omega -Y^{A}\theta_{A}\Omega \Big)+\mathcal{Q}_{0}, \label{extchrd4d}
\end{align}
where $T$ and $X$ are set of supertranslations and $Y^{A}$ denotes superrotation. The explicit details can be found in \cite{PhysRevLett.116.091101,Donnay2016,PhysRevD.98.124016}. Next we consider the memory effect analysis for non-extreme and extreme considerations.

\subsection{Non-Extremal Case}\label{4des}

Let us examine the non-extreme case first. We again consider the fixed temperature configuration for the full metric. The deviation equation for the metric (\ref{m}) is
\begin{align}
\frac{2\Omega}{(1+\zeta\bar{\zeta})^{2}}(\partial^{2}_{v}S^{\bar{\zeta}}-\kappa\partial_{v}S^{\bar{\zeta}}) = -(R_{\zeta v\bar{\zeta} v} S^{\bar{\zeta}}+R_{\zeta v\zeta v} S^{\zeta}), \label{ddn21n}
\end{align}
where $S^{\bar{\zeta}}$ and $S^{\zeta}$ depict the $\bar{\zeta}$ and $\zeta$ components of the deviation vector and $\Omega$ is function of ($\zeta,\bar{\zeta}$). The  corresponding Riemann tensor components to the linear order in $\kappa$ and $\rho$ are given by
\begin{align}
R_{\zeta v\bar{\zeta}v} =& \frac{1}{2} \rho (\partial_{v}\partial_{\bar{\zeta}}\theta_{\zeta}+\partial_{v}\partial_{\zeta}\theta_{\bar{\zeta}}-\kappa (\partial_{\bar{\zeta}}\theta_{\zeta}-\theta_{\zeta} \theta_{\bar{\zeta}}+\partial_{\zeta}\theta_{\bar{\zeta}})+\kappa\partial_{v}\lambda_{\zeta\bar{\zeta}}-\partial^{2}_{v}\lambda_{\zeta\bar{\zeta}})+ \mathcal{O}(\rho^{2})\\
R_{\zeta v\zeta v} =& \frac{\rho}{2 (\zeta\bar{\zeta}+1) \Omega} (-2 (\zeta\bar{\zeta}+1) \partial_{\zeta}\Omega \partial_{v}\theta_{\zeta}+2\kappa ((\zeta\bar{\zeta}+1) \partial_{\zeta}\Omega-2 \bar{\zeta} \Omega) \theta_{\zeta}+ \nonumber \\
& \kappa (\zeta\bar{\zeta}+1) \Omega \theta_{\zeta}^2+\Omega(-2\kappa (1 + \zeta\bar{\zeta}) \partial_{\zeta}\theta_{\zeta}+4 \bar{\zeta} \partial_{v}\theta_{\zeta}+2 (1+\zeta\bar{\zeta}) \partial_{v}\partial_{\zeta}\theta_{\zeta}+\nonumber \\
& \kappa (1+\zeta\bar{\zeta})\partial_{v}\lambda_{\zeta\zeta}-(1+\zeta\bar{\zeta})\partial^{2}_{v}\lambda_{\zeta\zeta}))+\mathcal{O}(\rho^{2}). \label{rmn4dfl}
\end{align}
Therefore, the change in the deviation vector can now be written as
\begin{align}
\frac{2\Omega}{(1+\zeta\bar{\zeta})^{2}}\Big(1-\kappa\Delta v\Big)\Delta S^{\bar{\zeta}} = -\int\int(R_{\zeta v\zeta v}S^{\zeta}+R_{\zeta v\bar{\zeta}v}S^{\bar{\zeta}})dv dv . \label{gnd2}
\end{align}
Note that with fall-off $\mathcal{O}(\rho^{2})$
\begin{align}
    \int\int R_{\zeta v\bar{\zeta}v} dv dv = \frac{\rho}{2}\Big((\Delta (\partial_{\bar{\zeta}}\theta_{\zeta})+\Delta (\partial_{\zeta}\theta_{\bar{\zeta}}))\Delta v-\kappa\mathcal{H}(\zeta,\bar{\zeta})(\Delta v)^{2}+\kappa\Delta\lambda_{\zeta\bar{\zeta}}\Delta v-\Delta\lambda_{\zeta\bar{\zeta}}\Big), \label{rmn4dfull}
\end{align}
with $\mathcal{H}(\zeta,\bar{\zeta})=(\partial_{\bar{\zeta}}\theta_{\zeta}-\theta_{\zeta} \theta_{\bar{\zeta}}+\partial_{\zeta}\theta_{\bar{\zeta}})$. Similarly, one can now get an expression for integral of $R_{\zeta v\zeta v}$. One can also compute the $\Delta S^{\zeta}$ component of the displacement vector adopting the similar approach. We are concentrating on $\Delta S^{\bar{\zeta}}$ component throughout the draft. In general, the Eq. (\ref{gnd2}) will contain changes `$\Delta$' of $\theta_{\zeta}$, $\lambda_{\zeta\zeta}$ and $\lambda_{\zeta\bar{\zeta}}$ together with the changes in their derivatives with respect to $v$. We can see that all metric parameters would contribute in the displacement memory. These changes in the metric parameters can be determined using Einstein field equations. The relevant component of the field equation is given by
\begin{align}
8\pi T^{M}_{vv}=& \frac{\rho}{2 \Omega^3} ((\zeta\bar{\zeta}+1)^2 \Omega^2 (-\partial_{v}\theta_{\zeta}\theta_{\bar{\zeta}}-\kappa \partial_{\bar{\zeta}}\theta_{\zeta} +\partial_{v}\partial_{\bar{\zeta}}\theta_{\zeta}+\theta_{\zeta}(2 \kappa  \theta_{\bar{\zeta}}-\partial_{v}\theta_{\bar{\zeta}})- \kappa \partial_{\zeta}\theta_{\bar{\zeta}}+ \nonumber \\
& \partial_{v}\partial_{\zeta}\theta_{\bar{\zeta}}+\kappa \partial_{v}\lambda_{\zeta\bar{\zeta}}-\partial^{2}_{v}\lambda_{\zeta\bar{\zeta}})-\kappa (\Omega^2 (3 (\zeta\bar{\zeta}+1)^2 \theta_{\zeta} \theta_{\bar{\zeta}}-2 ((\zeta\bar{\zeta}+1)^2 \partial_{\bar{\zeta}}\theta_{\zeta} \nonumber \\
& +(\zeta\bar{\zeta}+1)^2 \partial_{\zeta}\theta_{\bar{\zeta}}- 2 ((\zeta\bar{\zeta}+1)^2 \partial_{v}\lambda_{\zeta\bar{\zeta}}-1))+4 \kappa (\zeta\bar{\zeta}+1)^2 \lambda_{\zeta\bar{\zeta}})- \nonumber \\
& 2 (\zeta\bar{\zeta}+1)^2 \partial_{\bar{\zeta}}\Omega \partial_{\zeta}\Omega+2 (\zeta\bar{\zeta}+1)^2 \Omega \partial_{\zeta}\partial_{\bar{\zeta}}\Omega))+\mathcal{O}(\rho^{2}). \label{efe4dfl}
\end{align}
Considering again the shockwave profile of the stress-tensor as $T_{vv}=\lim_{\rho\to 0}\frac{8\pi T^{M}_{vv}}{\rho}$, we could determine the required changes for the Eq. (\ref{gnd2}). Now, using the $vA$-component of the Einstein field equations, we get the following conditions to the $\mathcal{O}(\rho^{0})$  
\begin{align}
    \partial_{v}\theta_{A} = 0 \hspace{2mm} \Rightarrow \hspace{2mm} \theta_{A} = \mathcal{C}(x^{A}). \label{thtacondn}
\end{align}
Further, for computational simplification, we also consider $\Omega$ to be unity. Under these considerations, the calculation reduces significantly. The Riemann tensor components are given by
\begin{align}
\int\int R_{\zeta v\zeta v} dv dv = \frac{\rho}{2}\Big(\Delta\lambda_{\zeta\zeta} (\kappa\Delta v-1)-\kappa \mathcal{G}(\zeta,\bar{\zeta})(\Delta v)^{2}\Big), \label{4dr1}
\end{align}
where $ \mathcal{G}(\zeta,\bar{\zeta})=2\partial_{\zeta}\theta_{\zeta}+\theta_{\zeta}^{2}+\frac{4\bar{\zeta}\theta_{\zeta}}{(1+\zeta\bar{\zeta})} $, and
\begin{align}
\int\int R_{\zeta v\bar{\zeta}v} dv dv = \frac{\rho}{2}\Big(\Delta\lambda_{\zeta\bar{\zeta}}(\kappa\Delta v-1)-\kappa\mathcal{H}(\zeta,\bar{\zeta})(\Delta v)^{2}\Big). \label{4dr2}
\end{align}
This enables us to rewrite the GDE solution (\ref{gnd2}) in a compact form, 
\begin{align}
    \Delta S^{\bar{\zeta}} = \frac{\rho (1+\zeta\bar{\zeta})^{2}}{4}\Big((\Delta\lambda_{\zeta\bar{\zeta}}S^{\bar{\zeta}}+\Delta\lambda_{\zeta\zeta}S^{\zeta})-\kappa (\Delta v)^{2}(\mathcal{H}S^{\bar{\zeta}}+\mathcal{G}S^{\zeta})\Big)+\mathcal{O}(\rho^{2}). \label{ngd3n}
\end{align}
Thus we now only need to find the changes in $\lambda_{AB}$ in order to solve the Eq. (\ref{ngd3n}). This requires to solve the EFE as we did for three dimensional case. The $vv$-component EFE is
\begin{align}
8\pi T^{M}_{vv} = \frac{\rho}{2}  (\zeta\bar{\zeta}+1)^2 (\kappa (-3 \partial_{v}\lambda_{\zeta\bar{\zeta}}+\partial_{\bar{\zeta}}\theta_{\zeta}-\theta_{\zeta} \theta_{\bar{\zeta}}+\partial_{\zeta}\theta_{\bar{\zeta}})-\partial^{2}_{v}\lambda_{\zeta\bar{\zeta}}+\frac{4\kappa }{(\zeta\bar{\zeta}+1)^2})+\mathcal{O}(\rho^{2}). \label{nefe4d}
\end{align}
Now again defining $T_{vv}=\lim_{\rho\to 0}\frac{8\pi T^{M}_{vv}}{\rho}=aQ(\zeta,\bar{\zeta})\delta(v-v_{0})$, with some constant $a$, we get
\begin{align}
\Delta\lambda_{\zeta\bar{\zeta}} = \Big(\kappa\tilde{G}\Delta v-\frac{2aQ}{(1+\zeta\bar{\zeta})^{2}}\Big)\Delta v,
\end{align}
where $\tilde{G}(\zeta,\bar{\zeta})=\mathcal{H}(\zeta,\bar{\zeta})+\frac{4}{(1+\zeta\bar{\zeta})^{2}}+\frac{6a\Delta v}{(1+\zeta\bar{\zeta})^{2}}Q(\zeta,\bar{\zeta})$. Further, for determining $\Delta\lambda_{\zeta\zeta}$, we use $\zeta\zeta$-component of the field equation. We get an $\mathcal{O}(\rho^{0})$ term together with a linear order correction in $\rho$. The $\mathcal{O}(\rho^{0})$ term gives us
\begin{align}
\partial_{v}\lambda_{\zeta\zeta}+\kappa\lambda_{\zeta\zeta}=\frac{2\bar{\zeta}\theta_{\zeta}}{(1+\zeta\bar{\zeta})}-\frac{1}{2}\theta_{\zeta}^{2}. \label{nefe4dr}
\end{align}
This generates the change in $\lambda_{\zeta\zeta}$ with $\tilde{A}= \frac{2\bar{\zeta}\theta_{\zeta}}{(1+\zeta\bar{\zeta})}-\frac{1}{2}\theta_{\zeta}^{2}$.
\begin{align}
    \lambda_{\zeta\zeta} = \frac{1}{\kappa}\tilde{A}+C_{1}(\zeta,\bar{\zeta})(1-v\kappa) \Rightarrow \Delta\lambda_{\zeta\zeta} = \kappa \tilde{B}(\zeta,\bar{\zeta})\Delta v ,
\end{align}
where $C_{1}(\zeta,\bar{\zeta})$ appears as a integration constant. As $\Delta\lambda_{\zeta\zeta}$ and $\Delta\lambda_{\zeta\bar{\zeta}}$ are contained in Eq. (\ref{ngd3n}), we can replace these changes to obtain an explicit form of the memory. 
\begin{align}
    \Delta S^{\bar{\zeta}} = \frac{\rho (1+\zeta\bar{\zeta})^{2}}{4}\Big(\Big((\kappa\tilde{G}\Delta v-\frac{2aQ}{(1+\zeta\bar{\zeta})^{2}}) S^{\bar{\zeta}}+\kappa \tilde{B}S^{\zeta}\Big)\Delta v-\kappa (\Delta v)^{2}(\mathcal{H}S^{\bar{\zeta}}+\mathcal{G}S^{\zeta})\Big)+\mathcal{O}(\rho^{2}). \label{ngd3nn}
\end{align}
This is the four-dimensional version of displacement memory near the horizon of black holes similar to the one discussed in the previous section. We notice that the displacement memory $\Delta S^{\bar{\zeta}}$ has now been induced by the both components of the deviation vector ($S^{\zeta},S^{\bar{\zeta}}$). We also observe that there is a direct relation between $\lambda_{AB}$ and $\theta_{A}$  from Eq. (\ref{nefe4d}) or Eq. (\ref{nefe4dr}). This also suggests that the displacement memory is ultimately restored in the changes of metric parameters $\theta_{A}$. This completes our analysis of achieving the displacement memory for near the horizon of non-extremal black holes. \\

\textbf{Relation with BMS Symmetry}\\ \\
Let us now establish the connection between displacement memory and BMS symmetries in the vicinity of black hole horizon generated by shock wave profile.  
We have considered here the transformations in $\lambda_{\zeta\bar{\zeta}}$, $\lambda_{\zeta\zeta}$ and $\lambda_{\bar{\zeta}\bar{\zeta}}$. Hence, the general solution for $\lambda_{\zeta\bar{\zeta}}$, $\lambda_{\zeta\zeta}$ and $\lambda_{\bar{\zeta}\bar{\zeta}}$ can be written as 
\begin{align}
    \lambda_{\zeta\bar{\zeta}} =& A(\zeta,\bar{\zeta})+B(\zeta,\bar{\zeta})v+\kappa\tilde{\mathcal{K}}(\zeta,\bar{\zeta})v\Delta v \\
  	\lambda_{\zeta\zeta} =& \tilde{A}(\zeta,\bar{\zeta})+v\tilde{B}(\zeta,\bar{\zeta}) \kappa \\
  	 \lambda_{\bar{\zeta}\bar{\zeta}} =& \tilde{W}(\zeta,\bar{\zeta})+v\kappa\tilde{Z}(\zeta,\bar{\zeta}). \label{nd4}
\end{align}
Therefore, as before the changes in various components of $\lambda_{AB}$ are given by
\begin{align}
\Delta \lambda_{\zeta\bar{\zeta}}= B(\zeta,\bar{\zeta})\Delta v+\kappa \tilde{\mathcal{K}}(\zeta,\bar{\zeta})(\Delta v)^{2} \hspace{1mm} ; \hspace{1mm} \Delta\lambda_{\zeta\zeta}=\kappa\tilde{B}(\zeta,\bar{\zeta})\Delta v \hspace{1mm} ; \hspace{1mm} \Delta\lambda_{\bar{\zeta}\bar{\zeta}}=\kappa\tilde{Z}(\zeta,\bar{\zeta})\Delta v . \label{dlmbda2} 
\end{align}
Recall $\kappa$ is small, and for a small $\Delta v$ we can disregard the term $\kappa (\Delta v)^2$. In this case, we are  considering only the part of $\Delta S^{\bar{\zeta}}$ in (\ref{ngd3nn}) that is induced by the changes in $\lambda_{AB}$. \par

Now consider the variation of $\lambda_{AB}(v,\zeta,\bar{\zeta})$ along Killing direction $\chi$ given by \cite{PhysRevD.98.124016} 
\begin{align}
{\cal L}_{\chi}\lambda_{AB} = f \partial_{v}\lambda_{AB}-\lambda_{AB}\partial_{v}f+\mathcal{L}_{Y}\lambda_{AB}+\theta_{A}\partial_{B}f+\theta_{B}\partial_{A}f-2\nabla_{A}\nabla_{B}f , \label{chi2}
\end{align}
where $\chi$ is the Killing vector. For non-extreme case, the $v$ component of the Killing vector for the fixed temperature configuration generates two sets of supertranslations $T(\phi)$ and $X(\phi)$,
\begin{align}
    f(v,\zeta,\bar{\zeta}) = T(\zeta,\bar{\zeta})+(1-\kappa v)X(\zeta,\bar{\zeta})+\mathcal{O}(\kappa^{2}). \label{k4d}
\end{align}
In order to make the variation of different components of $\lambda_{AB}$ along the Killing direction independent of $v$, we again set the $v$ coefficients to be zero. It is apparent from Eq. (\ref{nd4}) and Eq. (\ref{k4d}) the terms dependent on $v$ will only be linear in $v$. Considering $\mathcal{L}_{\chi}\lambda_{\zeta\bar{\zeta}}$, we get the following condition
\begin{align}
    \partial_{C}\Big((B+\tilde{\mathcal{K}}\kappa\Delta v)Y^{C}\Big)-(\theta_{\zeta}\partial_{\bar{\zeta}}X+\theta_{\bar{\zeta}}\partial_{\zeta}X)\kappa+2\kappa\nabla_{\zeta}\nabla_{\bar{\zeta}}X = 0. \label{bms4d1}
\end{align}
Similarly, considering $\mathcal{L}_{\chi}\lambda_{\zeta\zeta}$, we obtain
\begin{align}
    \kappa\Big( Y^{C}\partial_{C}\tilde{B}+2\tilde{B}\partial_{\zeta}Y^{\zeta}-2\theta_{\zeta}\partial_{\zeta}X+2\nabla_{\zeta}\nabla_{\zeta}X\Big) = 0, \label{bms4d2}
\end{align}
and $\mathcal{L}_{\chi}\lambda_{\bar{\zeta}\bar{\zeta}}$ gives
\begin{align}
    \kappa\Big(Y^{C}\partial_{C}\tilde{Z}+2\partial_{\bar{\zeta}}Y^{\bar{\zeta}}\tilde{Z}-2\theta_{\bar{\zeta}}\partial_{\bar{\zeta}}X-\nabla_{\bar{\zeta}}\nabla_{\bar{\zeta}}X\Big)=0. \label{bms4dnxt3}
\end{align}
In principle Eq. (\ref{bms4d1}), Eq. (\ref{bms4d2}) and Eq. (\ref{bms4dnxt3}) can be solved for $Y^{\zeta}$, $Y^{\bar{\zeta}}$ and $X$. These equations appear as conditions which make sure that the variation of $\lambda_{AB}$ will be independent of $v$. Further, if one freezes the supertranslation $X$\footnote{Since $X$ is absent in the leading order charge Eq.(\ref{chrg1d4}), this is a legitimate assumption. }, this reduces the above conditions in the following form
\begin{align}
    \partial_{C}\Big((B+\tilde{\mathcal{K}}\kappa\Delta v)Y^{C}\Big) = 0 \hspace{2mm};\hspace{2mm}
    \kappa( Y^{C}\partial_{C}\tilde{B}+2\tilde{B}\partial_{\zeta}Y^{\zeta}) = 0 \hspace{2mm};\hspace{2mm}
    \kappa\Big(Y^{C}\partial_{C}\tilde{Z}+2\partial_{\bar{\zeta}}Y^{\bar{\zeta}}\tilde{Z}\Big)=0 .
\end{align}
Using these set of equations one can obtain the solution for $Y^{\zeta}$
\begin{align}
    Y^{\zeta}(\zeta) = \tilde{a}e^{-\int\frac{\tilde{w}}{\tilde{p}}d\zeta} , \label{intnxt4d}
\end{align}
where $\tilde{w}$ and $\tilde{p}$ are functions of ($\zeta,\bar{\zeta}$) and $\tilde{a}$ appears as an integration constant. $\tilde{a}$ will be a function of $\bar{\zeta}$ only. Similarly, one can also find the solution for $Y^{\bar{\zeta}}(\bar{\zeta})$. 
Therefore, we can explicitly find some $Y^{A}$ that would induce the desired shift in the deviation vector components.
\subsection{Extremal Case}\label{nextml2}
Next, we analyze the extremal case in which the $g_{vv}$ component of the metric becomes $\mathcal{N}(\zeta,\bar{\zeta})\rho^{2}+\mathcal{O}(\rho^{3})$. We obtain the GDE as
\begin{align}
\frac{2}{(1+\zeta\bar{\zeta})^{2}}\partial^{2}_{v}S^{\bar{\zeta}}_{E} = -(R_{\zeta v\bar{\zeta} v} S^{\bar{\zeta}}_{E}+R_{\zeta v\zeta v} S^{\zeta}_{E}), \label{ddn21}
\end{align}
where $S^{\bar{\zeta}}_{E}$ and $S^{\zeta}_{E}$ depict the $\bar{\zeta}$ and $\zeta$ components of the deviation vector respectively for extremal case. The relevant components of the Riemann tensor are
\begin{align}
R_{\zeta v\bar{\zeta} v} = -\frac{\rho}{2}  \partial^{2}_{v}\lambda_{\zeta\bar{\zeta}} + \mathcal{O}(\rho^{2}) \hspace{2mm} ; \hspace{2mm} R_{\zeta v\zeta v} = -\frac{\rho}{2} \partial^{2}_{v}\lambda_{\zeta\zeta} + \mathcal{O}(\rho^{2}). \label{R1}
\end{align}
Therefore, the change in the deviation vector can now be written as
\begin{align}
\Delta S^{\bar{\zeta}}_{E} =\frac{\rho}{4}(1+\zeta\bar{\zeta})^{2}(\Delta\lambda_{\zeta\bar{\zeta}} S^{\bar{\zeta}}_{E}+\Delta\lambda_{\zeta\zeta} S^{\zeta}_{E})+\mathcal{O}(\rho^{2}). \label{ngd2}
\end{align}
This is similar to the results obtained in (\ref{ngd3nn}). Here, derivative of $\lambda_{AB}$ mimics the News tensor obtained in the far region case. Further, we again consider the shock wave profile of stress tensor as mentioned in the non-extreme case.  The relevant components of the EFE can be obtained by setting $\kappa=0$ in Eq. (\ref{ngd3n}) and Eq. (\ref{nefe4dr}) to give,
\begin{align}
    \Delta\lambda_{\zeta\bar{\zeta}} = \mathcal{P}(\zeta,\bar{\zeta})\Delta v \hspace{2mm} ; \hspace{2mm} \Delta\lambda_{\zeta\zeta} = \tilde{C}(\zeta,\bar{\zeta})\Delta v . \label{ext4ddlbda}
\end{align}
Eq. (\ref{ngd2}) and Eq. (\ref{ext4ddlbda}) indicate that we get a finite jump in the deviation vector depicting displacement memory. 

Here also, we may seek for BMS symmetry that induces the shift in $S^{A}_{E}$. This is again in principle be determined from (\ref{chi2}) by choosing $\lambda_{AB}$ linear in $v$. The entire change in $\lambda_{AB}$ then would be determined by functions that are dependent on $\zeta,\bar \zeta$ only. Further, for extreme case, the $\chi^{v}$ component contains two set of supertranslations, i.e. $f(v,\zeta,\bar{\zeta})=T(\zeta,\bar{\zeta})+vX(\zeta,\bar{\zeta})$. Let us take the general solution of $\lambda_{AB}$ for extremal consideration of the following form 
\begin{align}
    \lambda_{\zeta\bar{\zeta}} = \mathcal{A}(\zeta,\bar{\zeta})+v\mathcal{P}(\zeta,\bar{\zeta}) \hspace{2mm} ; \hspace{2mm} \lambda_{\zeta\zeta} = \mathcal{B}(\zeta,\bar{\zeta})+v\tilde{C}(\zeta,\bar{\zeta}). \label{lbdsol4d}
\end{align}
The $\bar{\zeta}\bar{\zeta}$-component of the EFE will generate a $\lambda_{\bar{\zeta}\bar{\zeta}}$, similar to $\lambda_{\zeta\zeta}$. Again, we set $v$ coefficient to be zero in order to make the variation of $\lambda_{AB}$ along the Killing direction to be independent of $v$. For $\lambda_{\zeta\bar{\zeta}}$, we obtain
\begin{align}
    \partial_{C}(Y^{C}\mathcal{P})-(\theta_{\zeta}\partial_{\bar{\zeta}}X+\theta_{\bar{\zeta}}\partial_{\zeta}X)-2\nabla_{\zeta}\nabla_{\bar{\zeta}} X = 0. \label{cndn2n4d}
\end{align}
and for $\lambda_{\zeta\zeta}$ variation
\begin{align}
    Y^{C}\partial_{C}\tilde{C}+2\tilde{C}\partial_{\zeta}Y^{\zeta}-2\theta_{\zeta}\partial_{\zeta}X-2\nabla_{\zeta}\nabla_{\zeta}X = 0. \label{cndn4d2n}
\end{align}
Similarly, variation in $\lambda_{\bar{\zeta}\bar{\zeta}}$ also generates a equation similar to Eq. (\ref{cndn4d2n}). These conditions should serve the purpose to obtain explicit solutions for $X$, $Y^{\zeta}$ and $Y^{\bar{\zeta}}$ that can generate variations in $\lambda_{AB}$ independent of $v$. However, obtaining an explicit solution for $X$ or $Y$ seems to be difficult in this case. Some insights can be gained considering a reduced form of the metric which we study in the next section.


\subsection{Memory for a less generic metric}\label{rdcdm}

In this section, we consider a reduced form of the metric (\ref{m}) where we set $g_{vA}=0$. This can be regraded as an asymptotic form of a metric near the horizon of a spherically symmetric black hole which is deformed in the spatial sector. The metric now reads
\begin{align}
  ds^{2} =& g_{vv} dv^{2}+2k dv d\rho+g_{AB}dx^{A}dx^{B}. \label{m1}
\end{align}

\subsubsection{Non-Extremal Case}\label{rdcdnxtm}

We consider the non-extreme case first for the above form of the metric. We shall be adopting the similar approach for computing the displacement memory as we have shown in section (\ref{4des}). We consider again the fixed temperature configuration, i.e. $\kappa$ to be constant. Let us write down the GDE for the metric (\ref{m1})
\begin{align}
\frac{2}{(1+\zeta\bar{\zeta})^{2}}(\partial^{2}_{v}S^{\bar{\zeta}}-\kappa\partial_{v}S^{\bar{\zeta}}) = -(R_{\zeta v\bar{\zeta} v} S^{\bar{\zeta}}+R_{\zeta v\zeta v} S^{\zeta}), \label{ddn21nn}
\end{align}
This again mimics the expression (\ref{ngd3nn}) where both components ($S^{\zeta},S^{\bar{\zeta}}$) of the deviation vector are contributing. Therefore, the change in the deviation vector can now be written as
\begin{align}
\Delta S^{\bar{\zeta}} =\frac{\rho}{4}(1+\zeta\bar{\zeta})^{2}(\Delta\lambda_{\zeta\bar{\zeta}} S^{\bar{\zeta}}+\Delta\lambda_{\zeta\zeta} S^{\zeta})+\mathcal{O}(\rho^{2}). \label{ngd2nn}
\end{align}
The $vv$-component of the EFE is given by
\begin{align}
- \frac{1}{2}\left(\zeta\bar{\zeta} + 1\right)^{2} \Big(\left( 3\kappa \partial_{v} \lambda_{\zeta\bar{\zeta}} + \partial^{2}_{v} \lambda_{\zeta\bar{\zeta}} \right) + 2 \kappa\Big)\rho  - T^{M}_{vv}{\left(v,\zeta, \bar{\zeta} \right)} = 0. \label{En1n}
\end{align}
With a shockwave profile $T_{vv}=\lim_{\rho\to 0}\frac{8\pi T^{M}_{vv}}{\rho}=aQ(\zeta,\bar{\zeta})\delta(v-v_{0})$, for some constant $a$, we get
\begin{align}
\Delta \lambda_{\zeta\bar \zeta} = -\frac{2\, a}{(1+\zeta\bar \zeta)^2}\Delta v\,(1- 3\,\kappa\,\Delta v)\, Q(\zeta,\bar \zeta)+\frac{4\,\kappa\,(\Delta v)^2}{(1+\zeta\bar \zeta)^2}. \label{RR4nn}
\end{align}
Further, the $\zeta\zeta$-component of the EFE one gets
\begin{align}
\Delta\lambda_{\zeta\zeta}=\kappa\mathcal{W}(\zeta,\bar{\zeta})\Delta v .\label{lbda2rdcn}    
\end{align}
Hence, from Eq. (\ref{ngd2nn}), Eq. (\ref{RR4nn}) and Eq. (\ref{lbda2rdcn}), we find that the memory can be associated with $\Delta\lambda_{\zeta\bar{\zeta}}$ and $\Delta\lambda_{\zeta\zeta}$ to the linear order in $\rho$. \\

\textbf{Relation with BMS Symmetry}\\ \\
To relate it with BMS symmetry, we have the general solutions of $\lambda_{AB}$ of the same form as it appears for non-extreme case in section (\ref{4des}). 
\begin{align}
    \lambda_{\zeta\bar{\zeta}} = A(\zeta,\bar{\zeta})+B(\zeta,\bar{\zeta})v+\kappa\mathcal{C}(\zeta,\bar{\zeta})v\Delta v \hspace{2mm} ; \hspace{2mm} \lambda_{\zeta\zeta} = \tilde{A}(\zeta,\bar{\zeta})+v\tilde{B}(\zeta,\bar{\zeta}) \kappa \label{lbdardcd4d}
\end{align}
This generates the same conditions as appeared in section (\ref{4des}). We obtain the similar form of the solution as mentioned in Eq. (\ref{intnxt4d}).

\subsubsection{Extremal Case}\label{xnt4nd}
Let us now consider the extremal case. As we have seen the extremal analysis for the full metric in (\ref{nextml2}), the $g_{vv}$ component would change by a function of ($\zeta,\bar{\zeta}$) to the $\rho^{2}$ order. Under this consideration, the deviation equation for the metric (\ref{m1}) reads
\begin{align}
\frac{2}{(1+\zeta\bar{\zeta})^{2}}\partial^{2}_{v}S^{\bar{\zeta}}_{E} = -(R_{\zeta v\bar{\zeta} v} S^{\bar{\zeta}}_{E}+R_{\zeta v\zeta v} S^{\zeta}_{E}), \label{dd21}
\end{align}
The change in the deviation vector can now be written as
\begin{align}
\Delta S^{\bar{\zeta}}_{E} =\frac{\rho}{4}(1+\zeta\bar{\zeta})^{2}(\Delta\lambda_{\zeta\bar{\zeta}} S^{\bar{\zeta}}_{E}+\Delta\lambda_{\zeta\zeta} S^{\zeta}_{E})+\mathcal{O}(\rho^{2}). \label{gd2}
\end{align}
From the $vv$-component of EFE we get 
\begin{align}
-\frac{\rho}{2}(1+\zeta\bar{\zeta})^{2}\partial^{2}_{v}\lambda_{\zeta\bar{\zeta}} = 8\pi T^{M}_{vv}. \label{ffd2}
\end{align}
Using the same definition of stress-energy tensor as mentioned in non-extreme case, we get 
\begin{align}
\Delta \lambda_{\zeta\bar{\zeta}} = H(\zeta,\bar{\zeta})\Delta v . \label{n4dnew}
\end{align}
From $\zeta\zeta$-component of the EFE to the $\mathcal{O}(\rho^{0})$, $\lambda_{\zeta\zeta}$ turns out to be a function of $(\zeta,\bar{\zeta})$ only. Hence, there will be no change induced by $\lambda_{\zeta\zeta}$, and Eq. (\ref{gd2}) becomes
\begin{align}
    \Delta S^{\bar{\zeta}}_{E} =\frac{\rho}{4}(1+\zeta\bar{\zeta})^{2}H(\zeta,\bar{\zeta}) \Delta v S^{\bar{\zeta}}_{E}+\mathcal{O}(\rho^{2}). \label{mxt4d}
\end{align}
It matches closely to the expression (5.1.4) of far region analysis as oppose to the results of non-extreme cases, where two components of the deviation vector contribute. Let us now look at the relation with BMS symmetry for extremal consideration. It is to note that $\mathcal{O}(\rho^{0})$ term of $\bar{\zeta}\bar{\zeta}$-component of the EFE also turns out to be function of $(\zeta,\bar{\zeta})$ only. Hence, the change in diagonal components of $\lambda_{AB}$ with respect to $v$ does not contribute in the memory. However, general solutions of $\lambda_{\zeta\zeta}$ and $\lambda_{\bar{\zeta}\bar{\zeta}}$ can still contribute while relating to BMS symmetry. The general solution for the off-diagonal component of $\lambda_{AB}$ can be written as $\lambda_{\zeta\bar{\zeta}}(v,\zeta,\bar{\zeta}) = \tilde{P}(\zeta,\bar{\zeta})+vH(\zeta,\bar{\zeta})$. While $\lambda_{\zeta\zeta}$ and $\lambda_{\bar{\zeta}\bar{\zeta}}$ are of the form $\lambda_{\zeta\zeta}=\tilde{E}(\zeta,\bar{\zeta})$ and $\lambda_{\bar{\zeta}\bar{\zeta}}=\tilde{W}(\zeta,\bar{\zeta})$. To have the desired change as depicted in Eq. (\ref{n4dnew}), we obtain the following condition for $\lambda_{\zeta\bar{\zeta}}$
\begin{align}
    \partial_{C}(Y^{C}H)-2\nabla_{\zeta}\nabla_{\bar{\zeta}}X = 0. \label{diff4d2}
\end{align}
This can further be written as
\begin{align}
    \partial_{C}(Y^{C}H)-2\Big(\partial_{\zeta}\partial_{\bar{\zeta}}X-(\Gamma^{\zeta}{}_{\zeta\bar{\zeta}}\partial_{\zeta}X+\Gamma^{\bar{\zeta}}{}_{\zeta\bar{\zeta}}\partial_{\bar{\zeta}}X)\Big) = 0, \label{diff4d22}
\end{align}
where
\begin{align}
    \Gamma^{\zeta}{}_{\zeta\bar{\zeta}}=\frac{\rho}{4} (1+\zeta\bar{\zeta})^{2}\partial_{\zeta}\lambda_{\bar{\zeta}\bar{\zeta}} +\mathcal{O}(\rho^{2}) \hspace{2mm} ; \hspace{2mm} \Gamma^{\bar{\zeta}}{}_{\zeta\bar{\zeta}} = \frac{\rho}{4} (1+\zeta\bar{\zeta})^{2}\partial_{\bar{\zeta}}\lambda_{\zeta\zeta}+\mathcal{O}(\rho^{2}). \label{chrst}
\end{align}
To the order $\mathcal{O}(\rho^{0})$, the Eq. (\ref{diff4d22}) becomes
\begin{align}
    \partial_{C}(Y^{C}H)-2\partial_{\zeta}\partial_{\bar{\zeta}}X = 0. \label{rd4d}
\end{align}
This looks to be a simpler condition to be satisfied by $X$ and $Y^{A}$. Still, getting explicit expressions is not an easy task. However, if we freeze the superrotation of the Eq. (\ref{rd4d}) reduces to,  $\partial_{\bar{\zeta}}\partial_{\zeta}X(\zeta,\bar{\zeta})=0$ \footnote{This is legitimate as the 4D charge expression contains supertranslation and superrotation separately Eq. (\ref{extchrd4d})}. For which, we can obtain an exact solution, i.e., $X(\zeta,\bar{\zeta})=X_{1}(\zeta)+X_{2.}(\bar{\zeta})$. Hence, there is a  supertranslation $X(\zeta,\bar{\zeta})$ that can induce the displacement memory effect depicted in Eq. (\ref{mxt4d}). We also notice here, the form of the equation (\ref{ffd2}) is quite similar that was obtained in the far region in \cite{strominger2018lectures}. The extremal case for this reduced metric mimics the memory or shift in the far region  \cite{PhysRevD.98.124016, 2020n1}.
 
Finally, we would like to mention here that the metric considered here generates gravitational waves which can be captured in the Newman-Penrose (NP) scalar $\psi_4.$ This is displayed in the appendix (\ref{cds}). Although $\lambda_{AB}$ are not explicitly present in the charge  Eq. (\ref{extchrd4d}) but $\psi_4$ contains $\lambda_{AB}$ and their derivatives, ensuring the physicality and relevance of $\lambda_{AB}$. 

\section{Discussion}\label{dscsn}
The primary objective of this report is to study how GWs affect inertial test detectors near the horizon of black hole spacetimes. This analysis could serve as a model for estimating the deviation between the test masses of a GW detector. We also provide an explicit connection between BMS symmetries and memory effect near the horizon of black holes.\par
There are crucial differences between memory effect and asymptotic symmetries obtained in the future null infinities and the same obtained near the horizon of a black hole. As the algebra of asymptotic symmetries contains two copies of Virasoro algebra, we have two sets of supertranslation and a set of superrotation generators near the horizon of a four-dimensional black hole. The extra set of supertranslation generator is something that we do not recover at the null infinities. Another crucial difference seems to be in the structure of the geodesic deviation equation (GDE). The near-horizon GDE contains a linear derivative term of the deviation vector with respect to time apart from a double-time derivative, whereas the far region GDE contains only a double-time derivative of deviation vector to the leading order. The function $\lambda(\lambda _{AB}$ in four dimensions) mimics the derivative of News tensor $C_{zz}$ in the far region.  The Riemann tensor in GDE also contains a double derivative and a single derivative of $\lambda$, which is absent in the GDE of far region (\ref{asymdev}). These differences make the analyses of near-horizon memory significantly non-trivial than the far region case. We also see the memory in the near region of black holes is due to the dual effect of supertranslation and superrotation generators. For the sake of getting an explicit expression of BMS generators that would generate the displacement memory, we have considered one type of generator (either superrotations or supertranslations) at a time for extremal cases. Although it was not easy to obtain an explicit solution but   considering a slightly less generic metric, we could recover closed form solutions for BMS parameters. For this extremal case, the GDE governing the displacement memory becomes almost identical to the structure of the same in the far region.  The full non-extremal case seems to be analytically difficult to analyze. We have provided an analytic expression for the memory effect in small $\kappa$ approximation. We wish to report an analysis without such approximation in future.     \par 

In the near future, the observational aspects of GW will serve a great purpose in measuring the memory effect, and this could allow us to examine BMS symmetries in more depth as a direct evidence \cite{Khera:2020mcz}. The detection of near-horizon memory in the far region may be possible if the test detectors or particles near the black hole region interact with some matter or radiation that can be detected in the far region and encode the near-horizon physics including memory. It will be interesting to investigate the existence of such a phenomenon.

\chapter{Prospects of detecting supertranslation hair}\label{CH6}

\ifpdf
    \graphicspath{{Chapter6/Figs/Raster/}{Chapter6/Figs/PDF/}{Chapter6/Figs/}}
\else
    \graphicspath{{Chapter6/Figs/Vector/}{Chapter6/Figs/}}
\fi

The possibility of detecting the signatures of asymptotic symmetries near the event horizon of a black hole has generated a lot of activities, and there are several ways by which BMS symmetries have been recovered near the black hole horizon \cite{PhysRevLett.116.091101, Blau2016, PhysRevD.98.104009, Chandrasekaran2018, Ashtekar2018, Compere2016, Compere:2016hzt}. Finding the signatures of such symmetries is, in a way analogous to those which are considered in the far region should be a natural approach. We have already dealt with the memory effect for test geodesics near the horizon of black holes in the context of stitching of two black hole spacetimes and for the study analogous to the far region case; we have also established the connection with BMS symmetries. In this chapter, we include some more notions of memories of BMS like symmetries. We attempt to show a new kind of gravitation memory, i.e., \textit{spin memory effect} near the horizon of black holes which mimics the one originally obtained at null infinity \cite{Pasterski2016}. The spin memory appears as a sub-leading effect with respect to the displacement memory effect and there are difficulties involved in the detection of such memory. Next, we take a jump on the \emph{Wald's memory tensor} and introduce a memory tensor for the dynamical horizon. Finally, we address the issue of detection of a supertranslated black hole. This is motivated from the prospect of resolving the information loss puzzle as already discussed in chapter (\ref{CH1}). In this direction, it will be interesting to see if standard tests of GR can determine a configuration with a supertranslation field. We have attempted to provide a possible scenario to address this issue through the study of photon sphere for a dynamically evolving black hole carrying a supertranslation hair \cite{2021detect}. This study should offer a first step towards determining the shadow of such black hole spacetimes which are endowed with supertranslation fields.  

\section{Spin memory effect: A new gravitational memory}

We know that the displacement memory is induced by the radiative energy flux, and it has been shown that there exists a new kind of gravitational memory- \textit{spin memory effect} which is sourced by angular momentum flux. It was Pasterski, Strominger and Zhiboedov who proposed and showed the idea of a new type of gravitational memory in 2015, and established that the Bondi angular momentum aspect is the source of this information \cite{Pasterski2016}. It provides a new memory associated with superrotations. The original idea of the detection mechanism relies on the light rays that orbit a circle ($\mathcal{C}$) with a radius $L$ near the future null infinity. In this system, photons can go clockwise or counterclockwise. Here, we also assume that at early and late times, no null radiation travels through the system. The setup experiences the passage of radiation carrying angular momentum aspect $N_{A}$ during a finite interval of retarded time $(u_{i}, u_{f})$. The finite interval can be denoted as $\Delta u$ which defines the relative time delay between counter-orbiting light rays. However, our configuration is set in such a way that time separation $\Delta u$ is zero for $u<u_{i}$. The spin memory effect is caused by the fact that the $\Delta u$ is no longer zero at late times after the passage of null radiation. 

For the general form of the asymptotically flat metric (\ref{asymfd}), The total relative time delay integrated over orbits of light rays has been shown in \cite{Pasterski2016}, given as
\begin{align}
    \Delta u = \frac{1}{2\pi L} \int du \int_{\mathcal{C}} D^{A}C_{AB}dx^{B} 
\end{align}
As a result, it leads to the difference in the periods of two counter-orbiting massless particles in a circular orbit. Further, authors in \cite{Pasterski2016} have also shown that this new gravitational memory can be computed by an integration involving the local asymptotic angular momentum flux between the interval $(u_{i}, u_{f})$ which carries an explicit dependence over change in angular momentum aspect $N_{A}$.

In this direction, we show that one can recover spin memory effect near the horizon of black holes as well which mimics the one originally obtained at null infinity. This treatment of the memory might help us to catch on to the Hawking information paradox. We consider the similar setup as mentioned above for near-horizon asymptotic form of the four-dimensional metric (\ref{m}). The total relative time delay for two counter-orbiting massless particles in a circular orbit can be written as
\begin{align}
    \Delta v = \frac{\rho_{0}}{2\pi L} \int dv \int_{\mathcal{C}} \theta_{A}(v, x^{A})dx^{A}
\end{align}
Here, we again have: for $v<v_{i}\Rightarrow \Delta v=0$, and very late time after the passage of radiation, $\Delta v\neq 0$. This shows the restoration of spin memory effect near the horizon of black holes. One can also show that the near-horizon spin memory will be induced by the  angular momentum flux, and will be associated with extended form of the near-horizon BMS symmetry superrotation. We have not reported this connection in the thesis. We hope to report it elsewhere in the future.   

It is worth to mention that the conventional spin memory effect is sub-leading with respect to the displacement memory and inaccessible to the present-day detectors from observational perspectives \cite{PhysRevD.95.084048}. However, the future advanced detectors like LISA might be able to capture such effect. The detection of near-horizon form of the spin memory will also be difficult at the same time. The spin memory near the horizon of black holes has not been reported in the literature yet. Our attempt shows a motivation to understand the spin memory and its possible connection with extended asymptotic symmetries. 

\section{Wald's Memory tensor}\label{memt}

A memory tensor for Killing horizons has been proposed in \cite{PhysRevD.101.124010}, and it may be used as an useful device to detect supertranslation memory. It is expressed in terms of the expansion and shear tensor corresponding to the null congruence that generate the Killing horizon or a null boundary \cite{PhysRevD.101.124010, doi:10.1142/S0218271820430063, adami2021null}. If the null horizon is slightly perturbed from stationarity between two values of the null coordinate of the Killing horizon, we may expect to obtain a memory tensor in the form of expansion and shear. This construction is similar to the $B$-tensor introduced in Chapter (\ref{CH3}) and (\ref{CH4}) with certain differences. To define the $B$-tensor, we considered null congruences orthogonally crossing the null hypersurface whereas in the case of Wald's memory tensor, we deal with the horizon generating null congruence. Further, for $B$-tensor the null hypersurface considered to be the event horizon of a static black hole; however, the Wald's memory tensor can be defined for a more generic situation, namely for dynamical horizons. Let us now consider that the null horizon which becomes non-stationary between two values $v<v_0$ and $v>v_1$ of the null coordinate of the Killing horizon. The memory tensor takes the following form \cite{PhysRevD.101.124010}, 
\begin{align}\label{memoryt}
	\Delta_{AB}={1\over 2}\int_{v_0}^{v_1}\vartheta^{(k)}q_{AB}dv+2\int_{v_0}^{v_1}\sigma_{AB}^{(k)} dv,
\end{align}
where $k$ is the horizon generating null vector \footnote{In this chapter, we have used $a, b,\cdots$ for spacetime indices, $i,j,k\cdots$ for any 3-hypersurface, and $A,B, \cdots$ for a codimension-2 surface.}, and $\sigma_{AB}$ is symmetric traceless term known as \textit{shear}. $\vartheta^{(k)}$ denotes \textit{expansion}, and $q_{AB}$ is the codimension-1 metric to the null surface. Here, we make sure that the geodesics are hypersurface orthogonal, i.e, rotation or twist $\omega_{\alpha\beta}$ is zero. In the case when twist will not be zero, we shall include an additional term in the above equation. This consideration has been examined in section(\ref{stmt}). The \textit{second fundamental form} $K_{ij}$ given by,
\begin{align}\label{sff}
    K^{(k)}_{ij} = \frac{1}{2}\mathcal{L}_{k}q_{ij}
\end{align}
where the geodesic equation for the vector $k^{a}$ with the non-affinity parameter $\kappa$ is given as
\begin{align}
    k^{a}\nabla_{a}k^{b} = \kappa k^{b}.
\end{align}
It is to note here that the tensor field $q_{ij}$ satisfies the following condition on the null surface $\Sigma$: $k^{j}q_{ij} = k^{j}q_{ji} = 0$. Since, we know that the vector $k^{a}$ is null hypersurface generator, therefore, on the horizon: $k\cdot k=0$. In our consideration with the spatial metric, the second fundamental form is related to expansion and shear in the following way,
\begin{align}\label{mt}
K^{(k)}_{AB} = \frac{1}{(D-2)}\vartheta^{(k)}q_{AB}+\sigma^{(k)}_{AB}
\end{align}
where $\vartheta^{(k)}$ is expansion on the null hypersurface $\mathcal{\Sigma}$ for the geodesics defined by $k^{a}$, given by
\begin{align}\label{expn}
\vartheta^{(k)} = D_{i}k^{i}
\end{align}
With this framework, given a spacetime metric, we determine the form of the memory tensor. In this context, it will be interesting to investigate the memory tensor on the horizon for a spacetime given by (\ref{m}); we further also consider the spacetimes which will be important for detecting the signatures of supertranslations by ground based detectors. 

\subsection{Memory tensor: near-horizon 4-dimensional metric} 	
In this section, we consider Donnay's near-horizon asymptotic form of the metric described by (\ref{m}). We know that the horizon of the metric is given by $\rho=0$, hence, the null hypersurface is defined as: $\Sigma\equiv \rho=0$. The contravariant component of the vector $k^{\mu}$ is
\begin{align}
k^{\mu} = \Big(1,2 \rho  \kappa ,-\frac{\rho  (\zeta\bar{\zeta}+1)^2 \theta_{\bar{\zeta}}}{2 \Omega},-\frac{\rho  (\zeta\bar{\zeta}+1)^2 \theta_{\zeta}}{2 \Omega}\Big)
\end{align}
Here, $\Omega$, $\kappa$ and $\theta_{A}$ are functions of ($v,x^{A}$) with $x^{A}\equiv (\zeta,\bar{\zeta})$. The vector $k^{\mu}$ is a null vector at $\rho=0$ as $k^{\mu}k_{\mu}\vert_{\rho=0} = 0$, thus, becomes the generator of the null hypersurface. 

Using the definition (\ref{expn}), we obtain the expansion on the horizon of the following form,
\begin{align}\label{thta}
\vartheta^{(k)}\vert_{\rho=0} = \frac{\partial_{v}\Omega(v,x^{A})}{\Omega (v,x^{A})}+\kappa 
\end{align}
where the non-affinity parameter is the surface gravity $\kappa$. The definition (\ref{sff}) gives the non-vanishing second fundamental form as,
\begin{align}
    K^{(k)}_{\zeta\bar{\zeta}}\vert_{\rho=0} = \frac{\partial_{v}\Omega}{(1+\zeta\bar{\zeta})^{2}}
\end{align}
Since, our analysis is based on 4-dimensional spacetime metric, therefore from relation (\ref{mt}), the non-vanishing shear $\sigma_{AB}$ is given by,
\begin{align}
    \sigma_{\zeta\bar{\zeta}}\vert_{\rho=0} = \frac{-\kappa\Omega}{(1+\zeta\bar{\zeta})^{2}}
\end{align}
As a result, the memory tensor $\Delta_{AB}$ becomes,
\begin{align}\label{memoryt1}
	\Delta_{AB}={1\over 2}\int_{v_0}^{v_1}\Big(\frac{\partial_{v}\Omega(v,x^{A})}{\Omega (v,x^{A})}+\kappa\Big)q_{AB}dv+2\int_{v_0}^{v_1}\sigma_{AB}^{(k)} dv,
\end{align}
As a result, there is a non-vanishing contribution of ($\zeta,\bar{\zeta}$) component of the memory tensor on the horizon, i.e.,
\begin{align} \label{wldsm}
    \Delta_{\zeta\bar{\zeta}} = {1\over 2}\int_{v_0}^{v_1}\frac{2\Omega}{(1+\zeta\bar{\zeta})^{2}}\Big(\frac{\partial_{v}\Omega}{\Omega }+\kappa\Big)dv-2\int_{v_0}^{v_1}\frac{\kappa\Omega}{(1+\zeta\bar{\zeta})^{2}} dv
\end{align}
The difference between the memory tensor obtained above and $B$-memory discussed in Chapter (\ref{CH3}) and (\ref{CH4}) is the following- the $B$-memory emerges due to the jumps in expansion and shear of the orthogonal null congruence crossing the thin shell whereas the Wald's memory appears due to the permanent relative displacement in the null generators of a dynamical horizon whose early and late states can be considered as approximately stationary. Further, Eq. (\ref{wldsm}) shows the memory obtained at the horizon of a near-horizon asymptotic form of the black hole spacetime. In order to relate it with asymptotic symmetries, we notice that the memory tensor contains integration over $v$, i.e., the form of the end result of the $\Delta_{AB}$ will ultimately be independent of $v$ variable. Therefore, one can take the transformations of functions $\Omega$ and $\kappa$ along the killing direction, and set the $v$ coefficient to be independent of $v$ as we did in the Chapter (\ref{CH5}). This will establish the explicit connection between memory tensor and asymptotic symmetries.

\subsection{Memory tensor: for dynamical horizon}\label{stmt}
 	Here we consider a metric containing a dynamical horizon, the supertranslated Vaidya black hole (STVBH). Such a configuration can result from a Vaidya solution subjected to a linear perturbation resulting from a shock-wave like energy-momentum flux \cite{strominger2018lectures, PhysRevLett.116.231301, Chu2018}. As the perturbations need not be spherically-symmetric, the metric is in general not spherically symmetric as well, viz., \cite{2021detect}  
	\begin{align}
		d s^2 = -g_{vv} dv^2 + 2 dv dr + g_{v \theta} dv d\theta + r^2 \tilde{g}_{\theta \theta} d\theta^2 + r^2 sin^2 \theta \tilde{g}_{\phi \phi} d\phi^2, \label{STVBH}
	\end{align}
	where
	\begin{align} 
		g_{vv} & = 1-\frac{2 M(v)}{r}-\frac{M(v)}{r^2}f''-\dfrac{M(v)\cot \theta}{r^2}f'-\frac{2 \dot{M}(v)}{r} f \hspace{1mm} ; \hspace{1mm} g_{\theta \theta} =  1 + \dfrac{f''}{r} -\dfrac{\cot \theta}{r } f',\\
		g_{v \theta} & = \csc^2 \theta f' - 2\left(1-\dfrac{2 M(v)}{r}\right)f' - \cot \theta f'' -f''' \hspace{2mm} ; \hspace{2mm} g_{\phi \phi} = 1 - \dfrac{f''}{r} +\dfrac{\cot \theta}{r } f'.
	\end{align}
	Here, prime denotes the derivative of supertranslation field $f$ with respect to $\theta$, and dot denotes the derivative with respect to $v$. The supertranslation $f$ is assumed to be small and it depends only on the $\theta$ coordinate. This metric has a dynamical horizon ($H$), located at 
	\begin{align}
		r_H=2 M(v) +2 f \dot{M}(v)+ \frac{1}{2}(f''+f' \cot \theta) +\mathcal{O}(f^2).    
	\end{align}
	In the double null foliation of $H$, there will be two null normals $(k,l)$ to the codimension one hypersurface $H$. In this case, the expansion $\tilde{\vartheta}^{(k)}$ will vanish as $k $ is the null-normal to the horizon $H$. The other expansion $\tilde{\vartheta}^{(l)}$ will not be zero and will have a negative value \cite{Chu2018}. Therefore $H$ is a supertranslated version of a Future Outer Trapping Horizon (FOTH) \cite{PhysRevD.68.104030, Ashtekar2004, PhysRevD.103.024046, PhysRevD.102.124061}. Since $\tilde{\vartheta}^{(k)}$ is zero, and $\tilde{\vartheta}^{(l)}$ is independent of supertranslation hair on $r_H$, we don't get any memory. In this regard, there could be an equivalent way to define the memory tensor. We can consider the congruence generated by the spacelike normal to the codimension one hypersurface of $H$ given by $r^a=\frac{1}{\sqrt{2}}(k^a-l^a)$ \footnote{$k^a=\{1,\frac{g_{vv}}{2},0,0\};~~  l^a=\{0,-1,0,0\}$} . As $H$ is not null, we would have a slight change in the expression of memory tensor written above in (\ref{memoryt}). 
	\begin{align}\label{memorytn}
		\Delta^{(DH)}_{ij}={1\over 3}\int_{v_0}^{v_1}\vartheta^{(r)}\gamma_{ij}dv+2\int_{v_0}^{v_1}\sigma_{ij}^{(r)}dv +\int_{v_0}^{v_1}\omega^{(r)}_{ij}dv.
	\end{align}
	where $\Delta^{DH}$ denotes the memory tensor for dynamical horizon. The expansion for the congruence generated by $r^a$ is shown below and it includes the supertranslation field \cite{2021detect}. 
	\begin{align}
		\vartheta =& {Q(r,\theta)}^{-1} \left( 2 r^3 \sin ^2\theta (3 r-f \dot{M}(v))-    \right. 
		 \left. 3 r^2(\cos \theta  f'-\sin \theta  f'')^2+M(v) (f''+  \cot \theta  f') \nonumber \right. \\ 
		&\left. \times  (\sin \theta  f''-\cos \theta  f')^2-   2 r^3 \sin ^2\theta \right),
	\end{align}
	where $Q(r,\theta)=2 \sqrt{2} r^3 (r^2 \sin ^2\theta -(\cos \theta  f'-\sin \theta  f'')^2)$. We need to replace $r_H$ for $r$ in the above equation in order to get the $\vartheta^{(r)}$. Note that the memory tensor $\Delta^{(DH)}_{ij}$ contains a rotation part also. The only non-vanishing component of the rotation would be $\omega_{v\theta}$, given as
	\begin{align}
\omega_{v\theta} = \frac{1}{4 \sqrt{2} r^2}\Big(M(v) \left(f'''(\theta )+\cot (\theta ) f''-\csc ^2(\theta ) f'\right)-2 r f' M'(v)\Big)
\end{align}
	There will be a non-zero shear part as well. Interestingly, only the rotation vanishes if we set $f$ to be a constant on $H$ (a non-supertranslated horizon) \cite{PhysRevD.103.024046, PhysRevD.102.124061}. We also notice that the memory tensor will automatically carry supertranslations at $r=r_{H}$ as the metric is endowed with supertranslation hair. The motion along the vector $r^a$ can be regarded as time evolution of $H$ for a distant observer.
	
	Now,  a  pertinent  question  would  be  whether  such kinds of memories can be detectable through experiments performed   near earth. Investigating   the   memories discussed above would require us to place the detectors near  the  horizon  of  a  black  hole.   This  seems  to  be  a far fetched idea.  The obvious way to go ahead would be to see if a supertranslated configuration can be detected through the standard GR tests like gravitational lensing or  by  studying  the  black  hole  shadow.  Of course, if gravity waves are generated by such a source then it would be possible to detect supertranslation through memory effect, but we are interested in detecting the hair through standard tests of GR like light bending etc. For  any  static (or stationary) black hole with a supertranslation hair,it  is  not  practically  possible  to  detect  any  such  effect if  one  considers  standard  tests  of  GR  \cite{Comp_re_2016}. Since, for  example,  a  supertranslated  Schwarzschild  metric  is locally  diffeomorphic  to  a  Schwarzschild  metric,   any possible  deviation  from  the  Schwarzschild  metric  can always be nullified by a coordinate transformation (gauge choice)  in  a  local  patch. In  fact  one  needs  to  set an  array  of  detectors  around  the  entire  black  object to  be  able  to  estimate  the  charge  corresponding  to the  supertranslation  hair  for  this  kind  of  configuration \cite{Comp_re_2016}. Therefore we need to consider a dynamical black hole  which  will  not  be  plagued  with  such  limitations. When   a (isolated)   supertranslated   black   hole   gets perturbed  due  to  the  presence  of  matter  around  it, the  perturbed  black hole  needs not to  obey  the  same symmetries  before  it was  perturbed.   In  most  general  case, such  a  perturbed configuration may settle down to a completely different kind of a configuration,  for example a static state may settle  down  to  a  stationary  state  \cite{PhysRevD.101.124010}.  Therefore  it is plausible to expect an observable change can be seen in such perturbed  phase  of  a  black  hole  and  this change can not be undone by a diffeomorphism.
	
\section{Photon sphere: supertranslated Vaidya black hole}

As a first step in this direction, we study the evolution of the photon sphere of a slightly \textit{perturbed} STVBH (pSTVBH)  following  the  method  elucidated  in  \cite{PhysRevD.99.104080}.   The metric for this pSTVBH takes the following form \cite{2021detect},
\begin{align}\label{pSTVBH}
		ds^2 =& -g_{vv} dv^2 + 2 dv dr - 2a \psi(r,v) dr d\phi - 2 a \chi(r,v) dv d\phi + 2\epsilon \xi(r,v) d\theta d\phi + \nonumber \\ 
		& g_{v \theta} dv d\theta + r^2 \tilde{g}_{\theta \theta} d\theta^2 + r^2 \sin^2 \theta \tilde{g}_{\phi \phi} d\phi^2. 
	\end{align}
The additional terms appearing in the above metric are due to perturbations sourced from a suitable stress-energy tensor. They can be modeled in such a way that they are absent at $v=0$ and grows gradually, eventually saturating to a small value fixed by the constant $a$ and $\epsilon$\footnote{The evolution of the photon sphere is actually not sensitive to small deformation $\epsilon \xi(v,r) d\theta d\phi$  since we will take $\theta = \pi/2$.}. The choice made here indicates that the initially static configuration will gradually become stationary or rotating. Here $a$ plays the role of the spin parameter. During the dynamic phase, this black hole is not diffeomorphic to the Vaidya solution. 

The \textit{photon sphere} of a black hole is an unstable null geodesic that forms a close orbit.  Moreover, orbits which lie  inside  the  photon sphere  spiral  into  the  black  hole while  those  lying  outside  can  escape  to  infinity.   This gives  rise  to  the  prospect  of  detecting  a  black  hole  by observing  its  so-called  shadow \cite{perlick2021calculating}. In  our  study,  we shall consider only those photon orbits which lie on the equatorial  plane,  and  therefore  set $\theta = \frac{\pi}{2}$.    This essentially amounts to projecting the photon sphere on to the equatorial plane and studying the photon circle.  For stationary black holes, the radius of the photon circular orbit $r_{0}$ is  constant  and  it  is,  for  example,  3$M$ for  a Schwarzschild  black  hole.    This  result  is  given  by  an algebraic equation but if we consider a dynamical black hole, $r_{0}$ is  no  longer  a  constant.   Instead, $r_{0}=r_{0}(v)$, as the radius of the photon circle grows with time, and its  evolution  is  governed  by  a  second  order  differential equation  which  we  now  derive. Assuming  that  the supertranslation field is sufficiently small, it is reasonable to expect that the final geometry of the spacetime is that of a slightly distorted slowly rotating black hole; then $r_{0}$ for  the  dynamic  black  hole  will  eventually  saturate  to $ r_0(v_0) \approx 3 M_0 \mp 2/\sqrt{3}a$ where $M_{0}$ is the final mass of the BH, $v_{0}$ is a sufficiently late instant of time, and $a$ is very small \cite{1972ApJ...178..347B}. Let us consider two separate cases with vanishing and finite rotation parameter.

\subsection{Photon Sphere: pSTVBH with $a=0$}

Let us first focus on how to derive the photon sphere equation for $a=0$; since we are considering null trajectories, we have $ds^2 =0$ which, for a dynamical BH, gives us
	\begin{align}
		\left(\frac{d \phi}{d \lambda}\right)^2 = \left(\dfrac{d v}{d \lambda}\right)^2 \left[ h_1(r_0(v), v) + \dot{r}_0 h_2(r_0(v),v)\right], \label{Heq}
	\end{align}
	where $\lambda$ is an affine parameter, a dot denotes derivative with respect to $v$, and $h_1$ and $h_2$ are functions whose exact form depends on the metric under consideration.
	
	Moreover, the relevant geodesic equations tell us
	\begin{align}
		\frac{d^2 v}{d \lambda^2} &= F_1 (r, v)\left( \frac{dv}{d\lambda}\right)^2 + F_2 (r, v)\left( \dfrac{d\phi}{d \lambda}\right)^2 \label{Feq} ,\\ 
		\dfrac{d^2 r}{ d \lambda^2} &=  G_1 (r, v)\left( \dfrac{d r}{d \lambda}\right)^2 + G_2 (r, v)\left( \dfrac{d \phi}{d \lambda}\right)^2 
		 + G_3 (r, v) \frac{d v}{d \lambda} \frac{dr}{d \lambda} + G_4 (r, v)\left( \frac{d v}{d \lambda}\right)^2, \label{Geq}
	\end{align}
	where the function $F_i$'s and $G_i$'s can be read off directly from the geodesic equation. The actual form of the geodesic equations can be seen in appendix(\ref{ch6}) which can be arranged in the form (\ref{Feq}) and (\ref{Geq}).
	
	Since $r_0=r_0(v)$, we can write,
	\begin{align}
	{d r_0(v)}  =  \frac{ dr_0(v)}{dv} d v = \dot{r}_0 d v  \implies \frac{d^2 r_0(v)}{ d \lambda^2}  = \dot{r}_0 \frac{d^2 v}{ d \lambda^2} + \ddot{r}_0\left( \frac{d v}{d \lambda}\right)^2. \label{psp1}
	\end{align}
	Now by substituting \eqref{Heq}, \eqref{Feq}, \eqref{Geq} in \eqref{psp1}, we arrive at the second order differential equation governing the evolution of the photon sphere, viz.,
	\begin{align}
		\ddot{r}_0(v) + \dot{r}_0(v)\left[ f_1 + f_2 h_1 -g_2 h_2 -g_3 \right] +  \dot{r}^2(v)[f_2 h_2 - g_1] -\left[g_2 h_1 + g_4\right] =0, \label{psp2}
	\end{align} 
	where $f_i(r_0(v), v) =  F_i(r,v)|_{r=r_0(v)}$ and $g_i(r_0(v), v) =  G_i(r,v)|_{r=r_0(v)}$.
	
	We choose to model the growth of the mass parameter using smoothly growing functions,
	\begin{align}\label{mtan}
		&	M(v) = \dfrac{M_0}{2} (1 + \tanh(v)) \hspace{2mm} ; \hspace{2mm}  M(v) = \dfrac{M_0}{(1 + sech (v))}.
	\end{align}
	For concreteness, we have assumed that the supertranslation field $f(\theta)$ is of the form given by 
	\begin{align}
		f(\theta) = q P_2 (\cos(\theta)) \label{P2},
	\end{align}
	where $P_2(\cos \theta)$ is the second Legendre polynomial and $q$ is a constant such that $0<q \leq 1$ (and $f(\theta)$ has been normalized with respect to $M_0$). Note that supertranslation field becomes constant on setting $\theta = \pi/2$, so $q$ essentially characterizes the \emph{strength} of the supertranslation field, hence, can be called the supertranslation hair. We restrict ourselves to small values of $q$.
	
\begin{figure}[h!]\centering
    \includegraphics[height=7.3cm, width=11.5cm]{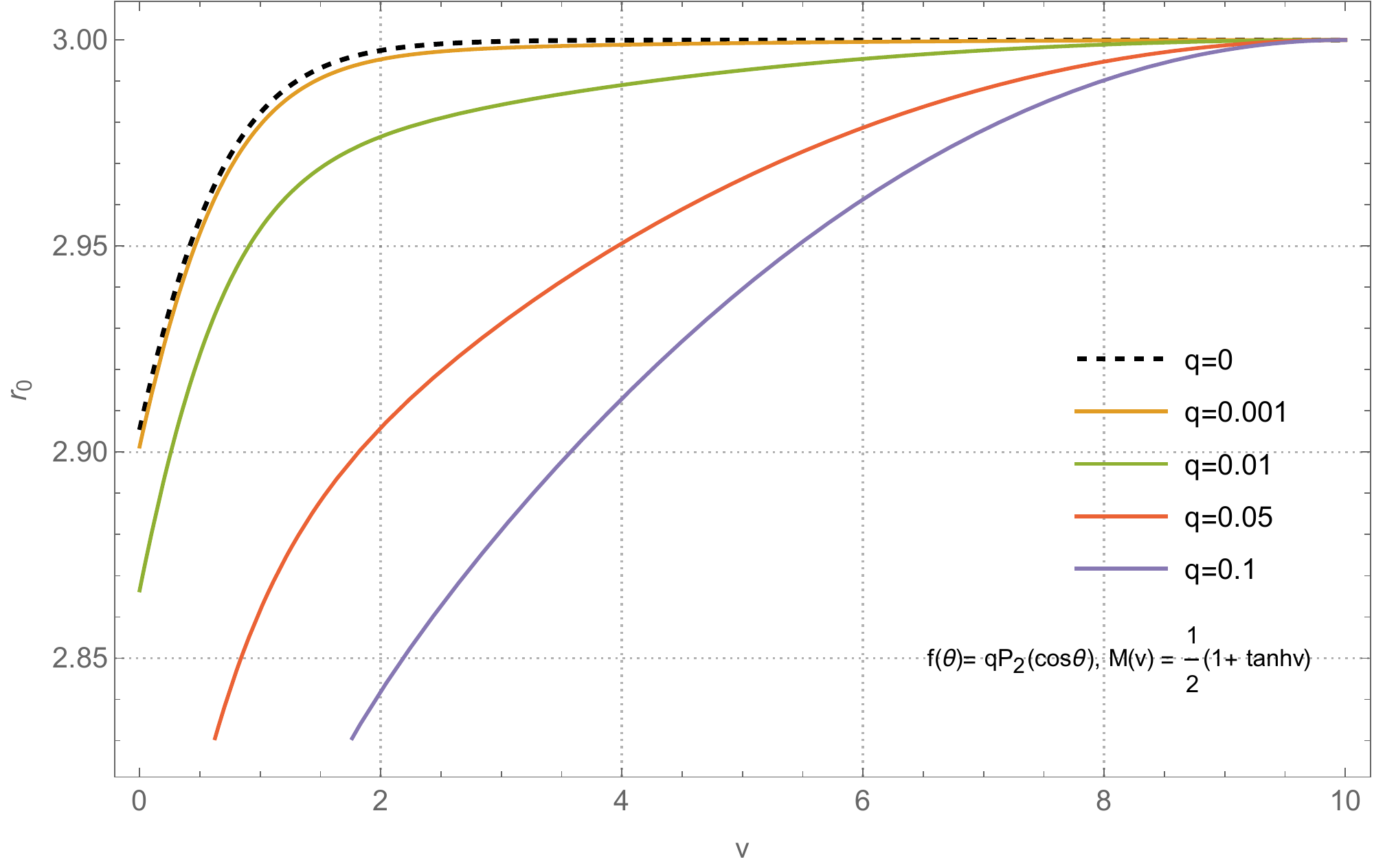} 
    \caption{$M(v)=1/2(1+ \tanh(v)), a=0, M_0=1, v_0 = 10.$}\label{svbh10}
\end{figure}

\begin{figure}[h!]\centering
    \includegraphics[height=7.3cm, width=11.5cm]{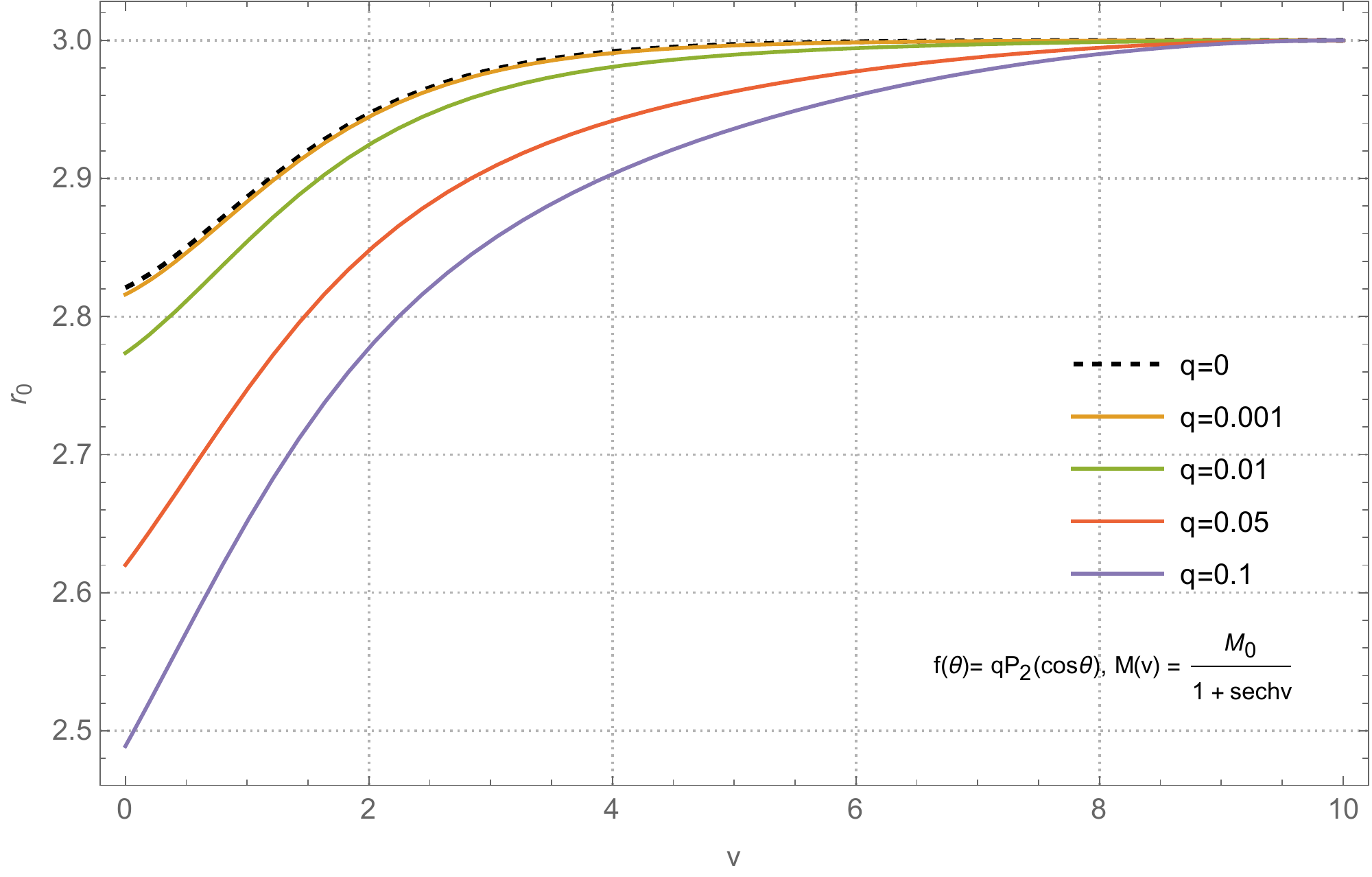} 
    \caption{$M(v)=1/(1+ sech(v)), a=0, M_0=1, v_0 = 10. $}\label{svbh20}
\end{figure}

	We plot the results for the pSTVBH in Fig.(\ref{svbh10}) and (\ref{svbh20}).  
	We observe that the photon sphere of a pSTVBH deviates from that of a Vaidya black hole ($q=0$). The impact of the supertranslation field on the evolution of photon sphere can be seen in the figures below.
	

\subsection{Photon Sphere: pSTVBH with $a\neq 0$}

Now, we study the signature of supertranslation through photon sphere for the dynamically evolving spacetime (\ref{pSTVBH}) with $a\neq 0$. With the case $a \neq 0$, the algebra is quite tedious but it will essentially give us an additional term
	\begin{align}
		&	\ddot{r}_0(v) + \dot{r}_0(v)\left[ f_1 + f_2 h_1 -g_2 h_2 -g_3 \right] + \dot{r}^2(v)[f_2 h_2 - g_1]
		  	-\left[g_2 h_1 + g_4\right]  \pm a H (r_0(v),v)=0, \label{psp3}
	\end{align} 
	where the $\pm$ corresponds to prograde and retrograde orbits respectively, and the exact form of $H(r_0(v),v)$ depends on the metric being studied (see \cite{PhysRevD.99.104080} for computational details). To study the behaviour of the photon sphere, again \eqref{psp3} must be solved numerically subjected to two boundary conditions. We shall focus on the prograde mode only.
	\begin{figure}[h!]\centering
    \includegraphics[height=7.3cm, width=11.5cm]{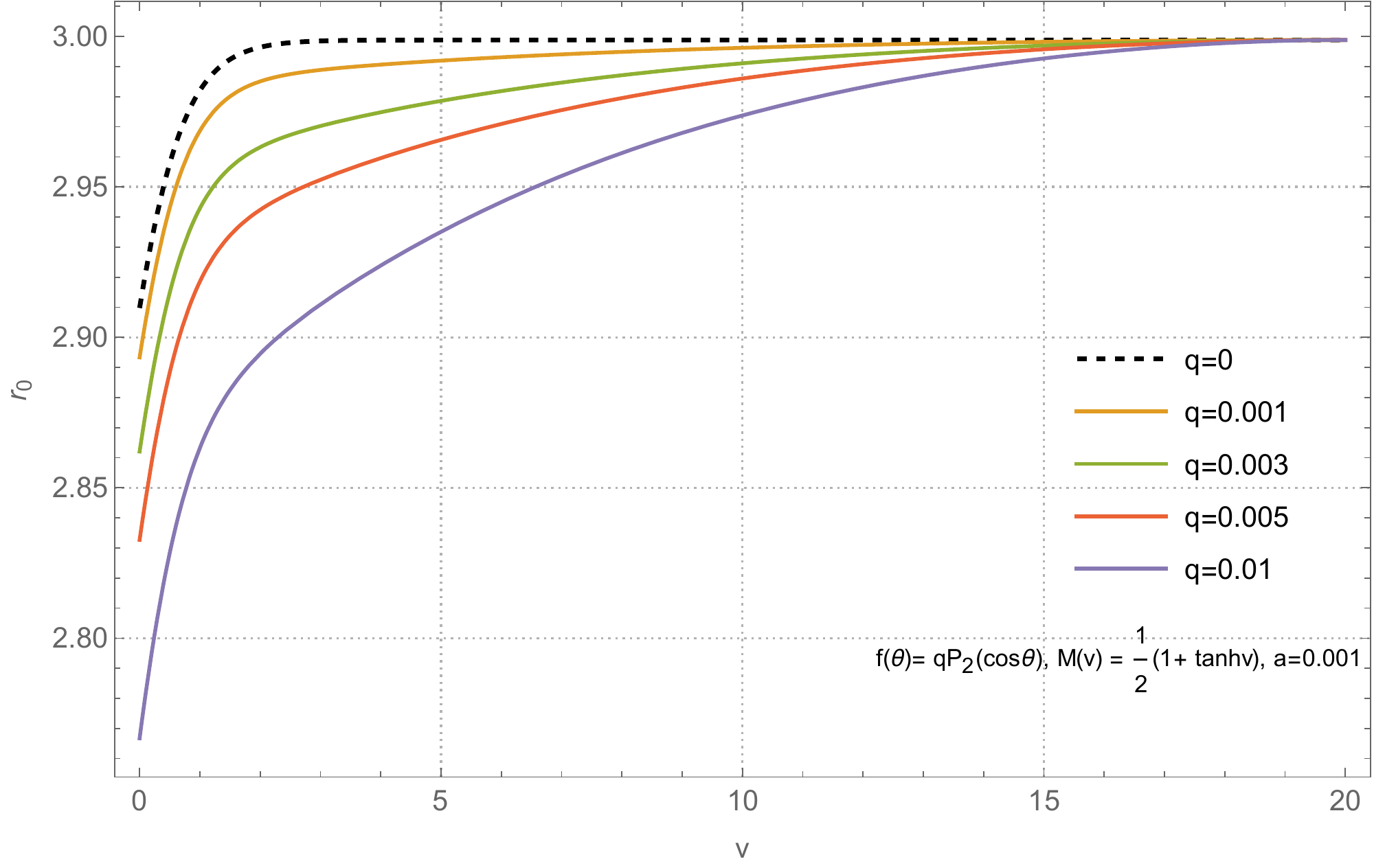} 
    \caption{$M(v)=1/2(1+ \tanh(v)), a=0.01, M_0=1, v_0 = 20$}\label{psvbh10}
\end{figure}
\begin{figure}[h!]\centering
    \includegraphics[height=7.3cm, width=11.5cm]{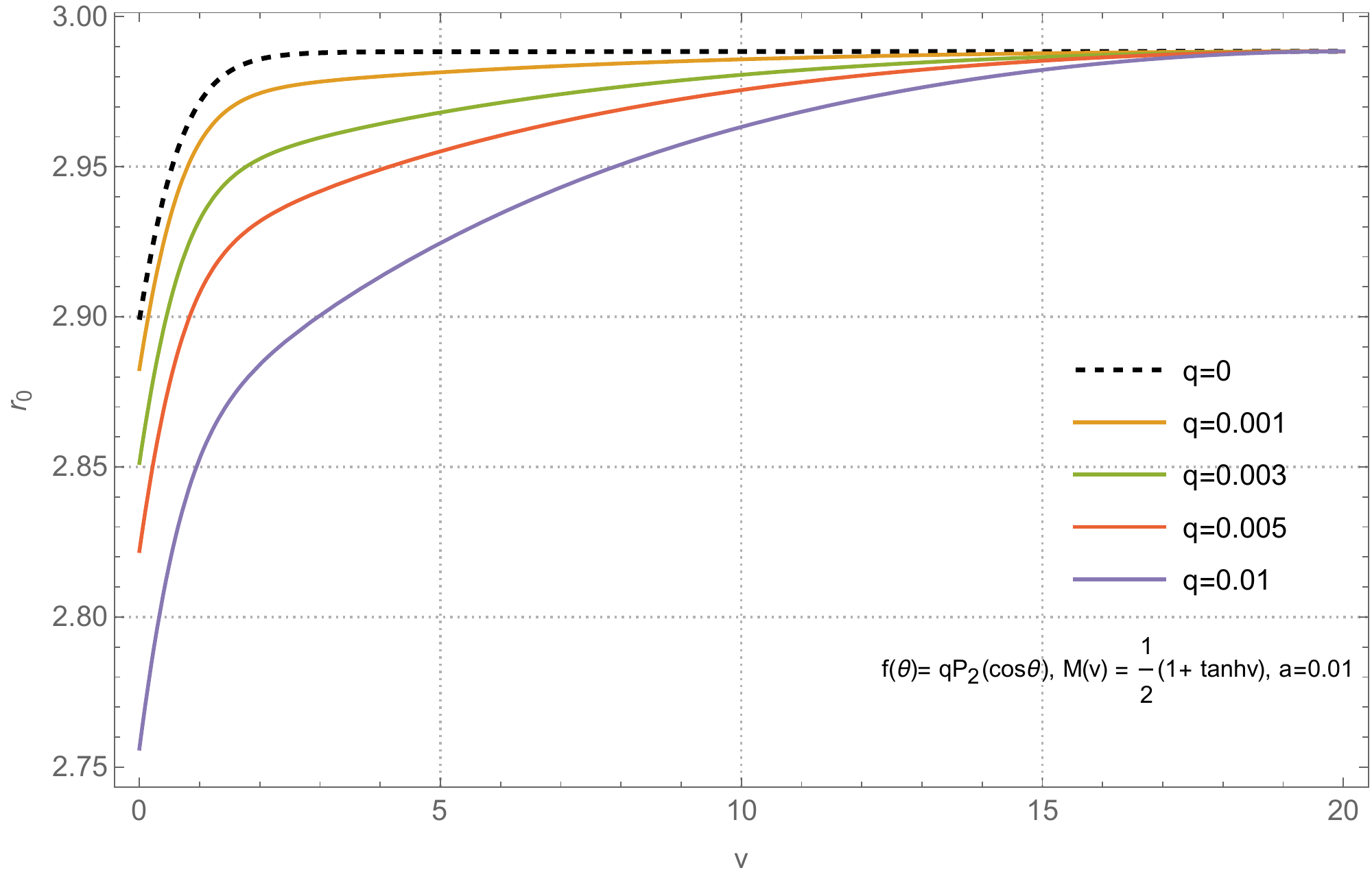} 
    \caption{$M(v)=1/2(1+ \tanh(v)), a=0.001, M_0=1, v_0 = 20$}\label{psvbh20}
\end{figure}
	It is physically reasonable to assume that at late times, the dynamical BH becomes stationary and stops growing. We therefore take the following two future boundary conditions at a sufficiently late time $v_0$, namely, $r_0(v_0)= 3M_0- 2/\sqrt{3}a$ and $\dot{r}_0(v_0)=0$.
	We further model the perturbations as follows
	\begin{align}
	    \psi(r,v) \equiv \tanh(v), ~ \xi(r,v) \equiv \dfrac{2 M(v)}{r} \tanh(v)
	\end{align}
	We observe that the evolution of photon sphere carrying supertranslation hair significantly deviates from the stationary configuration, and sets the motivation for detecting the supertranslated black holes. As an initial step in this direction, our study  open up a direction to investigate the signatures of asymptotic symmetries through black hole shadow.

\section{Discussion}

The observational signatures of asymptotic symmetries are yet to be made. Since such symmetries are directly related to the gravitational memory, hence, the detection of one feature might reveal intriguing aspects of the other. In this chapter, we have examined the spin memory effect near the horizon of black holes mimicking the obtained at null infinity and Wald's memory tensor for near-horizon asymptotic form of the metric as well as for dynamically evolving supertranslated black hole. We address the possibilities of detecting the signatures of supertranslation field through standard tests of general relativity. We find that the behaviour of the photon sphere of a dynamical black hole changes over time. The evolution of a photon sphere of a pSTVBH is distinct from that of a \emph{ordinary} Vaidya black hole. The models which we have considered here reduce to the scenario depicted in \cite{PhysRevD.99.104080} for $q=0$, that is, in the absence of the supertranslation field. Therefore it is reasonable to expect that a pSTVBH will give rise to a shadow that is distinguishable from a Vaidya black hole over time, and studying the evolution of the photon sphere is but a first step in this direction. A dynamical analogue of a supertranslated Schwarzschild blackhole formed as the final state of gravitational collapse is discussed in \cite{Compere:2016hzt}. A dynamic version of this metric also yields curious results where the photon sphere undergoes a growing phase followe by a decaying phase before it gets saturated \cite{2021detect}. We must admit that these considerations may still be regarded as some toy models to explore practical detection prospects of such configurations. However, we believe the framework proposed here can be more tightened up to get more robust results.

Usually, if the angular and radial parts of the first integral of motion for null geodesics are separable by introducing the Carter constant then one can study the shadow using a semi-analytic approach \cite{PhysRevD.99.104080}. The pSTVBH is not amenable to such a treatment. It is known that the shadow of static black holes with metrics having such cross terms \cite{PhysRevD.103.044057} are studied numerically using backward ray-tracing algorithms \cite{PhysRevD.94.104023}. At present, algorithms for studying dynamical black holes are not available to the best of our knowledge. Hence it would be fruitful to develop a numerical method that can tackle spacetimes which do not possess energy as a conserved quantity, that is, dynamical spacetimes, and generate the evolution of its shadow. We humbly leave this issue for the future.  
\chapter{Conclusion and Outlook}\label{CH7}




\ifpdf
    \graphicspath{{Chapter2/Figs/Raster/}{Chapter2/Figs/PDF/}{Chapter2/Figs/}}
\else
    \graphicspath{{Chapter2/Figs/Vector/}{Chapter2/Figs/}}
\fi

The idea of gravitational memory effect and its direct connection with asymptotic symmetries enable us to probe the near-horizon structure of black holes. In the light of discussion based on the previous chapters, it is clear that a detailed investigation of gravitational waves carrying memory from compact binary sources is worthwhile. In this direction, our studies put forward a way of comprehending the theoretical framework together with detection prospects of the BMS memory generated through cataclysmic astrophysical events. On a major note, as a model scenario for near the horizon of black holes, we provide two ways of estimating the memory signal; one is based on soldering of two black holes along a common null surface, and the second appears as a more realistic approach in the sense that it is analogous to the conventional memory originally obtained at asymptotic null infinity. The detection mechanism also addresses the starring role of near-horizon asymptotic symmetries and how such symmetries can be restored in the memory signal. As an alternative approach for detecting the astrophysical signatures of asymptotic symmetries, we have made an attempt to study the evolution of photon sphere of a dynamical supertranslated black hole. Let us take a look at the key findings of each chapter of the thesis and see how we can grasp and summarise the important concluding remarks. 

Keeping in mind the motivational ingredients of the chapter (\ref{CH1}), the chapter (\ref{CH2}) discusses the theoretical framework of understanding the null shell formalism and how asymptotic symmetries emerge in this context. We review the concept of gluing formalism and show that supertranslations and superrotations are recovered as a soldering freedom of the black hole spacetimes. The stitching analyses imply the generation of null shells which carry the impulsive lightlike signals such as null and matter and impulsive gravitational waves. In general, the impulsive lightlike signals carry both the null matter and IGWs, however, one can separate them out. In the chapter, the tensor field which carries impulsive lightlike signal is denoted by $\gamma_{ab}$, and can be investigated through the extrinsic curvature algorithm and off-shell extension of soldering transformation. This is the quantity whose examination implies the intrinsic properties of the null shell placed at the horizon: surface energy density, current and pressure. Interestingly, $\gamma_{ab}$ captures the asymptotic symmetries carried by impulsive signals. Further, the impact of such signals on test detectors has been examined in greater details in subsequent chapters.

In chapter (\ref{CH3}), we implement the basic methodology of gluing formalism on non-extreme black holes. Precisely, we study the restoration of asymptotic symmetries in the intrinsic formulation of Schwarzschild spacetimes. We extended the existing works concentrating on plane gravitational waves to spherical ones. We have shown that one can recover supertranslation-like transformations for Schwarzschild horizon shells. Further, it is shown that one can recover extended BMS transformations, i.e., superrotations for a null shell placed just outside the horizon $U=\epsilon$. We have estimated the finite changes in the measurable quantities for timelike and null congruence upon crossing the null shell placed near the null hypersurface. These changes are obtained in terms of BMS parameters associated with IGW which leave footprints on test particles or test geodesics upon passing through them. We have shown that the surface quantities like energy density ($\mu$), current ($j^{A}$) and pressure ($p$) carry asymptotic symmetry parameters. The relevant question comes into the picture whether one can detect this memory at a null surface positioned at a finite location of spacetime. As a result, we have analysed the effects of null shell, situated near the horizon of the Schwarzschild black hole, on the separation vector of two nearby timelike geodesics passing through the shell. The displacement between the geodesics was identified by supertranslation parameters which gave us Supertranslation memory effect. We also discussed the superrotational memory effect for constant curvature spacetimes. Due to non-zero surface current for constant curvature spacetime consideration, the test particles will be displaced off the initial 2-dimensional surface as the $V$-component of the deviation vector $X_{V}$ is non-zero. Whereas for Schwarzschild case, the surface current turns out to be zero. Thus, the test particles will remain on the spatial slice of the metric as $X_{V}$ is zero but with a relative change in the displacement. Further, we also calculated the measurable quantities such as expansion and shear for null geodesics indicating the supertranslational $B$-memory. Here, we used the fact that null geodesics are crossing orthogonally the null hypersurface supporting IGW. We have shown that null geodesics suffer a finite jump after passing through the null shell supporting IGW. This completed the BMS B-memory effect in the context of stitching of two Schwarzschild black hole spacetimes. This project might serve as a model scenario to obtain some observable effects for upcoming advanced detectors from the detection prospects of the asymptotic symmetries in gravitational memory.

Next, we accomplish the soldering analyses on extreme black holes in chapter (\ref{CH4}), and show the distinguishing features from Schwarzschild consideration. We computed the intrinsic properties of the extreme RN black hole in terms of BMS parameters and showed that the shell does not support pure IGW. The same effect is observed for extreme BTZ black holes as well. We have shown how conformal symmetries may appear as a soldering freedom when we place a null shell infinitesimally close to the event horizon of an extreme RN black hole as appears in the Schwarzschild case. We recovered superrotation-like symmetries for the null shell placed a little away from the horizon. Further, for extreme RN black holes, we derived deviation vector for timelike geodesics in terms of BMS parameter
crossing the null hypersurface and observed that the particles would get displaced off from the two dimensional spherical surface which opposes the Schwarzschild consideration. This concluded the memory effect for timelike geodesics. Next, we computed the memory effect on the horizon shell for null geodesics crossing the null hypersurface orthogonally and obtained finite jumps in measurable quantities such as expansion and shear. We also determined these measurable quantities off the horizon shell. This concludes the BMS $B$-memory effect similar to the non-extreme case, i.e., Schwarzschild for null geodesics crossing the null shell supporting IGW.

As a more realistic model scenario, we extend our analyses where we do not require the concept of soldering formalism. In this direction, our study provides a significant contribution in investigating the displacement memory effect near the horizon of black holes and its possible connection with asymptotic symmetries. The chapter (\ref{CH5}) discusses this consideration where two detector setups are being placed near the horizon of a black hole, and shows an impact of gravitational waves carrying memory on test detectors. We have shown a detailed study on memory effect for non-extremal and extremal black holes in three and four dimensions. In both cases, we estimated the change in the displacement vector between test detectors, and showed that the memory can be restored in terms of supertranslations and superrotations. We observe that the $\lambda_{AB}$ is the data available to be considered in the detection which mimics the data $C_{zz}$ present near the null infinity ($\mathcal{I}^{+}$). We also notice the crucial differences between memory effect and BMS symmetries obtained for the far region and near the horizon of a black hole. As a major distinguishing feature, one recovers two supertranslation parameters and a superrotation whereas we have only one supertranslation in order to relate it with displacement memory at null infinity. Further difference seems to be in the structure of GDE as well. These key differences make the analyses of near-horizon memory effect significantly non-trivial than the far region. This might serve as a model scenario and help us to understand the intriguing role of asymptotic symmetries in the context of gravitational memory. 

The chapter (\ref{CH6}) is the final contribution to the thesis which particularly focuses on detecting the astrophysical signatures of asymptotic symmetries. At first, we obtain a form of the memory, expansion and shear, for null congruences on the horizon $\rho=0$ for the near-horizon asymptotic form of the 4-dimensional metric. It is obvious that one can also obtain such memories to the linear order in $\rho$. Further, as a new gravitational memories, we were able to recover the spin memory effect for a 4-dimensional near-horizon asymptotic form of the metric which is analogous to the one obtained at asymptotic null infinity.  However, such memories are difficult to be detected in upcoming years. As an initial step in search of signatures of asymptotic symmetries, this has motivated us to investigate the behaviour of photon sphere for a dynamically evolving black hole spacetime. We have shown that the evolution of a photon sphere of a supertranslated Vaidya black hole is distinct from that of an ordinary Vaidya black hole. The models which we have considered reduce to the scenario where the strength of the supertranslation field vanishes, that is, $q = 0$. We have also shown the dynamical analogue of a static supertranslated Schwarzschild black hole formed as the final state of gravitational collapse \cite{2021detect}. The photon sphere of such a case is very interesting that it begins to shrink after reaching a certain maximum determined by the strength of the supertranslation field. If the supertranslation field is sufficiently strong, this behaviour might have observable consequences for the shadow of a supertranslated static black hole. This feature could therefore offer a novel way to understand whether a black hole carries a supertranslation hair. We have also addressed the mathematical complications arising in order to examine the shadow-analyses. It is clear from $g_{r\theta}$ component of the metric which causes difficulties to proceed for determining the shadow for a supertranslated Schwarzschild black hole \cite{2021detect}. A similar complication arises in the case of supertranslated Vaidya black hole which carries $\theta$-dependent terms in metric coefficients $g_{vv}$ and $g_{v\theta}$. These limitations strongly inspire us to investigate the nature of shadow for a dynamically evolving black hole spacetime. 


Further, in the sixth chapter, we provide the possibilities of detecting the signatures of supertranslation through standard GR tests. Detecting  signatures of asymptotic symmetries from gravitational lensing or similar tests might not be an appropriate direction for static black holes spacetimes as supertranslated geometry of these will not lead to any deviations from the known results. The reason behind this is, a supertranslated Schwarzschild black hole is diffeomorphic to a Schwarzschild black hole. Hence, anything we compute can be always undone by a diffeomorphism. Here one must also note, this is a little different context. The supertranslated black hole is a state with a supertranslation hair. The signature for such a hair could be detectable through the memory effect. However, we are interested to detect such a configuration via GR tests. This is done in the previous chapter considering a model where a supertranslated black hole is in dynamical phase. Further, black holes are not isolated, so a detection framework should be designed for realistic astrophysical situations when a black hole gets perturbed due to the matters present surrounding it. In this context, one can also define a memory tensor that will encode the supertranslation hair but will not be useful from an observational perspective. However, we still be hopeful to detect such hair in black hole shadows. The last chapter of this thesis provides a study of evolution of the photon sphere for a supertranslated Vaidya black hole and its perturbed version. We have found the photon sphere of such a black hole will evolve differently from that of a Vaidya black hole. This will have also interesting effect on the shadow of such black holes. Therefore, in the dynamical phase, there is a reasonable possibility of detecting such configurations. Further investigation in this line might set a stronger grounds for detecting the asymptotic symmetries through deflection angle approach and black hole shadows.

One of the major breakthroughs of gravitational wave astronomy in recent times is the direct detection of gravitational waves. It is expected that all sorts of gravitational wave sources will carry memory part along with them. Researchers actively involved are expecting to detect the memory signal experimentally in the coming decade. It is expected that present-day detectors like LIGO might not be able to play a crucial role. We hope that advanced detectors like aLIGO or LISA (Laser Interferometer Space Antenna) might be able to capture this effect as LISA operates in the lower frequency range than the LIGO, i.e., LISA is on the lookout for a relatively much longer wavelength to expand the detection domain to a wider length of gravitational wave sources. People are implementing wave extraction techniques to carry out the simulations for measuring the nonlinear GW memory generated by binary black hole mergers \cite{Khera:2020mcz, Pollney_2011, Hubner:2021amk}. Another study discusses the detectability of memory signals from a population of coalescing supermassive black hole binaries with pulsar timing arrays and LISA \cite{Islo:2019qht}. PN theory is also another most valuable and effective approach at present to compute the nonlinear memory effect \cite{PhysRevD.80.024002, Favata:2008ti, Favata:2009ii, Favata:2010zu}. These are strong attempts in order to have the possible detection of GW memory. This directly motivates us to look for simulation techniques which efficiently and effectively can help to extract the memory signal coming from black hole mergers.

Considering the current state of literature, there are a number of theoretical studies  which motivates us to investigate the signatures of supertranslation from an observational point of view. However, it is hard to find such signatures in the context of gravitational memory. Our last chapter of the thesis showed an alternative way of detecting supertranslation hair, but not from the memory picture. On the other hand, However, \textit{Marc Favata} has provided the study on limitations of numerical relativity in modelling the memory signal. He proposed the analytic modelling methods to search for the memory signal in a binary black hole merger due to unavailability of numerical simulations of the coalescence memory. He adopted the post-Newtonian (PN) approximation to show that the Christodoulou memory is derived from 2.5 post-Newtonian order multipole interactions and has an effect on the waveform at leading order. In building up the theoretical framework of PN formalism having relevance for asymptotic symmetries, one can post-Newtonian gravitational wave polarisations for a test particle circular orbit around a supertranslated black hole. It is expected that both polarisations, i.e., plus and cross should carry the supertranslation hair. This would surely provide a more direct approach to gravitational wave data analysts in order to understand the detection prospects of supertranslations, and could play a significant role in estimating memory signal.  

As I have been working on theoretical elements of the problems, and we know that Einstein field equations are highly nonlinear equations, the nonlinear effects of general relativity are explained by numerical relativity, however, it is not easy to compute the memory precisely in these simulations. But this helps to understand the complications involved in numerical relativity simulations in order to match the results from the analytical techniques of PN formalism. Moreover, the initial conditions for polarization modes in numerical simulations could be provided through PN technique. In addition to this, similar analyses for the search of extended BMS symmetry superrotations can also be carried out which would set a complete study of investigating the signatures of supertranslations and superrotations, and surely touch a crucial role for data analysts in coming years. Further, as a recent progress and also mentioned earlier in the direction of memory effect, Lasky et al. proposed to search for memory signal associated with the population of single BBH merger events like GW150914 \cite{PhysRevLett.117.061102}. He emphasized on the need of including higher order gravitational wave modes. This should leave a stronger impact on the memory detection. Nevertheless, the question of detecting BMS symmetries still remains open. On the other hand, the study by \textit{Michael Boyle} on transformations of asymptotic gravitational wave data investigates the role of asymptotic symmetries on GW data but not in the context of gravitational memory \cite{PhysRevD.93.084031}. However, a further extension to this has been carried out in \cite{PhysRevD.103.024031} which shows a relation between displacement memory and BMS balance laws. As both the fields are inter-linked, hence, it would be interesting to search for such symmetries in this context by analysing the gravitational wave amplitude of the memory signal. \vskip 0.10cm

The other alternative ideas for examining the role of BMS symmetries would be to work on tidal heating where late in the inspiral, individuals of a binary are highly sensitive to each other's tidal fields as the bodies approach their final plunge and merger. The generated tides by the objects would backreact on the orbit, and transfer energy together with angular momentum from their spin into the orbit; giving rise the effect known as tidal heating. Further, the post-Newtonian tidal environment analysis especially in terms of BMS symmetries can also be another direction which can be explored from BMS detection prospects. The connection with asymptotic symmetries in the context of binaries would be an important initiative from observational standpoints. The detection of such symmetries might also help in understanding the gravitational memory more at a precise level. 
\vskip 0.10cm
Furthermore, as we have been examining the problems from classical perspectives, recently, the quantum memory effect has also got considerable attention as quantum treatment might help us to understand the information loss puzzle. In this direction, the seminal works by Hawking-Perry-Strominger-Zhibodove (HPSZ) motivate us to dig into the near-horizon structure of black holes. We have shown an attempt for the spin memory effect recovered near the horizon of black holes which is analogous to the Pasterki's spin memory obtained at asymptotic null infinity. This provides a direction to search for signatures of asymptotic symmetries in the context of spin memory effect. In all these studies, the basic question arises, can we have a study of memory effect near the horizon of black holes which is analogous to the far region? In a different part of the story, one side, it is also a challenge to perform soldering transformation off the horizon shell in the context of stitching of two black hole spacetimes. How can we restore BMS-like symmetries in the memory signal? Later, on the other hand, it is again not very obvious and apparent to look at the displacement memory effect near the horizon of black holes. Can we restore memory and its relation with BMS symmetries for near the horizon of black holes? This set of questions motivated us to tackle the problems. To add more, the BMS algebra on null surface situated at a finite location of the manifold might disclose some fascinating role of symmetries on gravitational memory. Understanding the holographic aspect of quantum gravity in four-dimensional asymptotically flat spacetimes is also a main inspirational factor for infrared studies. The study of soft particles and IR divergences in various theories such as quantum electrodynamics (QED), Yang-Mills theory, and collider physics has a vast and central scope in quantum field theory. In relation with information loss puzzle, this dilemma is linked to the deep IR sector because the formation and evaporation of black holes produce an infinite number of soft gravitons and soft photons. Thus, the triangular relation for soft theorem, BMS symmetries and memory effect is itself one of the most intriguing field of research which might help in understanding the Hawking, Perry and Strominger's conjecture on hawking information paradox \cite{strominger2018lectures, Strominger2016, Strominger2014}. The quantum treatment in this line would serve a great purpose as an extension of current studies at a more precise level. \vskip 0.10cm

To sum up, the work presented in this thesis covers a number of model scenarios regarding the emergence of asymptotic symmetries and their detection prospects. These studies certainly probe the near horizon structure of black holes, and throw a light on the observational standpoints of BMS memories and understanding the information loss puzzle. How such near horizon symmetries will affect the profile of gravitational waveforms would certainly be an interesting direction of research. The observation of shadows of black holes with BMS hair would be an intriguing way to probe the question of the existence of soft BMS-like hairs. The studies depicted in this thesis, we believe, should serve a significant contribution towards understanding some of these aspects of black hole physics.

\begin{spacing}{0.9}


\bibliographystyle{unsrt} 
\bibliographystyle{plainnat} 
\cleardoublepage



\end{spacing}


\begin{appendices} 
\chapter{} \label{appen1}

\section{Null surface near Killing Horizon} 
Lastly, we present the off-shell extension of the symmetry transformations mentioned in (\ref{trans2}). keeping only those terms which are linear in $h(\zeta),\bar h(\bar \zeta)$ and $\epsilon$ we get, 
\begin{align}
\begin{split} \label{new3}
&U_+= (U-\epsilon) A, V_+= -\frac{a\,\tilde \Omega(\zeta,\bar \zeta)}{b\,\epsilon}+V(1-\tilde \Omega(\zeta,\bar \zeta))+(U-\epsilon) C, \\&
\zeta_+=\zeta+h(\zeta)+(U-\epsilon) B, \bar \zeta_+=\bar \zeta+\bar h(\bar \zeta)+(U-\epsilon) \bar B.
\end{split}
\end{align}
$a=4 m^2, b=-\frac{8 m^2}{e}$ and $\tilde \Omega(\zeta,\bar\zeta)$ appears as a conformal factor. Then demanding the continuity of the metric across the null shell upto $\mathcal{O}(U-\epsilon)$ yields,
\begin{align}
\begin{split} \label{new4}
&A=1+\frac{ \left((\zeta \bar \zeta+1) \left(h'(\zeta)+\bar h'(\bar \zeta)\right)-2 \bar \zeta h(\zeta)-2 \zeta \bar h(\bar \zeta)\right)}{1+\zeta \bar \zeta}, C=\mathcal{O}(h(\zeta)^2,\bar h(\bar \zeta)^2),\\&
B=-2 e^{1-m^2} \Big(2 h(\zeta)-(\zeta \bar \zeta+1) \left((\zeta \bar \zeta+1) \bar h''(\bar \zeta)-2 \zeta \bar h'(\bar \zeta)\right)-2 \zeta^2 \bar h(\bar \zeta)\Big),\\&
\bar B=-2 e^{1-m^2} \Big(2 \bar h(\bar \zeta)-(\zeta \bar \zeta+1) \left((\zeta \bar \zeta+1) h''(\zeta)-2 \bar \zeta h'(\zeta)\right)-2 \bar \zeta^2 h(\zeta)\Big).
\end{split}
\end{align}

\section{Jacobian of transformations}

Here we present the expression for \textit{Jacobian} matrix mentioned in (\ref{coordtt}) which is crucial for computation of the memory effect through (\ref{jump}). We use the expressions  (\ref{kerr4}) and (\ref{kerr5}) to get
\begin{align}
\begin{split}
\Big(\frac{\partial x_{+}^{\beta}}{\partial x^{\alpha}}\Big)^{-1}\Big|_
{U=0} = \frac{1}{2m^2\,e}\left(
\begin{array}{cccc}
2 m^2 e F_{V} & 0 & 0 & 0 \\
 \frac{1}{2 F_{V}}\Big(F_{\theta}^{2}+\frac{F_{\phi}^{2}}{\sin^{2}\theta}\Big) & \frac{2 m^2\,e}{F_{V}} & -\frac{2 m^2\,e\,F_{\theta}}{F_{V}} & -\frac{2 m^2\, e\, F_{\phi}}{F_{V}} \\
 -F_{\theta} & 0 & 2 m^2\,e & 0 \\
 -\frac{F_{\phi}}{\sin^{2}\theta} & 0 & 0 & 2 m^2 e\\
\end{array}
\right).\end{split}
\label
{Jacobian-1}
\end{align}
For the case of supertranslation, it further simplifies.
\begin{align}
\begin{split}
\Big(\frac{\partial x_{+}^{\beta}}{\partial x^{\alpha}}\Big)^{-1}\Big|_
{U=0} =\frac{1}{2m^2\,e}\left(
\begin{array}{cccc}
2 m^2 e  & 0 & 0 & 0 \\
 \frac{1}{2}\Big(T_{\theta}^{2}+\frac{T_{\phi}^{2}}{\sin^{2}\theta}\Big) & 2 m^2\,e & -2 m^2\,e\,T_{\theta} & -2 m^2\,e\, T_{\phi} \\
 -T_{\theta} & 0 & 2 m^2\,e & 0 \\
 -\frac{T_{\phi}}{\sin^{2}\theta} & 0 & 0 & 2 m^2 e\\
\end{array}
\right).\end{split}
\label{Jacobian-2}\end{align}

For off-shell, we already know that null congruence to the $-$ side is related to the coordinates on the surface via the following expression,
\begin{align*}
x^{\alpha} = x_{0}^{\alpha}+UN_{0}^{\alpha}
\end{align*}
The B-tensor in this mapping would give us the off shell extended version. The transformation can be written as,
\begin{align}
\tilde{B}_{AB} = \frac{\partial x_{0}^{M}}{\partial x^{A}}\frac{\partial x_{0}^{N}}{\partial x^{B}}B_{MN} \hspace{2mm} ; \hspace{2mm} with \hspace{2mm} ; \hspace{2mm}  \frac{\partial x_{0}^{M}}{\partial x^{A}} = \frac{\delta^{B}_{A}}{(\delta^{B}_{M}-U\frac{\partial N^{B}}{\partial x^{N}})}
=\frac{I}{J}.
\end{align} 
$B_{MN}$ is nothing but the components of $B-tensor$ calculated on the shell. We would be considering here only BMS case. Here, $J=(\delta^{B}_{M}-U\frac{\partial N^{B}}{\partial x^{N}})$, and inverse of the $J$ matrix is then given by, 
\begin{align}
\Big(\delta^{B}_{M}-U\frac{\partial T^{B}}{\partial x^{N}}\Big)^{-1} = \frac{1}{det(J)}\left(
\begin{array}{cc}
 1+\frac{2U}{e}T_{\phi\phi} & -\frac{2U}{e}T_{\theta\phi} \\
 -\frac{2U}{e}T_{\theta\phi} &  1+\frac{2U}{e}T_{\theta\theta}
\end{array}
\right)
\label{Jacobian4}
,\end{align}
where determinant of $J$ is given by,
\begin{align*}
det(J) =1+\frac{2U}{e}(T_{\theta\theta}+F_{\phi\phi})-\frac{4U^{2}}{e^{2}}(T_{\theta\phi}^{2}-T_{\theta\theta}T_{\phi\phi}) .
\end{align*}
 

Next, for the null surface near Killing horizon, using (\ref{new3}) and (\ref{new4}) we get ,
\begin{align}
\begin{split} \Big(\frac{\partial x_{+}^{\beta}}{\partial x^{\alpha}}\Big)^{-1}\Big|_
{U=\epsilon}=\left(
\begin{array}{cccc}
 1-A_{1}(\zeta,\bar{\zeta}) & 0 & 0 & 0 \\
 0 &1+\tilde \Omega (\zeta,\bar \zeta) & \frac{(a+b\, V\,  \epsilon ) \partial_{\zeta} \tilde \Omega(\zeta,\bar \zeta)}{b\, \epsilon } & \frac{(a+b\, V\, \epsilon ) \partial_{\bar \zeta}\tilde  \Omega(\zeta,\bar \zeta)}{b \, \epsilon } \\
 - B(\zeta,\bar \zeta) & 0 & 1- h'(\zeta) & 0 \\
 -\bar B(\zeta,\bar \zeta) & 0 & 0 & 1- \bar h'(\bar \zeta) \\
\end{array}
\right).
\end{split}
\label{Jb3}\end{align}
where, $A_{1}(\zeta,\bar{\zeta})=\frac{ \left((\zeta \bar \zeta+1) \left(h'(\zeta)+\bar h'(\bar \zeta)\right)-2 \bar \zeta h(\zeta)-2 \zeta \bar h(\bar \zeta)\right)}{1+\zeta \bar \zeta}$.

\chapter{}

\section{Soldering Freedom \& Intrinsic Quantities in Extreme BTZ Black Hole}\label{apnA}

In 3-dimensions, BTZ black hole provides a good model where we can study the horizon shells for black holes with rotation. In $4$-dimensions, the analysis becomes quite difficult.  We thus study the soldering freedom for rotating BTZ black holes and try to gain some insight  for the case when black hole possesses angular momentum.

First, we present the intrinsic quantities for extreme BTZ black hole. We solder two rotating extreme BTZ metrics with horizon situated at $r_{0}=r_{\pm}=l\sqrt{\frac{M}{2}}$. $l^2=-\frac{1}{\Lambda}$, and $\Lambda<0$ is the cosmological constant. $M$ is the mass of the BTZ black holes. For $\mathcal{M}_{-}$ manifold, in EF coordinate system, the metric takes the following form
\begin{align}
ds^{2} = -\frac{(r^{2}-r_{0}^{2})^{2}}{r^{2}l^{2}}dv^{2}+2dvdr+r^{2}(d\phi+N^{\phi}dv)^{2}, 
\end{align}
where, \hspace{3mm} $N^{\phi} = -\frac{J}{2r^{2}} \hspace*{3mm} ; \hspace*{3mm}  J = Ml$ \hspace{2mm} ; \hspace{2mm} $r_{\pm}^{2} = \frac{Ml^{2}}{2}\Big\lbrace 1\pm\Big[1-\Big(\frac{J}{Ml}\Big)^{2}\Big]^{\frac{1}{2}}\Big\rbrace .\label{horizon}$

Here, we see when the angular momentum $J$ equals $M l$, $r_{\pm}$ becomes $r_{0}$. Now, considering the supertranslation type transformations as
\begin{align}
v_{+}=v+T(\phi) \hspace*{3mm} ; \hspace*{3mm} r_{+} = r \hspace*{3mm}  ; \hspace*{3mm} \phi_{+}=\phi.
\end{align}
We obtain intrinsic quantities of the shell
\begin{align}
\mu =& -\frac{1}{8\pi}\sigma^{AB}[\kappa_{AB}] = -\frac{1}{8\pi r_{0}^{2}}\Big(\Gamma^{v_{+}}_{\phi\phi} + 2T_{\phi}\Gamma^{v_{+}}_{v_{+}\phi}+T_{\phi}^{2}\Gamma^{v_{+}}_{v_{+}v_{+}}+r_{0} \Big)\\
J^{A} =& \frac{1}{8\pi}\sigma^{AB}[\kappa_{vB}] = \frac{1}{8\pi r_{0}^{2}}\Big(\Gamma^{v_{+}}_{v_{+}\phi}+T_{\phi}\Gamma^{v_{+}}_{v_{+}\phi}\Big)\\
p =& -\frac{1}{8\pi} [\kappa_{vv}] = -\frac{1}{8\pi}(\Gamma^{v_{+}}_{v_{+}v_{+}}-\Gamma^{v}_{vv}) .
\end{align}
We again recover supertranslation in the shell's intrinsic quantities. We also observe, there can not be a shell without matter supporting pure IGW. We skip displaying the long expressions for Christofell symbols as those are not required to comprehend the appearance of BMS type symmetries at the horizon shell. It seems in $3$-dimensions, one can't make any shell that can induce something similar as superrotation-like symmetry.  Due to the periodic identification of angular coordinate, no construction may produce a feasible solution.

\section{Inverse Jacobian}\label{apnB}
 We provide here the inverse Jacobian of the coordinate transformation used in Eq. (\ref{N_0}) for the extreme RN case. The coordinate transformations are given in (\ref{ern-offc}). We just label the $'+'$ coordinates as $'-'$, and compute the components of $\mathcal{N}_{0}$ on the hypersurface. The Jacobian matrix reads
\begin{align}
\left[\Big(\frac{\partial x_{-}^{\beta}}{\partial x^{\alpha}}\Big)^{-1}\right] =  \left(
\begin{array}{cccc}
 \frac{\psi '(F) F_{V}}{\psi '(V)} & 0 & 0 & 0 \\
 \frac{ \psi '(F) \left(\csc ^2(\theta ) F_{\phi}^{2}+F_{\theta}^{2}\right)}{2 M^2 F_{V}} & \frac{1}{F_{V}} & -\frac{F_{\theta}}{F_{V}} & -\frac{F_{\phi}}{F_{V}} \\
 -\frac{\psi '(F) F_{\theta}}{M^2} & 0 & 1 & 0 \\
 -\frac{\csc ^2(\theta ) \psi '(F) F_{\phi}}{M^2} & 0 & 0 & 1 \\
\end{array}
\right).
\end{align}

\chapter{}\label{cds}

\section{Newman-Penrose Tetrad and Weyl Scalar} \label{exactan}

Here we provide the explicit expressions for Newman-Penrose (NP) tetrad together with Weyl scalar. The NP formalism  is a tetrad formalism based on a set of four null vectors. There are two real vectors denoted as $l^{a}$ and $n^{a}$, and two complex conjugate vectors $m^{a}$ and $\bar{m}^{a}$. For computational convenience, we are considering the reduced form of the metric (\ref{m1}), which is independent of $\theta_{A}(x^{A})$. We provide $\Psi_{4}$ component of the Weyl scalar for non-extreme and extreme cases separately by constructing the null tetrad. Let us mention the the normalizaiton conditions for NP tetrad are given by
\begin{align}
-l^{a}n_{a} =& m^{a}\bar{m}_{a}=1 \\
l^{a}m_{a} =& l^{a}\bar{m}_{a}=n^{a}m_{a}=n^{a}\bar{m}_{a} = 0\\
l^{a}l_{a} =& n^{a}n_{a}=m^{a}m_{a}=\bar{m}_{a}\bar{m}_{a} = 0
\end{align}
Here, $n^{\alpha}$ is nothing but $n^{\alpha}_{[NP]}$.

\subsection{Weyl scalar for the less generic metric: Non-Extremal Case}

We consider the metric (\ref{m1}) for non-extreme case. The null tetrad for this are given by
\begin{align}
l^{a} = \frac{1}{\sqrt{2}}(0,1,0,0) \hspace{3mm} ; \hspace{3mm} n^{a} = \frac{2}{\sqrt{2}}(-1,-2\kappa\rho,0,0)
\end{align}
and other two complex null tetrad
\begin{align}
m^{a} = \frac{1}{\sqrt{2}}(0,0,m_{1},m_{2}) \hspace{2mm} ; \hspace{2mm}
\bar{m}^{a} = \frac{1}{\sqrt{2}}(0,0,m_{1},\bar{m}_{2})
\end{align}
Here $m_{2}$ and $\bar{m_{2}}$ will be complex conjugate of each other. And
\begin{align}
m_{1}=-\frac{\sqrt{\rho } (\zeta\bar{\zeta}+1)^2 \sqrt{\lambda _{\bar{\zeta}\bar{\zeta}}}}{\sqrt{\rho ^2 (\zeta\bar{\zeta}+1)^4 \lambda _{\bar{\zeta}\bar{\zeta}} \lambda _{\zeta\zeta}-\left(\rho  (\zeta\bar{\zeta}+1)^2 \lambda _{\zeta\bar{\zeta}}+2\right){}^2}} 
\end{align}
\begin{align}
m_{2} =& \frac{i}{B_{1}(\rho,\zeta,\bar{\zeta})}\Big(-\sqrt{\left(\rho ^2 (\zeta\bar{\zeta}+1)^4 \lambda _{\bar{\zeta}\bar{\zeta}} \lambda _{\zeta\zeta}-\left(\rho  (\zeta\bar{\zeta}+1)^2 \lambda _{\zeta\bar{\zeta}}+2\right){}^2\right){}^3}\Big)+\nonumber \\
& \frac{1}{B_{1}(\rho,\zeta,\bar{\zeta})}\Big(\rho  (\zeta\bar{\zeta}+1)^2 (\rho  (\zeta\bar{\zeta}+1)^2 (\lambda _{\bar{\zeta}\bar{\zeta}} \lambda _{\zeta\zeta} (\rho  (\zeta\bar{\zeta}+1)^2 \lambda _{\zeta\bar{\zeta}}+2)+ \nonumber \\
& \lambda _{\zeta\bar{\zeta}}{}^2 (-\rho  (\zeta\bar{\zeta}+1)^2 \lambda _{\zeta\bar{\zeta}}-6))-12 \lambda _{\zeta\bar{\zeta}})-8\Big)
 \end{align}
where, 
\begin{align}
B_{1}(\rho,\zeta,\bar{\zeta}) = \sqrt{\rho } \sqrt{\lambda _{\bar{\zeta}\bar{\zeta}}} \left(\rho ^2 (\zeta\bar{\zeta}+1)^4 \lambda _{\bar{\zeta}\bar{\zeta}} \lambda _{\zeta\zeta}-\left(\rho  (\zeta\bar{\zeta}+1)^2 \lambda _{\zeta\bar{\zeta}}+2\right){}^2\right){}^{3/2}
\end{align}
Using the non-vanishing Weyl tensor components, we compute the desired Newman-Penrose scalar. The Weyl scalar $\Psi_{4}$ is given by
\begin{align}
\Psi_{4} = -C_{\alpha\beta\mu\nu}n^{\alpha}\bar{m}^{\beta}n^{\mu}\bar{m}^{\nu}.
\end{align}
The expression for Weyl scalar $\Psi_{4}$ is given by 

\begin{align}
\Psi_{4} = -\frac{1}{8}\Big((1+\zeta\bar{\zeta})^{2}\lambda_{\bar{\zeta}\bar{\zeta}}(\kappa\partial_{v}\lambda_{\zeta\zeta}+\partial^{2}_{v}\lambda_{\zeta\zeta})\Big)\rho^{2}+\mathcal{O}(\rho^{3}) .\label{psi4ext}
\end{align}
Similarly, one can compute the Weyl scalar $\Psi_{4}$ for extremal case, and get

\begin{align}
\Psi_{4} = -\frac{1}{8}\Big((1+\zeta\bar{\zeta})^{2}\lambda_{\bar{\zeta}\bar{\zeta}}\partial^{2}_{v}\lambda_{\zeta\zeta}\Big)\rho^{2}+\mathcal{O}(\rho^{3}). \label{psi4dextmn}
\end{align}

\subsection{Weyl Tensor for full metric}
For the full metric the expression for NP tetrads are quite large, however we  display here some Weyl tensor components that would be relevant for computing $\psi_4$.  
To the linear order in $\rho$ and $\kappa$ for the full metric (\ref{m}) with $\theta_{A}=\mathcal{C}(x^{A})$, the two shortest and relevant Weyl tensor components are
\begin{align}
   C_{\zeta v\zeta v} =& -\frac{\rho}{2}\partial^{2}_{v}\lambda_{\zeta\zeta}+\mathcal{O}(\rho^{2}) \\
   C_{\zeta v\bar{\zeta} v} =& \frac{\rho}{{6 (\zeta\bar{\zeta}+1)^2}}\Big(\kappa (2 (\zeta\bar{\zeta})^{2} \partial_{v}\lambda_{\zeta\bar{\zeta}}+4 \zeta\bar{\zeta} \partial_{v}\lambda_{\zeta\bar{\zeta}}+2 \partial_{v}\lambda_{\zeta\bar{\zeta}}-(\zeta\bar{\zeta})^{2} \partial_{\zeta}\theta_{\bar{\zeta}}- \nonumber \\
   & (\zeta\bar{\zeta}+1)^2 \partial_{\bar{\zeta}}\theta_{\zeta}-2 \zeta\bar{\zeta} \partial_{\zeta}\theta_{\bar{\zeta}}-\partial_{\zeta}\theta_{\bar{\zeta}}+4)\Big)+\mathcal{O}(\rho^{2}).
\end{align}
From these components it is apparent that the NP scalar would depend on $\lambda$, $\theta$ and their derivatives. 

\chapter{}\label{cds}

\section{Geodesic equations for perturbed supertranslated Vaidya black hole}\label{ch6}

The Lagrangian for the perturbed supertranslated Vaidya black hole can be written as
\begin{align}
L=-a \phi ' r' \psi (r,v)-a \left(\phi ' v'\right) \xi (r,v)-\frac{1}{2} v'^2 g_{vv}(r,v)+\frac{1}{2} r^2 g_{\phi\phi}(r) \phi '^2+r' v'
\end{align}
Here, $\phi, r$ and $v$ are all function of $\lambda$. Also, prime denotes the derivative with respect to $\lambda$. Once we have the Lagrangian, we can immediately find the various derivatives of the $L$, which result into the following geodesic equations,
\begin{align}
2 r'' =& 2 a \phi ' r' \partial_{r}\xi (r,v)-2 a \phi ' r' \partial_{v}\psi (r,v)+2 a \phi '' \xi (r,v)+2 r' v' \partial_{r}g_{vv}(r,v)+ \nonumber \\
& v'^2 g_{vv}^{(0,1)}(r,v)+2 v'' g_{vv}(r,v)
\end{align}
\begin{align}
2 a \phi ' v' \partial_{r}\xi(r,v)+ v'^2 \partial_{r}g_{vv}(r,v)+2 v'' =& 2 a \phi ' v' \partial_{v}\psi(r,v)+2 a \phi '' \psi (r,v)+ \nonumber \\
& r^2 \phi '^2 g_{\phi\phi}'(r)+2 r g_{\phi\phi}(r) \phi '^2
\end{align}
\begin{align}
r^2 \phi ' r' g_{\phi\phi}'(r)+2 r g_{\phi\phi}(r) \phi ' r'+r^2 g_{\phi\phi}(r) \phi '' =& a r'' \psi (r,v)+a r' v' \partial_{r}\xi (r,v)+ \nonumber \\
& a r' v' \partial_{v}\psi(r,v)+a r'^2 \partial_{r}\psi(r,v)+ \nonumber \\
& a v'' \xi (r,v)+a v'^2 \partial_{v}\xi(r,v)
\end{align}
These set of equations have been used to determine the behaviour of photon sphere for the given slightly perturbed spacetime geometry.
\end{appendices}

\printthesisindex 

\end{document}